%% file: automodeling-src.tex
\documentclass[12pt]{article}

\RequirePackage[OT1]{fontenc}
\usepackage{amsthm,amsmath,natbib, enumerate, amssymb, mathrsfs, graphicx, subfigure, multirow,soul,bbm}
\usepackage[linesnumbered,ruled,vlined]{algorithm2e}
\usepackage[normalem]{ulem}
\RequirePackage{hypernat}
\usepackage{amsfonts}
\usepackage{epic}
\usepackage{fancybox}
\usepackage[pdftex]{color}
\usepackage[font=small,labelfont=bf]{caption}
\usepackage{textcomp}
\usepackage{adjustbox}
\usepackage{tikz}

\usepackage{ulem}
\usepackage{hyperref}
\usepackage[sectionbib]{bibunits}

\RequirePackage{ifthen}

\newcommand{\blind}{0}
\theoremstyle{plain}

\newtheorem{theorem}{Theorem}
\newtheorem{lemma}{Lemma}
\newtheorem{proposition}{Proposition}

\newtheorem{corollary}{Corollary}

\newtheorem{remark}{Remark}
\newtheorem{definition}{Definition}

\newtheorem{alg}{Algorithm}

\newcommand{\red}{\color{red}}

\usepackage{ulem}
\usepackage{marginnote}
\usepackage{fontawesome5}

\newcommand{\CLdel}[1]{{\marginpar{\vspace{-0.15in}\color{red}\Large$\checkmark$}}{\color{red} \sout{#1}}}

\definecolor{gray}{rgb}{0.3,0.3,0.3}
\begin{document}

\def\spacingset#1{\renewcommand{\baselinestretch}%
{#1}\small\normalsize} \spacingset{1}

\newcommand{\bX}{{\mathbf X}}
\newcommand{\bY}{{\mathbf Y}}
%%%%%%%%%%%%%%%%%%%%%%%%%%%%%%%%%%%%%%%%%%%%%%%%%%%%%%%%%%%%%%%%%%%%%%%%%%%%%%

%\red

%\begin{frontmatter}
%\title{Point Estimation of Many Normal Means}
%\title{A New Approach to Estimation of the Mean of Multivariate Normal Distribution}
%\title{Many Normal Means: James-Stein Estimators and Inferential Models}
%\title{Elucidating Foundations of Statistical Inference with the Cauchy Distribution}
%\title{Predictive/Adaptive Generalized Ridge Linear Regression}
\if0\blind
{
\title{% Automodeling: A Look at Overparameterization \\ from a Model Building Perspective
%	Estimation of Overparametrized Models via Fitting Models to Future Observations
%	Fitting Overparametrized Models to Future Observations
%	Prediction via Fitting Models to Future Observations
	Estimation of Over-parameterized Models from an Auto-Modeling Perspective
	}
%\title{Elucidating Inferential Models with the Cauchy Distribution}
%\runtitle{Many Normal Means}

%\begin{aug}
\author{
	Yiran Jiang \quad and \quad
Chuanhai Liu\\
	Department of Statistics, Purdue University
%\footnote{{\color{red} Keep it as simple as possible, not simpler!}}
%  \fnms{Chuanhai} \snm{Liu}
%\thanksref{t2}\ead[label=e1]{chuanhai@purdue.edu}
}
	\date{}

\maketitle

% \vspace{-0.4in}
} \fi

\if1\blind
{
  \bigskip
  \bigskip
  \bigskip
  \begin{center}
    {\LARGE\bf Estimation of Over-parameterized Models from an Auto-Modeling Perspective}
\end{center}
  \medskip
} \fi

\bigskip

%\thankstext{t2}{Department of Statistics, Purdue University,
%    150 N. University Street, West Lafayette, IN 47907.}%  \printead{e1,u1}.}
%\runauthor{C. Liu}

%\affiliation{Academia Sinica and Purdue University}

%\address{Department of Statistics, Purdue University, 250 N. University Street,
%West Lafayette, IN 47907.  \printead{e1,u1}.}

%\end{aug}

% 	\vspace{-0.3in}
% {
% \color{blue}\input{tmp-comments}
% \clearpage
% }

% \defaultbibliographystyle{apalike}
% \defaultbibliography{amb}
% \input{automodeling-src.tex}
\input{automodeling-src-figures-in-body}
%\ifthenelse{1=0}{}{
%\setcounter{page}{1}
%\bibliographystyle{listbib}
\bibliographystyle{apalike}
\bibliography{amb}
% \pagebreak
% \input{appendices.tex}

%}

\end{document}

\def\spacingset#1{\renewcommand{\baselinestretch}%
{#1}\small\normalsize} \spacingset{1}

\newcommand{\bX}{{\mathbf X}}
\newcommand{\bY}{{\mathbf Y}}
%%%%%%%%%%%%%%%%%%%%%%%%%%%%%%%%%%%%%%%%%%%%%%%%%%%%%%%%%%%%%%%%%%%%%%%%%%%%%%
\useunder{\uline}{\ul}{}
\newcommand{\bm}[1]{\boldsymbol{#1}}

\newcommand{\dual}{duality}
% \input{automodeling-src.tex}
%\ifthenelse{1=0}{}{
%\setcounter{page}{1}
%\bibliographystyle{listbib}

% \pagebreak
% \input{automodeling-src-figures-in-body}

\useunder{\uline}{\ul}{}

\renewcommand{\bm}[1]{#1}
\newcommand{\bx}{\boldsymbol{x}}
\newcommand{\by}{\boldsymbol{y}}
% \spacingset{1.9} % DON'T change the spacing!

% \spacingset{1.9} % DON'T change the spacing!

\addtolength{\textheight}{.5in}%

\appendix
\renewcommand{\thesection}{S}
\renewcommand{\theequation}{S.\arabic{equation}} 
\setcounter{equation}{0}

\renewcommand{\thefigure}{S.\arabic{figure}}

% Set the figure counter to 0 initially so that the first figure is S1
\setcounter{figure}{0}

\renewcommand{\thetable}{S.\arabic{table}}
\setcounter{table}{0}

\renewcommand{\thealg}{S.\arabic{alg}}
\setcounter{alg}{0}

%%{
  \bigskip
  \bigskip
  \bigskip
  \begin{center}
    {\Large {\bf Estimation of Over-parameterized Models from an Auto-Modeling Perspective}:\\

\vspace{0.2in}

    \textit{\Large Supplementary Material: Derivations, Proofs, \\ Implementation Details and Additional Discussions}

\vspace{0.2in}
    \text{
	Yiran Jiang \quad and \quad
Chuanhai Liu
}

\vspace{0.1in}

\text{
	Department of Statistics, Purdue University
}
}

\end{center}
  \medskip
%%}

\tableofcontents
\clearpage
\input{appendices.tex}
\bibliographystyle{apalike}
\bibliography{amb-supp}
%}

%% file: automodeling-src-figures-in-body.tex
\useunder{\uline}{\ul}{}
\newcommand{\bm}[1]{\boldsymbol{#1}}

\newcommand{\dual}{duality}
\newcommand{\mostplau}{{most plausible}}

\begin{abstract}

From a model-building perspective, we propose a paradigm shift for fitting over-parameterized models. Philosophically, the mindset is to fit models to future observations rather than to the observed sample. Technically, given an imputation method to generate future observations, we fit over-parameterized models to these future observations by optimizing an approximation of the desired expected loss function based on its sample counterpart and an adaptive {\it \dual~function}. The required imputation method is also developed using the same estimation technique with an adaptive $m$-out-of-$n$ bootstrap approach. We illustrate its applications with the many-normal-means problem, $n < p$ linear regression, and neural network-based image classification of MNIST digits. The numerical results demonstrate its superior performance across these diverse applications. While primarily expository, the paper conducts an in-depth investigation into the theoretical aspects of the topic. It concludes with remarks on some open problems.  

\end{abstract}

%For example, for the many-normal-means problem, our method uniformly dominates James-Stein and Efron's $g$-modeling; for the linear regression, it is showcased by our improved estimate of model parameters over those produced using the leading-edge techniques; and for the MNIST image classification, it shows state-of-the-art performance in common model structures. %% DELETE IT TO SAVE SPACE

\noindent%
{\bf \it Keywords:}
%Bagging and Stacking,
Bootstrap,
Cross--Validation,
Future Observations,
Image Classification,
%Multiple Imputation,
%Penalty Function,
%Over-Dispersionn,
Resampling
%\CLadd{For the 5 key words, Bootstrap and Image Classification seem to be less relevant. Should we include over-parameterized models or model estimation? Your suggested are in the title already} 
%\footnote{Cross-Validation is removed.}
%Simplicity-Preference

%\vfill

\newpage
% \spacingset{1.9} % DON'T change the spacing!

\addtolength{\textheight}{.5in}%

\section{Introduction}\label{s:introduction}
Over-parameterized models such as neural networks play a crucial role in statistical analysis. Their primary benefit lies in their ability to flexibly and efficiently approximate non-linear functions across diverse structures. However, the application of over-parameterized models can present challenges ({\it c.f.}, \cite{nalisnick2018deep}). The primary challenge arises during model estimation, which typically involves minimizing a loss function based on observed data \citep{Vapnik1991}. In this case, seemingly optimistic performance on the observed data fails to generalize to the population data, leading to the well-recognized issue of \textit{overfitting} due to the existence of \textit{generalization gap}. This issue of overfitting also serves as the main characteristic for the definition of over-parameterization \citep[see, {\it e.g.},][%and references therein
]{oneto2023we}. %\CLmar{This sentence is added to address the comment on the definition of ``over-parameterization''.}

Traditionally, to enhance the effectiveness of over-parameterized models while mitigating their associated challenges, prediction-oriented model selection is essential. %a careful model selection process is essential. %A common model selection procedure is performed based on prediction methods. % \citep{box1980sampling, rubin1984bayesianly}. That is, the selected model has the lowest prediction error among a set of candidate models. 
Implementation of methods of prediction is typically based on
the simple and effective idea of \textit{cross-validation} with the key references
%Stone (1974, 1977), Allen (1974), Geisser (1975), and
%Efron and Tibshirani (1998, p. 255; and references therein).
%Stone (1974),Geisser (1975), and
%Efron and Tibshirani (1998, p. 255; and references therein).
\cite{stone1974cross,Stone1977},
\cite{geisser1975predictive}, and \citet[p.~255, and references therein]{efron1994introduction}. In the modern machine learning era, %for over-parameterized models, 
\textit{regularization} techniques \citep{buhlmann2011statistics} are commonly used for calibration by making the over-parameterized models ``simpler'' in order to prevent overfitting. The regularization process typically involves selecting \textit{hyper-parameters}, making the process of model selection fundamentally a task of hyper-parameter optimization. 

Despite its simplicity and overall effectiveness, the current framework for fitting over-parameterized models has certain limitations. First, the processes of model estimation and model selection are separated. This separation often requires limiting the number of candidate models to ensure computational feasibility. For instance, in  $L_1$ penalized models, a grid-search method is typically employed to explore a limited range of hyper-parameter values (denoted by $\lambda$ in what follows). This approach can result in inconsistencies in the estimated models, arising from variations in the choice of candidate sets. Second, the hyper-parameters tuning process may not adapt well to the observed data. For example,  in methods like $K$-fold cross-validation, the same tuned hyper-parameters are used across $K$ models, each fitted on different observations, as well as the final model using the full observation data \citep[see][and references therein]{tibshirani2009}. %However, the optimal hyper-parameters are inherently dependent on the specific observations used to fit the model, and may not even lead to a unique estimation.
Additionally, recent studies, such as \cite{Bates2023}, have highlighted issues with cross-validation, suggesting that it may not adequately estimate the prediction error.
Moreover, and perhaps most importantly, it appears that new methods are desirable when high-dimensional hyperparameters are used to fully harness the power of over-parameterized models.

Here, we take a model-building perspective and propose %\CLdel{a paradigm shift} 
a new framework, \textit{Auto-Modeling} (AM), for estimating over-parameterized models. Philosophically, the mindset is to fit models to future observations rather than to the observed sample.  Technically, given an imputation method to generate future observations, we fit over-parameterized models to these future observations by optimizing an approximation to the desired expected loss function. This optimization is based on the empirical counterpart and an adaptive \textit{\dual~function} that extends the penalty function with estimable hyper-parameters.
The required imputation method is also developed using the same estimation technique with an adaptive $m$-out-of-$n$ bootstrap approach.

The proposed estimation framework itself appears to be applicable and attractive for creating imputation models using bootstrap methods%\CLdel{. This is because we have }
, with the bootstrap population to serve as future observations and bootstrap samples to serve as observed data, respectively. 
However, due to the difficulty of the standard ($n$-out-of-$n$) bootstrap method %\CLdel{methods}
for high-dimensional problems \citep[\textit{c.f.},][and references therein]{Jiang2024}, we use an adaptive $m$-out-of-$n$ bootstrap-based imputation approach. %, 
%which is motivated by \cite{jiang2024finite-blind}. 
Notably, it follows that the proposed method for final estimation serves as a genuine approach to combining the resampling-based results in the context of over-parameterization; see Remark \ref{remark:combine}.

%\CLdel{The proposed method}
AM is illustrated through various applications, including the many-normal-means problem, $n<p$ linear regression, and neural network-based image classification of MNIST digits. The numerical results show that for the many-normal-means problem, 
%\CLdel{our method consistently outperforms other popular methods.} 
AM outperforms other popular methods in most scenarios.
%\CLdel{ James-Stein and Efron's $g$-modeling across all examples}
%\CLdel{In the application of $n<p$ linear regression, our method}
For the linear regression, AM yields model parameter estimates that result in greatly improved performance, demonstrated by both lower prediction errors and enhanced prediction interval coverage, compared with leading-edge techniques. In the MNIST image classification, %\CLdel{it}
AM significantly surpasses several commonly used regularization methods when applied to standard model structures.

In the rest of the paper, we comprehensively explore the proposed AM framework in Section~\ref{s:framework}. The required numerical algorithms are discussed in Section~\ref{s:numerical-methods}. %\ref{s:numerical-methods}.
%Section \ref{s:examples} provides simple illustrative examples.
Section~\ref{s:theoretical-results} provides relevant theoretical results. % supporting the proposed method.
Applications of the method are given in Section~\ref{s:applications} for the three different examples. 
Section~\ref{s:discussion} concludes with a few remarks.
%provide a brief discussion of the popular cross-validation methods in the context of the proposed method and concludes with a few open problems.

%\clearpage

\section{The General Framework}
\label{s:framework}

%\subsection{The Setting: Essential Guidelines for Model Estimation}
%\subsection{The setting and the objective of estimation}
\subsection{The Setting}
\label{ss:setting}

% The setting:
\renewcommand{\bm}[1]{#1}
\newcommand{\bx}{\boldsymbol{x}}
\newcommand{\by}{\boldsymbol{y}}

Consider a sample $(\bx, \by):=\{(x_i, y_i): i=1,...,n\}$ of measurements
$(X, Y) $ from an unknown population $\mathbb{P}$,
where $(X, Y) \in \mathbb{X}\times \mathbb{Y}$ with
$\mathbb{X} \subseteq \mathbb{R}^k$ 
and  $\mathbb{Y} \subseteq \mathbb{R}^q$.
The problem of interest is to build a model for predicting $y$ given $x$ based on the observed sample
\begin{equation}\label{eq:model-01}
Y|\{X=\bm{x}\} \sim \mathbb{P}_{\bm{x}, \theta}\qquad(Y\in\mathbb{Y}, \ \bm{x}\in \mathbb{X}, \ \theta \in \Theta^{(n)} \subseteq \mathbb{R}^p).
\end{equation}
That is, $\mathbb{P}_{\bm{x}, \theta}$ is used as an
approximation to the true underlying
conditional distribution %\CLdel{$\mathbb{P}_{\bm{x}}$}\CLadd{(The notation is not used)}
of $Y$ given $X=x$. For simplicity, we write $(X, Y) \sim \mathbb{P}_{\theta}$ as the distribution of the data sampled from the model.
In such a case, we assume that the distribution of
$X$ is independent of $\theta$ and that the problem of interest remains on $\mathbb{P}_{x,\theta}$ in (\ref{eq:model-01}).  The notation $\Theta^{(n)}$ is used to imply that the parameter space and hence its dimension $p$ depend on the sample size $n$. %More exactly, we model each $(x, y)$ by

\ifthenelse{1=0}{}{
\begin{remark}
	\label{remark:sample-size-specific}
% \hl{
%We assume the dimension of the parameter $\theta$ depends on $n$ in view of the fact that the models potentially changes with the sample size. %{\it i.e.} More complicated models are used when larger data set is available. 
% Ideally and importantly, for a \textit{valid} modeling strategy, it should be expected that when more data is available ($n$ grows), the model should be able to approximate $\mathbb{P}$ better. Such an asymptotic property is formally formulated later in Section~\ref{s:theoretical-results} as the model validity.
Besides over-parameterized modeling strategies, the fact that the model of choice changes potentially with the sample size has been recognized in the statistical literature,
	\it e.g., in non-parametric statistics
	underlying the sieve method; see \mbox{\cite{grenander1981abstract}},
	\mbox{\cite{geman1982nonparametric}},
	\mbox{\cite{shen1999random}},
	\mbox{\cite{wasserman2006all}},
	 and references therein.

% 	 }
% 	\st{Our discussion is more on its generality from modeling
% 	with small data, where this notion is implicit, 
% 	to modeling with big data, where simple models appear to be limited in
% 	capturing model information that data can provide.
% 	It is the latter that has motivated our use of the term auto-modeling
% 	in this paper. }
\end{remark}
}

% Following the above discussion, the model estimation procedure should be established with the belief that there is no \textit{true} $\theta$ for each model. Thus, it is worth discussing the final goal of model estimation.

We consider estimating the model $\{\mathbb{P}_{\theta}: \theta \in \Theta^{(n)}\}$ from the observed data $(\bx,\by)$.
% where the model and the observed sample size $n$ are given. 
% Thus, the superscript $(n)$ is omitted when possible for simplicity.
%write $\mathbb{P}_{\theta}^{(n)}$ as $\mathbb{P}_{\theta}$ for simplicity. 
%
Let $\ell(\theta|X, Y)$ be the log-likelihood function of $\theta \in \Theta^{(n)}$, given $(X, Y) \in \mathbb{X}\times \mathbb{Y}$. Take
\begin{equation}\label{eq:nll} L(\theta|X, Y) = -\ell(\theta|X, Y)
\end{equation}
to be the loss function.
From the modeling perspective elaborated in Section~\ref{s:introduction}, we define the optimal estimate of $\theta$ as a set of $\theta$-values
that minimizes the expected loss with respect to the population. That is, such $
\theta$-values form the set
\begin{equation}\label{eq:obj-01}
\Theta_*^{(n)} = \left\{\theta: 
\theta \in \Theta^{(n)};
E_{(X, Y)\sim\mathbb{P}}L(\theta|X, Y) =
	\min_{\tilde{\theta} \in \Theta^{(n)}} E_{(X, Y)\sim\mathbb{P}}L(\tilde{\theta}|X, Y)
\right\}.
\end{equation}
Thus, for optimal prediction, our primary objective of estimation is
to find some $\theta$ in $\Theta_*^{(n)}$. 

%\CLdel{Note that with (\ref{eq:nll}), the set can be seen as the counterpart of MLE of $\theta$, applied to the entire population. It also represents the set of $\{\mathbb{P}_{\theta}: \theta \in \Theta^{(n)}\}$}
Notably, the set $\{\mathbb{P}_{\theta}: \theta \in \Theta_*^{(n)}\}$ consists of the models that provide the closest approximation to the true population $\mathbb{P}$, as measured by the Kullback-Leibler (KL) divergence (\citeauthor{lehmann1983theory}, \citeyear{lehmann1983theory}; \citeauthor{Pardo2018}, \citeyear{Pardo2018}). A similar concept, viewed from the perspective of population risk minimization in learning theory, is discussed in \cite{Vapnik1991}.

\subsection{A New Method of Model Estimation}\label{ss:estimation}
Let $\hat{\mathbb{P}}$ denote the empirical distribution
defined by the sample $\{(x_i, y_i): i=1,..., n\}$. The method as an empirical counterpart of \eqref{eq:obj-01} is known as
empirical risk minimization (ERM). ERM aims at finding $\hat{\theta}_{ERM}$ that minimizes the empirical loss
\(R(\theta)
	= E_{(X, Y)\sim\hat{\mathbb{P}}} L(\theta|X, Y)
	= \frac{1}{n}\sum_{i=1}^{n}L(\theta|x_i,y_i).
\) 
However, for over-parameterized models, ERM suffers from generalization gap.
To reduce the generalization error, ERM is modified via the so-called technique of regularization.
A comprehensive review of regularization techniques
is given by \cite{buhlmann2011statistics}.  

As has been elaborated in Section \ref{s:introduction}, existing frameworks such as ERM can have certain limitations.
This motivates us to find an approximate solution to 
\begin{equation}
\label{eq:obj-02}
\theta^* = \underset{\theta}{\text{arg min}} \ E_{(X, Y)\sim\mathbb{P}} L(\theta|X, Y).
\end{equation}
In pursuit of this, we make use of
the sample counterpart of \eqref{eq:obj-02} and %thus %, this is equivalent to
decompose the objective function into the sum of
the empirical loss and the % corresponding
generalization gap
\[
\theta^* =
	\underset{\theta}{\text{arg min}} \left[ E_{(X, Y)\sim\hat{\mathbb{P}}}L(\theta|X, Y)  + \left(E_{(X, Y)\sim\mathbb{P}}L(\theta|X, Y) - E_{(X, Y)\sim\hat{\mathbb{P}}}L(\theta|X, Y) \right)\right].
\]
%It can be seen that the above objective function contains two parts: the empirical loss and the generalization gap. 
In order for the fitting to the empirical distribution to be
meaningful, here we introduce the
%\textit{simplicity-preference} function
\dual~function $ \pi(\theta, \lambda)$ ($\theta \in \Theta^{(n)}, \lambda \in \Lambda^{(n)}$) as an extension of the penalty function, where $\Lambda^{(n)}$ is the
%simplicity-preference
duality parameter space, %. % to be defined with the choice of $\pi(\theta,\lambda)$.
%By including this new function with new parameters denoted by $\lambda$, we
and write the target objective function as
\begin{equation}
\label{eq:obj-10}
%\begin{split}\label{eq:obj-03.01} E_{(X, Y)\sim\mathbb{P}} L(\theta|X, Y) = \ & E_{(X, Y)\sim\hat{\mathbb{P}}} L(\theta|X, Y)  + \pi(\theta, \lambda) \\&+ \left(E_{(X, Y)\sim\mathbb{P}} L(\theta|X, Y)  - E_{(X, Y)\sim\hat{\mathbb{P}}} L(\theta|X, Y)  - \pi(\theta, \lambda)\right), \end{split}
%	\label{eq:obj-03.01}
	E_{(X, Y)\sim\mathbb{P}} L(\theta|X, Y) =
	G_{\hat{\mathbb{P}}}(\theta, \lambda) + V_{\mathbb{P}, \hat{\mathbb{P}}}(\theta, \lambda),
\end{equation}
where $G_{\hat{\mathbb{P}}}(\theta, \lambda)$ is the penalized empirical loss
\begin{equation}\label{eq:obj-03.02}
G_{\hat{\mathbb{P}}}(\theta, \lambda)
= E_{(X, Y)\sim\hat{\mathbb{P}}} L(\theta|X, Y)
  + \pi(\theta, \lambda)
\end{equation}
%\CLadd{which is used to find $\hat{\theta}_\lambda$ given $\lambda \in\Lambda^{(n)}$,}
and $V_{\mathbb{P}, \hat{\mathbb{P}}}(\theta, \lambda)$ is the
modified
generalization gap
\begin{equation}\label{eq:obj-11}
V_{\mathbb{P}, \hat{\mathbb{P}}}(\theta, \lambda) = 
E_{(X, Y)\sim\mathbb{P}}L(\theta|X, Y)
 - E_{(X, Y)\sim\hat{\mathbb{P}}} L(\theta|X, Y)
 - \pi(\theta, \lambda).
\end{equation}
Accordingly, the duality function serves a dual purpose in equations (\ref{eq:obj-03.02}) and (\ref{eq:obj-11}) for model estimation and model selection, respectively.  

\ifthenelse{1=1}{}
{
\color{red}
\begin{enumerate}
    \item[1.] For any $\lambda \in\Lambda^{(n)}$, find
    \[
    \theta_\lambda = \arg\min_\theta G_{\hat{\mathbb{P}}}(\theta, \lambda)
    \]
    to make it $\theta$ estimable from the over-parameterized model.
    \item[2.] For given $\lambda$, \eqref{eq:obj-11} is thus a function of $\lambda$ alone.
    It follows that \eqref{eq:obj-10} is an only function of $\lambda$. In other words, we have introduced an expansion parameter $\lambda$ to be interested in $\theta$ values indexed by $\lambda$. That is, this induces the following constrained objective function \[      E_{(X, Y)\sim\mathbb{P}} L(\theta|X, Y),\quad \theta \in \{\theta_\lambda, \; \lambda\in\Lambda^{(n)}\}.     \]

\end{enumerate}
}

With the intended use of the duality function $\pi(\theta, \lambda)$, we estimate $\theta$ conceptually in two steps: %(a) 
find $\theta(\lambda)=\arg\min_{\tilde\theta} G_{\hat{\mathbb{P}}}(\tilde\theta, \lambda)$ for all $\lambda\in\Lambda^{(n)}$; and %(b)
take the estimate $\hat\theta = \theta(\hat{\lambda})$ with $\hat\lambda = \arg\min_{\lambda}E_{(X, Y)\sim\mathbb{P}} L(\theta(\lambda)|X, Y).$
Such estimation process is equivalent to finding the solution $(\hat{\theta}, \hat{\lambda})$ for the optimization problem
\ifthenelse{1=1}{
\begin{equation}\label{eq:am-objective}
\underset{\theta, \lambda}{\text{min}}
\; G_{\hat{\mathbb{P}}}(\theta, \lambda) + V_{\mathbb{P}, \hat{\mathbb{P}}}(\theta, \lambda),  \quad \text{subject to}  \quad
\theta = \underset{\tilde\theta}{\text{arg min}} \, G_{\hat{\mathbb{P}}}(\tilde\theta, \lambda).
\end{equation}
}{
\begin{equation}\label{eq:am-objective}
\begin{aligned}
& \underset{\theta, \lambda}{\text{minimize}}
& & G_{\hat{\mathbb{P}}}(\theta, \lambda) + V_{\mathbb{P}, \hat{\mathbb{P}}}(\theta, \lambda) \\
& \text{subject to}  
& & \theta = \underset{\tilde\theta}{\text{arg min}} \, G_{\hat{\mathbb{P}}}(\tilde\theta, \lambda).
\end{aligned}
\end{equation}
}
%\CLadd{Note that if a more conservative approach is desired, one can replace $\Lambda$ with, for example, $\Lambda_\epsilon=\{\lambda:\; \lambda\in \mathbb{R}^p, \lambda_i\geq \epsilon \mbox{ for $i=1,...,p$}\}$ for some pre-specified $\epsilon>0$.} {\color{red} (I think we can replace $\epsilon$ with 0 for simplicity.)}
An effective iterative algorithm for finding the solution $(\hat{\theta}, \hat{\lambda})$ is proposed in Section~\ref{s:numerical-methods}. More discussion on the theoretical properties of $(\hat{\theta}, \hat{\lambda})$ is given in Section~\ref{s:theoretical-results}. A detailed discussion of the efficiency of (\ref{eq:am-objective}) over the standard regularization is given in Supplementary~S.8. Moreover, the duality function in our framework is not limited to penalty functions with closed forms; for example, it can be related to model structures. An illustrative example involving tree models is provided in the Supplementary~S.13. In such cases, employing the grid-search method to determine the duality parameter $\lambda$ remains necessary, and developing methods capable of handling high-dimensional parameters becomes essential.  % \CLdel{(the ``substantial example'').}

Since $\mathbb{P}$ is not available in practice, we approximate it with an imputed population $\mathbb{Q}$. This leads to the following algorithm for point estimation of  the model parameter $\theta$. % \CLdel{\eqref{eq:model-01}}.

%\addtocounter{alg}{+2}
\begin{alg}[Estimation]\label{alg:estimation}
\rm
Let $\mathbb{Q}$ denote the imputation distribution.
The AM estimate of $\theta$, denoted by $\hat{\theta}_{AM}$, is determined by the AM objective (\ref{eq:am-objective}) with the empirical distribution $\hat{\mathbb{P}} := (\bx, \by)$ and the future population $\mathbb{P} = \mathbb{Q}$\;.
\end{alg}
An imputation method to create such an imputed population $\mathbb{Q}$ is proposed in Section \ref{ss:imputation} by applying this same estimation method with the required $\mathbb{P}$ and $\hat{\mathbb{P}}$, obtained via $K$-fold data splitting and adaptive $m$-out-of-$n$ bootsrap resampling. 

% \begin{remark}\CLdel{\raisebox{2mm}{Move this remark to Section 6?}}
%     In addition to the imputation method introduced in the subsequent Section \ref{ss:imputation}, recent advancements in synthetic data generation techniques have demonstrated their effectiveness across various tasks, particularly in handling complex models and high-dimensional data \citep{Liu2024, Shen2024, Tian2024}. Although these methods merit further exploration, the primary focus of this paper remains on the foundational aspects of modeling. Consequently, we have opted not to use these more complex models for imputation in our current study. For further discussion on this topic, including Box's perspective on modeling as an iterative process \citep{box1980sampling}, see Supplementary~S.12. 
% \end{remark}
% Nevertheless, exploring these methods for generating the imputation population $\mathbb{Q}$ could potentially enhance the robustness and applicability of AM.

% A weaker condition that is well-defined and can be implemented in practice is
% \begin{equation}\label{eq:obj-v-condition}
% \hat{\lambda} = 
% \min_{\lambda}\left\|\frac{ \partial V_{{\mathbb P}, \hat{\mathbb P}}(\theta, \lambda)}{\partial \theta}
% \right\|_2
% \end{equation}
%\subsection{An Imputation and Estimation Scheme}\label{ss:imputation-estimation}
\subsection{Imputation via Data Splitting and Adaptive Resampling}\label{ss:imputation-estimation}\label{ss:imputation}

%\CLdel{We solve the AM optimization problem (\ref{eq:am-objective}) by approximating $\mathbb{P}$ with an imputed population $\mathbb{Q}$.} 
In this paper, we primarily concern the imputation of $Y$ given the observed covariates $X = x$. More precisely, we generate future observations via multiple imputations
% \begin{equation}
% Y|X = x \sim \mathbb{Q}_{\theta}\qquad(\theta \in \Theta^{(n)} \subseteq \mathbb{R}^p).
% \end{equation}
\begin{equation}\label{eq:multiple-imputations}
Y|\{X = x \}\sim \mathbb{Q} := \frac{1}{B}\sum_{b = 1}^B \mathbb{Q}_{\hat{\theta}_b},
\end{equation}
where $x$ is the observed covariate and $\mathbb{Q}$ is the final imputation distribution formed as a mixture of $B$ imputation models with model parameters $\hat{\theta}_b \in \Theta^{(n)}$, $b = 1,\dots, B$, which are estimates of the model parameter $\theta$ in \eqref{eq:model-01}.

\ifthenelse{1=1}{
% \color{blue}
The imputation model $\mathbb{Q}_{\hat{\theta}_b}$ in \eqref{eq:multiple-imputations}
is constructed using a $K$-fold data partition along with adaptive $m$-out-of-$n$ bootstrap resampling, where $m$ serves as the adaptive parameter. For each fold, we set it aside as holdout and use its covariate data $x$ for imputation and the responses $y$ for estimating the adaptive parameter $m$. The model is then fit to the remaining data, serving as the required future population $\mathbb{P}$, and a bootstrap resample from these remaining data, representing the observed data $\hat{\mathbb{P}}$. Denote the $m$-out-of-$n$ resampling scheme as $r(\cdot)$. 
This imputation procedure is detailed by Algorithm \ref{alg:imputation}, followed by a discussion on estimating the adaptive parameter $m$ for the specification of $r(\cdot)$.

\color{black}
}{
\CLdel{The imputation model $\mathbb{Q}_{\hat{\theta}_b}$ in \eqref{eq:multiple-imputations} 
can be obtained using resampling-based methods. Denote by $r(\cdot)$ a resampling scheme such as the bootstrap method \citep{efron1979}. Since for any resampled data  %\CLdel{$r((x,y))$. CHANGE ALL $(x, y)$???} %\CLadd{$r((\bX,\bY))$}, 
$r((\bx,\by))$ 
the original observed sample $(\bx, \by)$ serves as an approximation to the future population, each imputation model can be obtained by solving (\ref{eq:am-objective}) with $\mathbb{P}=(\bx, \by)$ and $\hat{\mathbb{P}}=r((\bx,\by))$. 
% To validate the imputation model and to select an appropriate resampling scheme, 
%\CLdel{For simplicity here in this paper,}
To develop a simple way of creating effective resampling-based imputations,
we further require the imputation to be performed on $x$ that were not used in estimating the imputation model. In consequence, this approach ensures that the observed values of $y$ corresponding to these holdout $x$ can also be used to assess the effectiveness of the imputation models. The evaluation method will be detailed below in this section. This naturally leads us to use the simple $K$-fold data splitting and construct the imputation scheme summarized by Algorithm \ref{alg:imputation}.}
}

%\addtocounter{algorithm}{1}
\setcounter{algocf}{1}

\ifthenelse{1=1}{
%\addtocounter{algorithm}{1}
%\setcounter{algorithm}{1}
\begin{algorithm}% \SetAlgoNoLine%
% \color{blue}
\footnotesize
        Specify the resampling scheme $r(\cdot)$, $B$, and $K$\;
        
        \For{$b \gets 1$ \KwTo $B$} {
            
            Randomly split the observations $(\bx, \by)$ into $K$ equal subsets $\{(\bx_b^{(k)}, \by_b^{(k)})\}_{k=1}^K$\;

            \For{$k \gets 1$ \KwTo $K$}{
                %Split the observations into $(\bx_b^{(-k)}, \by_b^{(-k)})$ and the holdout set $(\bx_b^{(k)} ,\by_b^{(k)})$\;
                Estimate the imputation model $\mathbb{Q}^{(k)}_{\hat{\theta}_b}$ by solving the AM objective (\ref{eq:am-objective}) with %the future population 
                $\mathbb{P} = \hat{\mathbb{P}}^{(k)}_b:= (\bx_b^{(-k)}, \by_b^{(-k)})$
                and %the empirical distribution 
                $\hat{\mathbb{P}} = \tilde{\mathbb{P}}^{(k)}_b := r((\bx_b^{(-k)}, \by_b^{(-k)}))$                 \;
                
                Generate future observations $(\bx_b^{(k)}, \by_{*b}^{(k)})$ with $\bx_b^{(k)}$ using $\mathbb{Q}^{(k)}_{\hat{\theta}_b}$;
            }
        }
        Return the generated future observations $\mathbb{Q} := \{(\bx_b^{(k)}, \by_{*b}^{(k)}): b = 1,\dots, B \text{ and } k = 1,\dots,K\}$.
   \caption{The Imputation of Future Observations\label{alg:imputation}}
  
\end{algorithm}
}{
\begin{algorithm}% \SetAlgoNoLine%
\footnotesize
	    %\color{blue}
        Specify the resampling scheme $r(\cdot)$\;
        % $\gamma^{(0)} \gets \ln\frac{q^{(0)}}{1-q^{(0)}}$ [or $\ln m^{(0)}$]\;
        %$\ell_0 \gets \ell(\hat{\theta}, X, Y)$\;
        \For{$b \gets 1$ \KwTo $B$} {
            %$\vtheta^{(t+1)} = \vtheta^{(t)} -\alpha_t\nabla f(\vtheta^{(t)})$\;
            %Initialize the predictive set $\mathcal{P}$ that consists of a list of predictive distribution for each observation in $\{(x_i, y_i):\; i=1,..., n\}$\;
            Randomly split the observations $(\bx, \by)$ into $K$ equal partitions $\{(\bx_b^{(k)}, \by_b^{(k)})\}_{k=1}^K$\;

            % Sample  $\{(x_i^*, y_i^*):\; i=1,..., m\}$ or, in the matrix notation, $(X^*, Y^*)$ from $\{(x_i, y_i):\; i=1,..., n\}$ with replacement\;    
            % Compute $\hat{\theta}_*$, the MLE of $\theta$ with possible adjustment for finite sample performance\;

            \For{$k \gets 1$ \KwTo $K$}{
                Split the observations into $(\bx_b^{(-k)}, \by_b^{(-k)})$ and the holdout set $(\bx_b^{(k)} ,\by_b^{(k)})$\;
                Estimate the imputation model $\mathbb{Q}^{(k)}_{\hat{\theta}_b}$ by solving the AM objective (\ref{eq:am-objective}) with the empirical distribution $\hat{\mathbb{P}} = \tilde{\mathbb{P}}^{(k)}_b := r((\bx_b^{(-k)}, \by_b^{(-k)}))$ and the future population $\mathbb{P} = \hat{\mathbb{P}}^{(k)}_b:= (\bx_b^{(-k)}, \by_b^{(-k)})$ \;
                % $\{(x_i, y_i):\; i=1,..., n\}$
                % using generative sampling or the method of SIR with $\theta = \hat{\theta}_*$\;
                Generate future observations $(\bx_b^{(k)}, \by_{*b}^{(k)})$ with $\bx_b^{(k)}$ and $\mathbb{Q}^{(k)}_{\hat{\theta}_b}$;
            }
        }
        Return the generated future observations $\mathbb{Q} := \{(\bx_b^{(k)}, \by_{*b}^{(k)}): b = 1,\dots, B \text{ and } k = 1,\dots,K\}$.
   \caption{The Imputation of Future Observations\label{alg:imputation}}
   % {\color{red}[FIXME, to make it to work for the general case]}
\end{algorithm}
}
From Algorithm~\ref{alg:imputation}, the estimated data generation distribution (\ref{eq:multiple-imputations}) can be rewritten as%\footnote{\color{red}The notation $\mathbb{Q}$ is both a set in in Algorithm 1 and a distribution in (10).}\CLadd{, with a slightly abused notation,}
\begin{equation}\label{eq:multiple-imputations-K-fold}
Y|\{X = x\} \sim \mathbb{Q} := \frac{1}{B}\sum_{b = 1}^B \sum_{k = 1}^K \mathbb{Q}_{\hat{\theta}_b}^{(k)} \cdot \mathbbm{1}\left(x \in \{\bx_b^{(k)}\}\right), 
\end{equation}
where $\mathbbm{1}(\cdot)$ denotes the indicator function and $\{\bx_b^{(k)}\}$ denotes the holdout set of $x$ in the $k$-th fold. This process requires fitting $B\cdot K$ imputation models, and a practical choice is $B = 5$ and $K = 5$. According to our experiments for the examples in Section~\ref{s:applications}, increasing $B$ and $K$ does not show significant changes in the results. %\CLadd{See \cite{rubinmultiple imputation for nonresponse in surveys}} 
Incidentally, it is interesting to note that the use of a small number of imputations is also reported to be effective in a different imputation context of \cite{rubin1987multiple} and \cite{hopke2001multiple}.

%%%%\color{cyan}\CLdel{Move the following subsection to Appendix??? If yes, explain briefly in words here about what the procedure is.}
%\subsection{Search for $m$ in the Adapative $m$-out-of-$n$ Resampling Scheme}
To find a resampling scheme $r(\cdot)$ that can effectively estimate the data distribution for imputation, we recall that for the continuous true distribution function $F(\cdot | x)$, 
\begin{equation}\label{eq:generative-inference}
   F(y_i | x_i) \sim \text{Uniform}(0,1), \quad i = 1,\dots,n, 
\end{equation}
given \textit{i.i.d.} samples $(\bx, \by) := \{(x_i, y_i): i = 1,\dots, n\}$. Based on (\ref{eq:multiple-imputations-K-fold}) and (\ref{eq:generative-inference}), to better approximate the true data distribution,  we check against the condition that for $b = 1,\dots,B$, 
% \begin{equation}\label{eq:generative-inference-imputation}
% \frac{1}{K} \sum_{k=1}^{K}\hat{F}^{(k)}_{\hat{\theta}_b}(y^{(k)}_i | x^{(k)}_i) \sim \text{Uniform}(0,1), \quad \text{for } i = 1,\dots,n \text{ and } b = 1,\dots,B
% \end{equation}
\begin{equation}\label{eq:generative-inference-imputation}
\hat{F}^{(k)}_{\hat{\theta}_b}(y^{(k)}_{bi} | x^{(k)}_{bi}) \sim \text{Uniform}(0,1), \quad  i = 1,\dots,n - n_k \text{ and } k = 1,...,K,
\end{equation}
where $\hat{F}_{\hat{\theta}_b}^{(k)}(\cdot|x)$ denotes the distribution function of the imputation model $\mathbb{Q}^{(k)}_{\hat{\theta}_b}$. The term $n_k$ denotes the sample size of the observations $\hat{\mathbb{P}}^{(k)}_b:= (\bx^{(-k)}, \by^{(-k)})$ used to estimate $\mathbb{Q}^{(k)}_{\hat{\theta}_b}$, and $\{(x_{bi}^{(k)}, y_{bi}^{(k)}): i = 1,\dots,n - n_{k}\}$ denotes the holdout set in the $k$-th fold, as outlined in Algorithm~\ref{alg:imputation}.
% where $\hat{F}_{\hat{\theta}_b}^{(k)}(\cot|x)$ denotes the distribution function of the imputation model $\mathbb{Q}^{(k)}_{\hat{\theta}_b}$, $n_k$ denotes the sample size of $(x^{(-k)}, y^{(-k)})$ used to fit $\mathbb{Q}^{(k)}_{\hat{\theta}_b}$, and $\{(x_i^{(k)}, y_i^{(k)}): i = 1,\dots,n - n_{k}\}$ denotes the holdout set in $k$-th fold, described in Algorithm~\ref{alg:imputation}. 
This implementation has the desired property of finite-sample valid predictive coverage. From a model checking perspective, this serves as an important indicator of the imputation model's proficiency in efficiently approximating the true data distribution, and is further elaborated in Supplement~S.3.
When $y$ follows a discrete distribution, a simple randomization approach of \cite{dunn1996randomized} can be used to provide a surrogate value of $\hat{F}(y|x)$. This approach is discussed in detail in Section~\ref{s:nn}.

% that the distribution function of the imputation model $\mathbb{Q}$, denoted by $\hat{F}(\cdot|x)$, follows (\ref{eq:generative-inference}). 
Checking the condition (\ref{eq:generative-inference-imputation}) in a simple and efficient way guides us to use the Kolmogorov–Smirnov test (KS-test,  \citeauthor{Massey1951} (\citeyear{Massey1951}); see also \cite{liu2023reweighted}) for evaluating the effectiveness of the imputation model. Specifically, the $p$-value from the KS-test, comparing the imputation distribution outlined in (\ref{eq:generative-inference-imputation}) against the standard uniform distribution, serves as the measurement of effectiveness. Such a measurement, in turn, serves as a pivotal guide for selecting an appropriate resampling scheme to estimate imputation models. %\CLdel{, which will be elaborated following the remark.}
% The $K$-fold data-splitting-based imputation further provide hold-out sets that can be used for validating (\ref{eq:generative-inference-imputation}). 
\ifthenelse{1=1}{}{
\CLdel{\begin{remark}
    When $y$ follows a discrete distribution, a simple randomization approach of \cite{dunn1996randomized} can be used to provide a surrogate value of $\hat{F}(y|x)$. The approach is discussed in detail in Section~\ref{s:nn}.
\end{remark}}
}

As investigated in \cite{bickel2008choice}
 and \cite{Jiang2024}, by adaptively choosing the resampling data size $m$ (with replacement), the $m$-out-of-$n$ bootstrap can be a powerful tool in capturing the uncertainty of parameter estimation, especially in high-dimensional settings. In the context of AM, the value of $m$ is pivotal in regulating ``how much we know about the future population.'' The smaller the value of $m$, the greater the anticipated uncertainty.
More specifically, we consider setting $m = \lceil\tilde{\alpha} n\rceil$, where $\tilde{\alpha}$ is the ratio used to control the resampled data size, thereby influencing the effectiveness of the imputation model for achieving (\ref{eq:generative-inference-imputation}). The effect of varying $m$ on such effectiveness is demonstrated using an application example discussed in Section~\ref{s:applications}. Further illustrations of this are provided in Supplement~S.7.1. Based on the goal (\ref{eq:generative-inference-imputation}), the following algorithm provides an efficient and straightforward way to choose $\tilde{\alpha}$ using a small candidate set.

\ifthenelse{1=1}{ 
\begin{algorithm}% \SetAlgoNoLine%
% \color{blue}
\footnotesize
%\color{blue}
    Choose a candidate set of $\tilde{\alpha}$, such as $\{0.2,0.5,0.8,1.0,1.5\}$, denoted as $\tilde{\boldsymbol{\alpha}}$\;

            Randomly split the observations $(\bx, \by)$ into $K$ equal subsets $\{(\bx^{(k)}, \by^{(k)})\}_{k=1}^K$\;
    \For{$\tilde\alpha$ \textnormal{\textbf{in}} $\tilde{\boldsymbol{\alpha}}$} {

            \For{$k \gets 1$ \KwTo $K$}{
                %Split the observations into $(\bx^{(-k)}, \by^{(-k)})$ and the holdout set $(\bx^{(k)} ,\by^{(k)})$\;
                Obtain the empirical distribution $\tilde{\mathbb{P}}^{(k)}$ by resampling $m_k = \lceil \tilde{\alpha}n_k \rceil$ observations with replacement from the samples $(\bx^{(-k)}, \by^{(-k)})$ whose sample size is $n_k$\;
                Estimate the imputation model $\mathbb{Q}^{(k)}_{\hat{\theta}}$ by solving the AM objective (\ref{eq:am-objective}) with the empirical distribution $\hat{\mathbb{P}} = \tilde{\mathbb{P}}^{(k)}$ and the future population $\mathbb{P} = \hat{\mathbb{P}}^{(k)}:= (\bx^{(-k)}, \by^{(-k)})$ \;
            }
                Calculate the KS-test statistic of the imputation distribution described in (\ref{eq:generative-inference-imputation}) against the standard uniform distribution, denoted  by $P_{\tilde{\alpha}}$  the corresponding $p$-value \;
            
        }
        Return $\tilde{\alpha} = \underset{\tilde\alpha}{\text{arg max}} \ P_{\tilde\alpha}$.
        
   \caption{Resampling Scheme Selection\label{alg:choose-m}}
\end{algorithm}
}{

\begin{algorithm}% \SetAlgoNoLine%
\footnotesize
%\color{blue}
    Choose a candidate set of $\tilde{\alpha}$, such as $\{0.2,0.5,0.8,1.0,1.5\}$, denoted as $\tilde{\boldsymbol{\alpha}}$\;
        % $\gamma^{(0)} \gets \ln\frac{q^{(0)}}{1-q^{(0)}}$ [or $\ln m^{(0)}$]\;
        %$\ell_0 \gets \ell(\hat{\theta}, X, Y)$\;

            Randomly split the observations $(\bx, \by)$ into $K$ equal partitions $\{(\bx^{(k)}, \by^{(k)})\}_{k=1}^K$\;
    \For{$\tilde\alpha$ \textnormal{\textbf{in}} $\tilde{\boldsymbol{\alpha}}$} {
            %$\vtheta^{(t+1)} = \vtheta^{(t)} -\alpha_t\nabla f(\vtheta^{(t)})$\;
            %Initialize the predictive set $\mathcal{P}$ that consists of a list of predictive distribution for each observation in $\{(x_i, y_i):\; i=1,..., n\}$\;
            % Sample  $\{(x_i^*, y_i^*):\; i=1,..., m\}$ or, in the matrix notation, $(X^*, Y^*)$ from $\{(x_i, y_i):\; i=1,..., n\}$ with replacement\;    
            % Compute $\hat{\theta}_*$, the MLE of $\theta$ with possible adjustment for finite sample performance\;

            \For{$k \gets 1$ \KwTo $K$}{
                Split the observations into $(\bx^{(-k)}, \by^{(-k)})$ and the holdout set $(\bx^{(k)} ,\by^{(k)})$\;
                Obtain the empirical distribution $\tilde{\mathbb{P}}^{(k)}$ by resampling $m_k = \lceil \tilde{\alpha}n_k \rceil$ observations with replacement from the samples $(\bx^{(-k)}, \by^{(-k)})$ whose sample size is $n_k$\;
                Estimate the imputation model $\mathbb{Q}^{(k)}_{\hat{\theta}}$ by solving the AM objective (\ref{eq:am-objective}) with the empirical distribution $\hat{\mathbb{P}} = \tilde{\mathbb{P}}^{(k)}$ and the future population $\mathbb{P} = \hat{\mathbb{P}}^{(k)}:= (\bx^{(-k)}, \by^{(-k)})$ \;
            }
                % $\{(x_i, y_i):\; i=1,..., n\}$
                % using generative sampling or the method of SIR with $\theta = \hat{\theta}_*$\;
                Calculate the KS-test statistics of the imputation distribution described in (\ref{eq:generative-inference-imputation}) against the standard uniform distribution, denoted by $P_{\tilde{\alpha}}$\;
            
        }
        Return $\tilde{\alpha} = \underset{\tilde\alpha}{\text{arg max}} \ P_{\tilde\alpha}$.
        
   \caption{Resampling Scheme Selection\label{alg:choose-m}}
   % {\color{red}[FIXME, to make it to work for the general case]}
\end{algorithm}
}
In our experiments, the results are demonstrated to be robust with respect to the grid density of the candidate set, as long as the set covers a reasonable range, for example, from 0.2 to 1.5. Further evidence supporting Algorithm~\ref{alg:choose-m} is presented by the large $p$-values and the corresponding quantile-quantile (Q-Q) plots in Supplement~S.7.2, which indicate effective approximations to (\ref{eq:generative-inference-imputation}) and, thereby, satisfactory imputation.
%%%%%%%%%%\CLdel{END OF MOVING???}\color{black}

Once a satisfactory $\tilde{\alpha}$ is obtained, Algorithm~\ref{alg:imputation} can be readily applied to estimate imputation models and generate future observations, by choosing the resampling scheme $r(\cdot)$ as the $m$-out-of-$n$ bootstrap with $m = \lceil\tilde{\alpha} n\rceil$. See Supplementary~S.9 for a discussion of the limitations of the standard bootstrap imputation approach (fixing $m = n$).

\ifthenelse{1=1}{}{
\addtocounter{alg}{+2}
\begin{alg}[Estimation]\label{alg:estimation}
\rm
Let $\mathbb{Q}$ denote the imputation distribution.
The estimation of $\theta$ with the sample $(\bx, \by):=\{(x_i, y_i)\}_{i=1}^n$, denoted by $\hat{\theta}_{AM}$, is obtained
	by  solving the AM objective (\ref{eq:am-objective}) with the empirical distribution $\hat{\mathbb{P}} := (\bx, \by)$ and the future population $\mathbb{P} = \mathbb{Q}$\;.
\end{alg}
}

\renewcommand{\floatpagefraction}{.8}%
\ifthenelse{1=0}{
%       \centerline{--- Fig. \ref{fig:procedure} ---}
}{

\begin{figure}[!htbp]
\centering
\includegraphics[width=14cm]{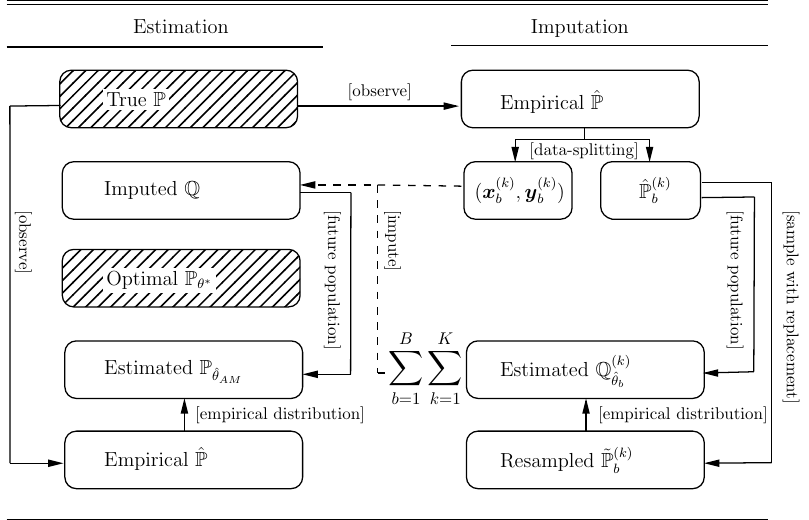}
\caption{%\footnotesize
The graphical illustration of the proposed imputation-estimation scheme
described in Algorithms~\ref{alg:imputation} and \ref{alg:estimation}.
% The left column
% is the list of true $(\mathbb{P})$, imputation $(\mathbb{Q})$, data or empirical $(\hat{\mathbb{P}})$,
% and estimated $(\mathbb{P}_{\hat{\theta}_{AM}}).$
% The right column corresponds to that of the left column,
% but for the bootstrap method of multiple imputations of
% the missing population.
The symbol $\sum_{b=1}^{B} \sum_{k = 1}^K$ between the two columns
stands for a mixture of $B \cdot K$ imputation models.
The notation $\mathbb{P}_{\theta^*}$ denotes the distribution obtained with the optimal estimate by (\ref{eq:obj-01}). The two distributions denoted by $\hat{\mathbb{P}}$ in both columns are identical. The unknown distributions are shaded.
 }
\label{fig:procedure}
\end{figure}
}

For clarity, a pictorial representation of the main components
of the proposed framework is given in Figure~\ref{fig:procedure}.
A simple illustrative example is provided in Supplement~S.1. 
 The clarity and computational efficiency of AM are elaborated further in the following remarks. 
 % Furthermore, the computational efficiency of the AM method, along with its advantages over the cross-validation framework, is detailed in Supplement~S.13. {\color{blue}}

\begin{remark}\label{remark:combine}
The estimation procedure (Algorithm \ref{alg:estimation}) can also be viewed as combining individual imputation models to obtain a single model, analogous to bootstrap and Bayesian averaging. This approach is particularly relevant to over-parameterized models, where the effectiveness of bootstrap and Bayesian-like averaging is questionable.
\end{remark}

\begin{remark}
    The proposed imputation method shares the conceptual similarities with existing ensemble techniques, notably Bagging and Stacking, which are well-established in the fields of statistics and machine learning \citep{breiman1996bagging, Wolpert1992}. 
    % \CLdel{
    % The essence of the ensemble approach is to develop multiple high-variance base models and integrate them to achieve a robust prediction system. Similarly, our method involves the generation of $B\cdot K$ base imputation models, each based on a different subset of the resampled data. This approach aligns with the principles of the bagging technique \citep{breiman1996bagging}, where multiple models are generated from varied data subsets.The integration of these $B\cdot K$ base models incorporates principles from the stacking technique \citep{Wolpert1992}, where the collective outputs of the base models are further leveraged. 
    % }
However, being different from the ensemble approach, our method is motivated by statistical modeling and only keeps a single model. Because of this key difference, our method contributes significantly to the ease of both interpreting and inferring the model.
\end{remark}

\begin{remark}
The imputation and estimation scheme is computationally efficient and does not exceed the computational cost typically associated with widely used cross-validation frameworks. 
% This efficiency stems primarily from the development of effective numerical optimization methods, detailed in Section~\ref{s:numerical-methods} for solving the AM objective (\ref{eq:am-objective}). 
The efficient numerical methods in Section~\ref{s:numerical-methods} ensure that the computational complexity of fitting a single AM model is comparable to that of standard model estimation procedures with fixed hyper-parameters, commonly employed in cross-validation. Given this efficiency, the entire AM imputation-estimation process, including Algorithms~\ref{alg:estimation}, \ref{alg:imputation}, and \ref{alg:choose-m}, aligns computationally with performing K-fold 
cross-validation across $(B-1) + |\tilde{\boldsymbol{\alpha}}|$  different hyper-parameter combinations. Here, $|\tilde{\boldsymbol{\alpha}}|$ denotes the number of candidate $\tilde{\alpha}$ values used in Algorithm~\ref{alg:choose-m}. Notably, in practice, we have $B + |\tilde{\boldsymbol{\alpha}}| \leq 10$. The term $B - 1$ arises because selecting $\tilde{\alpha}$ via Algorithm~\ref{alg:choose-m} inherently produces an imputation model for generating future observations. In the neural network model example in Section~\ref{s:dnn}, run on the same personal computer, AM with $B + |\tilde{\boldsymbol{\alpha}}| = 9$ takes approximately 18 minutes, while tuning a set of 6 hyperparameters using cross-validation takes around 10 minutes, reflecting a relative computational cost close to the theoretical analysis. Furthermore, similar to cross-validation, AM can be easily parallelizable. 
% \CLdel{\raisebox{2mm}{No new paragraph here is necessary?}}
Another notable computational advantage of the AM framework over cross-validation is its applicability for estimating extremely large models. In such cases, the AM framework allows the use of less resource-intensive yet effective models for imputation, substantially reducing the overall computational cost compared to cross-validation.

% {\red TODO: Briefly evaluate the computational cost.}
\end{remark}

\section{Numerical Optimization Methods}
\label{s:numerical-methods}

In this section, we develop efficient numerical optimization algorithms for the AM estimator. For ease of analysis, we assume that the regularity conditions for both the loss function and the duality function, as detailed in Supplement~S.2.1, are satisfied. Since the imputation step, implemented with Algorithm~\ref{alg:imputation} and \ref{alg:choose-m}, and the estimation step with Algorithm~\ref{alg:estimation} involve identical optimization problems, this section will focus on the estimation step. Specifically, we aim to solve 
% \eqref{eq:am-objective} \CLadd{by replacing the unknown $\mathbb{P}$ with the imputation population $\mathbb{Q}$,} that is,
\ifthenelse{1=1}{
\begin{equation}\label{eq:am-objective-Q}
\underset{\theta, \lambda}{\text{min}}
\; G_{\hat{\mathbb{P}}}(\theta, \lambda) + V_{\mathbb{Q}, \hat{\mathbb{P}}}(\theta, \lambda), \quad
\text{subject to}  \quad
\theta = \underset{\tilde\theta}{\text{arg min}} \, G_{\hat{\mathbb{P}}}(\tilde\theta, \lambda),
\end{equation}
}{
\begin{equation}\label{eq:am-objective-Q}
\begin{aligned}
& \underset{\theta, \lambda}{\text{minimize}}
& & G_{\hat{\mathbb{P}}}(\theta, \lambda) + V_{\mathbb{Q}, \hat{\mathbb{P}}}(\theta, \lambda) \\
& \text{subject to}  
& & \theta = \underset{\tilde\theta}{\text{arg min}} \, G_{\hat{\mathbb{P}}}(\tilde\theta, \lambda),
\end{aligned}
\end{equation}
}
% An intriguing problem here is to find a pair of solutions ($\hat\theta$, $\hat\lambda$) that satisfy the condition for both $G_{\hat{\mathbb P}}(\theta, \lambda)$ and $V_{{\mathbb Q}, \hat{\mathbb P}}(\theta, \lambda)$ at the same time. 
% The following simple coordinate descent-type algorithm 
% can be used.
where the $G$ function and the $V$ function take the form (\ref{eq:obj-03.02}) and (\ref{eq:obj-11}) respectively. Ideally, the solution $(\hat{\theta}, \hat{\lambda})$ can minimize both $G_{\hat{\mathbb{P}}}(\theta, \lambda)$ and $V_{\mathbb{Q}, \hat{\mathbb{P}}}(\theta, \lambda)$. This amounts to requiring
\begin{equation} \label{eq:obj-g-condition}
	\left.
    \frac{\partial G_{\hat{\mathbb{P}}}(\theta, \lambda)}{\partial \theta_j}
	\right|_{\theta=\hat\theta, \lambda=\hat\lambda} = \left.\left(\frac{1}{n}\sum^n_{ i=1} \frac{\partial L(\theta|x_{i}, y_{i})}{\partial \theta_j}
	+ \frac{\partial \pi(\theta, \lambda)}{\partial \theta_j}\right)\right|_{\theta=\hat\theta, \lambda=\hat\lambda}
	= 0
\end{equation}
for $j=1,...,p$, and $V_{\mathbb{Q}, \hat{\mathbb{P}}}(\theta, \lambda)$ satisfying
\begin{equation}\label{eq:obj-200}
	\left.
\frac{\partial V_{\mathbb{Q}, \hat{\mathbb{P}}}(\theta, \lambda)}{\partial \theta_j} \right|_{\theta=\hat\theta, \lambda=\hat\lambda} = 0
\end{equation}
for $j=1,...,p$. Subgradient methods \citep{Shor1985} 
% or proximal gradient descent methods \citep{Boyd2014} 
can be used if the duality function $\pi(\theta, \lambda)$ is non-differentiable.

However, the requirement \eqref{eq:obj-200} may not be achievable due to the constraints on the duality parameter space $\Lambda^{(n)}$ such as $\Lambda^{(n)}=\{\lambda:\; \lambda\in \mathbb{R}^p, \lambda_j\geq 0 \mbox{ for $j=1,...,p$}\}$. 
 % This indeed makes the simultaneous optimization of the two functions not equivalent to directly optimizing (\ref{eq:obj-03.01}), where the newly introduced terms cancel out. 
 %Philosophically, this ``forces" the function $\pi(\theta, \lambda)$ to take effect.
% for all $i=1,...,p$.
%As an alternative,
%we require the validation function to be approximately a constant in the neighborhood of $\hat\theta$.  This leads to the adaptive choice of $\lambda$ to be obtained by optimizing over coefficients of the Taylor expansion of the validation function $V_{\mathbb{P}, \hat{\mathbb{P}}}(\theta, \lambda)$ at $\theta=\hat\theta$.  
For a generalized method to overcome the difficulty,
we consider %the generalized objective function can be obtained in this way, where the components of the adaptive parameter $\lambda$ are given by
the values of $\lambda_j$'s that
minimize the magnitude of the quantity on the left-hand side of
\eqref{eq:obj-200} for $j=1,...,p$.
This can be done by considering
\begin{equation}\label{eq:obj-v-condition}
\hat{\lambda} = 
	\underset{\lambda}{\text{arg min}}\sum_{j=1}^p\left|\frac{\partial V_{\mathbb{Q}, \hat{\mathbb{P}}}(\theta, \lambda)}{\partial \theta_j}\right|_{\theta=\hat\theta},
\end{equation}
where
\[
\left|\frac{ \partial V_{{\mathbb Q}, \hat{\mathbb P}}(\theta, \lambda)}{\partial \theta_j}\right|
	= \left| E_{(X, Y)\sim\mathbb{Q}}\left[\frac{\partial L(\theta|X, Y)}{\partial \theta_j}\right]
- \frac{1}{n}\sum^n_{i=1}\frac{\partial L(\theta|x_{i}, y_{i})}{\partial \theta_j}
- \frac{\partial \pi(\theta, \lambda)}{\partial \theta_j}\right|
\]
for $j=1,...,p$. It is interesting to see that
for many types of the duality function, including those used in the numerical examples in this paper such as $\pi(\theta, \lambda)=\sum_{j=1}^p \lambda_j|\theta_j|$ with
$\Lambda=\{\lambda:\; \lambda\in \mathbb{R}^p, \lambda_j\geq 0 \mbox{ for $j=1,...,p$}\}$, $\hat\lambda$ in \eqref{eq:obj-v-condition} is given by
\begin{equation}\label{eq:obj-21}
    \hat{\lambda}_j = 
	\underset{\lambda_j}{\text{arg min}}\left|\frac{\partial V_{\mathbb{Q}, \hat{\mathbb{P}}}(\theta, \lambda)}{\partial \theta_j}\right|_{\theta=\hat\theta}
\end{equation}
for $j=1,...,p$.

% Recall that the equilibrium conditions are
% % \begin{enumerate}[{C}1]
% %     \item 
% %     \begin{equation}\label{eq:optim-102}
% %     0=
% % \frac{ \partial G_{\hat{\mathbb P}}(\theta, \lambda)}{\partial \theta}
% % = \frac{1}{n}\sum_{i=1} \frac{\partial L(\theta|x_i, y_i)}{\partial \theta}
% % 	+ \frac{\partial \pi(\theta, \lambda)}{\partial \theta}
% %     \end{equation}
% %     \item 
% %     \begin{equation}\label{eq:optim-103}
% % \left|\frac{ \partial V_{{\mathbb Q}, \hat{\mathbb P}}(\theta, \lambda)}{\partial \theta}\right|_i
% % 	= \left| E_{(X, Y)\sim\mathbb{Q}}\left[\frac{\partial L(\theta|X, Y)}{\partial \theta}\right]
% % - \frac{1}{n}\sum_{i=1}\frac{\partial L(\theta|x_i, y_i)}{\partial \theta}
% % - \frac{\partial \pi(\theta, \lambda)}{\partial \theta}\right|_i
% %     \end{equation}
% %     is minimized for $i=1,...,p$.  

% % \end{enumerate}

% \begin{equation}\label{eq:optim-102}
% \frac{ \partial G_{\hat{\mathbb P}}(\theta, \lambda)}{\partial \theta_i}
% = \frac{1}{n}\sum^n_{ j=1} \frac{\partial L(\theta|x_{j}, y_{j})}{\partial \theta_i}
% 	+ \frac{\partial \pi(\theta, \lambda)}{\partial \theta_i} = 0
% \end{equation}
% and that
% \begin{equation}\label{eq:optim-103}
% \left|\frac{ \partial V_{{\mathbb Q}, \hat{\mathbb P}}(\theta, \lambda)}{\partial \theta_i}\right|
% 	= \left| E_{(X, Y)\sim\mathbb{Q}}\left[\frac{\partial L(\theta|X, Y)}{\partial \theta}\right]
% - \frac{1}{n}\sum^n_{j=1}\frac{\partial L(\theta|x_{j}, y_{j})}{\partial \theta}
% - \frac{\partial \pi(\theta, \lambda)}{\partial \theta}\right|_i
% \end{equation}
% is minimized over $\lambda$ for $i=1,...,p$. 

To conclude, the solution $(\hat{\theta}, \hat{\lambda})$ is an equilibrium point that simultaneously satisfies (\ref{eq:obj-g-condition}) and (\ref{eq:obj-v-condition}). Intuitively, for obtaining a feasible solution $(\hat{\theta}, \hat{\lambda})$, first note that the objective (\ref{eq:obj-g-condition}) provides a constraint on $\theta$ given $\lambda$, from which $\theta$ can be seen as a function of $\lambda$,
denoted by $\theta_\lambda$. With such a constraint,
% the $\theta$ solution for (\ref{eq:obj-g-condition}) can be seen as a function of $\lambda$. 
the other objective (\ref{eq:obj-v-condition}) can be seen as a function solely depending on $\lambda$, given by
\(
	\zeta(\lambda) = \sum_{j=1}^p
	\left|\frac{\partial V_{\mathbb{Q}, \hat{\mathbb{P}}}(\theta, \lambda)}{\partial \theta_j}\right|_{\theta=\theta_{\lambda}}.
\)
%for $i=1,...,p$.
In this way, it remains to solve the single $\theta$-free objective function for 
\(
\hat{\lambda} = 
\underset{\lambda}{\text{arg min}}\zeta(\lambda).
\)
%for $i = 1,...,p$,
% \begin{eqnarray}
% \label{eq:optim-102}
% 0=
% \frac{ \partial G_{\hat{\mathbb P}}(\theta, \lambda)}{\partial \theta}
% &=& \frac{1}{n}\sum_{i=1} \frac{\partial L(\theta|x_i, y_i)}{\partial \theta}
% 	+ \frac{\partial \pi^{(n)}(\theta, \lambda)}{\partial \theta}
% 	\\
% \label{eq:optim-103}
% 0=
% \frac{ \partial V_{{\mathbb P}, \hat{\mathbb P}}(\theta, \lambda)}{\partial \theta}
% 	&=& E_{(X, Y)\sim\mathbb{P}}\left[\frac{\partial L(\theta|X, Y)}{\partial \theta}\right]
% - \frac{1}{n}\sum_{i=1}\frac{\partial L(\theta|x_i, y_i)}{\partial \theta}
% - \frac{\partial \pi^{(n)}(\theta, \lambda)}{\partial \theta}
% \end{eqnarray}
This naturally leads to Algorithm~\ref{alg:coordinate-descent}. For ease of implementation, the target (\ref{eq:obj-v-condition}) can also be implemented to minimize
\[
\left\|\frac{ \partial V_{{\mathbb Q}, \hat{\mathbb P}}(\theta, \lambda)}{\partial \theta}
\right\|_2^2
=
\left\|
	E_{(X, Y)\sim\mathbb{Q}}\left[\frac{\partial L(\theta|X, Y)}{\partial \theta}\right]
- \frac{1}{n}\sum_{i=1}^{n} \frac{\partial L(\theta|x_{i}, y_{i})}{\partial \theta}
- \frac{\partial \pi(\theta, \lambda)}{\partial \theta}
\right\|_2^2
\]
over $\lambda$, where $\| \cdot\|^2_2$ denotes the squared Euclidean norm.
% takes $\partial V_{{\mathbb Q}, \hat{\mathbb P}}(\theta, \lambda)/\partial \theta$  in \eqref{eq:optim-103} as the partial derivatives with respect to
% $\lambda$.
\setcounter{algocf}{3}
\begin{algorithm}[]
\footnotesize
Choose starting values $(\theta^{(0)},\lambda^{(0)})$ and set $t \gets 0$\;
\While{\textnormal{$\theta^{(t)}$ \textbf{has not converged}}}{
    Compute $\frac{1}{n}\sum_{i=1}^{n} \frac{\partial L(\theta|x_{i}, y_{i})}{\partial \theta}$ and $E_{(X, Y)\sim\mathbb{Q}}\left[\frac{\partial L(\theta|X, Y)}{\partial \theta}\right]$ at the current estimate $\theta^{(t)}$\;
    Fix $\lambda = \lambda^{(t)}$, update $\theta^{(t)}$ to $\theta^{(t+1)}$ toward the target \eqref{eq:obj-g-condition} with one gradient descent step\;
    Fix $\theta = \theta^{(t+1)}$, update $\lambda^{(t)}$ to $\lambda^{(t+1)}$ toward the target \eqref{eq:obj-v-condition} with one gradient descent step\;
    Set $t = t + 1$\;
}
Return converged values $\theta^{(t)}$.
\caption{Coordinate Descent Algorithm for Iterative Updates of $\theta$ and $\lambda$\label{alg:coordinate-descent}}
\end{algorithm}

Numerous variants of Algorithm~\ref{alg:coordinate-descent} can be developed. For example, a stochastic coordinate descent algorithm can be easily obtained
as a stochastic variant of Algorithm~\ref{alg:coordinate-descent} by replacing
the update steps for $\theta$ and $\lambda$ with, for example, the stochastic gradient descent (SGD) updates or the ADAM updates \citep{kingma2014adam}.
This variant is used in Section \ref{s:nn}
for the neural network 
application.
For all the numerical examples in this paper, Algorithm~\ref{alg:coordinate-descent} and its variants produce satisfactory convergence results.
Formal theoretical properties of these numerical methods will be
%investigated and
reported elsewhere.

%\clearpage

%\section{Theoretical Investigations}
\section{Theoretical Considerations}\label{s:theoretical-results}
\subsection{Model Validity and Estimation Validity}

The success of using modern over-parameterized models for big data has led us to believe that it is important to both consider modeling processes for increasing sample sizes %\CLadd{,} 
and introduce a new concept of validity or, perhaps more precisely, effectiveness regarding the underlying potential modeling strategies. It is within this context that our mathematical definition of validity is formulated below in this section to ensure clarity of the results presented in Section~\ref{ss:am-validity}; see Remark~\ref{remark:sample-size-specific} for similar ideas.

% \CLdel{The effective use of modern over-parameterized models in big data analysis highlights the need for models that accommodate increasing sample sizes. In these scenarios, traditional assumptions of correct model specification, and the related concepts like consistent estimators, might lose their relevance. This motivates us to formulate a new concept of validity specifically designed for such contexts. }

%\CLdel{The support for the proposed AM framework is based on the theory of the likelihood-based loss functions and the KL divergence.} %{\color{red} (Mentioned earlier in the end of Section 2.1.)} \CLdel{Kullback-Leibler divergence \CLadd{KL} ({\it e.g.}, \cite{lehmann1983theory}, \cite{pardo2006statistical}).}\CLdel{What is new in the proposed procedure is that it requires explicit reduction of the divergence between the fitted model $\mathbb{P}_{\hat{\theta}_{AM}}$ and the imputed population $\mathbb{Q}$. The additional requirement is obtained by the use ofthe % simplicity-preference \dual~function.}
%Clearly, the most important theoretical concern about the proposed framework 
%are the efficiency of the estimated model $\mathbb{P}_{\hat{\theta}}^{(n)}$.
Using the %\CLdel{Kullback-Leibler}
KL divergence denoted by $D_{KL}(\cdot, \cdot)$,
%this can be written as
%\[
%	D_{KL}(\mathbb{P}_{\hat{\theta}}^{(n)}, \mathbb{P})
%\]
we measure the optimal efficiency of 
the model $\{\mathbb{P}_{\theta}: \theta \in \Theta^{(n)}\}$
by the \textit{model error}
\begin{equation}\label{eq:model-error}
	D_{KL}(\mathbb{P}, \mathbb{P}_{\theta^*})
	=
	\min_{\theta\in\Theta^{(n)}}
	D_{KL}(\mathbb{P}, \mathbb{P}_{\theta})
\end{equation}
%{\color{red} What can be said about the model error, in terms of {\it validity}?}
%\subsection{Estimation errors}
%This is about additional errors due to the use of loss functions and randomness of data. 
with $\theta^* \in \Theta_*^{(n)}$, where $\Theta_*^{(n)}$ is the set of optimal estimate defined in (\ref{eq:obj-01}). It is noteworthy that for each $n$, the model error $D_{KL}(\mathbb{P}, \mathbb{P}_{\theta^*})$ remains the same for all $\theta^* \in \Theta_*^{(n)}$. Moreover, this model error is irrelevant to estimation methods and depends purely on the model. This leads to the following definition of model validity, which concerns the asymptotic performance of the underlying model as a way of summarizing a relaxed requirement in practice that has something to do with the cognition that ``all models are wrong'' \citep[][see also \cite{Tukey1954} and \cite{von1947mathematician}]{box1976science}. 
% 	\cite{box1976science}.
% \begin{equation}\label{eq:model-error-2}
%     \lim_{n \to \infty}D_{KL}(\mathbb{P}, \mathbb{P}_{\theta^*}^{(n)}) = 0
% \end{equation}

%[Large-Sample Validity]
\begin{definition}[Model Validity]\label{def:validity-model}
A model $\{\mathbb{P}_{\theta}: \theta \in \Theta^{(n)}\}$, or a sequence of models indexed by the sample size $n$, is said to be valid iff
%\begin{equation*}
\(
    \lim_{n \to \infty}D_{KL}(\mathbb{P}, \mathbb{P}_{\theta^*}) = 0
    \)
%\end{equation*}
for any $\theta^* \in \Theta_*^{(n)}$. 
\end{definition}
%\CLdel{\begin{remark}}
    % It is important to highlight that in the context of regression, the asymptotic validity of model $\mathbb{P}_{\theta}$ is contingent upon the asymptotic ``correctness" of both the mean and residual distribution. This requirement ensures the overall validity of the model. 
    Conceptually, Definition~\ref{def:validity-model} relaxes the canonical assumption of exact correct specification of the model with any finite sample sizes, which is typically used to establish the asymptotic ``correctness" of the model. The following Proposition~\ref{prop:1} and \ref{prop:2} illustrate how this definition of model validity relates to conventional statistical assumptions. To begin, we introduce the concept of model generality for future reference.
\begin{definition}[Model Generality]\label{def:nested-poputation-model}
A parametric model $\mathbb{P}_\theta$ with parameter $\theta \in \Theta$ is said to be %\CLdel{nested within (or a restricted version of)\CLdel{or equal to} }
a special case of another parametric model $\mathbb{P}_\phi$ with parameter  $\phi\in\Phi$, written as $\mathbb{P}_\theta \preceq \mathbb{P}_\phi$ or
$\Theta \preceq \Phi$, if there exists a continuous onto mapping $\psi: \Phi \mapsto \Theta$ such that for all $\theta\in\Theta$, there exists a $\phi\in \psi^{-1}(\theta) \subseteq \Phi$ satisfying %$\theta=\psi(\phi)$ and 
$\mathbb{P}_\theta = \mathbb{P}_\phi$.  If $\mathbb{P}_\theta \preceq \mathbb{P}_\phi$, $\mathbb{P}_\phi$ is said to be more general than or equal to $\mathbb{P}_\theta$, written as $\mathbb{P}_\phi \succeq \mathbb{P}_\theta$ or $\Phi \succeq \Theta$.
\end{definition}
    
%\CLdel{\end{remark}}
% \begin{example}\label{example:1}
%     Suppose $\mathbb{P} := \mathbb{P}_{\theta_0}$, $\theta_0 \in \Theta_0$, where $\Theta_0$ is the parameter space for the true parameter $\theta_0$. Then, the model $\{\mathbb{P}_{\theta}: \theta \in \Theta_0\}$ is valid by Definition~\ref{def:validity-model}. If there exists a positive integer $N$ such that $\Theta_0 \subseteq \Theta^{(n)}$ for all $n > N$, then the model $\{\mathbb{P}_{\theta}: \theta \in \Theta^{(n)}\}$ is also valid. Moreover, a consistent estimator such that $\hat{\theta}_0 \xrightarrow{p} \theta_0$ is valid.
% \end{example}

%\marginpar{\color{red} $:=$ is for definition not for a condition.}
\begin{proposition}\label{prop:1}
    Suppose $\mathbb{P} = \mathbb{P}_{\theta_0}$, $\theta_0 \in \Theta_0$, where $\Theta_0$ is the parameter space for the true model parameter $\theta_0$. Then, the model $\{\mathbb{P}_{\theta}: \theta \in \Theta_0\}$ is valid by Definition~\ref{def:validity-model}.
\end{proposition}
\begin{proposition}\label{prop:2}
     Suppose $\mathbb{P} = \mathbb{P}_{\theta_0}$, $\theta_0 \in \Theta_0$. If there exists a positive integer $N$ such that $\Theta_0 \preceq \Theta^{(n)}$ for all $n \geq N$, then the model $\{\mathbb{P}_{\theta}: \theta \in \Theta^{(n)}\}$ is valid by Definition~\ref{def:validity-model}. %%\footnote{\color{red} The condition can be relaxed to $\Theta_0 \cap \Theta^{(n)}\neq \emptyset$. No?}\CLadd{}
\end{proposition}

The proof of the two propositions is given in Supplement~S.2.2. As demonstrated in Proposition~\ref{prop:2}, Definition~\ref{def:validity-model} validates the applicability of some over-parameterized models whose parameter space consistently covers the true parameter space.  Accordingly, for a given data set $\{(x_i, y_i)\}_{i=1}^n$,
estimation methods strive for reducing
the non-negative \textit{estimation error} %\footnote{Better goal?}
% \begin{equation}\label{eq:estimation-error}
\(
	D_{KL}(\mathbb{P}, \mathbb{P}_{\hat{\theta}}) - D_{KL}(\mathbb{P}, \mathbb{P}_{\theta^*})
\)
% \end{equation}
with some parameter estimate $\hat{\theta} \in \Theta^{(n)}$, where $\theta^* \in \Theta_*^{(n)}$. This leads to the following definition of estimation validity, indicating an asymptotically vanishing estimation error.

\begin{definition}[Estimation Validity]\label{def:validity-estimation}
An estimator $\hat{\theta} \in \Theta^{(n)}$ is said to be valid iff
\begin{equation*}
D_{KL}(\mathbb{P}, \mathbb{P}_{\hat{\theta}}) - D_{KL}(\mathbb{P}, \mathbb{P}_{\theta^*})\xrightarrow{p} 0, \quad n\rightarrow \infty,
\end{equation*}
for any $\theta^* \in \Theta_*^{(n)}$. Such an estimator is said to provide a method of valid estimation. 
\end{definition}

\begin{proposition}\label{prop:3}
    Suppose that $\mathbb{P} = \mathbb{P}_{\theta_0}$, $\theta_0 \in \Theta_0 $ \, and that the likelihood function is continuous on $\Theta_0$. Any consistent estimator $\hat{\theta} \in \Theta_0$ such that $\hat{\theta}_0 \xrightarrow{p} \theta_0$, is valid by Definition~\ref{def:validity-estimation}.
\end{proposition}
The proof of Proposition~\ref{prop:3} is given in Supplement~S.2.2. Using the above definitions, it can be concluded that valid estimation (Definition~\ref{def:validity-estimation}) using a valid model (Definition~\ref{def:validity-model}) results in the asymptotic ``correctness'' in terms of $D_{KL}(\mathbb{P}, \mathbb{P}_{\hat{\theta}}) \xrightarrow{p} 0$.
 % Thus, an estimator is considered to be valid by Definition~\ref{def:validity-estimation} if it satisfies two requirements: $(i)$ it must be based on a valid model, and $(ii)$ it has a converging estimation error such that $D_{KL}(\mathbb{P}, \mathbb{P}_{\hat{\theta}}) - D_{KL}(\mathbb{P}, \mathbb{P}_{\theta^*})\xrightarrow{p} 0$ for any $\theta^* \in \Theta_*^{(n)}$.    

% Thus, a valid estimator requires both a valid model by Definition\ref{def:validity-model} and a converging estimation error such that $D_{KL}(\mathbb{P}, \mathbb{P}_{\hat{\theta}}) - D_{KL}(\mathbb{P}, \mathbb{P}_{\theta^*})\xrightarrow{p} 0$.
% and the efficiency of the estimated model is measured as
% % \begin{equation}\label{eq:estimation-efficiency} 
% % \end{equation}
% with the lower bound of .
% \hl{which is lower-bounded by the model error \mbox{(\ref{eq:model-error})}}.

\subsection{Estimation Validity of AM}\label{ss:am-validity}
%\CLdel{In this part, }
Here, we show that the proposed AM estimator is valid by Definition~\ref{def:validity-estimation} under mild conditions.
%\CLdel{we establish the asymptotic consistency of the proposed AM estimator.} 
%\CLdel{We start by simplifying the notations. }
Note that the AM objective (\ref{eq:am-objective}) involves two distributions: one for the future population and the other for the empirical distribution. The AM imputation-estimation scheme introduced in %Section~\ref{ss:imputation-estimation} 
Sections~\ref{ss:estimation} and \ref{ss:imputation}
requires utilizing different data samples to represent these two distributions. For simplicity, with a modified notation, we define the AM operator $\mathfrak{A}_{\theta}: \mathbb{P}_{fut} \times \mathbb{P}_{emp} \to \mathbb{R}^p$, where $\mathbb{P}_{fut}$ plays the role of the future population and $\mathbb{P}_{emp}$ plays the role of the empirical distribution. The operator results in a vector representing the parameter estimated by solving (\ref{eq:am-objective}). The AM imputation-estimation framework, utilizing these simplified notations, is illustrated in Figure~\ref{fig:notation}. For ease of theoretical analysis, it is assumed that each optimization process, as denoted by the AM operator $\mathfrak{A}_{\theta}$, gives the global optimal solution with respect to its objective function. Similar to other methods of model estimation, the pursuit of the global optimal solution for the AM objective function depends on the numerical optimization algorithm, particularly the optimizer. However, this aspect is beyond the scope of our current discussion.

\ifthenelse{1=0}{
%       \centerline{--- Fig. \ref{fig:procedure} ---}
}{

\begin{figure}[!htbp]
\centering
\includegraphics[width=10cm]{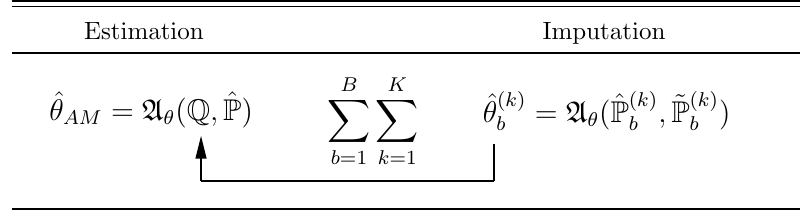}
\caption{The simplified AM imputation-estimation framework using the AM operator $\mathfrak{A}_\theta$. The involved distributions are described in Algorithm~\ref{alg:estimation} and \ref{alg:imputation}. \label{fig:notation}}
\end{figure}
}

With the above notation, the desired AM estimator is expressed as
\begin{equation}\label{eq:4.4}
    {\hat{{\theta}}^*_{AM}} = \mathfrak{A}_{\theta}(\mathbb{P}, \hat{\mathbb{P}})
\end{equation}
% Note that the estimation is related to the sample size $n$, as articulated in Section~\ref{s:framework}.
with ${\hat{{\theta}}^*_{AM}} \in \Theta^{(n)}$.
% Furthermore, 
Since the true distribution $\mathbb{P}$ is unknown in practice, (\ref{eq:4.4}) represents the optimal estimator achievable with AM  under the assumption of ``perfect" imputation. The following theorem and its corollary demonstrate the validity of the estimator (\ref{eq:4.4}) under mild conditions, with the proof given in Supplement~S.2.3 and S.2.4, respectively.

\begin{theorem} [Validity of AM] \label{theorem:1}
% For any $\theta^* \in \Theta_*^{(n)}$, define the active set as\label{theorem:1}
% \[
%  \mathcal{A}(\theta^*):=\{j: \theta^*_j \neq 0 \text{ for } j = 1,\dots,p\}.
% \]
% % If $|\mathcal{A}(\theta^*)| < \infty$ for any $\theta^* \in \Theta_*^{(n)}$, 
% If the limit parameter space on the active set denoted as $\Theta^{(\infty)}_\mathcal{A}$ exists, 
% For some $N \in \mathbb{Z}^+$,  suppose there exists a fixed parameter space $\Theta_{\mathcal{A}}$ such that $\Theta_{\mathcal{A}} \subseteq \Theta^{(n)}$ for all $n \geq N$, satisfying 
Suppose that there exists a model $\{\mathbb{P}_\phi: \phi \in \Phi \}$ with compact parameter space $\Phi$ such that $\Phi \preceq \Theta^{(n)}$ for every $n \geq N$, where $N$ is some positive integer, satisfying 
\begin{equation}\label{eq:thm1-condition}
	\min_{\phi \in \Phi}  D_{KL}(\mathbb{P}, \mathbb{P}_{\phi})
	=
	 \lim_{n \to \infty}D_{KL}(\mathbb{P}, \mathbb{P}_{\theta^*}) 
\end{equation}
% \begin{equation}\label{eq:thm1-condition}
%     \min_{\theta \in \Theta_{\mathcal{A}}} E_{(X, Y)\sim\mathbb{P}}L(\theta|X, Y) = \lim_{n\to \infty} \min_{\theta \in \Theta^{(n)}} E_{(X, Y)\sim\mathbb{P}}L(\theta|X, Y).
% \end{equation}
for any $\theta^* \in \Theta_*^{(n)}$. %%\CLadd{}\footnote{\color{red} Does the condition $\Theta_{\mathcal{A}} \subseteq \Theta^{(n)}$ imply that the dimension of $\Theta^{(n)}$ is fixed for all $n>N$?}
Assume that the loss function takes the form (\ref{eq:nll}) and %CLdel{some}
the regularity conditions %\CLdel{hold}
 in Supplement~S.2.1 hold. Then, the optimal AM estimator ${\hat{{\theta}}^*_{AM}} = \mathfrak{A}_{\theta}(\mathbb{P}, \hat{\mathbb{P}}) \in \Theta^{(n)}$ is valid by Definition~\ref{def:validity-estimation}.  %%\CLadd{}\footnote{\color{red}Do we need the condition \eqref{eq:thm1-condition} to be zero?}
% \[
% D_{KL}(\mathbb{P}, \mathbb{P}_{\theta^{*}_{AM}}) - D_{KL}(\mathbb{P}, \mathbb{P}_{\theta^{*}}) \xrightarrow{p}  0
% \]
% for any $\theta^* \in \Theta_*^{(n)}$. 
% Morevoer, if the model is valid by Definition~\ref{def:validity-model}, then $\theta^*_{AM}$ is a valid estimator by the Definition~\ref{def:validity-estimation}. i.e.,
% % if the model error \eqref{eq:model-error} goes to zero as $n \rightarrow \infty$,
% \(
% D_{KL}(\mathbb{P}, \mathbb{P}_{\theta^{*}_{AM}}) \xrightarrow{p} 0.
% \)

\end{theorem}
%\CLadd{}\footnote{\color{red} Keep it as a corollary if you want to add some statement similar to the conclusion of Theorem 1.}
\begin{corollary}\label{corollary:1}
    Suppose $\mathbb{P} = \mathbb{P}_{\theta_0}$, $\theta_0 \in \Theta_0$, where $\Theta_0$ is the parameter space for the true model parameter $\theta_0$. Assume $|\theta_{0j}| < \infty, \text{ for } j = 1,\dots,p$. If there exists a positive integer $N$ such that $\Theta_0 \preceq \Theta^{(n)}$ for every $n \geq N$, the condition (\ref{eq:thm1-condition}) holds.
\end{corollary}
% \subsection{Imputing True Population
In practice, Algorithm~\ref{alg:imputation} provides an imputation population $\mathbb{Q}$ as an approximation to the true population distribution $\mathbb{P}$, and the AM estimator takes the form
\(
        {\hat{\theta}_{AM}} = \mathfrak{A}_{\theta}(\mathbb{Q}, \hat{\mathbb{P}}).
\)
Lemma~\ref{Lemma:4.1} below demonstrates the validity of the AM estimator using the ``asymptotically correct'' imputation distribution, with the proof given in Supplement~S.2.5.

\begin{lemma}\label{Lemma:4.1}
%\CLdel{With}
Assume that
\(
 D_{KL}(\mathbb{P}, \mathbb{Q}) \xrightarrow{p} 0
\), and that the conditions in Theorem~\ref{theorem:1} hold. Then, the AM estimator \(
       {\hat{\theta}_{AM}} = \mathfrak{A}_{\theta}(\mathbb{Q}, \hat{\mathbb{P}}) \in \Theta^{(n)}
\) is valid by Definition~\ref{def:validity-estimation}. 
%if the negative log-likelihood function is used as the loss function and some regularity conditions hold as in Corollary~\ref{corollary:1}.
\end{lemma}
% The asymptotic correctness of the imputation distribution is supported by the well-established asymptotic theory of the bootstrap \citep{Kosorok2008, spencer2023}.
% Formally, we present the following theorem, with the proof given in Supplement~S.7. 
Next, we establish the asymptotic correctness of the imputation distribution. The result is summarized by the following theorem, with its proof provided in Supplement~S.2.6.

\begin{theorem}[Imputation Consistency] \label{theorem:2} 
Let the parameters $B$ and $K$,  which determine the number of imputation models, be fixed. Suppose that there exists a constant $\tilde{\alpha}_0 \in (0,1)$ such that the resampling ratio $\tilde{\alpha}$ used in the $m$-out-of-$n$ bootstrap has $\tilde{\alpha} \geq \tilde{\alpha}_0$ for every $n \geq N$, where $N$ is some positive integer. 
%Following the definition of Corollary~\ref{corollary:1}, assume $|\mathcal{A}(\theta^*)| < \infty$ for any $\theta^* \in \Theta_*^{(n)}$.
%CLdel{With }
Assume that the model is valid by Definition~\ref{def:validity-model}.
	If the
conditions in Theorem~\ref{theorem:1} hold, then
	\(
 D_{KL}(\mathbb{P}, \mathbb{Q}) \xrightarrow{p} 0
    \).
%if the negative log-likelihood function is used as the loss function and some regularity conditions hold as in Corollary~\ref{corollary:1}.
\end{theorem}
Finally, by synthesizing the results of Theorem~\ref{theorem:1} and Lemma~\ref{Lemma:4.1}, Theorem~\ref{theorem:2} establishes the validity of the AM estimator \(
        {\hat{\theta}_{AM}} = \mathfrak{A}_{\theta}(\mathbb{Q}, \hat{\mathbb{P}})
\) according to Definition~\ref{def:validity-estimation}.

\section{Applications}
\label{s:applications}
\label{s:dnn}

\subsection{Simultaneous Estimation of Many Normal Means}
\label{s:mnm}

The many-normal-means problem is about making inference
on the unknown means $\mu_1, ..., \mu_n$ from the sample $y_1,..., y_n$
with the model
%\[
$	Y_i|\{\mu_1,...,\mu_n\} \sim N(\mu_i, 1)$, $i=1,...,n$,
%\]
where $y_1,..., y_n$ are independent of each other
(\textit{c.f.}  a recent discussion by \cite{Qiu2021almond}
and references therein).
In the context of Section~\ref{s:framework}, where $y_1,...,y_n$ are assumed to be sampled from a 
single unknown population, an intriguing way of formulating this problem is to consider $\mu_1, ..., \mu_n$ as missing values taken from a different unknown
population. That is, at a high level, we consider the empirical Bayes approach similar to the $g$-modeling %\CLdel{\citep[][see \cite{Liu2022} for more discussion about different approaches]}
\citep{efron2016empirical}. More specifically, the model can be written as
\[
	\mu_i \sim {\mathbb P}_\theta\quad\mbox{ and }\quad
	Y_i|\mu_i \sim N(\mu_i, 1)\qquad(Y_i\in\mathbb{Y}, \ \theta \in \Theta^{(n)} \subseteq \mathbb{R}^p)
\]
for $i=1,...,n$.
Although more general and alternative non-parametric models for ${\mathbb P}_\theta$ can be considered,
 we opt for a simple approach by considering a discrete distribution
at $l$ $(l\ge n)$ unknown points $\eta_1\leq \eta_2 \leq \ldots\leq \eta_l$
with the probability mass % function
%\begin{equation} \label{eq:mnm02}
%	\Prob{\mu=\eta_k} =\alpha_k\quad (k=1,...,m)
%\end{equation}
$P(\mu=\eta_j) =\alpha_j$, $j=1,...,l.$
It follows that the observed-data likelihood based on $y_i$ is given by
%\begin{equation}
%	\label{eq:mnm03}
%	\frac{1}{\sqrt{2\pi}}
%	\sum_{k=1}^m\alpha_k e^{-\frac{(Y_i-\eta_k)^2}{2}} 
%\end{equation}
% In the following simulation study, 
% the conditional mean of $\mu_i$ given $Y_i$
% is taken as the point estimation of $\mu_i$.
%That is, \begin{equation} \label{eq:mnm05} \hat{\mu_i} = \end{equation}
$\frac{1}{\sqrt{2\pi}}\sum_{j=1}^l\alpha_j e^{-\frac{(y_i-\eta_j)^2}{2}}.$

% Firstly, consider the true model to be $\mu_i \sim N(M, A)$, where $M$ is the location parameter and $A$ is the scale parameter.
Unlike \cite{efron2016empirical} where $\eta_1,\dots,\eta_l$ are fixed, in our approach, $\eta_1,\dots,\eta_l$ are parameters to be estimated, resulting in an over-parameterized empirical Bayes model.
In the following simulation studies, the AM framework is used for estimating the discrete distribution model $\mathbb{P}_\theta$, which provides an approximation to the true distribution. By writing $\theta_1=\eta_1$, $\theta_j = \eta_{j} - \eta_{j-1}$ for $j=2,...,l$ and $\theta_{l+j} = \alpha_j$ for $j=1,...,l$, we define the %simplicity-preference 
\dual~function as
% For the simplicity-preference function,
% write $\theta_1=\eta_1$, $\theta_k = \eta_{k} - \eta_{k-1}$ for $k=2,...,m$ and $\theta_{m+k} = \alpha_k$ for $k=1,...,m$.
% We take the $L_1$-penalty on $(\theta_2,...,\theta_{m})'\in {\mathbb R}^{m-1}$, that is,
\[
	\pi(\theta,\lambda)
	=\sum_{j=2}^{l}\lambda_j |\theta_j| \qquad(\lambda_j\geq 0 \mbox{ for }j=2,\dots,l).
\]
% with $\lambda_k\ge 0$ for $k=2,...,m.$ 
% In this case, the simplicity of the model is obtained by restricting $(\theta_2,...,\theta_{m})$.
It is important to note that in this case, imposing $\lambda_2 = \dots = \lambda_l$ simplifies the duality function to $\pi(\theta,\lambda) = \lambda(\eta_l - \eta_1)$ where $\lambda\geq 0$.  Obviously, this reduced form is less ideal since it provides much less flexibility for parameter estimation; see Section \ref{s:discussion}. This suggests that CV with grid-search method, which is only feasible for tuning low-dimensional hyperparameters, fails in handling such model formulation. The technical and implementation details of the AM framework for this problem are provided in Supplement~S.4.

With the estimated parameter $\hat{\theta}_1,\dots,\hat{\theta}_{2l}$, the posterior mean of $\mu_i$ given $y_i$ is taken as the point estimate of $\mu_i$, namely
%\begin{equation}
	\(    \hat{\mu}_i = \hat{\mu}_i^{AM} = E_{\mu_i \sim \mathbb{P}_{\hat{\theta}}}(\mu_i | y_i)
	\).
%\end{equation}
The mean prediction error (MPE), equivalent to the average total squared error, is calculated as
% \begin{equation}\label{eq:mse}
\(
	\text{MPE} = \frac{1}{n}\sum_{i=1}^n ( \mu_i - \hat{\mu}_i )^2.
\)
% \end{equation}

%\CLdel{For comparison with other methods in addition to the $g$-modeling, we consider a more conceptually straightforward estimate of $\mu$ given by the maximum likelihood estimator (MLE)
%\begin{equation}
%	\(    \hat{\mu_i}^{MLE} = Y_i	\).
%\end{equation}
%\cite{Stein1956} first proved that such MLE is inadmissible when $n \geq 3$. Specifically, there exists at least an estimator that dominates MLE in terms of MPE in the form (\ref{eq:mse}) when $n \geq 3$. This phenomenon is thereafter known as Stein's paradox. \cite{Stein1956} and \cite{js1961} proposed the James-Stein estimator
%\[
%    \hat{\mu_i}^{JS} = \bar{Y} + \left[1 - \frac{n-3}{\sum_{i=1}^n (Y_i - \bar{Y})^2} \right] (Y_i %- \bar{Y})
%\]
%for $i = 1,\dots,n$, where $\bar{Y_i} = (1/n)\sum_{i=1}^n Y_i$ is the overall sample mean. The estimator shrinks the MLE estimator of each $\mu$ towards the overall sample mean of the observed data and has been proved to dominate the original MLE estimator when $n \geq 3$.}
It is important to note that although $g$-modeling treats $\eta_1,\dots,\eta_l$ to be fixed, this is primarily for computational feasibility. Theoretically, by allowing the density of knots to approach infinity, $g$-modeling can achieve the same flexibility as our proposed modeling approach. However, as demonstrated in Supplementary~S.10.4, increasing the knot density beyond a necessary level does not lead to performance improvement. Therefore, in this section, we present results for $g$-modeling with an optimally sufficient knot density.  

% {\red 1. For comparison with $g$-modeling, although we use fixed number of knots (later after comparison), it can theoretically use more density grid such $g$-modeling provides the equivalent flexibility when knots give  2. We consider DPMM}

Besides $g$-modeling, we consider the Dirichlet process mixture models (DPMM, \citeauthor{Li2019Dirichlet}, \citeyear{Li2019Dirichlet}; \citeauthor{Gordon2023}, \citeyear{Gordon2023}), 
% which uses the same structure of the data generating model in the simulation study. 
% % \color{magenta}--- How about we delete this sentence, because it is mentioned again in the end?}
% We also consider 
the canonical maximum likelihood estimator (MLE)
	\(    \hat{\mu}_i^{MLE} = y_i,	\)
 and the James-Stein estimator
\[
    \hat{\mu}_i^{JS} = \bar{y} + \left[1 - \frac{n-3}{\sum_{i=1}^n (y_i - \bar{y})^2} \right] (y_i - \bar{y}), \quad i = 1,\dots,n,
\]
where $\bar{y}_i = (1/n)\sum_{i=1}^n y_i$. The James-Stein (JS) estimator shrinks the MLE estimator of each $\mu$ towards the overall sample mean of the observed data and has been shown to dominate the  MLE estimator when $n \geq 3$ 
 \citep{Stein1956,js1961}. To ensure fair comparison in terms of modeling flexibility,  we also include the multiple-shrinkage generalization of the JS estimator \citep{George1987}, implemented adaptively using information derived from the DPMM results. Detailed implementation steps and insights regarding this estimator are provided in Supplementary~S.10 .

For numerical comparison, three simulation studies are considered. The first simulation study is on a simple case
in favor of the standard JS estimator: $\mu_i \sim N(0, A)$, where $A$ is the scale parameter. \cite{efron_hastie_2016} shows that the expected MPE of the JS estimator has the closed-form expression
\(
    E \left(\frac{1}{n}\sum_{i=1}^n ( \mu_i - \hat{\mu}_i^{JS} )^2\right) = \frac{A}{A+1} + \frac{3}{n}\left(1- \frac{A}{A+1}\right).
\)
Also note that for MLE, the expected MPE is
\(
    E \left(\frac{1}{n}\sum_{i=1}^n ( \mu_i - \hat{\mu}_i^{MLE} )^2\right) = 1. 
\)
This suggests that when $n \geq 3$ and $A = 0$, the James-Stein estimator performs the best such that it provides the greatest reduction of MPE compared to MLE. To maintain the randomness of $\mu$, we set $A = 0.01$, a value that is close to 0.
The second simulation study concerns the multi-modal case: %\CLdel{ when the James-Stein estimator in general does not provide satisfactory results:}
half of the $n$ unknown means are from $N(-2, 0.01)$ and the other half from $N(2, 0.01)$. This generates a bimodal distribution of the observed samples. 
%\CLdel{Ideally, the sample means should shrink towards the two modes, depending on their locations. Since the JS estimator can only shrink the sample means toward its overall mean, it will not provide satisfactory shrinkage in this case. %\CLdel{general.} %The problem is also addressed in \cite{Escobar1994}. More discussion along this line can also be found in \cite{Escobar1994}.}
The third simulation study, considers a normal model example presented in
{\cite{Narasimhan2020}}, which is used to introduce the developed statistical software for $g$-modeling, $\texttt{deconvolveR}$. Each $\mu$ is set to be $0$ with probability $0.9$ and is generated from $N(-3,1)$ with probability $0.1$. This simulates a scenario where the true model is zero-inflated and the true non-zeros $\mu$s are sparse.  

%In the last simulation study, we consider an exponential model $\mu \sim \exp(2)$ where $2$ is the rate.

For all simulation studies, three different sample sizes, $n = 10, \ 20, $ and $50$ are considered. For each case, $M = 500$ data sets are generated. MLE, the JS estimator, the multiple-shrinkage JS (MJS) estimator, DPMM, $g$-modeling, and AM are applied to each data set with MPE calculated. The mean of the MPE is taken over the corresponding
estimate in $M$ data sets. For the AM method, we set the number of unknown points to be $l = n$ and performed the imputation with $B = 5$ and $K = 5$ using Algorithm~\ref{alg:imputation}, with the resampling parameter $\tilde{\alpha}$ selected by Algorithm~\ref{alg:choose-m}. 

\ifthenelse{1=0}{
%       \centerline{--- Fig. \ref{fig:procedure} ---}
}{

\begin{table}[!htbp]
\caption{MPE results under three simulation settings with different methods. Each entry is taken as the average value obtained with $500$ repetitions and the standard deviation of each value (estimated with bootstrap) is given in parentheses. The best result in each setting is highlighted in boldface.\label{tb:1}}
\centering
\scriptsize
\def\arraystretch{1.2}
\begin{adjustbox}{max width=\textwidth}
\begin{tabular}{cccccccccccc}
\hline
\multirow{2}{*}{\textit{Method}} & \multicolumn{3}{c}{$\mu \sim N(0,0.01)$} & & \multicolumn{3}{c}{\begin{tabular}[c]{@{}c@{}}$\mu^1 \sim N(-2,0.01)$\\ $\mu^2 \sim N(2,0.01)$\end{tabular}} & & \multicolumn{3}{c}{\begin{tabular}[c]{@{}c@{}}$\mu^1 = 0$\\ $\mu^2 \sim N(-3,1)$\end{tabular}} \\ \cline{2-4} \cline{6-8} \cline{10-12} 
& {\ul $n = 10$} & {\ul $n = 20$} & {\ul $n = 50$} & & {\ul $n = 10$} & {\ul $n = 20$} & {\ul $n = 50$} & & {\ul $n = 10$} & {\ul $n = 20$} & {\ul $n = 50$} \\ 
% \hline
MLE & \def\arraystretch{0.5}\begin{tabular}[c]{@{}c@{}}1.009\\ (0.021)\end{tabular} & \def\arraystretch{0.5}\begin{tabular}[c]{@{}c@{}}0.990\\ (0.014)\end{tabular} & \def\arraystretch{0.5}\begin{tabular}[c]{@{}c@{}}0.987\\ (0.009)\end{tabular} & & \def\arraystretch{0.5}\begin{tabular}[c]{@{}c@{}}1.019\\ (0.021)\end{tabular} & \def\arraystretch{0.5}\begin{tabular}[c]{@{}c@{}}0.995\\ (0.014)\end{tabular} & \def\arraystretch{0.5}\begin{tabular}[c]{@{}c@{}}0.989\\ (0.009)\end{tabular} & & \def\arraystretch{0.5}\begin{tabular}[c]{@{}c@{}}0.973\\ (0.020)\end{tabular} & \def\arraystretch{0.5}\begin{tabular}[c]{@{}c@{}}1.005\\ (0.015)\end{tabular} & \def\arraystretch{0.5}\begin{tabular}[c]{@{}c@{}}1.001\\ (0.009)\end{tabular} \\ \vspace{-0.1in} \\
JS & \def\arraystretch{0.5}\begin{tabular}[c]{@{}c@{}}0.313\\ (0.017)\end{tabular} & \def\arraystretch{0.5}\begin{tabular}[c]{@{}c@{}}0.155\\ (0.007)\end{tabular} & \def\arraystretch{0.5}\begin{tabular}[c]{@{}c@{}}0.069\\ (0.003)\end{tabular} & & \def\arraystretch{0.5}\begin{tabular}[c]{@{}c@{}}0.899\\ (0.018)\end{tabular} & \def\arraystretch{0.5}\begin{tabular}[c]{@{}c@{}}0.850\\ (0.012)\end{tabular} & \def\arraystretch{0.5}\begin{tabular}[c]{@{}c@{}}0.812\\ (0.007)\end{tabular} & & \def\arraystretch{0.5}\begin{tabular}[c]{@{}c@{}}0.535\\ (0.016)\end{tabular} & \def\arraystretch{0.5}\begin{tabular}[c]{@{}c@{}}0.514\\ (0.012)\end{tabular} & \def\arraystretch{0.5}\begin{tabular}[c]{@{}c@{}}0.496\\ (0.007)\end{tabular} \\ \vspace{-0.1in} \\
% MJS & \def\arraystretch{0.5}\begin{tabular}[c]{@{}c@{}}0.226\\ (0.019)\end{tabular} & \def\arraystretch{0.5}\begin{tabular}[c]{@{}c@{}}0.110\\ (0.007)\end{tabular} & \def\arraystretch{0.5}\begin{tabular}[c]{@{}c@{}}0.049\\ (0.002)\end{tabular} & & \def\arraystretch{0.5}\begin{tabular}[c]{@{}c@{}}0.873\\ (0.018)\end{tabular} & \def\arraystretch{0.5}\begin{tabular}[c]{@{}c@{}}0.836\\ (0.012)\end{tabular} & \def\arraystretch{0.5}\begin{tabular}[c]{@{}c@{}}0.806\\ (0.007)\end{tabular} & & \def\arraystretch{0.5}\begin{tabular}[c]{@{}c@{}}0.449\\ (0.015)\end{tabular} & \def\arraystretch{0.5}\begin{tabular}[c]{@{}c@{}}0.463\\ (0.013)\end{tabular} & \def\arraystretch{0.5}\begin{tabular}[c]{@{}c@{}}0.464\\ (0.008)\end{tabular} \\ \vspace{-0.1in} \\
MJS & \def\arraystretch{0.5}\begin{tabular}[c]{@{}c@{}}0.345\\ (0.021)\end{tabular} & \def\arraystretch{0.5}\begin{tabular}[c]{@{}c@{}}0.180\\ (0.009)\end{tabular} & \def\arraystretch{0.5}\begin{tabular}[c]{@{}c@{}}0.092\\ (0.004)\end{tabular} & & \def\arraystretch{0.5}\begin{tabular}[c]{@{}c@{}}0.775\\ (0.027)\end{tabular} & \def\arraystretch{0.5}\begin{tabular}[c]{@{}c@{}}0.633\\ (0.022)\end{tabular} & \def\arraystretch{0.5}\begin{tabular}[c]{@{}c@{}}0.519\\ (0.015)\end{tabular} & & \def\arraystretch{0.5}\begin{tabular}[c]{@{}c@{}}0.614\\ (0.020)\end{tabular} & \def\arraystretch{0.5}\begin{tabular}[c]{@{}c@{}}0.576\\ (0.017)\end{tabular} & \def\arraystretch{0.5}\begin{tabular}[c]{@{}c@{}}0.535\\ (0.013)\end{tabular} \\ \vspace{-0.1in} \\
DPMM & \textbf{\def\arraystretch{0.5}\begin{tabular}[c]{@{}c@{}}0.131\\ (0.007)\end{tabular}} & \textbf{\def\arraystretch{0.5}\begin{tabular}[c]{@{}c@{}}0.080\\ (0.004)\end{tabular}} & \textbf{\def\arraystretch{0.5}\begin{tabular}[c]{@{}c@{}}0.043\\ (0.002)\end{tabular}} & & \def\arraystretch{0.5}\begin{tabular}[c]{@{}c@{}}0.782\\ (0.028)\end{tabular} & \def\arraystretch{0.5}\begin{tabular}[c]{@{}c@{}}0.531\\ (0.018)\end{tabular} & \def\arraystretch{0.5}\begin{tabular}[c]{@{}c@{}}0.388\\ (0.010)\end{tabular} & & \def\arraystretch{0.5}\begin{tabular}[c]{@{}c@{}}0.531\\ (0.020)\end{tabular} & \def\arraystretch{0.5}\begin{tabular}[c]{@{}c@{}}0.481\\ (0.015)\end{tabular} & \def\arraystretch{0.5}\begin{tabular}[c]{@{}c@{}}0.379\\ (0.008)\end{tabular} \\ \vspace{-0.1in} \\
$g$-modeling & \def\arraystretch{0.5}\begin{tabular}[c]{@{}c@{}}0.394\\ (0.014)\end{tabular} & \def\arraystretch{0.5}\begin{tabular}[c]{@{}c@{}}0.390\\ (0.010)\end{tabular} & \def\arraystretch{0.5}\begin{tabular}[c]{@{}c@{}}0.170\\ (0.004)\end{tabular} & & \def\arraystretch{0.5}\begin{tabular}[c]{@{}c@{}}0.771\\ (0.018)\end{tabular} & \def\arraystretch{0.5}\begin{tabular}[c]{@{}c@{}}0.731\\ (0.012)\end{tabular} & \def\arraystretch{0.5}\begin{tabular}[c]{@{}c@{}}0.729\\ (0.008)\end{tabular} & & \def\arraystretch{0.5}\begin{tabular}[c]{@{}c@{}}0.548\\ (0.015)\end{tabular} & \def\arraystretch{0.5}\begin{tabular}[c]{@{}c@{}}0.538\\ (0.013)\end{tabular} & \def\arraystretch{0.5}\begin{tabular}[c]{@{}c@{}}0.378\\ (0.007)\end{tabular} \\ \vspace{-0.1in} \\
Auto-modeling & \def\arraystretch{0.5}\begin{tabular}[c]{@{}c@{}}0.196\\ (0.013)\end{tabular} & \def\arraystretch{0.5}\begin{tabular}[c]{@{}c@{}}0.116\\ (0.007)\end{tabular} & \def\arraystretch{0.5}\begin{tabular}[c]{@{}c@{}}0.059\\ (0.003)\end{tabular} & & \textbf{\def\arraystretch{0.5}\begin{tabular}[c]{@{}c@{}}0.646\\ (0.029)\end{tabular}} & \textbf{\def\arraystretch{0.5}\begin{tabular}[c]{@{}c@{}}0.497\\ (0.019)\end{tabular}} & \textbf{\def\arraystretch{0.5}\begin{tabular}[c]{@{}c@{}}0.365\\ (0.010)\end{tabular}} & & \textbf{\def\arraystretch{0.5}\begin{tabular}[c]{@{}c@{}}0.437\\ (0.020)\end{tabular}} & \textbf{\def\arraystretch{0.5}\begin{tabular}[c]{@{}c@{}}0.400\\ (0.014)\end{tabular}} & \textbf{\def\arraystretch{0.5}\begin{tabular}[c]{@{}c@{}}0.324\\ (0.008)\end{tabular}} \\ 
\hline
\end{tabular}
\end{adjustbox}
\end{table}
}

%The results suggest that Auto-modeling method performs better than James-Stein in all cases. Specifically, the greatest reduction of MPE by Auto-modeling compared to the James-Stein estimator can be seen in the second simulation study example with bimodal distribution samples.  Figure~\ref{fig:bimodal} visualizes the shrinkage pattern of both methods for one of the simulated data set. It shows that Auto-modeling shrinks all of the sample means toward the correct direction. At the same time, the James-Stein estimator is only capable of shrinking the sample means toward the center, regardless of its location, which is not ideal in this case. 

The results, summarized in Table~\ref{tb:1}, indicate that while DPMM outperforms AM in the initial, simpler example, AM surpasses all other methods, including DPMM, in the two more challenging examples. These results demonstrate AM's capability to capture complex data-generating structures in the many-normal-means problem. The superior performance of AM over both $g$-modeling and DPMM in these challenging scenarios also suggests its estimation efficiency when similar model structures are applied. 
 %\CLdel{Besides} \CLdel{ Notably, DPMM assumes that $\mu$ is generated from the mixture of Gaussian distributions, aligning with the true underlying structure and potentially giving it an advantage over other methods without this assumption.} 
 Additional numerical results for the case where the underlying $\mu$ does not follow the normal distribution are provided in Supplementary~S.10.6.

\ifthenelse{1=1}{
%       \centerline{--- Fig. \ref{fig:procedure} ---}
}{

\begin{figure}[!htbp]
\centering
\includegraphics[width=10cm]{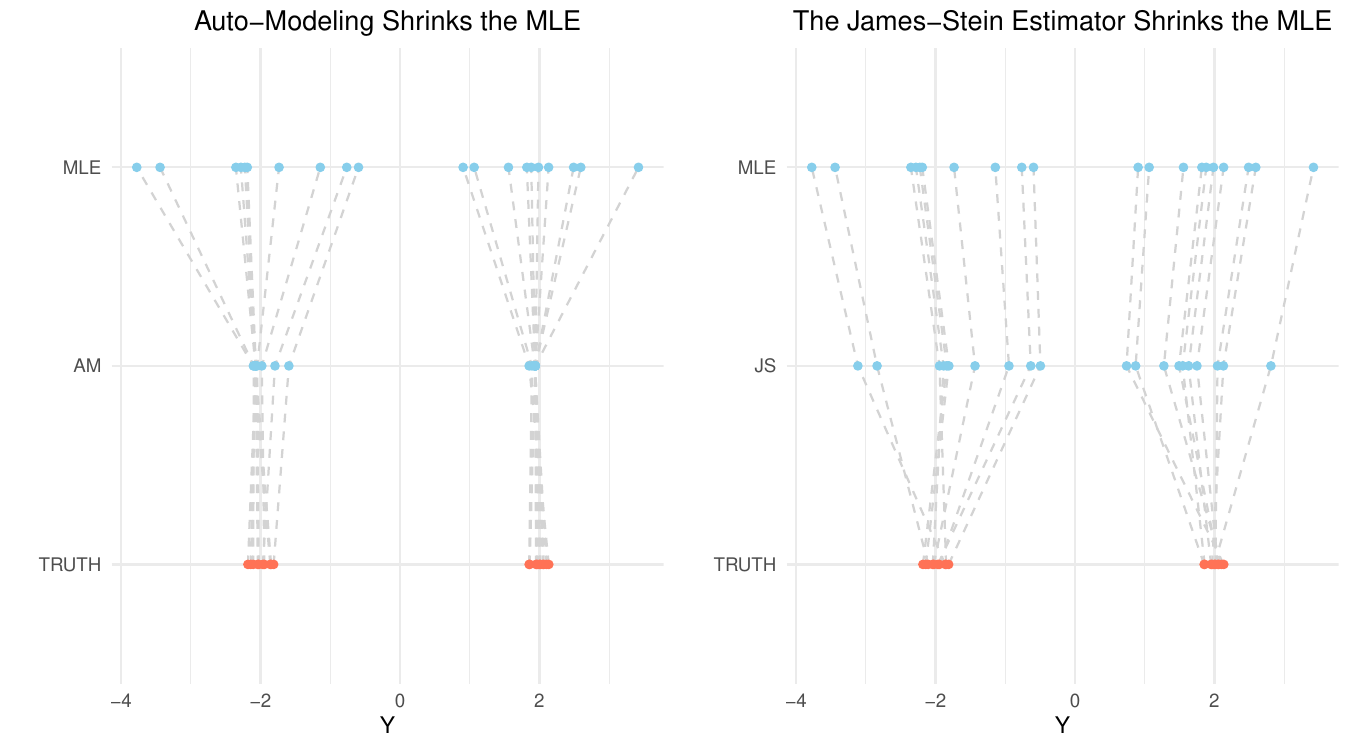}
\caption{{\red (We probabily need to remove the plot.) } Shrinkage pattern of the James-Stein estimator and AM in the bimodal simulation example with a typical data set.}\label{fig:bimodal}
\end{figure}
}

%\subsection{Linear Regression: the Case of $p > n$}\label{s:lr}
\subsection{The $n<p$ Linear Regression}\label{s:lr}
Consider the linear regression with $p$ predictors and $n$ observations $\{(x_i, y_i): i=1,..., n\}$:
\[
    y_i = x_i'\beta + \epsilon_i \qquad (y_i \in \mathbb{R}, \ x_i \in \mathbb{R}^p \mbox{ for } i = 1,\dots,n; \ \beta \in \mathbb{R}^p),
\]
where $x_i = (x_{i1}, \dots, x_{ip})'$ is a $p$-dimensional vector, $\epsilon_i \sim N(0,\sigma^2)$, and $\beta = (\beta_1, \dots, \beta_p)'$.
In the high-dimensional case of $n < p$, the estimation of the model parameters can be challenging due to the residuals approaching zero, and may rely on regularization techniques such as Lasso \citep{tibshirani1996regression} and ridge \citep{1970ridge}. However, these methods can not directly provide a good estimate %CLdel{estimation} 
of the variance term $\hat{\sigma}$, and additional techniques \citep{Reid2016, Yu2019} are required. Moreover, even with the estimated variance term, the finite-sample parametric prediction interval coverage can not be guaranteed.  Here with a simulation study, we show that the proposed AM can not only produce an efficient point estimate of future observations but also provide a satisfactory parametric prediction interval coverage.

For AM, we write the loss function as
\(
L(\beta, \sigma |X = x, Y = y) = \frac{1}{2}\log(\sigma^2) + \frac{1}{2\sigma^2n}\sum_{i=1}^n(y_i-x_i'\beta)^2
\)
and consider the weighted $L_1$ and $L_2$ \dual~function
% \[
% \pi(\beta, \sigma, \lambda) = \lambda_0 \log\left(\frac{1}{\sigma^2}\right) +  \sum_{j = 1}^p \lambda_j | \beta_j | \quad (\lambda_j \geq 0, j = 0, \ 1,\dots, p),
% \]
% the weighted $L_2$ \dual~function
% \[
% \pi(\beta, \sigma, \lambda) = \lambda_0 \log\left(\frac{1}{\sigma^2}\right) + \sum_{j = 1}^p \lambda_j \beta_j^2 \quad (\lambda_j \geq 0, j = 0, 1,\dots,p),
% \]
\[
\pi(\beta, \sigma, \lambda) = \lambda_0 \log\left(\frac{1}{\sigma^2}\right) +  \sum_{j = 1}^p \lambda_j | \beta_j |\quad \text{and} \quad  \pi(\beta, \sigma, \lambda) = \lambda_0 \log\left(\frac{1}{\sigma^2}\right) + \sum_{j = 1}^p \lambda_j \beta_j^2
\]
where $\lambda_j\geq 0 \mbox{ for } j=0,1,\dots,p$, as well as their unweighted versions obtained by imposing $\lambda_1 =\dots=\lambda_p$.
%\CLdel{which we call $L_1$-duality function and $L_2$-duality function, respectively. Moreover, we call the duality to be weighted or unweighted depending on whether we further impose $\lambda_1 =\dots=\lambda_p$.} 
% The technical and implementation details of the AM framework are provided in Supplement~S.10.
The implementation details of AM are provided in Supplement~S.5.

We consider the simulation setting in \cite{Reid2016} and \cite{Yu2019}. Let $p = 500$ and $n = 100$. Each row of the design matrix \( \bx \) is generated from a multivariate normal distribution $N(0, \Sigma)$. The covariance matrix $\Sigma$ is structured such that $\Sigma_{ii} = 1$ and $\Sigma_{ij} = \rho$ for $i \neq j $. The regression coefficient vector $\beta$ has $\lceil n^{\alpha} \rceil$ non-zero elements, where \( \alpha \) is a parameter that controls the sparsity of \( \beta \). These non-zero elements of \( \beta \) are randomly selected and follow a Laplace distribution with unit rate. The error variance in the model is determined by the parameter \( \tau \) and is calculated as \( \sigma^2 = \frac{1}{\tau} \beta^T \Sigma \beta \). We set $\rho = 0.5$ and vary $\alpha \in \{0.3,0.6,0.9\}$ and $\tau \in \{0.3,1,3\}$. 100 data sets are generated under each specific setting. For other methods to be compared with, we consider  Lasso \citep{tibshirani1996regression}, ridge \citep{1970ridge}, the elastic net \citep{Zou2005}, and the adaptive Lasso \citep{Zou2006}. These methods are applied with the cross-validation framework implemented in the R package $\texttt{glmnet}$ \citep{Friedman2010}. 

We evaluate the mean-squared error (ME) of different methods with
\(
\text{ME} = (\hat{\beta} - \beta)^T \Sigma (\hat{\beta} - \beta),
\)
a performance measure provided in \cite{tibshirani1996regression}. We also evaluate the $95\%$ prediction interval coverage with the additionally simulated data ($n = 1,000$). The prediction interval for the new observed covariate value $\hat{x}_{new}$ is given by 
\(
\left(\hat{x}_{new} \hat{\beta}- Z_{1-\alpha/2} \hat{\sigma}, \right.\\ \left.\hat{x}_{new} \hat{\beta} + Z_{1-\alpha/2}\hat{\sigma}\right),
\)
where $Z_{1-\alpha/2}$ denotes the CDF value of the standard normal distribution at $1- \alpha/2$, with $\alpha = 0.05$ in this case. The variance estimator of other methods is obtained with the effective canonical method described in \cite{Reid2016}. For AM, the imputation process is performed with the unweighted duality functions, and the final estimation is performed with both weighted and unweighted duality functions for comparison.

\ifthenelse{1=0}{
%       \centerline{--- Fig. \ref{fig:procedure} ---}
}{

\begin{table}[!htbp]
\caption{Summary of ME results for the linear regression examples with different methods with the simulation data ($p = 500, \  n = 100$). The parameters $\alpha$ and $\tau$ are those that control the sparsity of the true signals and the signal-to-noise ratio (SNR). Each entry is taken as the average value obtained with $100$ repetitions and the standard deviation of each value (estimated with bootstrap) is given in parentheses. The best result in each setting is highlighted in boldface. \label{tab:lr-1}}
\centering
% \tiny
\scriptsize
\def\arraystretch{0.8}
  \begin{adjustbox}{max width=\textwidth}

\begin{tabular}{cccclccclccc}
\hline
\multirow{2}{*}{\textit{Method}} & \multicolumn{3}{c}{$\alpha = 0.3$}                                                                                                                                                               & \multicolumn{1}{c}{} & \multicolumn{3}{c}{$\alpha = 0.6$}                                                                                                                                                                  & \multicolumn{1}{c}{} & \multicolumn{3}{c}{$\alpha = 0.9$}                                                                                                                                                                  \\ \cline{2-4} \cline{6-8} \cline{10-12} 
                                 & {\ul $\tau =0.3$}                                              & {\ul $\tau =1$}                                                & {\ul $\tau =3$}                                                & \multicolumn{1}{c}{} & {\ul $\tau =0.3$}                                               & {\ul $\tau =1$}                                                 & {\ul $\tau =3$}                                                 & \multicolumn{1}{c}{} & {\ul $\tau =0.3$}                                               & {\ul $\tau =1$}                                                 & {\ul $\tau =3$}                                 \\   \vspace{-0.05in}               \\
AM-weighted-$L_1$                & \begin{tabular}[c]{@{}c@{}}4.24\\ (0.33)\end{tabular}          & \begin{tabular}[c]{@{}c@{}}4.42\\ (0.47)\end{tabular}          & \begin{tabular}[c]{@{}c@{}}4.49\\ (0.50)\end{tabular}          &                      & \begin{tabular}[c]{@{}c@{}}19.08\\ (0.94)\end{tabular}          & \begin{tabular}[c]{@{}c@{}}16.34 \\ (0.87)\end{tabular}         & \begin{tabular}[c]{@{}c@{}}16.70\\ (1.11)\end{tabular}          &                      & \begin{tabular}[c]{@{}c@{}}70.79\\ (1.95)\end{tabular}          & \begin{tabular}[c]{@{}c@{}}66.48\\ (1.90)\end{tabular}          & \begin{tabular}[c]{@{}c@{}}67.39\\ (1.97)\end{tabular}       \\   \vspace{-0.05in}    \\
AM-unweighted-$L_1$              & \textbf{\begin{tabular}[c]{@{}c@{}}3.46\\ (0.27)\end{tabular}} & \textbf{\begin{tabular}[c]{@{}c@{}}2.79\\ (0.31)\end{tabular}} & \textbf{\begin{tabular}[c]{@{}c@{}}2.47\\ (0.34)\end{tabular}} &                      & \begin{tabular}[c]{@{}c@{}}17.68\\ (0.87)\end{tabular}          & \textbf{\begin{tabular}[c]{@{}c@{}}13.51\\ (0.72)\end{tabular}} & \textbf{\begin{tabular}[c]{@{}c@{}}12.71\\ (0.88)\end{tabular}} &                      & \begin{tabular}[c]{@{}c@{}}71.79\\ (2.12)\end{tabular}          & \begin{tabular}[c]{@{}c@{}}64.88\\ (2.05)\end{tabular}          & \textbf{\begin{tabular}[c]{@{}c@{}}61.19\\ (1.84)\end{tabular}} \\   \vspace{-0.05in}   \\
AM-weighted-$L_2$                & \begin{tabular}[c]{@{}c@{}}4.45\\ (0.34)\end{tabular}          & \begin{tabular}[c]{@{}c@{}}4.36\\ (0.46)\end{tabular}          & \begin{tabular}[c]{@{}c@{}}4.41\\ (0.48)\end{tabular}          &                      & \begin{tabular}[c]{@{}c@{}}20.07\\ (0.99)\end{tabular}          & \begin{tabular}[c]{@{}c@{}}16.17\\ (0.86)\end{tabular}          & \begin{tabular}[c]{@{}c@{}}16.34\\ (1.09)\end{tabular}          &                      & \begin{tabular}[c]{@{}c@{}}77.01\\ (2.35)\end{tabular}          & \begin{tabular}[c]{@{}c@{}}65.74\\ (1.85)\end{tabular}          & \begin{tabular}[c]{@{}c@{}}65.75\\ (1.84)\end{tabular}      \\   \vspace{-0.05in}    \\
AM-unweighted-$L_2$              & \begin{tabular}[c]{@{}c@{}}3.79\\ (0.30)\end{tabular}          & \begin{tabular}[c]{@{}c@{}}4.03\\ (0.43)\end{tabular}          & \begin{tabular}[c]{@{}c@{}}4.16\\ (0.46)\end{tabular}          &                      & \textbf{\begin{tabular}[c]{@{}c@{}}17.26\\ (0.85)\end{tabular}} & \begin{tabular}[c]{@{}c@{}}15.04\\ (0.79)\end{tabular}          & \begin{tabular}[c]{@{}c@{}}15.55\\ (1.04)\end{tabular}          &                      & \textbf{\begin{tabular}[c]{@{}c@{}}64.92\\ (1.80)\end{tabular}} & \textbf{\begin{tabular}[c]{@{}c@{}}61.45\\ (1.68)\end{tabular}} & \begin{tabular}[c]{@{}c@{}}62.82\\ (1.79)\end{tabular}     \\   \vspace{-0.05in}     \\
Adaptive Lasso                   & \begin{tabular}[c]{@{}c@{}}6.19\\ (0.46)\end{tabular}          & \begin{tabular}[c]{@{}c@{}}7.53\\ (0.92)\end{tabular}          & \begin{tabular}[c]{@{}c@{}}6.45\\ (0.80)\end{tabular}          &                      & \begin{tabular}[c]{@{}c@{}}26.22\\ (1.47)\end{tabular}          & \begin{tabular}[c]{@{}c@{}}25.40\\ (2.20)\end{tabular}          & \begin{tabular}[c]{@{}c@{}}25.76\\ (2.40)\end{tabular}          &                      & \begin{tabular}[c]{@{}c@{}}122.94\\ (8.14)\end{tabular}         & \begin{tabular}[c]{@{}c@{}}114.80\\ (8.65)\end{tabular}         & \begin{tabular}[c]{@{}c@{}}128.07\\ (8.73)\end{tabular}   \\   \vspace{-0.05in}       \\
Lasso                            & \begin{tabular}[c]{@{}c@{}}5.78\\ (0.44)\end{tabular}          & \begin{tabular}[c]{@{}c@{}}6.34\\ (0.82)\end{tabular}          & \begin{tabular}[c]{@{}c@{}}4.68\\ (0.67)\end{tabular}          &                      & \begin{tabular}[c]{@{}c@{}}25.56\\ (1.42)\end{tabular}          & \begin{tabular}[c]{@{}c@{}}23.93\\ (2.14)\end{tabular}          & \begin{tabular}[c]{@{}c@{}}22.68\\ (2.34)\end{tabular}          &                      & \begin{tabular}[c]{@{}c@{}}121.14\\ (8.04)\end{tabular}         & \begin{tabular}[c]{@{}c@{}}112.83\\ (8.43)\end{tabular}         & \begin{tabular}[c]{@{}c@{}}123.53\\ (8.39)\end{tabular}     \\   \vspace{-0.05in}     \\
Elastic Net                      & \begin{tabular}[c]{@{}c@{}}5.72\\ (0.43)\end{tabular}          & \begin{tabular}[c]{@{}c@{}}6.36\\ (0.82)\end{tabular}          & \begin{tabular}[c]{@{}c@{}}4.69\\ (0.67)\end{tabular}          &                      & \begin{tabular}[c]{@{}c@{}}25.33\\ (1.41)\end{tabular}          & \begin{tabular}[c]{@{}c@{}}23.91\\ (2.14)\end{tabular}          & \begin{tabular}[c]{@{}c@{}}22.65\\ (2.34)\end{tabular}          &                      & \begin{tabular}[c]{@{}c@{}}120.37\\ (8.03)\end{tabular}         & \begin{tabular}[c]{@{}c@{}}112.03\\ (8.43)\end{tabular}         & \begin{tabular}[c]{@{}c@{}}123.24\\ (8.40)\end{tabular}   \\   \vspace{-0.05in}    \\
Ridge                            & \begin{tabular}[c]{@{}c@{}}6.02\\ (0.45)\end{tabular}          & \begin{tabular}[c]{@{}c@{}}7.11\\ (0.85)\end{tabular}          & \begin{tabular}[c]{@{}c@{}}6.18\\ (0.74)\end{tabular}          &                      & \begin{tabular}[c]{@{}c@{}}25.83\\ (1.41)\end{tabular}          & \begin{tabular}[c]{@{}c@{}}25.11\\ (2.14)\end{tabular}          & \begin{tabular}[c]{@{}c@{}}25.17\\ (2.36)\end{tabular}          &                      & \begin{tabular}[c]{@{}c@{}}121.06\\ (8.03)\end{tabular}         & \begin{tabular}[c]{@{}c@{}}113.47\\ (8.41)\end{tabular}         & \begin{tabular}[c]{@{}c@{}}125.68\\ (8.36)\end{tabular}    \\ \hline
\end{tabular}
\end{adjustbox}
\end{table}
}

\ifthenelse{1=0}{
%       \centerline{--- Fig. \ref{fig:procedure} ---}
}{

\begin{table}[!htbp]
\caption{Summary of $95\%$ prediction interval coverage with different methods, under the same settings as Table~\ref{tab:lr-1}. The result closest to $95\%$ in each setting is highlighted in boldface.\label{tab:lr-2}}
\centering
% \tiny
\scriptsize
\def\arraystretch{0.8}
  \begin{adjustbox}{max width=\textwidth}
\begin{tabular}{cccccccccccc}
\hline
\multirow{2}{*}{\textit{Method}} & \multicolumn{3}{c}{$\alpha = 0.3$}                                                                                                                                          &  & \multicolumn{3}{c}{$\alpha = 0.6$}                                                                                                                                          &  & \multicolumn{3}{c}{$\alpha = 0.9$}                                                                                                                                          \\ \cline{2-4} \cline{6-8} \cline{10-12} 
                                 & {\ul $\tau =0.3$}                                       & {\ul $\tau =1$}                                         & {\ul $\tau =3$}                                         &  & {\ul $\tau =0.3$}                                       & {\ul $\tau =1$}                                         & {\ul $\tau =3$}                                         &  & {\ul $\tau =0.3$}                                       & {\ul $\tau =1$}                                         & {\ul $\tau =3$}                            \\   \vspace{-0.05in}              \\
AM-weighted-$L_1$                & \textbf{\begin{tabular}[c]{@{}c@{}}0.947\\ (0.002)\end{tabular}} & \textbf{\begin{tabular}[c]{@{}c@{}}0.950\\ (0.002)\end{tabular}} & \textbf{\begin{tabular}[c]{@{}c@{}}0.948\\ (0.002)\end{tabular}} &  & \begin{tabular}[c]{@{}c@{}}0.947\\ (0.002)\end{tabular} & \begin{tabular}[c]{@{}c@{}}0.943\\ (0.002)\end{tabular} & \begin{tabular}[c]{@{}c@{}}0.942\\ (0.002)\end{tabular} &  & \begin{tabular}[c]{@{}c@{}}0.947\\ (0.002)\end{tabular} & \begin{tabular}[c]{@{}c@{}}0.946\\ (0.002)\end{tabular} & \textbf{\begin{tabular}[c]{@{}c@{}}0.947\\ (0.002)\end{tabular}}  \\   \vspace{-0.05in} \\
AM-unweighted-$L_1$              & \begin{tabular}[c]{@{}c@{}}0.955\\ (0.002)\end{tabular} & \begin{tabular}[c]{@{}c@{}}0.970\\ (0.002)\end{tabular} & \begin{tabular}[c]{@{}c@{}}0.982\\ (0.002)\end{tabular} &  & \textbf{\begin{tabular}[c]{@{}c@{}}0.952\\ (0.002)\end{tabular}} & \begin{tabular}[c]{@{}c@{}}0.955\\ (0.002)\end{tabular} & \begin{tabular}[c]{@{}c@{}}0.966\\ (0.002)\end{tabular} &  & \textbf{\begin{tabular}[c]{@{}c@{}}0.951\\ (0.002)\end{tabular}} & \textbf{\begin{tabular}[c]{@{}c@{}}0.952\\ (0.002)\end{tabular}} & \begin{tabular}[c]{@{}c@{}}0.960\\ (0.002)\end{tabular}  \\   \vspace{-0.05in} \\
AM-weighted-$L_2$                & \begin{tabular}[c]{@{}c@{}}0.956\\ (0.002)\end{tabular} & \begin{tabular}[c]{@{}c@{}}0.958\\ (0.002)\end{tabular} & \begin{tabular}[c]{@{}c@{}}0.961\\ (0.002)\end{tabular} &  & \begin{tabular}[c]{@{}c@{}}0.960\\ (0.002)\end{tabular} & \textbf{\begin{tabular}[c]{@{}c@{}}0.954\\ (0.003)\end{tabular}} & \textbf{\begin{tabular}[c]{@{}c@{}}0.952\\ (0.003)\end{tabular}} &  & \begin{tabular}[c]{@{}c@{}}0.957\\ (0.002)\end{tabular} & \begin{tabular}[c]{@{}c@{}}0.958\\ (0.002)\end{tabular} & \begin{tabular}[c]{@{}c@{}}0.957\\ (0.002)\end{tabular}  \\   \vspace{-0.05in} \\
AM-unweighted-$L_2$              & \begin{tabular}[c]{@{}c@{}}0.960\\ (0.002)\end{tabular} & \begin{tabular}[c]{@{}c@{}}0.963\\ (0.002)\end{tabular} & \begin{tabular}[c]{@{}c@{}}0.966\\ (0.002)\end{tabular} &  & \begin{tabular}[c]{@{}c@{}}0.963\\ (0.002)\end{tabular} & \begin{tabular}[c]{@{}c@{}}0.958\\ (0.002)\end{tabular} & \begin{tabular}[c]{@{}c@{}}0.957\\ (0.002)\end{tabular} &  & \begin{tabular}[c]{@{}c@{}}0.961\\ (0.002)\end{tabular} & \begin{tabular}[c]{@{}c@{}}0.962\\ (0.002)\end{tabular} & \begin{tabular}[c]{@{}c@{}}0.962\\ (0.002)\end{tabular}  \\   \vspace{-0.05in} \\
Adaptive Lasso                   & \begin{tabular}[c]{@{}c@{}}0.923\\ (0.004)\end{tabular} & \begin{tabular}[c]{@{}c@{}}0.874\\ (0.007)\end{tabular} & \begin{tabular}[c]{@{}c@{}}0.769\\ (0.012)\end{tabular} &  & \begin{tabular}[c]{@{}c@{}}0.902\\ (0.009)\end{tabular} & \begin{tabular}[c]{@{}c@{}}0.873\\ (0.007)\end{tabular} & \begin{tabular}[c]{@{}c@{}}0.767\\ (0.010)\end{tabular} &  & \begin{tabular}[c]{@{}c@{}}0.916\\ (0.005)\end{tabular} & \begin{tabular}[c]{@{}c@{}}0.883\\ (0.010)\end{tabular} & \begin{tabular}[c]{@{}c@{}}0.833\\ (0.011)\end{tabular}  \\   \vspace{-0.05in} \\
Lasso                            & \begin{tabular}[c]{@{}c@{}}0.923\\ (0.004)\end{tabular} & \begin{tabular}[c]{@{}c@{}}0.908\\ (0.007)\end{tabular} & \begin{tabular}[c]{@{}c@{}}0.898\\ (0.009)\end{tabular} &  & \begin{tabular}[c]{@{}c@{}}0.898\\ (0.010)\end{tabular} & \begin{tabular}[c]{@{}c@{}}0.885\\ (0.007)\end{tabular} & \begin{tabular}[c]{@{}c@{}}0.854\\ (0.009)\end{tabular} &  & \begin{tabular}[c]{@{}c@{}}0.918\\ (0.006)\end{tabular} & \begin{tabular}[c]{@{}c@{}}0.885\\ (0.010)\end{tabular} & \begin{tabular}[c]{@{}c@{}}0.859\\ (0.010)\end{tabular}  \\   \vspace{-0.05in} \\
Elastic Net                      & \begin{tabular}[c]{@{}c@{}}0.915\\ (0.005)\end{tabular} & \begin{tabular}[c]{@{}c@{}}0.893\\ (0.008)\end{tabular} & \begin{tabular}[c]{@{}c@{}}0.888\\ (0.009)\end{tabular} &  & \begin{tabular}[c]{@{}c@{}}0.891\\ (0.010)\end{tabular} & \begin{tabular}[c]{@{}c@{}}0.875\\ (0.007)\end{tabular} & \begin{tabular}[c]{@{}c@{}}0.847\\ (0.009)\end{tabular} &  & \begin{tabular}[c]{@{}c@{}}0.906\\ (0.006)\end{tabular} & \begin{tabular}[c]{@{}c@{}}0.876\\ (0.010)\end{tabular} & \begin{tabular}[c]{@{}c@{}}0.855\\ (0.011)\end{tabular}  \\   \vspace{-0.05in} \\
Ridge                            & \begin{tabular}[c]{@{}c@{}}0.925\\ (0.004)\end{tabular} & \begin{tabular}[c]{@{}c@{}}0.871\\ (0.007)\end{tabular} & \begin{tabular}[c]{@{}c@{}}0.754\\ (0.010)\end{tabular} &  & \begin{tabular}[c]{@{}c@{}}0.907\\ (0.009)\end{tabular} & \begin{tabular}[c]{@{}c@{}}0.868\\ (0.007)\end{tabular} & \begin{tabular}[c]{@{}c@{}}0.764\\ (0.011)\end{tabular} &  & \begin{tabular}[c]{@{}c@{}}0.925\\ (0.005)\end{tabular} & \begin{tabular}[c]{@{}c@{}}0.891\\ (0.010)\end{tabular} & \begin{tabular}[c]{@{}c@{}}0.845\\ (0.011)\end{tabular} \\ \hline
\end{tabular}
\end{adjustbox}
\end{table}
}

The results are summarized in Table~\ref{tab:lr-1} and Table~\ref{tab:lr-2}. We can see that AM greatly outperforms all other methods in terms of ME, regardless of the duality function it uses. Moreover, AM can provide a satisfactory prediction interval coverage under all settings at the 95\% level, while all other methods show significant undercoverage.

\subsection{Image classification with Neural Networks}
\label{s:nn}

%The over-parameterized model in the previous subsection is effectively an example of what is called the shallow network model, a special case of deepnets.
For an application of the proposed method to the neural network model, we 
consider a numerical example of image classification with the
famous MNIST data set \citep{Lecun1998}.
%Our results show that over-dispersed imputation has a great promise in data analysis with deepnets for validity, efficiency, and robustness.  The sparsity issue that has been emphasized in the statistical literature is resolved as a simple consequence.
The MNIST %(National Institute of Standards and Technology)
database is a large database of handwritten digits that is commonly used for training various image processing systems.
%See \href{https://en.wikipedia.org/wiki/MNIST_database}{\color{blue}\underline{https://en.wikipedia.org/wiki/MNIST\_database}} for details.
The training and test sample sizes are 60,000 and 10,000, respectively. For each handwritten digit (0-9),
the image size is $28\times 28$ pixels,
with pixel values measured at gray-scale levels from 0 to 255.
Thus, for each observation, $x_i$ stands for the image and $y_i$ stands for the
label or digit.
The classification problem is to predict $y_i$ given $x_i$. 

The imputation-estimation scheme of AM proposed in Section~\ref{ss:imputation-estimation}, and applied in this example is concisely summarized here for better clarity. The imputation process (Algorithm~\ref{alg:imputation}) involves estimating models that predict new labels for training digit images. These training images, along with their newly predicted labels, form the imputed future observations used in the final estimation process with Algorithm~\ref{alg:estimation}. Conceptually, each image in the dataset is associated with multiple, potentially varying labels, which helps effectively prevent model overfitting to a single label.

% \footnote{
% \color{red} %\scriptsize %\footnotesize

% }

Two different neural network structures are used to investigate the efficiency of the proposed method. The first structure is a feed-forward neural network with two fully connected layers. The number of nodes in the two hidden layers is chosen to be 400, 800, and 1600. The final layer 
uses the multivariate logistic link and returns the probabilities for the 10 labels. This classic structure 
 % containing 478,410 parameters 
is commonly used in the literature for evaluating model training strategies. The second structure to be considered is described in \cite{Jarrett2009}. It is obtained by using the output from a convolutional neural networks (CNN) feature extractor as the input to the first structure. The feature extractor is constructed with two convolutional layers with 32 and 64 channels respectively, each followed by a $2\times 2$ max-pooling layer. The size of the filters in each CNN layer is chosen to be $5\times5$ and the number of hidden nodes in the fully connected layers is chosen to be 200. The rectified linear unit (ReLU) activation function is used in all structures.
% This results in 299,306 parameters. 
% ReLU activation function is used for both structures, after each convolutional layer and each fully connected layer.

% In our auto-modeling setting, the 60,000 images in the training set are treated as the imputed population $\mathbb{Q}$ and a bootstrap resampled set of images are treated as the empirical distribution $\hat{\mathbb{P}}$ in Algorithm~\ref{alg:estimation}.

% In the imputation step of auto-modeling, the standard Algorithm~\ref{alg:2.1} is used to create 5 imputation models.  

For each model, denote all the trainable parameters in the model as $\theta_1, \dots, \theta_p$. The \dual~function of AM is defined as the counterpart of the $L_1$ and $L_2$ penalty functions, 
% \begin{equation}\label{eq:nn-duality-l1}
% \pi(\theta,\lambda) = \sum_{j}^p \lambda_j|\theta_j|
% \qquad(\lambda_j\geq 0 \mbox{ for }  j=1,\dots,p)
% \end{equation}
% and
% \begin{equation}\label{eq:nn-duality-l2}
% \pi(\theta,\lambda) = \sum_{j}^p \lambda_j\theta_j^2
% \qquad(\lambda_j\geq 0 \mbox{ for } j=1,\dots,p).
% \end{equation}
\begin{equation}\label{eq:nn-duality}
\pi(\theta,\lambda) = \sum_{j=1}^p \lambda_j|\theta_j|\quad \text{or} \quad  \pi(\theta,\lambda) = \sum_{j=1}^p \lambda_j\theta_j^2 \qquad(\lambda_j\geq 0 \mbox{ for } j=1,\dots,p).
\end{equation}
When applying Algorithm~\ref{alg:choose-m} to choose $m$ for the $m$-out-of-$n$ resampling scheme, a continuous CDF value of the data distribution within the range of $0$ to $1$ is necessary. Here, we adopt and extend the randomization approach proposed by \cite{dunn1996randomized}. Denote by $p_0, p_1, \dots,$ and $ p_9$ the predicted probabilities for the 10 digits obtained from the final layer for an input image $x$, and let $y$ denote the true label of the image. We randomly select $F(y |x)$ from a uniform distribution over the interval $\left[\sum_{k=0}^{y-1} p_k, \sum_{k=0}^y p_k\right]$, where $\sum_{k=0}^{-1} p_k$ is defined as $0$. Comprehensive details and theoretical justifications for this approach are provided in Supplement~S.6.4. We use $B = 5$ and $K = 2$ in the imputation process with Algorithm~\ref{alg:imputation}. 
Further implementation details are given in Supplement~S.6.

For comparison with other methods, the same model structures are trained on the full data set accompanied with fixed scalar (or unweighted) $L_1$ and $L_2$ penalties, which are commonly used in neural network training. We record the best outcomes across a reasonable range of penalty values. 
% In favor of the existing methods, the scalar $\lambda$ that yields the lowest test error is chosen. This can be seen as the potential best results obtainable by cross-validated choice of $\lambda$ in practice. 
Additionally, we incorporate the early-stopping technique and the widely-used dropout method \citep{Srivastava2014}. 
% Each method, including the estimation step of AM is repeated for 5 times. 
To isolate the effectiveness of our framework from the impact of using weighted duality functions, we perform an additional estimation procedure. In this procedure, we remove this confounding factor by setting $\lambda_1 = \dots =\lambda_p$ in (\ref{eq:nn-duality}), thereby creating what we refer to as the unweighted duality functions.

\ifthenelse{1=0}{

}{
\begin{table}[!htbp]
\caption{The test error ($\%$) of different methods on the MNIST data set with different structures. FC X-X in the header of the table denotes the fully-connected network with two hidden layers where the number of hidden nodes is X, and the values in the parentheses denote the number of parameters in the corresponding structure. The reported value is the mean taken over 5 repetitions (with standard deviations given in parentheses). The best two results in each structure are highlighted in boldface.
\label{tab:nn}}
\centering
% \tiny
\scriptsize
\begin{tabular}{cclclcll}
\hline
\textit{Method}       & \begin{tabular}[c]{@{}c@{}}FC 400-400\\ (478,410)\end{tabular} &  & \begin{tabular}[c]{@{}c@{}}FC 800-800\\ (127,6810)\end{tabular} &  & \begin{tabular}[c]{@{}c@{}}FC 1600-1600\\ (383,3610)\end{tabular} &  & \multicolumn{1}{c}{\begin{tabular}[c]{@{}c@{}}CNN\\ (299,306)\end{tabular}} \\ \cline{2-2} \cline{4-4} \cline{6-6} \cline{8-8} 
AM-weighted-$L_1$     & 1.222 (0.040)                                                  &  & 1.244 (0.042)                                                   &  & 1.202 (0.052)                                                     &  & \textbf{0.584 (0.043)}                                                      \\
AM-unweighted-$L_1$   & \multicolumn{1}{l}{\textbf{1.180 (0.040)}}                     &  & \multicolumn{1}{l}{\textbf{1.172 (0.064)}}                      &  & \multicolumn{1}{l}{\textbf{1.093 (0.059)}}                        &  & 0.586 (0.038)                                                               \\
AM-weighted-$L_2$     & \multicolumn{1}{l}{\textbf{1.178 (0.040)}}                     &  & \multicolumn{1}{l}{\textbf{1.184 (0.040)}}                      &  & \multicolumn{1}{l}{\textbf{1.129 (0.046)}}                        &  & \textbf{0.569 (0.040)}                                                      \\
AM-unweighted-$L_2$   & 1.246 (0.046)                                                  &  & 1.204 (0.025)                                                   &  & 1.146 (0.020)                                                     &  & 0.608 (0.016)                                                               \\
$L_1$- regularization & 1.476 (0.046)                                                  &  & 1.442 (0.052)                                                   &  & 1.386 (0.044)                                                     &  & 0.668 (0.068)                                                               \\
$L_2$- regularization & 1.408 (0.018)                                                  &  & 1.386 (0.066)                                                   &  & 1.446 (0.065)                                                     &  & 0.602 (0.039)                                                               \\
Early-stopping        & 1.770 (0.084)                                                  &  & 1.608 (0.121)                                                   &  & 1.602 (0.067)                                                     &  & 0.870 (0.093)                                                               \\
Dropout              & \multicolumn{1}{l}{1.512 (0.141)}                              &  & \multicolumn{1}{l}{1.478 (0.087)}                               &  & \multicolumn{1}{l}{1.494 (0.068)}                                 &  & 0.592 (0.050)                                                               \\ \hline
\end{tabular}
\end{table}

}

% \ifthenelse{1=0}{
% %       \centerline{--- Fig. \ref{fig:procedure} ---}
% }{

% \begin{figure}[!htbp]
% \centering
% \includegraphics[width=14cm]{plot/feedforward-plot.pdf}
% \caption{The learning curves in the estimation process and the histograms of the estimated parameters using AM (weighted $L_1$ %simplicity-preference 
% 	\dual~function) with different model structures.}\label{fig:convergence}
% \end{figure}
% }

% \begin{table}[!h]
% \centering
% \footnotesize
% \begin{tabular}{c|c|c}
% \hline
% \textit{Method}                                                 & Fully-connected                                         & \begin{tabular}[c]{@{}c@{}}CNN + \\ Fully-connected\end{tabular} \\ \hline
% \begin{tabular}[c]{@{}c@{}}AM Estimation ($L_1$)\end{tabular} & \begin{tabular}[c]{@{}c@{}}1.20\\ (0.0359)\end{tabular} & \begin{tabular}[c]{@{}c@{}}0.54\\ (0.0301)\end{tabular}          \\
% \begin{tabular}[c]{@{}c@{}}AM Estimation ($L_2$)\end{tabular} & \begin{tabular}[c]{@{}c@{}}1.20\\ (0.0397)\end{tabular} & \begin{tabular}[c]{@{}c@{}}0.59\\ (0.0350)\end{tabular}          \\ \hline
% \end{tabular}
% \caption{The test error of the final estimation model of auto-modeling.\label{tab:4}}
% \end{table}

 The test error results for all the methods, using four different models, are presented in Table~\ref{tab:nn}.
 % The results suggest that the final estimation model provides further improvements of test accuracy for the average performance of the 5 imputation models. 
 A significant performance improvement can be seen when AM is used. Notably, the performance of AM exceeds that of the Dropconnect \citep{Li2013, Mobiny2021}, a current state-of-the-art regularization technique, with identical model structures as reported in \cite{Li2013} and  \cite{Mobiny2021}. Compared to their method,  AM not only demonstrates faster convergence but also offers a more straightforward implementation. 
 % Remarkably, AM consistently reaches convergence within 100 epochs across all experiments. 
 The details of the estimation process of AM and the resultant parameters are depicted in Supplement~S.6.4. As an ancillary benefit of AM, it can detect incorrectly labeled data (detailed in Supplement~S.6.5).
 % Figure~\ref{fig:convergence}. 

 % An intriguing aspect of the AM estimator is its tendency to achieve nearly, but not quite, 100\% prediction accuracy on training images. By aggregating the training images that were incorrectly predicted across all experiments (as detailed in Table~\ref{tab:nn}), we identified 62 unique images. Closer examination of these images revealed that many seem to possess incorrect labels. A selection of these images is showcased in Figure~\ref{fig:questionable-labels-1}, with the complete set of 62 images detailed in Supplement~S.11.5.  Notably, some of the labels we identified as questionable also align with findings from recent research by \cite{Northcutt2021}, which focused on detecting label inaccuracies. Our method, however, uncovered a greater number of potential errors. Therefore, AM shows promise as a tool for effectively detecting incorrectly labeled data, serving as an ancillary benefit to its primary purpose.

    \ifthenelse{1=1}{
    %       \centerline{--- Fig. \ref{fig:procedure} ---}
    }{

    \begin{figure}[!htbp]
    \centering
    \includegraphics[width=14cm]{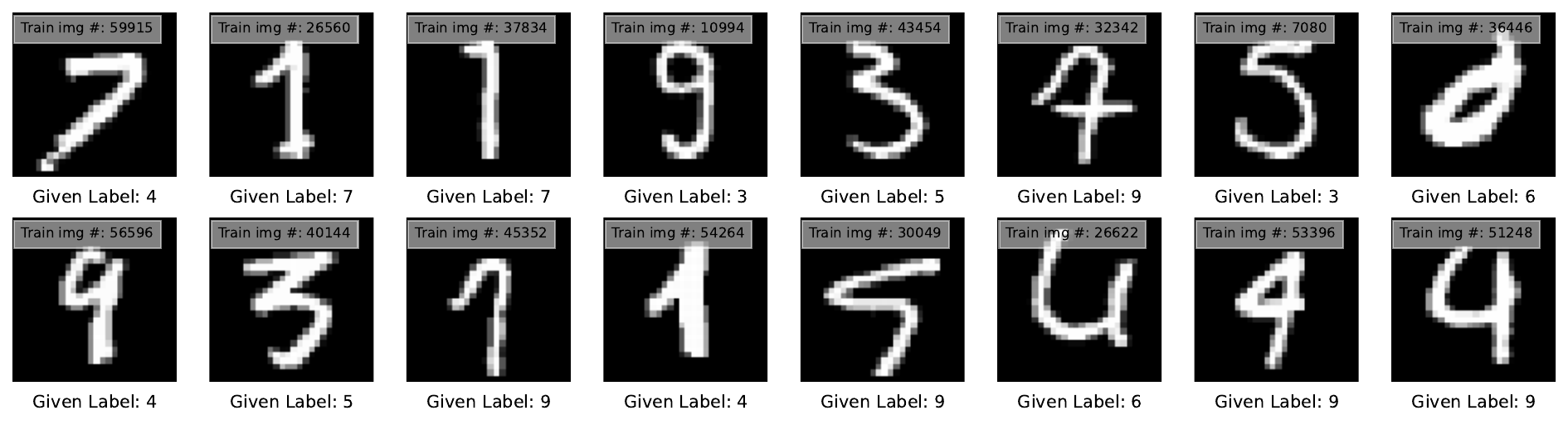}
    \caption{Selection of images detected by AM as having questionable labels, displayed in a left-to-right, top-to-bottom sequence, sorted by the decreasing frequency of their identification.}\label{fig:questionable-labels-1}
    \end{figure}
    }

As a remark, the imputation algorithm introduced in this paper does not create new images ($x$), and performance could improve with an imputation algorithm that does. For example, data augmentation methods like random deformation \citep{simard2003best} which is shown to be effective in practice, can be used to create more images. A popular technique in recent years called the generative adversarial network (GAN, \citeauthor{goodfellow2020generative}, \citeyear{goodfellow2020generative}) also provides a potential strategy towards this direction. %\CLdel{Perhaps a more conceptually easy way for imputing $x$ is by fitting a Gaussian mixture model on the images with reduced dimension and sampling from such distribution \citep{banijamali2018deep}}. 
% This is also in the same sense as ``creating over-dispersed population with currently observed data'', which is an important idea of the auto-modeling framework. 
For simplicity, such extended experiments have not been conducted here and will be reported elsewhere. 

% In addition to the above discussion, we highlight the reproducibility problem, which becomes one of the most challenging issue in deep learning models \citep{Chen2022}. One of the major cause of the reproducibility problem is the inconsistent choice of hyper-parameters. Indeed, for the most common $L_1$ and $L_2$ regularization, as well as the state-of-the-art method Dropconnect \citep{Li2013}, the results are seen to be very sensitive to the choice of regularization-related hyper-parameters. Unlike these existing methods, AM does not require any such hyper-parameters to be tuned with uncertainty, but automatically learn a model with high generalization power. This reduces the risk of being failure to reproduce the results due to the inconsistent choice of hyper-parameters. Besides, as is mentioned in \cite{Ahn2022}, the ensemble approach, which resembles AM as discussed in Section~\ref{s:theoretical-results}, is a promising technique to improve reproducibility in practice. Empirical evidence in Table~\ref{tab:3} shows that AM gives results with low variability in many cases.

%\section{Discussion}
\section{Concluding Remarks}
\label{s:discussion}

This paper proposed a %\CLdel{new} 
promising method for the estimation of over-parameterized models from a model-building perspective. %\CLdel{The theoretical results and applications have shown that the proposed method is promising. Future}
It is expected that future developments could focus on its applications, enhancing both over-parameterized and non-overparameterized models in machine learning and statistics. 
For instance, the numerical results from the many-normal-means example demonstrate that over-parameterization, when combined with the proposed estimation method, effectively enhances model flexibility and applicability. This, in turn, improves the efficiency in predicting future observations. These insights align with the widespread consensus on the success of deep neural networks. We believe that further investigations of this phenomenon for %\CLdel{the many-normal-means problem, as well as other well-known}
practically popular statistical models %\CLdel{ such as linear regression, factor analysis, mixed-effects models, mixture models, and Student-$t$ models,} 
could yield intriguing and valuable theoretical advancements. These advancements could enhance and extend beyond traditional likelihood-based inference methods. 

Technically, further improvements could be explored in the specification of the \dual~function and imputation methods. Since the primary focus of this paper is on the foundational aspects of modeling, we have opted to implement an adaptive bootstrapping approach emphasizing on model checking. Although this data-driven implementation has shown promise, it may still encounter unexpected limitations in robustness and efficiency, which are common in resampling methods, particularly in high-dimensional problems \citep{Liu2024}. Therefore, exploring alternative imputation methods could be worthwhile. For example, synthetic data generation techniques could be especially useful, as they have demonstrated effectiveness across various tasks, particularly in handling complex models and high-dimensional data \citep{Liu2024, Shen2024, Tian2024}. Additionally, it could also be interesting to investigate how the proposed method might, in turn, improve such models, considering the image classification example of Section \ref{s:nn} in particular and Box's perspective on modeling as an iterative process \citep{box1980sampling} in general; see Supplementary~S.12.

Progressively, the development of more efficient computational techniques would enhance the success of our proposed method, particularly in enabling more effective and valid analysis of large datasets.
Finally, methods of statistical inference, including conformal prediction  (\textit{c.f.}  \cite{imconformal} and references therein), can be effectively applied and developed within our proposed framework.

%\pagebreak
%\color{black}

\section*{Acknowledgement}
The authors are grateful to the editor, the associate editor, and anonymous referees for their insightful, critical, and constructive comments on earlier versions of the paper. In particular, their detailed and constructive feedback significantly shaped and improved the development of this work.
\if0\blind
{Jiang is supported in part by U.S. National Institutes of Health grants R01HG010171 and R01MH116527 and National Science Foundation grant DMS-2112711. Liu is supported in part by U.S. National Science Foundation grant DMS-2412629.
}\fi

%% Add Bruce Craig later (not this version) for discussion on checking discrete response models.
%%\clearpage

\ifthenelse{1=1}{
%       \centerline{--- Fig. \ref{fig:procedure} ---}
}{

\begin{table}[!htbp]
\centering
\scriptsize
\begin{tabular}{cccccccccccc}
\hline
\multirow{2}{*}{\textit{Method}} & \multicolumn{3}{c}{$\mu \sim N(0,0.01)$}                                                                                                                                                               &  & \multicolumn{3}{c}{\begin{tabular}[c]{@{}c@{}}$\mu^1 \sim N(-2,0.01)$\\ $\mu^2 \sim N(2,0.01)$\end{tabular}}                                                                                           &  & \multicolumn{3}{c}{\begin{tabular}[c]{@{}c@{}}$\mu^1 = 0$\\ $\mu^2 \sim N(-3,1)$\end{tabular}}                                                                                                         \\ \cline{2-4} \cline{6-8} \cline{10-12} 
                                 & {\ul $n = 10$}                                                   & {\ul $n = 20$}                                                   & {\ul $n = 50$}                                                   &  & {\ul $n = 10$}                                                   & {\ul $n = 20$}                                                   & {\ul $n = 50$}                                                   &  & {\ul $n = 10$}                                                   & {\ul $n = 20$}                                                   & {\ul $n = 50$}                                                   \\
MLE                              & \begin{tabular}[c]{@{}c@{}}0.996\\ (0.305)\end{tabular}          & \begin{tabular}[c]{@{}c@{}}0.977\\ (0.454)\end{tabular}          & \begin{tabular}[c]{@{}c@{}}1.012\\ (0.643)\end{tabular}          &  & \begin{tabular}[c]{@{}c@{}}0.936\\ (0.279)\end{tabular}          & \begin{tabular}[c]{@{}c@{}}1.016\\ (0.464)\end{tabular}          & \begin{tabular}[c]{@{}c@{}}1.020\\ (0.716)\end{tabular}          &  & \begin{tabular}[c]{@{}c@{}}1.015\\ (0.317)\end{tabular}          & \begin{tabular}[c]{@{}c@{}}0.999\\ (0.457)\end{tabular}          & \begin{tabular}[c]{@{}c@{}}0.998\\ (0.671)\end{tabular}          \\
James-Stein                      & \begin{tabular}[c]{@{}c@{}}0.302\\ (0.221)\end{tabular}          & \begin{tabular}[c]{@{}c@{}}0.161\\ (0.235)\end{tabular}          & \begin{tabular}[c]{@{}c@{}}0.064\\ (0.205)\end{tabular}          &  & \begin{tabular}[c]{@{}c@{}}0.836\\ (0.248)\end{tabular}          & \begin{tabular}[c]{@{}c@{}}0.857\\ (0.382)\end{tabular}          & \begin{tabular}[c]{@{}c@{}}0.829\\ (0.604)\end{tabular}          &  & \begin{tabular}[c]{@{}c@{}}0.548\\ (0.284)\end{tabular}          & \begin{tabular}[c]{@{}c@{}}0.497\\ (0.377)\end{tabular}          & \begin{tabular}[c]{@{}c@{}}0.499\\ (0.536)\end{tabular}          \\
$g$-modeling                     & \begin{tabular}[c]{@{}c@{}}0.375\\ (0.192)\end{tabular}          & \begin{tabular}[c]{@{}c@{}}0.383\\ (0.299)\end{tabular}          & \begin{tabular}[c]{@{}c@{}}0.180\\ (0.313)\end{tabular}          &  & \begin{tabular}[c]{@{}c@{}}0.722\\ (0.257)\end{tabular}          & \begin{tabular}[c]{@{}c@{}}0.738\\ (0.385)\end{tabular}          & \begin{tabular}[c]{@{}c@{}}0.758\\ (0.692)\end{tabular}          &  & \begin{tabular}[c]{@{}c@{}}0.571\\ (0.247)\end{tabular}          & \begin{tabular}[c]{@{}c@{}}0.528\\ (0.350)\end{tabular}          & \begin{tabular}[c]{@{}c@{}}0.384\\ (0.509)\end{tabular}          \\
Auto-Modeling                    & \textbf{\begin{tabular}[c]{@{}c@{}}0.204\\ (0.218)\end{tabular}} & \textbf{\begin{tabular}[c]{@{}c@{}}0.120\\ (0.241)\end{tabular}} & \textbf{\begin{tabular}[c]{@{}c@{}}0.061\\ (0.234)\end{tabular}} &  & \textbf{\begin{tabular}[c]{@{}c@{}}0.544\\ (0.392)\end{tabular}} & \textbf{\begin{tabular}[c]{@{}c@{}}0.482\\ (0.579)\end{tabular}} & \textbf{\begin{tabular}[c]{@{}c@{}}0.384\\ (0.915)\end{tabular}} &  & \textbf{\begin{tabular}[c]{@{}c@{}}0.446\\ (0.309)\end{tabular}} & \textbf{\begin{tabular}[c]{@{}c@{}}0.383\\ (0.447)\end{tabular}} & \textbf{\begin{tabular}[c]{@{}c@{}}0.328\\ (0.630)\end{tabular}} \\ \hline
\end{tabular}
\caption{Summarized MPE results under three simulation settings with different methods. The best result in each setting is highlighted in boldface.\label{tb:1}}
\end{table}
}

\ifthenelse{1=1}{
%       \centerline{--- Fig. \ref{fig:procedure} ---}
}{

\begin{table}[!htbp]
\centering
\scriptsize
\begin{tabular}{cccccccccccc}
\hline
\multirow{2}{*}{\textit{Method}} & \multicolumn{3}{c}{$\alpha = 0.3$}                                                                                                                                                               &  & \multicolumn{3}{c}{$\alpha = 0.6$}                                                                                                                                                                  &  & \multicolumn{3}{c}{$\alpha = 0.9$}                                                                                                                                                                  \\ \cline{2-4} \cline{6-8} \cline{10-12} 
                                 & {\ul $\tau =0.3$}                                              & {\ul $\tau =1$}                                                & {\ul $\tau =3$}                                                &  & {\ul $\tau =0.3$}                                               & {\ul $\tau =1$}                                                 & {\ul $\tau =3$}                                                 &  & {\ul $\tau =0.3$}                                               & {\ul $\tau =1$}                                                 & {\ul $\tau =3$}                                                 \\
AM-weighted-$L_1$                & \begin{tabular}[c]{@{}c@{}}4.26\\ (0.33)\end{tabular}          & \begin{tabular}[c]{@{}c@{}}4.42\\ (0.47)\end{tabular}          & \begin{tabular}[c]{@{}c@{}}4.51\\ (0.50)\end{tabular}          &  & \begin{tabular}[c]{@{}c@{}}19.09\\ (0.93)\end{tabular}          & \begin{tabular}[c]{@{}c@{}}16.39\\ (0.87)\end{tabular}          & \begin{tabular}[c]{@{}c@{}}16.77\\ (1.11)\end{tabular}          &  & \begin{tabular}[c]{@{}c@{}}71.06\\ (1.93)\end{tabular}          & \begin{tabular}[c]{@{}c@{}}67.23\\ (1.93)\end{tabular}          & \begin{tabular}[c]{@{}c@{}}67.66\\ (1.97)\end{tabular}          \\
AM-unweighted-$L_1$              & \textbf{\begin{tabular}[c]{@{}c@{}}3.52\\ (0.26)\end{tabular}} & \textbf{\begin{tabular}[c]{@{}c@{}}2.84\\ (0.31)\end{tabular}} & \textbf{\begin{tabular}[c]{@{}c@{}}2.59\\ (0.35)\end{tabular}} &  & \begin{tabular}[c]{@{}c@{}}18.02\\ (0.88)\end{tabular}          & \textbf{\begin{tabular}[c]{@{}c@{}}13.75\\ (0.72)\end{tabular}} & \textbf{\begin{tabular}[c]{@{}c@{}}13.11\\ (0.88)\end{tabular}} &  & \begin{tabular}[c]{@{}c@{}}72.66\\ (2.13)\end{tabular}          & \begin{tabular}[c]{@{}c@{}}66.42\\ (2.13)\end{tabular}          & \textbf{\begin{tabular}[c]{@{}c@{}}62.35\\ (1.86)\end{tabular}} \\
AM-weighted-$L_2$                & \begin{tabular}[c]{@{}c@{}}4.54\\ (0.34)\end{tabular}          & \begin{tabular}[c]{@{}c@{}}4.43\\ (0.47)\end{tabular}          & \begin{tabular}[c]{@{}c@{}}4.42\\ (0.48)\end{tabular}          &  & \begin{tabular}[c]{@{}c@{}}20.03\\ (1.00)\end{tabular}          & \begin{tabular}[c]{@{}c@{}}16.13\\ (0.86)\end{tabular}          & \begin{tabular}[c]{@{}c@{}}16.37\\ (1.09)\end{tabular}          &  & \begin{tabular}[c]{@{}c@{}}77.04\\ (2.40)\end{tabular}          & \begin{tabular}[c]{@{}c@{}}65.73\\ (1.80)\end{tabular}          & \begin{tabular}[c]{@{}c@{}}65.94\\ (1.87)\end{tabular}          \\
AM-unweighted-$L_2$              & \begin{tabular}[c]{@{}c@{}}3.79\\ (0.30)\end{tabular}          & \begin{tabular}[c]{@{}c@{}}4.02\\ (0.43)\end{tabular}          & \begin{tabular}[c]{@{}c@{}}4.16\\ (0.46)\end{tabular}          &  & \textbf{\begin{tabular}[c]{@{}c@{}}17.25\\ (0.85)\end{tabular}} & \begin{tabular}[c]{@{}c@{}}15.04\\ (0.79)\end{tabular}          & \begin{tabular}[c]{@{}c@{}}15.59\\ (1.04)\end{tabular}          &  & \textbf{\begin{tabular}[c]{@{}c@{}}65.13\\ (1.81)\end{tabular}} & \textbf{\begin{tabular}[c]{@{}c@{}}61.52\\ (1.68)\end{tabular}} & \begin{tabular}[c]{@{}c@{}}62.83\\ (1.79)\end{tabular}          \\
Adaptive Lasso                   & \begin{tabular}[c]{@{}c@{}}6.19\\ (0.46)\end{tabular}          & \begin{tabular}[c]{@{}c@{}}7.53\\ (0.92)\end{tabular}          & \begin{tabular}[c]{@{}c@{}}6.45\\ (0.80)\end{tabular}          &  & \begin{tabular}[c]{@{}c@{}}26.22\\ (1.47)\end{tabular}          & \begin{tabular}[c]{@{}c@{}}25.4\\ (2.20)\end{tabular}           & \begin{tabular}[c]{@{}c@{}}25.76\\ (2.40)\end{tabular}          &  & \begin{tabular}[c]{@{}c@{}}122.94\\ (8.14)\end{tabular}         & \begin{tabular}[c]{@{}c@{}}114.8\\ (8.65)\end{tabular}          & \begin{tabular}[c]{@{}c@{}}128.07\\ (8.73)\end{tabular}         \\
Lasso                            & \begin{tabular}[c]{@{}c@{}}5.78\\ (0.44)\end{tabular}          & \begin{tabular}[c]{@{}c@{}}6.34\\ (0.82)\end{tabular}          & \begin{tabular}[c]{@{}c@{}}4.68\\ (0.67)\end{tabular}          &  & \begin{tabular}[c]{@{}c@{}}25.56\\ (1.42)\end{tabular}          & \begin{tabular}[c]{@{}c@{}}23.93\\ (2.14)\end{tabular}          & \begin{tabular}[c]{@{}c@{}}22.68\\ (2.34)\end{tabular}          &  & \begin{tabular}[c]{@{}c@{}}121.14\\ (8.04)\end{tabular}         & \begin{tabular}[c]{@{}c@{}}112.83\\ (8.43)\end{tabular}         & \begin{tabular}[c]{@{}c@{}}123.53\\ (8.39)\end{tabular}         \\
Elastic Net                      & \begin{tabular}[c]{@{}c@{}}5.72\\ (0.43)\end{tabular}          & \begin{tabular}[c]{@{}c@{}}6.36\\ (0.82)\end{tabular}          & \begin{tabular}[c]{@{}c@{}}4.69\\ (0.67)\end{tabular}          &  & \begin{tabular}[c]{@{}c@{}}25.33\\ (1.41)\end{tabular}          & \begin{tabular}[c]{@{}c@{}}23.91\\ (2.14)\end{tabular}          & \begin{tabular}[c]{@{}c@{}}22.65\\ (2.34)\end{tabular}          &  & \begin{tabular}[c]{@{}c@{}}120.37\\ (8.03)\end{tabular}         & \begin{tabular}[c]{@{}c@{}}112.03\\ (8.43)\end{tabular}         & \begin{tabular}[c]{@{}c@{}}123.24\\ (8.40)\end{tabular}         \\
Ridge                            & \begin{tabular}[c]{@{}c@{}}6.02\\ (0.45)\end{tabular}          & \begin{tabular}[c]{@{}c@{}}7.11\\ (0.85)\end{tabular}          & \begin{tabular}[c]{@{}c@{}}6.18\\ (0.74)\end{tabular}          &  & \begin{tabular}[c]{@{}c@{}}25.83\\ (1.41)\end{tabular}          & \begin{tabular}[c]{@{}c@{}}25.11\\ (2.14)\end{tabular}          & \begin{tabular}[c]{@{}c@{}}25.17\\ (2.36)\end{tabular}          &  & \begin{tabular}[c]{@{}c@{}}121.06\\ (8.03)\end{tabular}         & \begin{tabular}[c]{@{}c@{}}113.47\\ (8.41)\end{tabular}         & \begin{tabular}[c]{@{}c@{}}125.68\\ (8.36)\end{tabular}         \\ \hline
\end{tabular}
\caption{Summary of ME results with different methods with the simulation data ($p = 500, n = 100$). $\alpha$ and $\tau$ are the parameters that control the sparsity of the true signals and the signal-to-noise ratio (SNR). Each value is taken as the average value obtained with $100$ repetitions and the standard deviation of each value (estimated with bootstrap) is given in parentheses. The best result in each setting is highlighted in boldface. \label{tab:lr-1}}
\end{table}
}

\ifthenelse{1=1}{
%       \centerline{--- Fig. \ref{fig:procedure} ---}
}{

\begin{table}[!htbp]
\centering
\scriptsize
\begin{tabular}{cccccccccccc}
\hline
\multirow{2}{*}{\textit{Method}} & \multicolumn{3}{c}{$\alpha = 0.3$}                    &  & \multicolumn{3}{c}{$\alpha = 0.6$}                    &  & \multicolumn{3}{c}{$\alpha = 0.9$}                    \\ \cline{2-4} \cline{6-8} \cline{10-12} 
                                 & {\ul $\tau =0.3$} & {\ul $\tau =1$} & {\ul $\tau =3$} &  & {\ul $\tau =0.3$} & {\ul $\tau =1$} & {\ul $\tau =3$} &  & {\ul $\tau =0.3$} & {\ul $\tau =1$} & {\ul $\tau =3$} \\
AM-weighted-$L_1$                & 0.950             & 0.956           & 0.954           &  & 0.953             & 0.948           & 0.947           &  & 0.952             & 0.951           & 0.952           \\
AM-unweighted-$L_1$              & 0.958             & 0.973           & 0.984           &  & 0.958             & 0.960           & 0.968           &  & 0.956             & 0.957           & 0.963           \\
AM-weighted-$L_2$                & 0.965             & 0.969           & 0.969           &  & 0.967             & 0.964           & 0.962           &  & 0.961             & 0.963           & 0.962           \\
AM-unweighted-$L_2$              & 0.968             & 0.972           & 0.973           &  & 0.970             & 0.967           & 0.966           &  & 0.965             & 0.966           & 0.966           \\
Adaptive Lasso                   & 0.923             & 0.874           & 0.769           &  & 0.902             & 0.873           & 0.767           &  & 0.916             & 0.883           & 0.833           \\
Lasso                            & 0.923             & 0.908           & 0.898           &  & 0.898             & 0.885           & 0.854           &  & 0.918             & 0.885           & 0.859           \\
Elastic Net                      & 0.915             & 0.893           & 0.888           &  & 0.891             & 0.875           & 0.847           &  & 0.906             & 0.876           & 0.855           \\
Ridge                            & 0.925             & 0.871           & 0.754           &  & 0.907             & 0.868           & 0.764           &  & 0.925             & 0.891           & 0.845           \\ \hline
\end{tabular}
\caption{Summary of the $95\%$ prediction interval coverage with different methods. The settings are the same as Tabel~\ref{tab:lr-1}. \label{tab:lr-2}}
\end{table}
}

\ifthenelse{1=1}{
%       \centerline{--- Fig. \ref{fig:procedure} ---}
}{

\begin{figure}[!htbp]
\centering
\resizebox {\textwidth} {!} {
\begin{tikzpicture}[x=0.75pt,y=0.75pt,yscale=-1,xscale=1]
%uncomment if require: \path (0,519); %set diagram left start at 0, and has height of 519

%Rounded Rect [id:dp34747255008486877] 
\draw   (65.9,90.78) .. controls (65.9,85.78) and (69.95,81.73) .. (74.95,81.73) -- (243.95,81.73) .. controls (248.95,81.73) and (253,85.78) .. (253,90.78) -- (253,117.95) .. controls (253,122.95) and (248.95,127) .. (243.95,127) -- (74.95,127) .. controls (69.95,127) and (65.9,122.95) .. (65.9,117.95) -- cycle ;
%Straight Lines [id:da10563954117042107] 
\draw    (24,30) -- (623,29) ;
%Straight Lines [id:da5424643971979787] 
\draw    (25,63.5) -- (273,63.5) ;
%Rounded Rect [id:dp9984023942620416] 
\draw   (381.9,90.78) .. controls (381.9,85.78) and (385.95,81.73) .. (390.95,81.73) -- (559.95,81.73) .. controls (564.95,81.73) and (569,85.78) .. (569,90.78) -- (569,117.95) .. controls (569,122.95) and (564.95,127) .. (559.95,127) -- (390.95,127) .. controls (385.95,127) and (381.9,122.95) .. (381.9,117.95) -- cycle ;
%Rounded Rect [id:dp10964234865043132] 
\draw   (67.9,162.78) .. controls (67.9,157.78) and (71.95,153.73) .. (76.95,153.73) -- (245.95,153.73) .. controls (250.95,153.73) and (255,157.78) .. (255,162.78) -- (255,189.95) .. controls (255,194.95) and (250.95,199) .. (245.95,199) -- (76.95,199) .. controls (71.95,199) and (67.9,194.95) .. (67.9,189.95) -- cycle ;
%Rounded Rect [id:dp9801851490706541] 
\draw   (383.9,162.78) .. controls (383.9,157.78) and (387.95,153.73) .. (392.95,153.73) -- (459.95,153.73) .. controls (464.95,153.73) and (469,157.78) .. (469,162.78) -- (469,189.95) .. controls (469,194.95) and (464.95,199) .. (459.95,199) -- (392.95,199) .. controls (387.95,199) and (383.9,194.95) .. (383.9,189.95) -- cycle ;
%Rounded Rect [id:dp48872599591013166] 
\draw  [pattern=_abhpn2cb5,pattern size=6pt,pattern thickness=0.75pt,pattern radius=0pt, pattern color={rgb, 255:red, 0; green, 0; blue, 0}] (67.9,231.78) .. controls (67.9,226.78) and (71.95,222.73) .. (76.95,222.73) -- (245.95,222.73) .. controls (250.95,222.73) and (255,226.78) .. (255,231.78) -- (255,258.95) .. controls (255,263.95) and (250.95,268) .. (245.95,268) -- (76.95,268) .. controls (71.95,268) and (67.9,263.95) .. (67.9,258.95) -- cycle ;
% %Rounded Rect [id:dp7850116150905193] 
% \draw  [pattern=_xryfpj67y,pattern size=6pt,pattern thickness=0.75pt,pattern radius=0pt, pattern color={rgb, 255:red, 0; green, 0; blue, 0}] (383.9,231.78) .. controls (383.9,226.78) and (387.95,222.73) .. (392.95,222.73) -- (561.95,222.73) .. controls (566.95,222.73) and (571,226.78) .. (571,231.78) -- (571,258.95) .. controls (571,263.95) and (566.95,268) .. (561.95,268) -- (392.95,268) .. controls (387.95,268) and (383.9,263.95) .. (383.9,258.95) -- cycle ;
%Rounded Rect [id:dp8095417732021417] 
\draw   (69.9,303.78) .. controls (69.9,298.78) and (73.95,294.73) .. (78.95,294.73) -- (247.95,294.73) .. controls (252.95,294.73) and (257,298.78) .. (257,303.78) -- (257,330.95) .. controls (257,335.95) and (252.95,340) .. (247.95,340) -- (78.95,340) .. controls (73.95,340) and (69.9,335.95) .. (69.9,330.95) -- cycle ;
%Rounded Rect [id:dp18800559719932952] 
\draw   (385.9,303.78) .. controls (385.9,298.78) and (389.95,294.73) .. (394.95,294.73) -- (563.95,294.73) .. controls (568.95,294.73) and (573,298.78) .. (573,303.78) -- (573,330.95) .. controls (573,335.95) and (568.95,340) .. (563.95,340) -- (394.95,340) .. controls (389.95,340) and (385.9,335.95) .. (385.9,330.95) -- cycle ;
%Rounded Rect [id:dp6660708758851814] 
\draw   (69.9,374.78) .. controls (69.9,369.78) and (73.95,365.73) .. (78.95,365.73) -- (247.95,365.73) .. controls (252.95,365.73) and (257,369.78) .. (257,374.78) -- (257,401.95) .. controls (257,406.95) and (252.95,411) .. (247.95,411) -- (78.95,411) .. controls (73.95,411) and (69.9,406.95) .. (69.9,401.95) -- cycle ;
%Rounded Rect [id:dp6389790523121001] 
\draw   (385.9,374.78) .. controls (385.9,369.78) and (389.95,365.73) .. (394.95,365.73) -- (563.95,365.73) .. controls (568.95,365.73) and (573,369.78) .. (573,374.78) -- (573,401.95) .. controls (573,406.95) and (568.95,411) .. (563.95,411) -- (394.95,411) .. controls (389.95,411) and (385.9,406.95) .. (385.9,401.95) -- cycle ;
%Straight Lines [id:da5124077289626672] 
\draw    (26.65,109.95) -- (66.5,109.5) ;
%Straight Lines [id:da5541996887694914] 
\draw    (26.65,109.95) -- (26.75,391) ;
%Straight Lines [id:da9762269567353284] 
\draw    (26.75,391) -- (66.75,391) ;
\draw [shift={(68.75,391)}, rotate = 180] [fill={rgb, 255:red, 0; green, 0; blue, 0 }  ][line width=0.08]  [draw opacity=0] (12,-3) -- (0,0) -- (12,3) -- cycle    ;
%Straight Lines [id:da20493694157127385] 
\draw    (374,62.5) -- (622,62.5) ;
%Straight Lines [id:da6823905983680469] 
\draw    (24,27) -- (623,26) ;
%Straight Lines [id:da9127041988292843] 
\draw    (25,435) -- (624,434) ;
%Straight Lines [id:da785689008827375] 
\draw    (254.75,178) -- (295,178) ;
%Straight Lines [id:da5692551149383286] 
\draw    (295,178) -- (294.75,321) ;
%Straight Lines [id:da9681794212604404] 
\draw    (259,321) -- (294.75,321) ;
\draw [shift={(257,321)}, rotate = 0] [fill={rgb, 255:red, 0; green, 0; blue, 0 }  ][line width=0.08]  [draw opacity=0] (12,-3) -- (0,0) -- (12,3) -- cycle    ;
%Straight Lines [id:da13460265567384677] 
\draw    (164,366) -- (164,342) ;
\draw [shift={(164,340)}, rotate = 90] [fill={rgb, 255:red, 0; green, 0; blue, 0 }  ][line width=0.08]  [draw opacity=0] (12,-3) -- (0,0) -- (12,3) -- cycle    ;
%Straight Lines [id:da13773219999214814] 
\draw    (480,366) -- (480,342) ;
\draw [shift={(480,340)}, rotate = 90] [fill={rgb, 255:red, 0; green, 0; blue, 0 }  ][line width=0.08]  [draw opacity=0] (12,-3) -- (0,0) -- (12,3) -- cycle    ;
%Straight Lines [id:da8401313922639767] 
\draw    (571, 170) -- (626,170) ;
%Straight Lines [id:da9753705553508778] 
\draw    (626,170) -- (626,390.51) ;
%Straight Lines [id:da17874920929786187] 
\draw    (575,390.98) -- (626,390.51) ;
\draw [shift={(573,391)}, rotate = 360] [fill={rgb, 255:red, 0; green, 0; blue, 0 }  ][line width=0.08]  [draw opacity=0] (12,-3) -- (0,0) -- (12,3) -- cycle    ;
%Straight Lines [id:da33991457463602115] 
\draw    (571,182) -- (612.5,181.5) ;
%Straight Lines [id:da39420943178845913] 
\draw    (612.5,181.5) -- (612,319) ;
%Straight Lines [id:da9336717568335511] 
\draw    (574,319.95) -- (612,319) ;
\draw [shift={(572,320)}, rotate = 360] [fill={rgb, 255:red, 0; green, 0; blue, 0 }  ][line width=0.08]  [draw opacity=0] (12,-3) -- (0,0) -- (12,3) -- cycle    ;
%Straight Lines [id:da8507922734547805] 
\draw  [dash pattern={on 4.5pt off 4.5pt}]  (257,172) -- (316,172) ;
\draw [shift={(255,172)}, rotate = 0] [fill={rgb, 255:red, 0; green, 0; blue, 0 }  ][line width=0.08]  [draw opacity=0] (12,-3) -- (0,0) -- (12,3) -- cycle    ;
%Straight Lines [id:da9873980596024335] 
\draw  [dash pattern={on 4.5pt off 4.5pt}]  (316,172) -- (316,320) ;
%Straight Lines [id:da21017490258371652] 
\draw  [dash pattern={on 4.5pt off 4.5pt}]  (316,320) -- (325,320) ;
%Straight Lines [id:da12904547152465218] 
\draw    (253,110) -- (378,110) ;
\draw [shift={(380,110)}, rotate = 180] [fill={rgb, 255:red, 0; green, 0; blue, 0 }  ][line width=0.08]  [draw opacity=0] (12,-3) -- (0,0) -- (12,3) -- cycle    ;

%Straight Lines [id:da16731506969435772]
\draw    (479,127) -- (479,136.5) ;
%Straight Lines [id:da04700526148652573]
\draw    (424.25,136.5) -- (530.25,136.5) ;
%Straight Lines [id:da6946749154051068]
\draw    (424.25,136.5) -- (424.25,151.5) ;
\draw [shift={(424.25,154)}, rotate = 270] [fill={rgb, 255:red, 0; green, 0; blue, 0 }  ][line width=0.08]  [draw opacity=0] (12,-3) -- (0,0) -- (12,3) -- cycle    ;
%Straight Lines [id:da45259382607753973]
\draw    (530,136.5) -- (530,151.5) ;
\draw [shift={(530,154)}, rotate = 270] [fill={rgb, 255:red, 0; green, 0; blue, 0 }  ][line width=0.08]  [draw opacity=0] (12,-3) -- (0,0) -- (12,3) -- cycle    ;
%Rounded Rect [id:dp20946591042572504]
\draw   (491.5,162.78) .. controls (491.5,157.78) and (495.55,153.73) .. (500.55,153.73) -- (560.95,153.73) .. controls (565.95,153.73) and (570,157.78) .. (570,162.78) -- (570,188.95) .. controls (570,193.95) and (565.95,199) .. (560.95,199) -- (500.55,199) .. controls (495.55,198) and (491.5,193.95) .. (491.5,188.95) -- cycle ;
%Straight Lines [id:da9445123693112751]
\draw  [dash pattern={on 4.5pt off 4.5pt}]  (316,172) -- (383.5,173) ;

% Text Node
\draw (77.1,40.81) node [anchor=north west][inner sep=0.75pt]   [align=left] {Estimation};
% Text Node
\draw (405.13,40.81) node [anchor=north west][inner sep=0.75pt]   [align=left] {Resampling Imputation};
% Text Node
\draw (99.38,95.86) node [anchor=north west][inner sep=0.75pt]   [align=left] {True $\mathbb{P}$};
% Text Node
\draw (410.82,95.86) node [anchor=north west][inner sep=0.75pt]   [align=left] {Empirical $\hat{\mathbb{P}}$};
% Text Node
\draw (99.38,167.86) node [anchor=north west][inner sep=0.75pt]   [align=left] {Imputed $\mathbb{Q}$};
% Text Node
\draw (385.82,167.86) node [anchor=north west][inner sep=0.75pt]   [align=left] {$(X^{(k)}, Y^{(k)})$};
% Text Node
\draw (99.38,236.86) node [anchor=north west][inner sep=0.75pt]  [fill={rgb, 255:red, 255; green, 255; blue, 255 }  ][align=left] {Optimal $\mathbb{P}_{\theta^*}$};
% % Text Node
% \draw (410.82,236.86) node [anchor=north west][inner sep=0.75pt]  [fill={rgb, 255:red, 255; green, 255; blue, 255 }  ][align=left]  [align=left] {Optimal $\hat{\mathbb{P}}_{\theta^*}$(ERM)};
% Text Node
\draw (99.38,308.86) node [anchor=north west][inner sep=0.75pt]   [align=left] {Estimated $\mathbb{P}_{\hat{\theta}_{AM}}$};
% Text Node
\draw (410.82,308.86) node [anchor=north west][inner sep=0.75pt]   [align=left] {Estimated $\mathbb{Q}^{(k)}_{\hat{\theta}_b}$};
% Text Node
\draw (99.38,378.86) node [anchor=north west][inner sep=0.75pt]   [align=left] {Empirical $\hat{\mathbb{P}}$};
% Text Node
\draw (410.82,378.86) node [anchor=north west][inner sep=0.75pt]   [align=left] {Resampled $\tilde{\mathbb{P}}^{(k)}_b$};
% Text Node
\draw (288.5,190.5) node [anchor=north west][inner sep=0.75pt]  [font=\footnotesize,rotate=-90] [align=left] {[future observations]};
% Text Node
\draw (605.5,190.5) node [anchor=north west][inner sep=0.75pt]  [font=\footnotesize,rotate=-90] [align=left] {[future observations]};
% Text Node
\draw (649.5,190.5) node [anchor=north west][inner sep=0.75pt]  [font=\footnotesize,rotate=-90] [align=left] {[sample with replacement]};
% % Text Node
% \draw (459,130) node [anchor=north west][inner sep=0.75pt]  [font=\small] [align=left] {$\displaystyle S$};
% Text Node
\draw (171,345) node [anchor=north west][inner sep=0.75pt]  [font=\footnotesize] [align=left] {[current observations]};
% Text Node
\draw (488,345) node [anchor=north west][inner sep=0.75pt]  [font=\footnotesize] [align=left] {[current observations]};
% Text Node
\draw (323,285) node [anchor=north west][inner sep=0.75pt]  [font=\large] [align=left] {$\displaystyle \sum _{b=1}^{B}\sum _{k=1}^{K}$};
% Text Node
\draw (336,190.5) node [anchor=north west][inner sep=0.75pt]  [font=\footnotesize,rotate=-90] [align=left] {[impute]};
% Text Node
\draw (45.5,190.5) node [anchor=north west][inner sep=0.75pt]  [font=\footnotesize,rotate=-90] [align=left] {[observe]};
% Text Node
\draw (291,91) node [anchor=north west][inner sep=0.75pt]  [font=\footnotesize] [align=left] {[observe]};
\draw (434,137.5) node [anchor=north west][inner sep=0.75pt]  [font=\footnotesize] [align=left] {[data-splitting]};
% Text Node
\draw (520.32,167.98) node [anchor=north west][inner sep=0.75pt]  [align=left] {$\hat{\mathbb{P}}^{(k)}_b$};

\end{tikzpicture}
}
\caption{The graphical illustration of the proposed imputation-estimation scheme described in Algorithm~\ref{alg:imputation} and \ref{alg:estimation}.
% The left column
% is the list of true $(\mathbb{P})$, imputation $(\mathbb{Q})$, data or empirical $(\hat{\mathbb{P}})$,
% and estimated $(\mathbb{P}_{\hat{\theta}_{AM}}).$
% The right column corresponds to that of the left column,
% but for the bootstrap method of multiple imputations of
% the missing population.
The symbol $\sum_{b=1}^{B} \sum_{k = 1}^K$ between the two columns
stands for a mixture of $B \cdot K$ imputation models.
$\mathbb{P}_{\theta^*}$ denotes the optimal distribution obtained with the optimal estimate by (\ref{eq:obj-01}).
 }
\label{fig:procedure}
\end{figure}
}

\ifthenelse{1=1}{
%       \centerline{--- Fig. \ref{fig:procedure} ---}
}{

\begin{figure}[!htbp]
\centering
\begin{tikzpicture}[x=0.75pt,y=0.75pt,yscale=-1,xscale=1]
%uncomment if require: \path (0,198); %set diagram left start at 0, and has height of 198

%Curve Lines [id:da5717385208119037] 
\draw    (420.48,68.71) .. controls (294.75,70.69) and (312.11,60.91) .. (310.53,105.35) ;
\draw [shift={(310.48,106.71)}, rotate = 272.49] [color={rgb, 255:red, 0; green, 0; blue, 0 }  ][line width=0.75]    (10.93,-3.29) .. controls (6.95,-1.4) and (3.31,-0.3) .. (0,0) .. controls (3.31,0.3) and (6.95,1.4) .. (10.93,3.29)   ;
%Straight Lines [id:da4079007657156444] 
\draw    (183,4) -- (604,4) ;
%Straight Lines [id:da6961815335336228] 
\draw    (183,6) -- (604,6) ;
%Straight Lines [id:da04895947447978766] 
\draw    (183,38) -- (604,38) ;
%Straight Lines [id:da8459184645874684] 
\draw    (182,168) -- (603,168) ;
%Shape: Rectangle [id:dp05925126000308545] 
\draw   (420,44.71) -- (602.48,44.71) -- (602.48,160.71) -- (420,160.71) -- cycle ;

% Text Node
\draw (198,106) node [anchor=north west][inner sep=0.75pt]   [align=left] {
$
\hat{\theta }_{AM} \ =\ \mathfrak{A}_{\theta}(\hat{\mathbb{P}} ,\ \mathbb{Q})
$};
% Text Node
\draw (430,47) node [anchor=north west][inner sep=0.75pt]   [align=left] {
$
\hat{\theta }_{b}^{( 1)} \ =\ \mathfrak{A}_{\theta }\left(\tilde{\mathbb{P}}_{b}^{( 1)} ,\hat{\mathbb{P}}_{b}^{( 1)}\right)
$};
% Text Node
\draw (503,95) node [anchor=north west][inner sep=0.75pt]   [align=left] {...};
% Text Node
\draw (197,16) node [anchor=north west][inner sep=0.75pt]   [align=left] { Estimation};
% Text Node
\draw (428,16) node [anchor=north west][inner sep=0.75pt]   [align=left] {Resampling Imputation};
% Text Node
\draw (431,116) node [anchor=north west][inner sep=0.75pt]   [align=left] {
$
\hat{\theta }_{b}^{( K)} \ =\ \mathfrak{A}_{\theta }\left(\tilde{\mathbb{P}}_{b}^{( K)} ,\hat{\mathbb{P}}_{b}^{( K)}\right)
$};
% Text Node
\draw (383,78) node [anchor=north west][inner sep=0.75pt]   [align=left] {$\displaystyle \sum _{b\ =\ 1}^{B}$};

\end{tikzpicture}
\caption{The simplified AM imputation-estimation framework using AM operator $\mathfrak{A}_\theta$. The involved distributions are described in Algorithm~\ref{alg:imputation} and \ref{alg:estimation}. \label{fig:notation}}
\end{figure}
}

\ifthenelse{1=1}{
%       \centerline{--- Fig. \ref{fig:procedure} ---}
}{

\begin{figure}[!htbp]
\centering
\includegraphics[width=14cm]{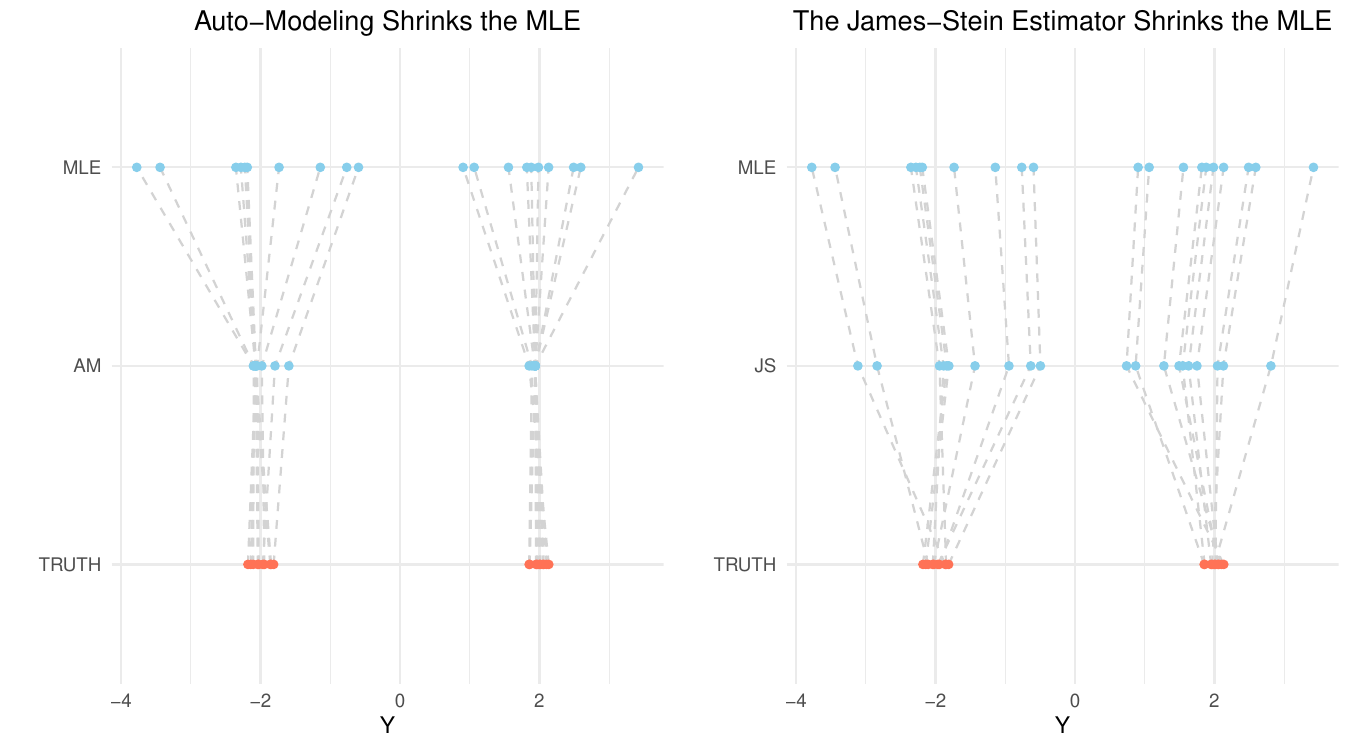}
\caption{Shrinkage pattern of the James-Stein estimator and AM in the bimodal simulation example with a typical data set.}\label{fig:bimodal}
\end{figure}
%\ifthenelse{1=0}{}{
% \begin{figure}[!htb]
% \centering
% \begin{subfigure}{}
%   \centering
%   \includegraphics[width=9cm]{plot/js_bimodal.png}
% \end{subfigure}%
% \begin{subfigure}{}
%   \centering
%   \includegraphics[width=9cm]{plot/am_bimodal.png}
% \end{subfigure}
% \caption{Shrinkage pattern of the James-Stein estimator and AM in the bimodal simulation study example.}\label{fig:bimodal}
% \end{figure}
%}
}

\ifthenelse{1=1}{

}{
\begin{table}[!htbp]
\centering
\footnotesize
\begin{tabular}{ccccl}
\hline
\textit{Method}       & \begin{tabular}[c]{@{}c@{}}FC 400-400\\ (478,410)\end{tabular} & \begin{tabular}[c]{@{}c@{}}FC 800-800\\ (127,6810)\end{tabular} & \begin{tabular}[c]{@{}c@{}}FC 1600-1600\\ (383,3610)\end{tabular} & \multicolumn{1}{c}{\begin{tabular}[c]{@{}c@{}}CNN\\ (299,306)\end{tabular}} \\ \hline
Early-stopping        & 1.514 (0.049)                                                  & 1.494 (0.037)                                                   & 1.466 (0.053)                                                     & 0.630 (0.047)                                                               \\
Drop-out              & \multicolumn{1}{l}{1.512 (0.141)}                              & \multicolumn{1}{l}{1.478 (0.087)}                               & \multicolumn{1}{l}{1.494 (0.068)}                                 & 0.592 (0.050)                                                               \\
$L_1-$ regularization & 1.476 (0.046)                                                  & 1.442 (0.052)                                                   & 1.386 (0.044)                                                     & 0.668 (0.068)                                                               \\
AM-weighted-$L_1$     & 1.222 (0.040)                                                  & 1.244 (0.042)                                                   & 1.202 (0.052)                                                     & \textbf{0.584 (0.043)}                                                      \\
AM-unweighted-$L_1$   & \multicolumn{1}{l}{\textbf{1.180 (0.040)}}                     & \multicolumn{1}{l}{\textbf{1.172 (0.064)}}                      & \multicolumn{1}{l}{\textbf{1.093 (0.059)}}                        & 0.586 (0.038)                                                               \\
$L_2-$ regularization & 1.408 (0.018)                                                  & 1.386 (0.066)                                                   & 1.446 (0.065)                                                     & 0.602 (0.039)                                                               \\
AM-weighted-$L_2$     & \multicolumn{1}{l}{\textbf{1.178 (0.040)}}                     & \multicolumn{1}{l}{\textbf{1.184 (0.040)}}                      & \multicolumn{1}{l}{\textbf{1.129 (0.046)}}                        & \textbf{0.569 (0.040)}                                                      \\
AM-unweighted-$L_2$   & 1.246 (0.046)                                                  & 1.204 (0.025)                                                   & 1.146 (0.020)                                                     & 0.608 (0.016)                                                               \\ \hline
\end{tabular}
\caption{The test error ($\%$) of different methods on the MNIST data set with different structures. FC X-X in the header of the table denotes the fully-connected network with two hidden layers where the number of hidden nodes is X. The values in the parentheses denote the number of parameters in the corresponding structure. The reported value is the mean taken over 5 repetitions (with standard deviations given in parentheses). The best two results in each structure are highlighted in boldface.
\label{tab:nn}}
\end{table}

}

\ifthenelse{1=1}{
%       \centerline{--- Fig. \ref{fig:procedure} ---}
}{

\begin{figure}[!htbp]
\centering
\includegraphics[width=14cm]{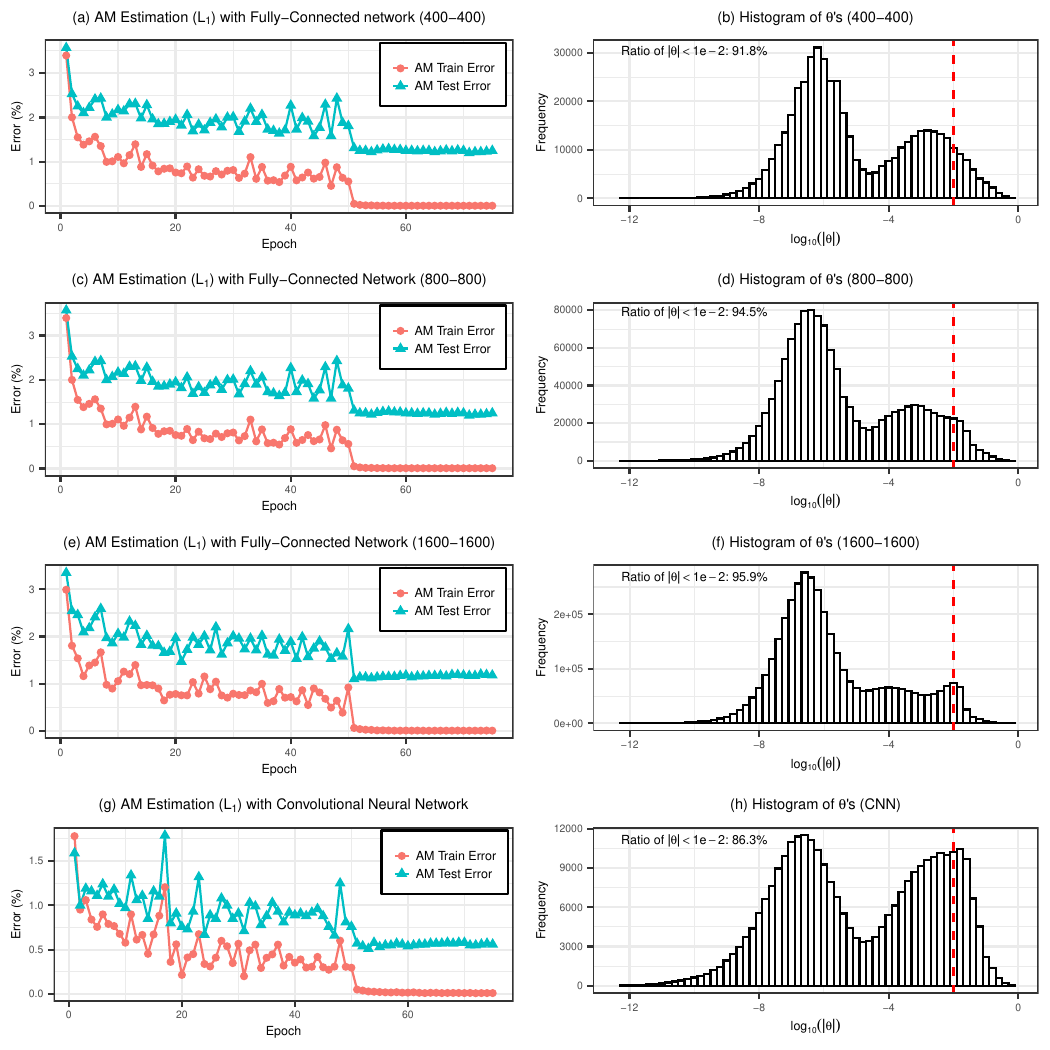}
\caption{The learning curves in the estimation process and the histograms of the estimated parameters using AM (weighted $L_1$ %simplicity-preference 
	\dual~function) with different model structures.}\label{fig:convergence}
\end{figure}
}

\ifthenelse{1=1}{
%       \centerline{--- Fig. \ref{fig:procedure} ---}
}{

\begin{figure}[!htbp]
\centering
\includegraphics[width=14cm]{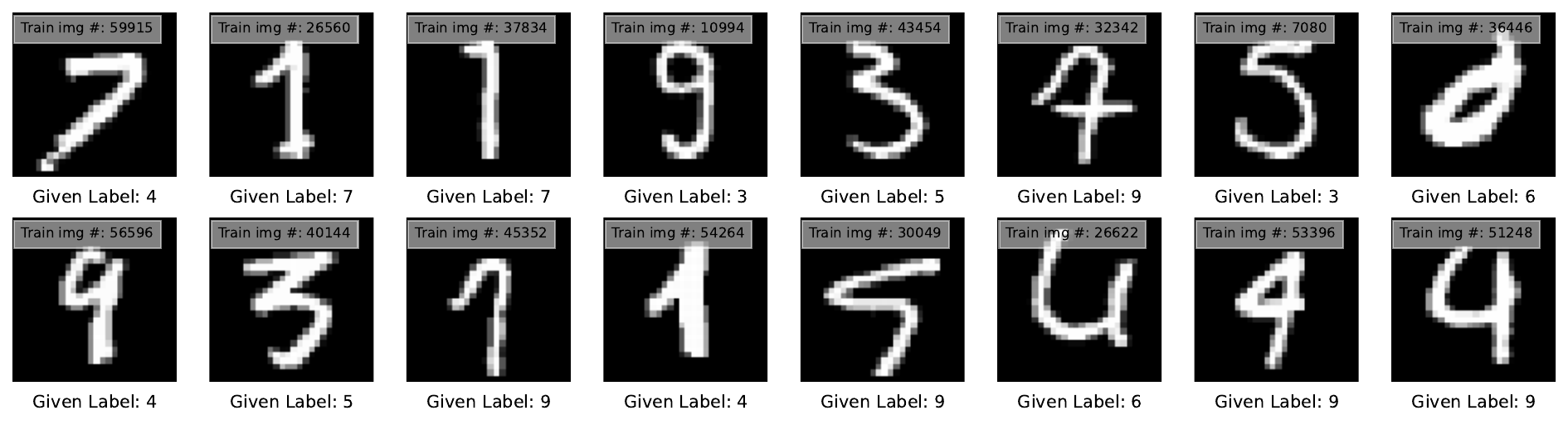}
\caption{Selection of images detected by AM for questionable labels, displayed in a left-to-right, top-to-bottom sequence, sorted by the decreasing frequency of their identification.}\label{fig:questionable-labels-1}
\end{figure}
}

%%% END OF FIGURES %%%

%\clearpage

% \appendix
% \renewcommand{\thesection}{A}
% \renewcommand{\theequation}{A\arabic{equation}} 
% \setcounter{equation}{0}

%% file: appendices.tex
%%\section*{Appendices}
\section*{}
%\subsubsection*{Proof for Theorem~\ref{theorem:1}}
% \CLdel{\Large Should this be created using a separated latex file? The references will be different if we do. The references to subections here will also be changed.}

\subsection{A Simple Example for Continuous Data with $p = 1$}\label{s:appendices-simple}
\subsubsection{Settings}
\label{s:example-01-B}

Suppose that $y_1 ,..., y_n$ is a sample of size $n$ from the model:
$y = \theta + z$ where $z \sim N(0,1)$. Take the AM \dual~function as
$\pi(\theta, \lambda) = \lambda|\theta|$ with $\lambda \in \mathbb{R}_+$, {\it i.e.}, $\lambda$ is a positive scalar.

\subsubsection{Imputation with Resampling}\label{s:appendices-1}
We demonstrate Algorithm~2 with an arbitrary $k = 1,\dots,K$ and $b = 1, \dots, B$. Let $\tilde{y}_1,...,\tilde{y}_{m}$, denoted by $\tilde{\mathbb{P}}^{(k)}_b$, be a re-sampled data drawn from
the sample data $y_1,...,y_{n_k}$ denoted by $\hat{\mathbb{P}}^{(k)}_b$ when leaving out the $k$-th fold, where $m$ denotes the sample size of $\tilde{\mathbb{P}}^{(k)}_b$ and $n_k$ denotes the sample size of $\hat{\mathbb{P}}^{(k)}_b$. 
In this case, we have
\[
G_{\tilde{\mathbb{P}}^{(k)}_b}(\theta, \lambda) = 
\frac{1}{2{m}}\sum_{i=1}^{m}(\theta-\tilde{y}_i)^2
+\lambda|\theta|
\]
and 
\[
V_{\hat{\mathbb{P}}^{(k)}_b,\tilde{\mathbb{P}}^{(k)}_b}(\theta, \lambda) = 
\frac{1}{2{n_k}}\sum_{i=1}^{n_k}(\theta-y_i)^2
-\frac{1}{2{m}}\sum_{i=1}^{m}(\theta-\tilde{y}_i)^2
-\lambda|\theta|.
\]
For illustrative purposes in the subsequent analysis, the notation $\mbox{sign}(\theta)$ is used as a substitute for the sub-gradient of $\theta$. The derivative of $V_{\hat{\mathbb{P}}^{(k)}_b,\tilde{\mathbb{P}}^{(k)}_b}(\theta, \lambda)$ is given by
\begin{align*}
\frac{\partial V_{\hat{\mathbb{P}}^{(k)}_b,\tilde{\mathbb{P}}^{(k)}_b}(\theta, \lambda)}{\partial\theta}
&= \theta - \bar{y}^{(k)} - \theta + \bar{\tilde{y}}^{(k)} - \lambda \cdot \mbox{sign}(\theta)\\
&= \bar{\tilde{y}}^{(k)} - \bar{y}^{(k)} - \lambda \cdot \mbox{sign}(\theta),
\end{align*}
where $\bar{\tilde{y}}^{(k)} = \frac{1}{m}\sum_{i = 1}^m \tilde{y}_i, \  \bar{y}^{(k)} = \frac{1}{n_k}\sum_{i = 1}^{n_k}y_i$. It is required to find
\[
\underset{\lambda}{\text{arg min}}\displaystyle\left\lvert\frac{\partial V_{\hat{\mathbb{P}}^{(k)}_b,\tilde{\mathbb{P}}^{(k)}_b}({\theta}, \lambda)}{\partial {\theta}} \displaystyle\right\lvert
= \underset{\lambda}{\text{arg min}}\displaystyle\left\lvert \bar{\tilde{y}}^{(k)} - \bar{y}^{(k)} - \lambda \cdot \mbox{sign}({\theta})  \displaystyle\right\lvert.\\
\]
With the constraint on $\lambda$ $\in \mathbb{R}_+$, it follows that
\[
\hat{\lambda}^{(k)}_b = 
\left\{
\begin{array}{lcl}
(\bar{\tilde{y}}^{(k)}-\bar{y}^{(k)}) \cdot \mbox{sign}({\theta})
&& \mbox{if
$(\bar{\tilde{y}}^{(k)}-\bar{y}^{(k)}) \cdot \mbox{sign}({\theta}) >0$,
}\\
0 && \mbox{otherwise}.
\end{array}
\right.
\]
Hence, in order to minimize $G_{\tilde{\mathbb{P}}^{(k)}_b}(\theta, \lambda) $ over 
$\theta$, we have
\[
\hat{\theta}^{(k)}_b
=
\left\{
\begin{array}{lcl}
\underset{\theta}{\text{arg min}} \ 
\frac{1}{2{m}}\sum_{i=1}^{m}(\theta-\tilde{y}_i)^2
 && \mbox{if
$(\bar{\tilde{y}}^{(k)}-\bar{y}^{(k)}) \cdot \mbox{sign}(\hat{\theta}^{(k)}_b) \leq 0$,
}\\
%\\
\underset{\theta}{\text{arg min}}\left(
\frac{1}{2{m}}\sum_{i=1}^{m}(\theta-\tilde{y}_i)^2 + 
|\bar{\tilde{y}}^{(k)} -\bar{y}^{(k)}| \cdot |\theta|
\right)&& \mbox{otherwise}.
\end{array}
\right.
\]
% This corresponds to solving for a global minimum solution of a \textit{segmented convex function}\footnote{I am not sure the word to be used here} where 0 is the knot. \\
One can further write the problem as
\[
\min_{\theta} \frac{1}{2{m}}\sum_{i=1}^{m}(\theta-\tilde{y}_i)^2
%\]
%\[
\qquad
s.t. \ \ |\theta| \cdot \mathbbm{1}\left((\bar{\tilde{y}}^{(k)}-\bar{y}^{(k)}) \cdot \mbox{sign}(\theta) > 0 \right)  \leq | \bar{\tilde{y}}^{(k)}-\bar{y}^{(k)} |, 
\]
where $\mathbbm{1}(\cdot)$ denotes the indicator function. This can be seen as the canonical convex optimization problem with the inequality constraint known as the halved Lasso constraint. 
%\CLdel{The solution depends on the three cases that are depicted in Figure~\ref{fig:appendices-1}.}

% \begin{figure}[!h]
% \centering
% \begin{subfigure}{}
%   \centering
%   \includegraphics[width=5cm]{plot/case1.png}
%   % \caption{A subfigure}
%   \label{fig:sub1}
% \end{subfigure}%
% \begin{subfigure}{}
%   \centering
%   \includegraphics[width=5cm]{plot/case2.png}
%   % \caption{A subfigure}
%   \label{fig:sub2}
% \end{subfigure}
% \begin{subfigure}{}
%   \centering
%   \includegraphics[width=5cm]{plot/case3.png}
%   % \caption{A subfigure}
%   \label{fig:sub3}
% \end{subfigure}
% \caption{Illustration of the global optimal solution in 3 different cases. The parameter $\theta$ is extended to the second dimension for a better visualization. {\color{red} --- HOW IT WAS EXTENDED OR HOW TO INTERPRET $\theta_1$ AND $\theta_2?$}}
% \label{fig:appendices-1}
% \end{figure}

Consider the following strategy to solve for $\theta$. Firstly, one can solve for the non-constrained optimization problem to get $\theta^{*}_{mle}$. Then in the case of $(\bar{\tilde{y}}^{(k)}-\bar{y}^{(k)})\cdot \mbox{sign}(\theta^{*}_{mle}) \leq 0$, we have $\hat{\theta}^{(k)}_b= \theta^{*}_{mle}$. Otherwise, we solve the constrained optimization problem (lasso) to get $\hat{\theta}^{(k)}_b = \theta^{*}_{lasso}$.
% \begin{remark}
% One can see that in this strategy $\theta^{*} = 0$ can be solved by lasso but does not lead to a positive penalty term. However, one can show that $\theta = 0$ is still the global optimal by the KKT condition (and of course with the fact that $\theta^{*}_{mle}$ is not achievable in this case).
% \end{remark}\\
% First, we initialize $\tilde{\theta}^{(0)}$ s.t. $(\bar{\tilde{y}}-\bar{y})\mbox{sign}(\tilde{\theta}^{(0)}) > 0$ to keep our penalty term $\tilde{\tau} = |\bar{\tilde{y}}-\bar{y}|$. Then we repeatedly update $\tilde{\theta}^{(t)}$ based on $\tilde{\theta}^{(t-1)}$ for $t = 1,2...$ until convergence. After each update, check if $(\bar{\tilde{y}}-\bar{y})\mbox{sign}(\tilde{\theta}^{(t)}) > 0$. This is to see whether we can still keep the penalty term. If it is not satisfied, we simply remove the penalty term.\begin{remark}
% This can be proved by taking $\tilde{\tau} = \left((\bar{\tilde{y}}-\bar{y})\mbox{sign}(\theta)\right)_+$ and setting first-order derivative of $\tilde{H}_\tilde{\lambda}(\theta)$ to 0 in each iteration. The convexity is naturally preserved and each update is a lasso shrinkage. One can prove that the result "converges" in at most 3 steps.
% \end{remark} 
In this simple example, one can show that the closed form solution of $\hat{\theta}^{(k)}_b$ is given by
\[
\hat{\theta}^{(k)}_b
=
\left\{
\begin{array}{lcl}
\bar{\tilde{y}}^{(k)} && \mbox{if} \ |\bar{\tilde{y}}^{(k)}| \leq |\bar{y}^{(k)}| \ \ \mbox{and} \ \ \mbox{sign}(\bar{\tilde{y}}^{(k)}) = \mbox{sign}(\bar{y}^{(k)}),\\
\bar{y}^{(k)} && \mbox{if} \ |\bar{\tilde{y}}^{(k)}| > |\bar{y}^{(k)}| \ \ \mbox{and} \ \ \mbox{sign}(\bar{\tilde{y}}^{(k)}) = \mbox{sign}(\bar{y}^{(k)}),\\
0 && \mbox{if} \ \mbox{sign}(\bar{\tilde{y}}^{(k)}) \neq \mbox{sign}(\bar{y}^{(k)}).\\
\end{array}
\right.
\]
It should be noted that $|\hat{\theta}^{(k)}_b| \le |\bar{\tilde{y}}^{(k)} |$ in all cases, indicating shrinkage of the estimate toward zero.

\subsubsection{Estimation of $\theta$}
In the estimation step, we have
\[
G_{\hat{\mathbb{P}}}(\theta, \lambda) = 
\frac{1}{2n}\sum_{i=1}^n(\theta-y_i)^2
+\lambda|\theta|
\]
and 
\[
V_{{\mathbb{Q}},\hat{\mathbb{P}}}(\theta, \lambda) = 
\frac{1}{2n}E\left[\sum_{i=1}^n(\theta-\mathbf{y_i})^2\right]
-\frac{1}{2n}\sum_{i=1}^n(\theta-y_i)^2
-\lambda|\theta|.
\]
Let $\bar{\mathbf{y}} = E[\mathbf{y}]$. Notice that the imputation data generator is $\mathbf{y} \sim \hat{\theta}^{(k)}_b + \mathbf{z}$, where $\mathbf{z} \sim N(0,1)$. Thus, we have 
\(
    \bar{\mathbf{y}} = E\left[\mathbf{y}\right] = E\left[\hat{\theta}^{(k)}_b\right],
\)
which concludes the derivation.

To obtain an approximate analytical solution, we assume that the sampling distribution can be approximated by $\bar{\tilde{y}}^{(k)} \sim N(\bar{y} , 1/n)$.  In such a case, $E\left[\hat{\theta}^{(k)}_b\right]$ takes an explicit form as
\[
E\left[\hat{\theta}^{(k)}_b\right] =  \left( 1- \Phi\left(-\sqrt{n} | \bar{y}|\right) \right) \bar{y} - \frac{1}{\sqrt{n}}\left\lvert \phi \left(-\sqrt{n} |\bar{y}|\right) - \phi(0)\right\rvert \cdot \text{sign}(\bar{y}),
\]
where $\Phi(\cdot)$ and $\phi(\cdot)$ denote the cumulative distribution function (CDF) and the probability distribution function (PDF) of the standard normal distribution, respectively. With arguments similar to those in Section~\ref{s:appendices-1}, one has the closed-form expression of the final estimate of $\theta$ as
\(
    \hat{\theta}_{AM} = E\left[\hat{\theta}^{(k)}_b\right].
\)

% \begin{remark}
% This can be shown with the fact $|\bar{\mathbf{y}}|= |E[\tilde{\theta}]| \le |\bar{y}|$ and $\text{sign}(E[\tilde{\theta}]) = \text{sign}(\bar{y})$
% \end{remark}

% \pagebreak
\subsubsection{The Shrinkage Effect and Sampling Distribution of $\hat{\theta}$}

\begin{figure}[!htbp]
    \centering
    \includegraphics[width=0.60\linewidth]{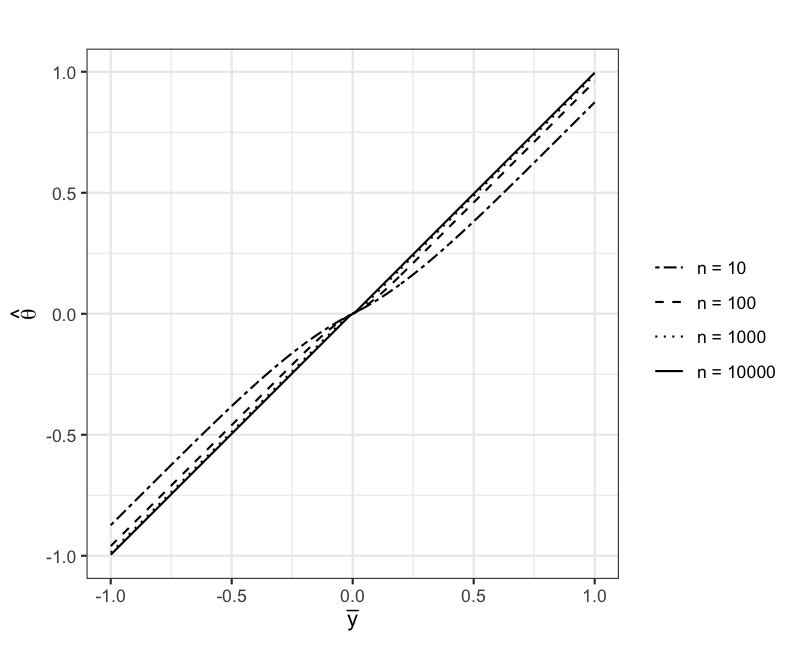}
    \caption{The shrinkage effect of the AM estimator in the simple example in Section~\ref{s:appendices-simple}.}
    \label{fig:simple-1}%
\end{figure}

\begin{figure}[!htbp]
\centering
\includegraphics[width=0.71\linewidth]{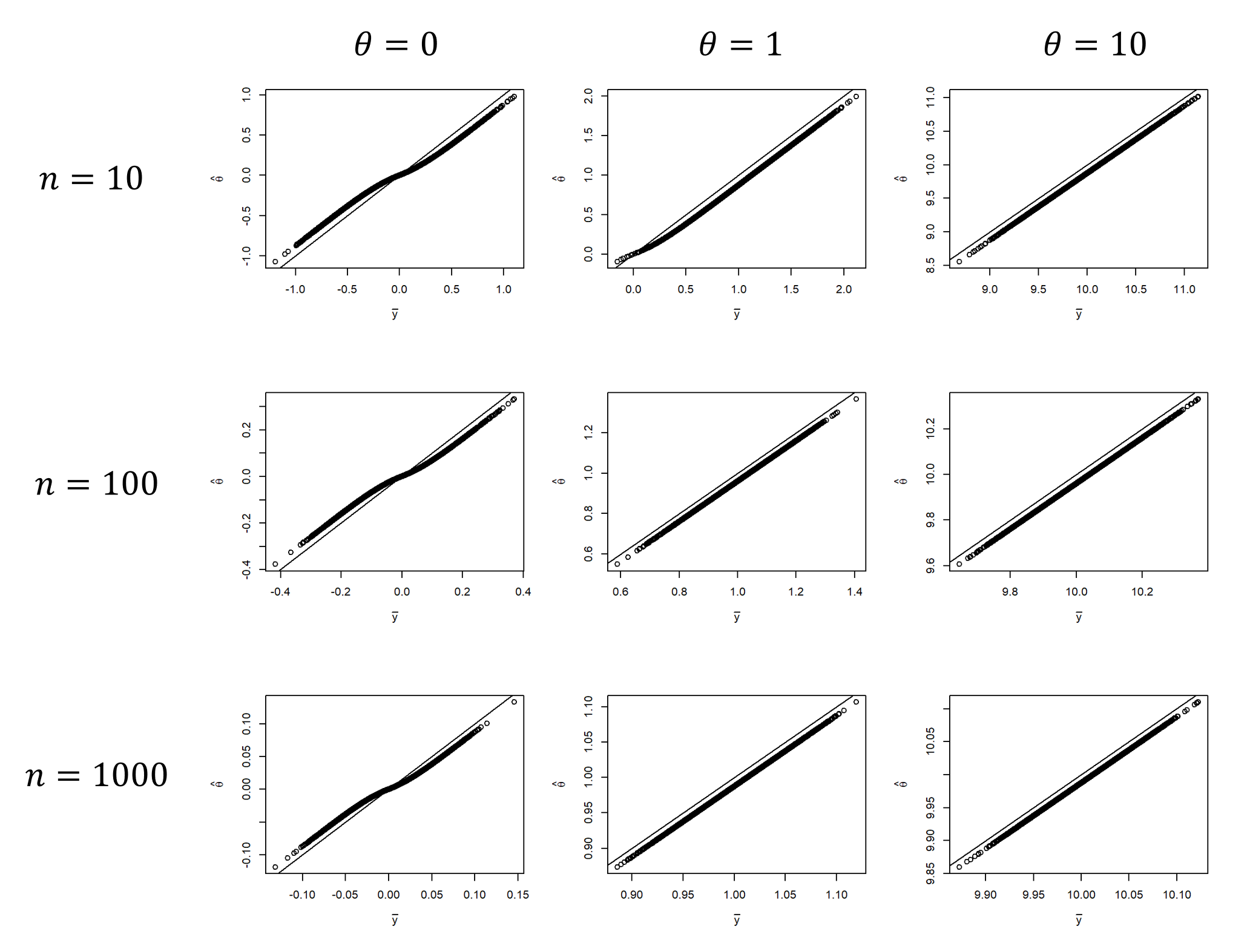}
\caption{Q-Q plots of the AM estimate versus MLEs %MLEs versus the AM estimator 
under different settings in the simple example in \ref{s:appendices-simple}. The plots show that the AM estimate are close to the MLEs but slightly shrunk towards zero.
\label{fig:simple-2}}
\end{figure}

The shrinkage effect of AM in this simple example is shown in Figure~\ref{fig:simple-1} for different sample size $n$. It can be seen that the AM estimator effectively shrinks the MLE of $\theta$ towards zero. We simulated $B=10,000$ data sets for different choices of $\theta$ and $n$ to study the sampling distribution of $\hat{\theta}_{AM}$. The pattern of $\hat{\theta}_{AM}$'s sampling distribution is shown in the Quantile-Quantile plots (Q-Q plots) in Figure~\ref{fig:simple-2}, where it can be seen that the AM estimate are close to the MLEs but slightly shrunk towards zero.

\subsubsection{Additional Notes on the Use of Lambda}

Suppose that shrinkage is not required, and $\lambda$ can take negative values. In this case, we provide the following theorem, which shows that the use of the duality function is equivalent to fitting models directly to future observations. 

\begin{theorem}[Fitting Simple Model to Future Observations] \label{theorem:S.1}
Suppose that $y_1 ,..., y_n$ is a sample of size $n$ from the model:
$y = \theta + z$ where $z \sim N(0,1)$. Take the AM \dual~function as
$\pi(\theta, \lambda) = \lambda|\theta|$ with $\lambda \in \mathbb{R}$, and let $\tilde{y}_1, \dots, \tilde{y}_m$ represent future observations. Then, the AM estimator is $\hat{\theta}_{AM} = \bar{\tilde{y}}$, where $\bar{\tilde{y}} = \frac{1}{m}\sum_{i=1}^m\tilde{y}_i.$
\end{theorem}

\begin{proof}
%\CLadd{Before I read your proof, I want to see that there exist a $\hat\lambda$ such that bothEquations (14) and (15) hold for $\lambda=\hat\lambda$ and $\theta=\hat\theta$, where $\hat\theta= \bar{\tilde{y}}$. It seems that you have shown that Equation (14) holds but not Equation (15). The logic has to be clear, even if the result holds. Please check the same logic for the theorem on LR.}
We first show that for any $\theta \in \mathbb{R}$, there exists some $\lambda = \hat{\lambda} \in \mathbb{R}$ such that $\frac{\partial G_{\hat{\mathbb{P}}}(\theta, \lambda)}{\partial \theta} = 0$ at $\lambda = \hat{\lambda}$. Using the sub-gradient notation and setting the derivative of $G_{\hat{\mathbb{P}}}(\theta, \lambda)$ with respect to $\theta$ to zero, we have
\[
\frac{\partial G_{\hat{\mathbb{P}}}(\theta, \lambda)}{\partial \theta} = \theta - \bar{y} + \lambda \cdot s(\theta) = 0
\]
where $s(\theta)  = \mbox{sign}(\theta)$ if  $\theta \neq 0$ and $s(\theta) \in [-1,1]$ otherwise.
Thus, we can choose $\hat{\lambda}$ such that:
\begin{equation*}
\hat{\lambda} =
\left\{
\begin{array}{lcl}
\bar{y} - \theta && \mbox{if 
%$ \bar{y} \le \bar{\tilde{y}} \le 0$ or $ \bar{y} \ge \bar{\tilde{y}} \ge 0$, {\it i.e.},
$\theta > 0$}\\
\theta - \bar{y} && \mbox{if 
%$ \bar{y} \le \bar{\tilde{y}} \le 0$ or $ \bar{y} \ge \bar{\tilde{y}} \ge 0$, {\it i.e.},
 $\theta < 0$}\\
\infty && \mbox{if 
%$ \bar{y} \le \bar{\tilde{y}} \le 0$ or $ \bar{y} \ge \bar{\tilde{y}} \ge 0$, {\it i.e.},
 $\theta = 0$.}\\
 
\end{array}
\right.
\end{equation*}

Let $\theta = \bar{\tilde{y}}$, and $\lambda (\bar{\tilde{y}})$ be the corresponding $\lambda$ value such that $\frac{\partial G_{\hat{\mathbb{P}}}(\theta, \lambda (\bar{\tilde{y}}))}{\partial \theta} = 0$ at $\theta = \bar{\tilde{y}}$. Then,
\begin{align*}
\frac{\partial V_{\hat{\mathbb{P}},\tilde{\mathbb{P}}}(\theta, \lambda (\bar{\tilde{y}}))} {\partial\theta}
&= \frac{\partial}{\partial\theta}\left(\frac{1}{2m} \sum_{i=1}^m(\tilde{y}_i - \theta)^2\right) -  
\frac{\partial G_{\hat{\mathbb{P}}}(\theta, \lambda (\bar{\tilde{y}}))}{\partial \theta}\\
&= 0.
\end{align*}
Also note that $\theta = \bar{\tilde{y}}$ is the unique solution to 
\[
\frac{\partial}{\partial\theta}\left(\frac{1}{2m} \sum_{i=1}^m(\tilde{y}_i - \theta)^2\right) = 0
\]
Therefore, the final AM estimator is given by $\hat{\theta}_{AM} = \bar{\tilde{y}}.$
% In AM, $\lambda$ is selected such that the resulting $\theta$, obtained by minimizing $G_{\hat{\mathbb{P}}}(\theta, \lambda)$, maximizes the likelihood of the future observations $\tilde{y}_1, \dots, \tilde{y}_m$. The above result shows that the MLE of the future observations $\tilde{y}_1, \dots, \tilde{y}_m$, specifically $\bar{\tilde{y}} \in \mathbb{R}$, can be achieved by selecting an appropriate $\lambda$. Therefore, the final AM estimator is given by $\hat{\theta}_{AM} = \bar{\tilde{y}}.$
\end{proof}

Such a conclusion from the simple model can also be generalized to more complex models, though additional work is required. For illustrative purposes, we present the results for the linear regression model below.

\begin{theorem}[Fitting Linear Regression Model to Future Observations] \label{theorem:S.2}
Suppose that $(x_1, y_1), \dots, (x_n, y_n)$ are $n$ observations from the linear regression model:
\[
y = x^\top \beta + z,
\]
where $z \sim N(0, \sigma^2)$ and $x \in \mathbb{R}^p$. Assume $\sigma$ is known and let the AM duality function be $\pi(\beta, \lambda) = \lambda^\top |\beta|$ with $\lambda \in \mathbb{R}^p$ and $|\beta|=(|\beta_1|, ..., |\beta|_p)'$, and let $(\tilde{x}_1, \tilde{y}_1), \dots, (\tilde{x}_m, \tilde{y}_m)$ represent future observations. Then, the AM estimator is $\hat{\beta}_{AM} = \hat{\tilde{\beta}}_{LS}$, where $\hat{\tilde{\beta}}_{LS}$ is the least squares solution for the future observations.
\end{theorem}

\begin{proof}
We first show that for any $\beta \in \mathbb{R}^p$, there exists some $\lambda = \hat{\lambda} \in \mathbb{R}^p$ such that $\frac{\partial G_{\hat{\mathbb{P}}}(\beta, \lambda)}{\partial \beta} = 0$ at $\lambda = \hat{\lambda}$. Using the sub-gradient notation and setting the derivative of $G_{\hat{\mathbb{P}}}(\beta, \lambda)$ with respect to $\beta$ to zero, we have:
\[
\frac{\partial G_{\hat{\mathbb{P}}}(\beta, \lambda)}{\partial \beta} = \frac{1}{n}\bx^\top (\bx \beta - \by) + \lambda \cdot s(\beta) = 0,
\]
where $s(\beta_i)  = \mbox{sign}(\beta_i)$ if  $\beta_i \neq 0$ and $s(\beta_i) \in [-1,1]$ otherwise, for $i = 1, \dots, p$. The notation $\bx$ and $\by$ are the design matrix and response vector for the observed data. For any given $\beta \in \mathbb{R}^p$, we can choose $\hat{\lambda}$ such that:
\begin{equation*}
\hat{\lambda}_j =
\begin{cases}
\frac{1}{n}[\bx^\top (\by - \bx \beta)]_j & \text{if } \beta_j > 0, \\
-\frac{1}{n}[\bx^\top (\by - \bx \beta )]_j & \text{if } \beta_j < 0, \\
\infty & \text{if } \beta_j = 0,
\end{cases}
\end{equation*}
for each component $j = 1, \dots, p$, which results in $\frac{\partial G_{\hat{\mathbb{P}}}(\beta, \lambda)}{\partial \beta} = 0$.

Let $\beta = \hat{\tilde{\beta}}_{LS}$, where $\hat{\tilde{\beta}}_{LS}$ is the least square solution such that when $\beta = \hat{\tilde{\beta}}_{LS}$,
\[
\frac{\partial}{\partial\beta}\left(\frac{1}{2m} (\tilde{\by} - \tilde{\bx} \beta)^T (\tilde{\by} - \tilde{\bx} \beta) \right) = 0.
\]
Let $\lambda (\hat{\tilde{\beta}}_{LS})$ be the corresponding $\lambda$ value such that $\frac{\partial G_{\hat{\mathbb{P}}}(\beta, \lambda (\hat{\tilde{\beta}}_{LS}))}{\partial \beta} = 0$ at $\beta = \hat{\tilde{\beta}}_{LS}$. Then,
\begin{align*}
\frac{\partial V_{\hat{\mathbb{P}},\tilde{\mathbb{P}}}(\beta, \lambda (\hat{\tilde{\beta}}_{LS}))} {\partial\beta}
&= \frac{\partial}{\partial\beta}\left(\frac{1}{2m} (\tilde{\by} - \tilde{\bx} \beta)^T (\tilde{\by} - \tilde{\bx} \beta) \right) -  
\frac{\partial G_{\hat{\mathbb{P}}}(\beta, \lambda (\hat{\tilde{\beta}}_{LS}))}{\partial \beta}\\
&= 0.
\end{align*}
Therefore, the final AM estimator is given by $\hat{\beta}_{AM} = \hat{\tilde{\beta}}_{LS}.$

\end{proof}

A similar conclusion can be drawn when $\sigma$ is unknown, as the estimation of $\beta$ does not depend on the value of $\sigma$. Additionally, the duality function can include the term $\lambda_{p+1} \log \left(\frac{1}{\sigma^2}\right)$, where $\lambda_{p+1} \in \mathbb{R}$. Relevant theoretical results will be reported elsewhere.

\clearpage
\subsection{Technical Proofs and Extended Theoretical Discussions}
\subsubsection{Regularity Conditions}\label{appendix:regularity-conditions}
Let $ (X, Y)$ take values in $\mathbf{W} \subseteq \mathbb{R}^{k + q}$ and $\theta \in \Theta^{(n)} \subseteq \mathbb{R}^p$, $\lambda \in \Lambda^{(n)} \subseteq \mathbb{R}^p$ with $p$ dependent on $n$. Assume that the loss function $L(\theta|X, Y)$ and the duality function $\pi(\theta, \lambda)$ satisfies the regularity conditions:
\begin{enumerate}[{C}1{.}]
    \item for each $n$ and $\theta \in \Theta^{(n)}$, $L (\theta | \cdot)$ is Borel Measurable on $\mathbf{W}$,
    \item for each $n$ and $(X,Y) \in \mathbf{W}$, $L(\cdot| X,Y)$ is continuous on $\Theta^{(n)}$,
    \item for every $n$ and for all $\theta \in \Theta^{(n)}$, the inequality $|L(\theta|X,Y)| < b(X,Y)$ is satisfied uniformly, where $b$ is a non-negative function on $\mathbf{W}$ such that $E|b(X,Y)| < \infty$,
    \item for each $n$, there exists $\theta_0 = \arg \min_{\theta \in \Theta^{(n)}} E_{(X,Y) \sim \mathbb{P}}L(\theta|X,Y)$,
    \item for each $n$, there exists $\delta\in(0,1)$ such that $ \mathbb{E}\left(\sup _{\theta \in \Theta^{(n)}}\left|L\left(\theta| X, Y\right)\right|^{\frac{1}{1-\delta}}\right)<\infty$ (required for Theorem~2),
    \item for each $n$ and $\theta \in \Theta^{(n)}$, $\pi(\theta,\lambda)$ is continuous on $\lambda \in \Lambda^{(n)}$,
    \item for each $n$ and $\lambda  \in \Lambda^{(n)}$, $\pi(\theta,\lambda)$ is continuous on $\theta \in \Theta^{(n)}$, and
    % \item for each $(X,Y) \in \mathbf{W}$, $L(\cdot | X, Y)$ is continuously differentiable on $\Theta^{(n)}$.
    \item for each $n$, $\lambda_0 = \boldsymbol{0} \in \Lambda^{(n)}$ and $\pi (\theta, \lambda_0) = C$ for any $\theta \in \Theta^{(n)}$, where $C \in \mathbb{R}$ is some constant.
\end{enumerate}

\subsubsection{Proof of Propositions 1 to 3}\label{proof:p1-p3}
To prove Proposition~1, by the definition of KL divergence, we have
\[
        D_{KL}(\mathbb{P}, \mathbb{P}_{\theta}) = E_{(X,Y) \sim \mathbb{P}}\left(\log p(X,Y) - \log p'(\theta | X,Y)\right),
\]
where $p(X,Y)$ denotes the true unknown population distribution and $p'(\theta | X,Y)$ denotes the likelihood function of the model with parameter $\theta \in \Theta_0$. By assumption, $\mathbb{P} = \mathbb{P}_{\theta_0}$, $\theta_0 \in \Theta_0$, resulting in
\(
        D_{KL}(\mathbb{P}, \mathbb{P}_{\theta}) =  D_{KL}(\mathbb{P}_{\theta_0}, \mathbb{P}_{\theta})
\)
and, thereby, $D_{KL}(\mathbb{P}_{\theta_0}, \mathbb{P}_{\theta}) = 0$ holds by taking $\theta = \theta_0$. Note that
\begin{equation}\label{eq:appendix-prop-1}
    E_{(X,Y) \sim \mathbb{P}}\log p'(\theta_0 | X,Y) \geq E_{(X,Y) \sim \mathbb{P}}\log p'(\tilde{\theta} | X,Y) 
\end{equation}
for any $\tilde{\theta} \in \Theta_0$, as the KL divergence is non-negative. By the definition of the optimal estimate in (3), it follows that $D_{KL}(\mathbb{P}, \mathbb{P}_{\theta^*}) = 0$ for $\theta^* \in \Theta_0^*$, where $\Theta_0^*$ denotes the set of $\theta$ that maximizes the expected log-likelihood function. This concludes the model validity since the parameter space is not dependent on $n$.

For Proposition~2, it is assumed that there exists a positive integer $N$ such that $\Theta_0 \preceq \Theta^{(n)}$ for all $n \geq N$. By the definition of model generality, for each $n \geq N$, there exists $\Theta_{\mathcal{A}}^{(n)} \subseteq \Theta^{(n)}$ such that 
\[
\{\mathbb{P}_{\theta} : \theta \in \Theta_0\} = \{\mathbb{P}_{\theta} : \theta \in \Theta_{\mathcal{A}}^{(n)}\}.
\]
As a result, we have
% \begin{equation*}\label{eq:appendix:prop-1}
%      \min_{\tilde{\theta} \in \Theta^{(n)}} E_{(X, Y)\sim\mathbb{P}}L(\tilde{\theta}|X, Y) \leq \min_{\tilde{\theta} \in \Theta_0} E_{(X, Y)\sim\mathbb{P}}L(\tilde{\theta}|X, Y)
% \end{equation*}
\begin{align*}
\min_{\tilde{\theta} \in \Theta^{(n)}} E_{(X, Y)\sim\mathbb{P}}L(\tilde{\theta}|X, Y) &= \min \left( \min_{\tilde{\theta} \in \Theta_{\mathcal{A}}^{(n)}} E_{(X, Y)\sim\mathbb{P}}L(\tilde{\theta}|X, Y), \ \min_{\tilde{\theta} \in \Theta^{(n)} \setminus \Theta_{\mathcal{A}}^{(n)}} E_{(X, Y)\sim\mathbb{P}}L(\tilde{\theta}|X, Y)  \right) \\
&\leq \min_{\tilde{\theta} \in \Theta_{\mathcal{A}}^{(n)}} E_{(X, Y)\sim\mathbb{P}}L(\tilde{\theta}|X, Y)\\
&= \min_{\tilde{\theta} \in \Theta_0} E_{(X, Y)\sim\mathbb{P}}L(\tilde{\theta}|X, Y)
\end{align*}
holds for all $n \geq N$. Let $L(\theta|X,Y)$ denote the negative log-likelihood function. We obtain from the proof of Proposition~1 that
\[
\min_{\tilde{\theta} \in \Theta_0} E_{(X, Y)\sim\mathbb{P}}L(\tilde{\theta}|X, Y) = -E_{(X,Y) \sim \mathbb{P}}\log p'(\theta_0 | X,Y)
\]
with the true parameter $\theta_0 \in \Theta_0$ and (\ref{eq:appendix-prop-1}) holds for any $\tilde{\theta} \in \Theta^{(n)}$. As a result, we can conclude that $\lim_{n\to\infty} D_{KL}(\mathbb{P}, \mathbb{P}_{\theta^*}) = 0$, where $\theta^* \in \Theta^{(n)}_*$.

For Proposition~3, given that the likelihood function is continuous and  $\hat{\theta}_0 \xrightarrow{p} \theta_0$ for $\hat{\theta} \in \Theta_0$, by the continuous mapping theorem (CMT), we have
\[
   E_{(X,Y) \sim \mathbb{P}}\log p'(\hat{\theta} | X,Y)  \xrightarrow{p} E_{(X,Y) \sim \mathbb{P}}\log p'(\theta_0 | X,Y).
\]
Since $\theta_0$ is the true parameter, this is equivalent to $D_{KL}(\mathbb{P}, \mathbb{P}_{\hat{\theta}}) \xrightarrow{p} 0$, completing the proof. 

\qed

\subsubsection{Proof of Theorem~1}\label{proof:t1}

% Recall that the AM objective functions consist of
% \begin{align}\label{eq:appendix-g-function}
% \begin{split}
% G_{\hat{\mathbb{P}}}(\theta, \lambda) &=
% E_{(X, Y)\sim\hat{\mathbb{P}}}L(\theta|X, Y)
% +\pi(\theta, \lambda)\\
% &= 
%     \frac{1}{n}\sum_{i=1}^{n}L(\theta|x_i, y_i)
%     + \pi(\theta, \lambda)
% \end{split}
% \end{align}
% and
% \begin{align*}
%     V_{\mathbb{P},\hat{\mathbb{P}}}(\theta, \lambda) &=
%     E_{(X, Y)\sim\mathbb{P}}L(\theta|X, Y)
%     -
%     E_{(X, Y)\sim\hat{\mathbb{P}}}L(\theta|X, Y)
%     -\pi(\theta, \lambda) \\
%     &= E_{(X, Y)\sim\mathbb{P}}L (\theta|X, Y)
%     -
%     \frac{1}{n}\sum_{i=1}^{n}L(\theta|x_i, y_i)
%     -\pi(\theta, \lambda).
% \end{align*}
By assumption, for some $N > 0$,  there exists a model $\{\mathbb{P}_\phi: \phi \in \Phi \}$ with compact parameter space $\Phi$ such that $\Phi \preceq \Theta^{(n)}$, satisfying 
\(
	\min_{\phi \in \Phi}  D_{KL}(\mathbb{P}, \mathbb{P}_{\phi})
	=
	 \lim_{n \to \infty}D_{KL}(\mathbb{P}, \mathbb{P}_{\theta^*}) 
\). By the definition of model generality, for every $n \geq N$, there exists $\Theta_{\mathcal{A}}^{(n)} \subseteq \Theta^{(n)}$ such that 
\begin{equation}\label{eq:appendix:restricted-parameter-space}
\{\mathbb{P}_{\phi} : \phi \in \Phi\} = \{\mathbb{P}_{\theta} : \theta \in \Theta_{\mathcal{A}}^{(n)}\}.
\end{equation}
Since the mapping $\psi$ is continuous by definition, the set $\Theta_{\mathcal{A}}^{(n)}$ is also compact for every $n \geq N$. Furthermore, we have 
\[
\min_{\theta \in \Theta_{\mathcal{A}}^{(n)}}  D_{KL}(\mathbb{P}, \mathbb{P}_{\theta})
	=
	 \lim_{n \to \infty}D_{KL}(\mathbb{P}, \mathbb{P}_{\theta^*}) 
\]
for any $\theta^* \in \Theta_*^{(n)}$, as defined in (3), and $n \geq N$.
% there exists a fixed parameter space $\Theta_{\mathcal{A}}$ such that $\Theta_{\mathcal{A}} \subseteq \Theta^{(n)}$ for all $n \geq N$.\CLadd{}\footnote{\color{red}Does this imply that the dimension of $\Theta_{\mathcal{A}} \subseteq \Theta^{(n)}$ is independent of $n$ for $n >N$?}
% Such $\Theta_{\mathcal{A}}$ satisfies
% \[
% 	\min_{\theta \in \Theta_{\mathcal{A}}}  D_{KL}(\mathbb{P}, \mathbb{P}_{\theta})
% 	=
% 	 \lim_{n \to \infty}D_{KL}(\mathbb{P}, \mathbb{P}_{\theta^*}) 
% \]
% for any $\theta^* \in \Theta_*^{(n)}$, which 
This further implies
\begin{align}\label{eq:appendix-thm1-1}
\begin{split}
    \min_{\theta \in \Theta_{\mathcal{A}}^{(n)}} E_{(X, Y)\sim\mathbb{P}}L(\theta|X, Y) &= \lim_{n\to \infty} \min_{\theta \in \Theta^{(n)}} E_{(X, Y)\sim\mathbb{P}}L(\theta|X, Y). \\
    &= \lim_{n\to \infty} E_{(X, Y)\sim\mathbb{P}}L(\theta^*|X, Y) \\
\end{split}
\end{align}
for some optimal estimate $\theta^* \in \Theta^{(n)}_*$ and $n \geq N$.

For simplicity in the asymptotic analysis, all subsequent expressions involving n implicitly assume $n \geq N$. With the regularity conditions C1---C4 for the loss function in Section~\ref{appendix:regularity-conditions}, the uniform law of large numbers (ULLN) can be applied \citep[see, {\it e.g.},][]{newey1994} to the restricted parameter space $\Theta_{\mathcal{A}}^{(n)}$ due to its compactness. Thus, 
\begin{equation}\label{eq:appendix-ULLN}
\sup_{\theta \in \Theta_{\mathcal{A}}^{(n)}} \displaystyle\left\lvert  E_{(X,Y) \sim \mathbb{P}}L(\theta|X,Y) - \frac{1}{n}\sum_{i=1}^n L(\theta |x_i,y_i)\displaystyle\right\rvert \xrightarrow{p} 0.
\end{equation}
% or equivalently,
% \[
% \sup_{\theta \in \Theta_{\mathcal{A}}} \displaystyle\left\lvert V_{\mathbb{P},\hat{\mathbb{P}}}(\theta, \lambda) + \pi(\theta, \lambda) \displaystyle\right\rvert \xrightarrow{p} 0.
% \]
The result in (\ref{eq:appendix-ULLN}) further implies
\[
\sup_{\theta \in \Theta_{\mathcal{A}}^{(n)}} \displaystyle\left\lvert  E_{(X,Y) \sim \mathbb{P}}L(\theta|X,Y) + \pi(\theta, \lambda) - \left(\frac{1}{n}\sum_{i=1}^n L(\theta |x_i,y_i) + \pi(\theta, \lambda)\right) \displaystyle\right\rvert \xrightarrow{p} 0
\]
for any $\lambda \in \Lambda^{(n)}$ and
\[
\frac{1}{n}\sum_{i=1}^n L(\theta |x_i,y_i) + \pi(\theta, \lambda) \xrightarrow{p} E_{(X,Y) \sim \mathbb{P}}L(\theta|X,Y) + \pi(\theta, \lambda)
\]
for all $\theta \in \Theta_{\mathcal{A}}^{(n)}$ and $\lambda \in \Lambda^{(n)}$.  
% This further implies
% \[
% \frac{1}{n}\sum_{i=1}^n L\left((\hat{\theta}^{*}_{\mathcal{A}})_\lambda|x_i,y_i\right) + \pi\left((\hat{\theta}^{*}_{\mathcal{A}})_\lambda , \lambda\right) \xrightarrow{p} E_{(X,Y) \sim \mathbb{P}}L\left((\hat{\theta}^{*}_{\mathcal{A}})_\lambda |X,Y\right) + \pi\left((\hat{\theta}^{*}_{\mathcal{A}})_\lambda , \lambda\right)
% \]
% where 
We define the $\lambda$-dependent AM estimator in the restricted parameter space as
\begin{equation}\label{eq:appendix:theta}
    (\hat{\theta}^{*}_{\mathcal{A}})_\lambda = \underset{\theta \in \Theta_{\mathcal{A}}^{(n)}}{\text{arg min}} \, \left( \frac{1}{n}\sum_{i=1}^n L(\theta |x_i,y_i) + \pi(\theta , \lambda) \right)
\end{equation}
for any given $\lambda \in \Lambda^{(n)}$. By the global optimal assumption of the AM estimator elaborated in Section~4.2, the duality parameter $\lambda$ is chosen at $\lambda = \hat{\lambda}$ such that
\begin{equation}\label{eq:appendix:lambda}
    \hat{\lambda} = \underset{\lambda \in \Lambda^{(n)}}{\text{arg min}} \, E_{(X,Y) \sim \mathbb{P}}L\left((\hat{\theta}^{*}_{\mathcal{A}})_\lambda |X,Y\right).
\end{equation}
Note that $\pi(\theta , \lambda) \geq 0$ for any $\theta \in \Theta_{\mathcal{A}}^{(n)}$, $\lambda \in \Lambda^{(n)}$. 
% Denote the AM estimate as $(\hat{\theta}^*_{AM}, \hat{\lambda})$. We have
% \[
% G_{\hat{\mathbb{P}}}(\hat{{\theta}}^*_{AM}, \hat{\lambda}) \xrightarrow{p} E_{(X,Y) \sim \mathbb{P}}L(\hat{{\theta}}^*_{AM}|X,Y) .
% \]
% and 
We then consider $\lambda_0 = \boldsymbol{0}$, which is assumed that $\lambda_0 \in \Lambda^{(n)}$ by the regularity condition C8 for the duality function in Section~\ref{appendix:regularity-conditions}. Furthermore, by the same regularity condition, we have
\[
\frac{1}{n}\sum_{i=1}^n L\left((\hat{\theta}^{*}_{\mathcal{A}})_{\lambda_0}|x_i,y_i\right) + \pi\left((\hat{\theta}^{*}_{\mathcal{A}})_{\lambda_0} , \lambda_0\right) \xrightarrow{p} E_{(X,Y) \sim \mathbb{P}}L\left((\hat{\theta}^{*}_{\mathcal{A}})_{\lambda_0} |X,Y\right) + C,
\]
where $C \in \mathbb{R}$ is a constant. By (\ref{eq:appendix:theta}), $(\hat{\theta}^{*}_{\mathcal{A}})_{\lambda_0}$ is the empirical loss minimizer in the parameter space $\Theta_{\mathcal{A}}^{(n)}$. The uniform convergence of the empirical loss in (\ref{eq:appendix-ULLN}) indicates the convergence of its minimizer to the true population risk minimizer \citep{shalev-shwartz2010}. By (\ref{eq:appendix-thm1-1}) and the definition of the optimal estimate in (3), the asymptotic optimality
\begin{equation}\label{eq:appendix:estimation-convergence}
    E_{(X,Y) \sim \mathbb{P}}L\left((\hat{\theta}^{*}_{\mathcal{A}})_{\lambda_0}|X,Y\right)  \xrightarrow{p} E_{(X,Y) \sim \mathbb{P}}L(\theta^*|X,Y)
\end{equation}
is achievable for some optimal estimate $\theta^* \in \Theta_{*}^{(n)}$. By the AM objective function (\ref{eq:appendix:lambda}), we know that for any $\hat{\lambda} \in \Lambda^{(n)}$,
\[
E_{(X,Y) \sim \mathbb{P}}L(\theta^*|X,Y) \leq E_{(X,Y) \sim \mathbb{P}}L\left((\hat{\theta}^{*}_{\mathcal{A}})_{\hat\lambda} |X,Y\right) \leq E_{(X,Y) \sim \mathbb{P}}L\left((\hat{\theta}^{*}_{\mathcal{A}})_{\lambda_0} |X,Y\right)
\]
for each $(X,Y) \in \mathbf{W}$. Now, we consider the unrestricted model with the full parameter space $\Theta^{(n)}$ and $\Lambda^{(n)}$. By the AM objective functions (\ref{eq:appendix:theta}) and (\ref{eq:appendix:lambda}) again, the optimal AM estimator $\hat{{\theta}}^*_{AM} \in \Theta^{(n)}$ satisfies
\begin{equation}\label{eq:appendix:bounded-estimation}
E_{(X,Y) \sim \mathbb{P}}L(\theta^*|X,Y) \leq 
E_{(X,Y) \sim \mathbb{P}}L\left(\hat{{\theta}}^*_{AM}|X,Y\right)
\leq E_{(X,Y) \sim \mathbb{P}}L\left((\hat{\theta}^{*}_{\mathcal{A}})_{\hat\lambda} |X,Y\right)
\end{equation}
for each $(X,Y) \in \mathbf{W}$. As a result, for any $\epsilon > 0$, the two events
\begin{align*}
    A &= \{E_{(X,Y) \sim \mathbb{P}}L\left(\hat{{\theta}}^*_{AM}|X,Y\right) - E_{(X,Y) \sim \mathbb{P}}L(\theta^*|X,Y) \geq \epsilon\} \\
    B &= \{E_{(X,Y) \sim \mathbb{P}}L\left((\hat{\theta}^{*}_{\mathcal{A}})_{\lambda_0}|X,Y\right) - E_{(X,Y) \sim \mathbb{P}}L(\theta^*|X,Y) \geq \epsilon \},
\end{align*}
defined using $\hat{{\theta}}^*_{AM}$ and $(\hat{\theta}^{*}_{\mathcal{A}})_{\lambda_0}$,
satisfy $A \subseteq B$. With (\ref{eq:appendix:estimation-convergence}) and \eqref{eq:appendix:bounded-estimation}, this implies for any $\epsilon > 0$, 
\[
\lim_{n \to \infty}P\left(\left|E_{(X,Y) \sim \mathbb{P}}L\left(\hat{{\theta}}^*_{AM}|X,Y\right) - E_{(X,Y) \sim \mathbb{P}}L(\theta^*|X,Y)\right| \geq \epsilon \right) = 0
\]
and, thereby, we have
\[
   E_{(X,Y) \sim \mathbb{P}}L\left(\hat{{\theta}}^*_{AM}|X,Y\right)  \xrightarrow{p} E_{(X,Y) \sim \mathbb{P}}L(\theta^*|X,Y)
\]
for $\hat{{\theta}}^*_{AM} \in \Theta^{(n)}$ and some optimal estimate $\theta^* \in \Theta_{*}^{(n)}$. For the KL divergence distance measure denoted as $D_{KL}(\cdot, \cdot)$, by definition,
\begin{align}\label{eq:ap4}
\begin{split}
        D_{KL}(\mathbb{P}, \mathbb{P}_{\hat{{\theta}}^*_{AM}}) &= E_{(X,Y) \sim \mathbb{P}}\left(\log p(X,Y) - \log p'\left(\hat{{\theta}}^*_{AM} | X,Y\right)\right)\\
        D_{KL}(\mathbb{P}, \mathbb{P}_{\theta^{*}}) &= E_{(X,Y) \sim \mathbb{P}}\left(\log p(X,Y) - \log p'(\theta^{*} | X,Y)\right),
\end{split}
\end{align}
where $p(X,Y)$ is the true unknown population distribution and $p'(\cdot | X,Y)$ is the likelihood function of the model with given parameters. When the negative log-likelihood function is used as the loss function, by (\ref{eq:appendix:estimation-convergence}), we conclude that
\[
D_{KL}(\mathbb{P}, \mathbb{P}_{\hat{{\theta}}^*_{AM}}) \xrightarrow{p}  D_{KL}(\mathbb{P}, \mathbb{P}_{\theta^{*}}),
\]
% It can be seen that if the model error satisfies $\lim_{n \to \infty} D_{KL}(\mathbb{P}, \mathbb{P}_{\theta^{*}}) = 0$ by the validity assumption in Definition~\ref{def:validity-model}, we have
% \[
% D_{KL}(\mathbb{P}, \mathbb{P}_{\hat{{\theta}}^*_{AM}}) \xrightarrow{p} 0,
% \]
completing the proof.
\qed

\subsubsection{Proof of Corollary~1}\label{proof:c1}
By assumption, $\mathbb{P} = \mathbb{P}_{\theta_0}$, $\theta_0 \in \Theta_0$, %  \subseteq \mathbb{R}^p$, 
where $\Theta_0$ is the parameter space for the true model parameter $\theta_0$. Moreover,  $|\theta_{0j}| < \infty, \text{ for } j = 1,\dots,p$. We construct a compact parameter space 
\[
    \Theta_{0\mathcal{A}} := \{\theta:\; \theta \in \Theta_0 \text{ and } \theta_j = 0 \text{ if } |\theta_j| > |\theta_{0j}| + \epsilon , \text{ for } j = 1,\dots, p\},
\]
where $\epsilon > 0$ is a small constant. Since $|\theta_{0j}| < \infty, \text{ for } j = 1,\dots,p$, the set $\Theta_{0\mathcal{A}}$ is compact. Furthermore, we have $\theta_0 \in \Theta_{0\mathcal{A}}$. By assumption, there exists a positive integer $N$ such that $\Theta_0 \preceq \Theta^{(n)}$ for every $n \geq N$. This implies that $ \Theta_{0\mathcal{A}} \preceq \Theta^{(n)}$ for all $n \geq N$. By the definition of model generality, there exists $\Theta_{\mathcal{A}}^{(n)} \in \Theta^{(n)}$ for all $n \geq N$ such that
\[
\{\mathbb{P}_{\theta} : \theta\in \Theta_{0\mathcal{A}}\} = \{\mathbb{P}_{\theta} : \theta \in \Theta_{\mathcal{A}}^{(n)}\}.
\]

With similar arguments used in Section~\ref{proof:p1-p3}, the expected negative log-likelihood function denoted by 
\(
E_{(X,Y) \sim \mathbb{P}} L(\tilde{\theta} | X,Y)\) is minimized at $\tilde{\theta} = \theta_0$. Since $\theta_0 \in \Theta_{0\mathcal{A}}$, we have 
% \begin{equation}\label{eq:appendix-c1-1}
%         \lim_{n \to \infty} \min_{\theta \in \Theta^{(n)}} E_{(X, Y)\sim\mathbb{P}}L(\theta|X, Y) \leq \min_{\theta \in \Theta_{\mathcal{A}}} E_{(X, Y)\sim\mathbb{P}}L(\theta|X, Y) = E_{(X, Y)\sim\mathbb{P}}L(\theta_0|X, Y).
% \end{equation}
\begin{align}\label{eq:appendix-c1-1}
    \begin{split}
        & \ \ \ \lim_{n \to \infty} \min_{\theta \in \Theta^{(n)}} E_{(X, Y)\sim\mathbb{P}}L(\theta|X, Y) \\ &= \min \left( \min_{\theta \in \Theta_{\mathcal{A}}^{(n)}} E_{(X, Y)\sim\mathbb{P}}L(\theta|X, Y), \ \lim_{n \to \infty}\min_{\theta \in \Theta^{(n)} \setminus \Theta_{\mathcal{A}}^{(n)}} E_{(X, Y)\sim\mathbb{P}}L(\theta|X, Y)  \right) \\
        &\leq \min_{\theta \in \Theta_{\mathcal{A}}^{(n)}} E_{(X, Y)\sim\mathbb{P}}L(\theta|X, Y)\\
        &= \min_{\theta \in \Theta_{0\mathcal{A}}} E_{(X, Y)\sim\mathbb{P}}L(\theta|X, Y)\\
        &= E_{(X, Y)\sim\mathbb{P}}L(\theta_0|X, Y).
    \end{split}
\end{align}
The first equality in (\ref{eq:appendix-c1-1}) follows since $ \Theta_{\mathcal{A}}^{(n)} \subseteq \Theta^{(n)}$ for all $n \geq N$. Given that
\[
        D_{KL}(\mathbb{P}, \mathbb{P}_{\theta}) = E_{(X, Y)\sim\mathbb{P}}L(\theta|X, Y) - E_{(X, Y)\sim\mathbb{P}}L(\theta_0|X, Y) \geq 0
\]
for any $\theta \in \Theta^{(n)}$, it follows that
\[
\lim_{n \to \infty} \min_{\theta \in \Theta^{(n)}} E_{(X, Y)\sim\mathbb{P}}L(\theta|X, Y) \geq E_{(X, Y)\sim\mathbb{P}}L(\theta_0|X, Y),
\]
resulting in 
\[
\lim_{n \to \infty} \min_{\theta \in \Theta^{(n)}} E_{(X, Y)\sim\mathbb{P}}L(\theta|X, Y) = \min_{\theta \in \Theta_{\mathcal{A}}^{(n)}} E_{(X, Y)\sim\mathbb{P}}L(\theta|X, Y) = E_{(X, Y)\sim\mathbb{P}}L(\theta_0|X, Y)
\]
from (\ref{eq:appendix-c1-1}). This further implies
\[
	\min_{\theta \in \Theta_{\mathcal{A}}^{(n)}}  D_{KL}(\mathbb{P}, \mathbb{P}_{\theta})
	=
	 \lim_{n \to \infty}D_{KL}(\mathbb{P}, \mathbb{P}_{\theta^*}) 
\]
for any $\theta^* \in \Theta_*^{(n)}$, by the definition of the optimal estimate in (3), completing the proof.
\qed

\subsubsection{Proof of Lemma~1} \label{proof:l1}
By assumption, $D_{KL}(\mathbb{P}, \mathbb{Q})$ converges to zero in probability as $n \to \infty$. By Pinsker's inequality \citep[Lemma 2.5]{Tsybakov2009}, it implies the convergence in total variation distance denoted by $d_{TV}(\cdot,\cdot)$ between $\mathbb{P}$ and $\mathbb{Q}$. That is, for any $\epsilon > 0$,
\begin{equation}\label{eq:appendix-total-variation}
    \lim_{n \to \infty} P (d_{TV}(\mathbb{P}, \mathbb{Q}) \geq \epsilon) = 0.
\end{equation}
The total variation distance is defined as
\[
d_{TV}(\mathbb{P}, \mathbb{Q}) = \sup_{A} \vert\mathbb{P}(A) - \mathbb{Q}(A)\vert,
\]
where the supremum is taken over all measurable sets $A$.

  By regularity condition C3 in Section~\ref{appendix:regularity-conditions}, the loss function is bounded by a non-negative function $b$ such that $|L(\theta|X,Y)| < b(X,Y)$ for all $\theta \in \Theta^{(n)}$ and $E|b(X,Y)| < \infty$. Thus, we have
%\begin{align*}
\begin{eqnarray}
    \Big\vert E_{(X,Y) \sim \mathbb{P}}L(\theta|X,Y) - E_{(X,Y) \sim \mathbb{Q}}L(\theta|X,Y) \Big\vert &=& \Big\vert \int L(\theta|X, Y) d\mathbb{P} - \int L(\theta|X, Y) d\mathbb{Q} \Big\vert \nonumber\\
    &\leq& \int b(X,Y) d|\mathbb{P} -  \mathbb{Q}| \label{eq:b-inequality}
\end{eqnarray}
%\end{align*}
for any $\theta \in \Theta^{(n)}$. The condition $E|b(X,Y)| < \infty$ implies that the function $b(X, Y)$ is almost surely bounded. %\CLadd{This is because $b(X, Y)$ is non-negative, and any positive measure where $|b(X, Y)| = \infty$ results in $E|b(X,Y)| = \infty$. } 
Thus, it can be approximated by simple functions. As a result, by the property of total variation distance, for any $\epsilon > 0$, there exists a $\delta > 0$ such that if $d_{TV}(\mathbb{P}, \mathbb{Q}) < \delta$, then $\int b(X,Y) d|\mathbb{P} -  \mathbb{Q}| < \epsilon$. Thus, by (\ref{eq:appendix-total-variation}), we have for any $\epsilon > 0$, 
\[
\lim_{n \to \infty} P (\int b(X,Y) d|\mathbb{P} -  \mathbb{Q}|\geq \epsilon) = 0. %%%%,
\]
With \eqref{eq:b-inequality}, this further implies
\begin{equation}\label{eq:appendix-ULLN-2}
\sup_{\theta \in \Theta^{(n)}} \Big\vert  E_{(X,Y) \sim \mathbb{P}}L(\theta|X,Y) - E_{(X,Y) \sim \mathbb{Q}}L(\theta|X,Y) \Big\vert \xrightarrow{p} 0.
\end{equation}
% As in \ref{proof:t1}, the uniform convergence in (\ref{eq:appendix-ULLN-2}) implies the convergence of the minimizer. That is,
% \[
% E_{(X,Y) \sim \mathbb{Q}}L(\theta|X,Y) \xrightarrow{p} E_{(X,Y) \sim \mathbb{P}}L(\theta|X,Y).
% \]
The uniform convergence in (\ref{eq:appendix-ULLN-2}) indicates the convergence of the empirical risk minimizer to the true population risk minimizer \citep{shalev-shwartz2010}, that is,
\[
\min_{\theta \in \Theta^{(n)}} E_{(X, Y)\sim\mathbb{Q}}L(\theta|X, Y) \xrightarrow{p} \lim_{n\to \infty} \min_{\theta \in \Theta^{(n)}} E_{(X, Y)\sim\mathbb{P}}L(\theta|X, Y)
\]
or, by definition,
\begin{equation}\label{eq:appendix-convergence-Q}
       E_{(X,Y) \sim \mathbb{Q}}L\left(\hat{\theta}^{*}|X,Y\right)  \xrightarrow{p} E_{(X,Y) \sim \mathbb{P}}L(\theta^*|X,Y)
\end{equation}
for some optimal estimate $\theta^* \in \Theta_{*}^{(n)}$, and $\hat{\theta}^{*} \in \Theta^{(n)}$ is some minimizer of the expected loss function with regard to $(X,Y) \sim \mathbb{Q}$.
 
Following the same arguments in Section~\ref{proof:t1}, we construct the restricted parameter space $\Theta_{\mathcal{A}}^{(n)} \subseteq \Theta^{(n)}$ specified in (\ref{eq:appendix:restricted-parameter-space}). We further define the $\lambda$-dependent AM estimator in the restricted parameter space $\Theta_{\mathcal{A}}^{(n)} \subseteq \Theta^{(n)}$ such that
\begin{equation}\label{eq:appendix:theta-2}
    (\hat{\theta}^{*}_{\mathcal{A}})_\lambda = \underset{\theta \in \Theta_{\mathcal{A}}^{(n)}}{\text{arg min}} \, \left( \frac{1}{n}\sum_{i=1}^n L(\theta |x_i,y_i) + \pi(\theta , \lambda) \right)
\end{equation}
for any given $\lambda \in \Lambda^{(n)}$ and the duality parameter $\lambda$ is chosen at $\lambda = \hat{\lambda}$ with
\begin{equation}\label{eq:appendix:lambda-2}
    \hat{\lambda} =  \underset{\lambda \in \Lambda^{(n)}}{\text{arg min}} \, E_{(X,Y) \sim \mathbb{Q}}L\left((\hat{\theta}^{*}_{\mathcal{A}})_\lambda |X,Y\right).
\end{equation}
We then consider $\lambda_0 = \boldsymbol{0}$, which is assumed that $\lambda_0 \in \Lambda^{(n)}$ by the regularity condition C8 for the duality function in Section~\ref{appendix:regularity-conditions}. Such $\lambda_0$ results in $\pi(\theta, \lambda_0) = C$ for any $\theta \in \Theta_{\mathcal{A}}^{(n)}$, where $C \in \mathbb{R}$ is a constant, by the same regularity condition. By (\ref{eq:appendix-ULLN-2}), we have 
\begin{equation}\label{eq:appendix:theta-intermediate}
E_{(X,Y) \sim \mathbb{Q}}L\left((\hat{\theta}^{*}_{\mathcal{A}})_{\lambda_0} |X,Y\right) \xrightarrow{p}  E_{(X,Y) \sim \mathbb{P}}L\left((\hat{\theta}^{*}_{\mathcal{A}})_{\lambda_0} |X,Y\right).
\end{equation}
By (\ref{eq:appendix:theta-2}), $(\hat{\theta}^{*}_{\mathcal{A}})_{\lambda_0}$ is the empirical risk minimizer in the parameter space $\Theta_{\mathcal{A}}^{(n)}$. Since $\Theta_{\mathcal{A}}^{(n)} \subseteq \Theta^{(n)}$ for $n \geq N$, the uniform convergence of the empirical risk in (\ref{eq:appendix-ULLN-2}) indicates the convergence of its minimizer to the true population risk minimizer \citep{shalev-shwartz2010}. As a result, it follows from (\ref{eq:appendix:theta-intermediate}) and (\ref{eq:appendix-thm1-1}) that
\[
   E_{(X,Y) \sim \mathbb{Q}}L\left((\hat{\theta}^{*}_{\mathcal{A}})_{\lambda_0} |X,Y\right) \xrightarrow{p} E_{(X,Y) \sim \mathbb{P}}L(\theta^*|X,Y)
\]
for some $\theta^* \in \Theta_*^{(n)}$. Thus, with the duality parameter estimated in the parameter space $\hat{\lambda} \in \Lambda^{(n)}$, it follows from (\ref{eq:appendix:lambda-2}) that
\[
E_{(X,Y) \sim \mathbb{Q}}L\left((\hat{\theta}^{*}_{\mathcal{A}})_{\lambda_0} |X,Y\right) \geq E_{(X,Y) \sim \mathbb{Q}}L\left((\hat{\theta}^{*}_{\mathcal{A}})_{\hat{\lambda}} |X,Y\right)
\]
for each $(X,Y) \in \mathbf{W}$. It follows from (\ref{eq:appendix-ULLN-2}) that
\[
  E_{(X,Y) \sim \mathbb{Q}}L\left((\hat{\theta}^{*}_{\mathcal{A}})_{\hat{\lambda}} |X,Y\right) \xrightarrow{p} E_{(X,Y) \sim \mathbb{P}}L\left((\hat{\theta}^{*}_{\mathcal{A}})_{\hat{\lambda}} |X,Y\right)
\]
with
\[
E_{(X,Y) \sim \mathbb{P}}L\left((\hat{\theta}^{*}_{\mathcal{A}})_{\hat{\lambda}} |X,Y\right) \geq E_{(X,Y) \sim \mathbb{P}}L(\theta^*|X,Y)
\]
for each $(X,Y) \in \mathbf{W}$ and some optimal estimate $\theta^* \in \Theta_{*}^{(n)}$. For simplifying the notations, we define the random variables $A, C, D$ and the value $B$ dependent on $n$ as follows:
\begin{align*}
    A &:=  E_{(X,Y) \sim \mathbb{Q}}L\left((\hat{\theta}^{*}_{\mathcal{A}})_{\lambda_0} |X,Y\right) &   B &:= E_{(X,Y) \sim \mathbb{P}}L(\theta^*|X,Y) \\
    C &:= E_{(X,Y) \sim \mathbb{Q}}L\left((\hat{\theta}^{*}_{\mathcal{A}})_{\hat{\lambda}} |X,Y\right) &   D &:= E_{(X,Y) \sim \mathbb{P}}L\left((\hat{\theta}^{*}_{\mathcal{A}})_{\hat{\lambda}} |X,Y\right).
\end{align*}
It has been shown that $A \xrightarrow{p} B$, $C \xrightarrow{p} D$, and $\forall \omega \in \Omega, A(\omega) \geq C(\omega)$ and $D(\omega) \geq B(\omega)$, where $\Omega$ is the sample space. Here we prove that $C \xrightarrow{p} B$. For any $\epsilon > 0$, we have
\begin{align*}
    P(|C - B| \geq \epsilon) &\leq P(|C - A| + |A - B| \geq \epsilon)\\
    &\leq P( A - C \geq \epsilon/2) + P( |A - B| \geq \epsilon/2)
\end{align*}
by the union bound and the fact that $\forall \omega \in \Omega, A(\omega) \geq C(\omega)$. From $A \xrightarrow{p} B$ we know that $\lim_{n\to \infty} P( |A - B| \geq \epsilon/2) = 0$ for any $\epsilon > 0$. It remains to show that $\lim_{n\to \infty} P( A - C \geq \epsilon/2) = 0$ for any $\epsilon > 0$. We notice that
\begin{align*}
    P( A - C \geq \epsilon/2) &= P( (A - B) + (B - D) + (D - C) \geq \epsilon/2)\\
    &\leq P( (A - B) + (D - C) \geq \epsilon/2)\\
    &\leq P( |A - B| \geq \epsilon/4) + P( |D - C| \geq \epsilon/4)
\end{align*}
by the union bound, which concludes $\lim_{n\to \infty} P( A - C \geq \epsilon/2) = 0$ given $A \xrightarrow{p} B$, $C \xrightarrow{p} D$. The first inequality above follows by the given condition $\forall \omega \in \Omega, D(\omega) \geq B(\omega)$. Thus, we have proved that
\begin{equation}\label{eq:appendix-convergence-Q-2}
    E_{(X,Y) \sim \mathbb{Q}}L\left((\hat{\theta}^{*}_{\mathcal{A}})_{\hat{\lambda}} |X,Y\right) \xrightarrow{p} E_{(X,Y) \sim \mathbb{P}}L(\theta^*|X,Y).
\end{equation}

The estimation performed in the full parameter space $\Theta^{(n)}$ and $\Lambda^{(n)}$ with the unrestricted model results in the AM estimator $\hat{\theta}_{AM} \in \Theta^{(n)}$. We have
\[
 E_{(X,Y) \sim \mathbb{Q}}L\left(\hat{\theta}^{*}|X,Y\right) \leq  E_{(X,Y) \sim \mathbb{Q}}L\left(\hat{\theta}_{AM} |X,Y\right) \leq E_{(X,Y) \sim \mathbb{Q}}L\left((\hat{\theta}^{*}_{\mathcal{A}})_{\hat{\lambda}} |X,Y\right)
\]
for each $(X,Y) \in \mathbf{W}$, where $\hat{\theta}^* \in \Theta^{(n)}$ is some optimal estimate with regard to $(X,Y)\sim \mathbb{Q}$ defined in (\ref{eq:appendix-convergence-Q}). Given the convergence results in (\ref{eq:appendix-convergence-Q}) and (\ref{eq:appendix-convergence-Q-2}) for the lower bound random variable and upper bound random variable, by the sandwich theorem, it follows that
\begin{equation}\label{eq:lemma1-proof201}
E_{(X,Y) \sim \mathbb{Q}}L\left(\hat{\theta}_{AM} |X,Y\right) \xrightarrow{p} E_{(X,Y) \sim \mathbb{P}}L(\theta^*|X,Y),
\end{equation}
which further implies
\begin{equation}\label{eq:lemma1-proof202}
E_{(X,Y) \sim \mathbb{P}}L\left(\hat{\theta}_{AM} |X,Y\right) \xrightarrow{p} E_{(X,Y) \sim \mathbb{P}}L(\theta^*|X,Y)
\end{equation}
by (\ref{eq:appendix-ULLN-2}). Following the definition of the KL divergence elaborated in Section~\ref{proof:t1}, it can be concluded from \eqref{eq:lemma1-proof201} and \eqref{eq:lemma1-proof202} that
\[
D_{KL}(\mathbb{P}, \mathbb{P}_{\hat{\theta}_{AM}}) \xrightarrow{p}  D_{KL}(\mathbb{P}, \mathbb{P}_{\theta^{*}})
\]
for any $\theta^* \in \Theta^{(n)}$, 
% and the model validity further implies
% \[
% D_{KL}(\mathbb{P}, \mathbb{P}_{\hat{\theta}_{AM}}) \xrightarrow{p}  0,
% \]
completing the proof.
\qed
% Thus, using the same argument as \ref{proof:t1}, for given $\hat{\mathbb{P}}, \ \theta$ and $\lambda$, we have
% \[
% V_{\mathbb{Q},\hat{\mathbb{P}}}(\theta,\lambda) \xrightarrow{p} V_{\mathbb{P},\hat{\mathbb{P}}}(\theta,\lambda)
% \]
% and thus
% \begin{equation*}
%         {\hat{\theta}_{AM}} = \mathfrak{A}_{\theta}(\hat{\mathbb{P}}, \mathbb{Q}) \xrightarrow{p} \mathfrak{A}_{\theta}(\hat{\mathbb{P}}, \mathbb{P}) =  {{\theta}^{*}_{AM}}.
% \end{equation*}
% The conclusion follows by Theorem~\ref{theorem:1}.

\subsubsection{Proof of Theorem~2} \label{proof:t2}
By Algorithm~2, in each imputation step at iteration $(b, k)$ for arbitrary $b = 1,\dots,B$ and $k = 1,\dots, K$,  the samples $\tilde{\mathbb{P}}^{(k)}_b:= \{(\tilde{x}_i, \tilde{y}_i) : i = 1,\dots,m\}$ are treated as the current observations and the samples $\hat{\mathbb{P}}^{(k)}_b:= \{(x_i, y_i) : i = 1,\dots,n_k\}$ are treated as the future observations, where the sample sizes in $\hat{\mathbb{P}}^{(k)}_b$ and $\tilde{\mathbb{P}}^{(k)}_b$ are denoted as $n_k$ and $m$, respectively. The index $(b, -k)$ on each indexed observation $(x, y)$ are omitted for simplicity.
 Recall that estimating the imputation model $\mathbb{Q}_{\hat{\theta}_b}^{(k)}$ aims to minimize the objective function 
 \[
E_{(X, Y)\sim{\hat{\mathbb{P}}^{(k)}_b}}L(\theta|X, Y)
= % \frac{1}{n}\sum_{i=1}^nL(\theta|X_i, y_i) + \pi(\theta)
G_{{\tilde{\mathbb{P}}^{(k)}_b}}(\theta, \lambda) 
+ 
V_{{\hat{\mathbb{P}}^{(k)}_b}, {\tilde{\mathbb{P}}^{(k)}_b}}(\theta, \lambda).
 \]
% Similar to the reasoning in Section~\ref{proof:t1}, we construct the restricted parameter space denoted by $\Theta_{\mathcal{A}}$, as specified in (\ref{eq:appendix:restricted-parameter-space}).  
By regularity conditions C1---C4 for the loss function in Section~\ref{appendix:regularity-conditions}, with the same ULLN argument in Section~\ref{proof:t1}, we have
\begin{equation}\label{eq:appendix-ULLN-4}
\sup_{\theta \in \Theta_{\mathcal{A}}^{(n)}} \displaystyle\left\lvert  E_{(X,Y) \sim \hat{\mathbb{P}}^{(k)}_b}L(\theta|X,Y) - E_{(X,Y) \sim \mathbb{P}}L(\theta|X,Y) \displaystyle\right\rvert \xrightarrow{p} 0
\end{equation}
where $\Theta_{\mathcal{A}}^{(n)}$ is the restricted parameter space specified in (\ref{eq:appendix:restricted-parameter-space}).
The (strong) ULLN for the $m$-out-of-$n$ bootstrap sample mean is proved in \cite[Theorem~1]{spencer2023}, requiring the additional conditions: there exists $\delta \in [0, 1)$ such that
\begin{equation}\label{eq:appendix-SULLN-condition}
    \lim _{n_k \rightarrow \infty} \frac{n_k^{1-\delta} \log (n_k)}{m_{n_k}}=0 \quad \text { and } \quad \mathbb{E}\left(\sup _{\theta \in \Theta_{\mathcal{A}}^{(n)}}\left|L\left(\theta| X, Y\right)\right|^{\frac{1}{1-\delta}}\right)<\infty,
\end{equation}
where $n_k$ is the sample size of the future observations used to obtain the estimated model $\mathbb{Q}_{\hat{\theta}_b}^{(k)}$. The resampling sample size used to formulate current observations is denoted by $m_{n_k} = \lceil \tilde{\alpha} n_k\rceil$, where $\tilde{\alpha}$ denotes the resampling ratio used in the $m$-out-of-$n$ bootstrap. By assumption, there exists a constant $\tilde{\alpha}_0 \in (0,1)$ such that the resampling ratio $\tilde{\alpha}$ has $\tilde{\alpha} \geq \tilde{\alpha}_0$ for every $n \geq N$, where $N$ is some positive integer. Thereby, for any $\delta \in (0,1)$, we have 
% Given the parameter $K$ used for data-splitting is fixed, we have for any $\delta \in (0,1)$
\begin{align}
\begin{split}\label{eq:appendix-SULLN-condition-2}
   \lim _{n_k \rightarrow \infty} \frac{n_k^{1-\delta} \log (n_k)}{m_{n_k}} &\leq  \lim _{n_k \rightarrow \infty} \frac{n_k^{1-\delta} \log (n_k)}{\tilde{\alpha}_0 n_k} \\
   &= \lim _{n_k \rightarrow \infty} \frac{C \cdot \log (n_k)}{ n_k^{\delta}} \\
   &= 0,
   \end{split}
\end{align}
where $C = 1/\tilde{\alpha}_0$ is a constant. Thus, the condition specified in (\ref{eq:appendix-SULLN-condition}) is satisfied given the regularity condition C5 in Section~\ref{appendix:regularity-conditions}. By assumption, the parameter $K$ used for data-splitting is fixed, implying $n_k \to \infty$ as $n \to \infty$. Together with \cite[Theorem~1]{spencer2023} and (\ref{eq:appendix-SULLN-condition-2}), this validates the ULLN for the bootstrap samples obtained by AM:
\begin{equation}\label{eq:appendix-ULLN-5}
\sup_{\theta \in \Theta_{\mathcal{A}}} \displaystyle\left\lvert  E_{(X,Y) \sim \tilde{\mathbb{P}}^{(k)}_b}L(\theta|X,Y) - E_{(X,Y) \sim \mathbb{P}}L(\theta|X,Y)\displaystyle\right\rvert \xrightarrow{p} 0.
\end{equation}
% Given (\ref{eq:appendix-ULLN-4}), using the same triangular inequality argument in \ref{proof:l1}, we have
% \[
% \sup_{\theta \in \Theta^{(n)}} \displaystyle\left\lvert E_{(X,Y) \sim \hat{\mathbb{P}}^{(k)}_b}L(\theta|X,Y) -  E_{(X,Y) \sim \tilde{\mathbb{P}}^{(k)}_b}L(\theta|X,Y)  \displaystyle\right\rvert \xrightarrow{p} 0,
% \]
% The assumption $D_{KL}(\mathbb{P}, \hat{\mathbb{P}}) \xrightarrow{p} 0$ implies $D_{KL}(\mathbb{P}, \hat{\mathbb{P}}^{(k)}_b) \xrightarrow{p} 0$.
% The regularity condition C1---C4 can be used to show the ULLN on the unrestricted parameter space, that is,
% \[
% \sup_{\theta \in \Theta^{(n)}} \displaystyle\left\lvert  \frac{1}{n_k}\sum_{i=1}^{n_k} L(\theta |x_i,y_i) - E_{(X,Y) \sim \mathbb{P}}L(\theta|X,Y)\displaystyle\right\rvert \xrightarrow{p} 0
% \]
% which further implies the convergence of the empirical risk minimizer such that
% \begin{equation}\label{eq:appendix-convergence-Q-2}
%        E_{(X,Y) \sim \hat{\mathbb{P}}^{(k)}_b}L(\hat{\theta}^{*}|X,Y)  \xrightarrow{p} E_{(X,Y) \sim \mathbb{P}}L(\theta^*|X,Y)
% \end{equation}
% for some optimal estimate $\theta^* \in \Theta_{*}^{(n)}$, and $\hat{\theta}^{*} \in \Theta^{(n)}$ is some minimizer of the expected loss function with regard to $(X,Y) \sim \hat{\mathbb{P}}^{(k)}_b$. 
The uniform convergence in (\ref{eq:appendix-ULLN-5}) indicates the convergence of the empirical risk minimizer to the true population risk minimizer \citep{shalev-shwartz2010}, that is,
\begin{equation}\label{eq:appendix-convergence-Q-3}
       E_{(X,Y) \sim \hat{\mathbb{P}}^{(k)}_b}L(\hat{\theta}^{*}|X,Y)  \xrightarrow{p} E_{(X,Y) \sim \mathbb{P}}L(\theta^*|X,Y)
\end{equation}
for some optimal estimate $\theta^* \in \Theta_{*}^{(n)}$, and $\hat{\theta}^{*} \in \Theta^{(n)}$ is some minimizer of the expected loss function with regard to $(X,Y) \sim \hat{\mathbb{P}}^{(k)}_b$. 

Following the same arguments in Section~\ref{proof:t1}, we construct the restricted parameter space $\Theta_{\mathcal{A}}^{(n)} \subseteq \Theta^{(n)}$ specified in (\ref{eq:appendix:restricted-parameter-space}). We further define the $\lambda$-dependent AM estimator in the restricted parameter space $\Theta_{\mathcal{A}}^{(n)} \subseteq \Theta^{(n)}$ as
\begin{equation}\label{eq:appendix:theta-3}
    (\hat{\theta}^{*}_{\mathcal{A}})_\lambda =\underset{\theta \in \Theta_{\mathcal{A}}}{\text{arg min}} \, \left( \frac{1}{m}\sum_{i=1}^m L(\theta |\tilde{x}_i,\tilde{y}_i) + \pi(\theta , \lambda) \right)
\end{equation}
for any given $\lambda \in \Lambda^{(n)}$, and the duality parameter $\lambda$ is chosen at $\lambda = \hat{\lambda}$ with
\begin{equation}\label{eq:appendix:lambda-3}
    \hat{\lambda} =  \underset{\lambda \in \Lambda^{(n)}}{\text{arg min}} \, E_{(X,Y) \sim \hat{\mathbb{P}}^{(k)}_b}L\left((\hat{\theta}^{*}_{\mathcal{A}})_\lambda |X,Y\right).
\end{equation}
We then consider $\lambda_0 = \boldsymbol{0}$, which is assumed that $\lambda_0 \in \Lambda^{(n)}$ by the regularity condition C8 for the duality function in Section~\ref{appendix:regularity-conditions}. Such $\lambda_0$ results in $\pi(\theta, \lambda_0) = C$ for any $\theta \in \Theta_{\mathcal{A}}^{(n)}$, where $C \in \mathbb{R}$ is a constant, by the same regularity condition. By (\ref{eq:appendix-ULLN-4}), we have 
\begin{equation}\label{eq:appendix-theta-intermediate-2}
E_{(X,Y) \sim \hat{\mathbb{P}}^{(k)}_b}L\left((\hat{\theta}^{*}_{\mathcal{A}})_{\lambda_0} |X,Y\right) \xrightarrow{p}  E_{(X,Y) \sim \mathbb{P}}L\left((\hat{\theta}^{*}_{\mathcal{A}})_{\lambda_0} |X,Y\right).
\end{equation}
By (\ref{eq:appendix:theta-3}), $(\hat{\theta}^{*}_{\mathcal{A}})_{\lambda_0}$ is the empirical risk minimizer in the parameter space $\Theta_{\mathcal{A}}^{(n)}$. The uniform convergence of the sample loss in (\ref{eq:appendix-ULLN-5}) indicates the convergence of its minimizer to the true population risk minimizer \citep{shalev-shwartz2010}. As a result, it follows from (\ref{eq:appendix-theta-intermediate-2}) and (\ref{eq:appendix-thm1-1}) that
\begin{equation}\label{eq:appendix:estimation-convergence-2}
   E_{(X,Y) \sim \hat{\mathbb{P}}^{(k)}_b}L\left((\hat{\theta}^{*}_{\mathcal{A}})_{\lambda_0} |X,Y\right) \xrightarrow{p} E_{(X,Y) \sim \mathbb{P}}L(\theta^*|X,Y).
\end{equation}
With (\ref{eq:appendix-convergence-Q-3}) and \eqref{eq:appendix:estimation-convergence-2}, the exact same argument in the relevant part of Section~\ref{proof:t2} can be used by simply replacing $\mathbb{Q}$ with $\hat{\mathbb{P}}^{(k)}_b$, which concludes that
\[
D_{KL}(\mathbb{P}, \mathbb{Q}_{\hat{\theta}_b}^{(k)}) \xrightarrow{p}  D_{KL}(\mathbb{P}, \mathbb{P}_{\theta^{*}})
\]
for the arbitrary pair $(b, k)$. Thus, when the model is valid,
\[
D_{KL}(\mathbb{P}, \mathbb{Q}_{\hat{\theta}_b}^{(k)}) \xrightarrow{p} \lim_{n \to \infty} D_{KL}(\mathbb{P}, \mathbb{P}_{\theta^*}) = 0.
\]
for the arbitrary pair $(b, k)$. Denote the final imputation distribution as the mixture distribution
\[
\mathbb{Q} := \frac{1}{B \cdot K}\sum_{b=1}^B \sum_{k=1}^K\mathbb{Q}_{\hat{\theta}_b}^{(k)}
\]
for fixed $K > 0$ and $B > 0$. The convexity of the KL divergence implies
\[
D_{KL}(\mathbb{P}, \mathbb{Q}) \leq \frac{1}{B \cdot K}\sum_{b=1}^B \sum_{k=1}^K D_{KL}(\mathbb{P}, \mathbb{Q}_{\hat{\theta}_b}^{(k)}).
\]
Since the KL divergence is non-negative, for any $\epsilon > 0$, we have
\[
\lim_{n \to \infty}P\left(D_{KL}(\mathbb{P}, \mathbb{Q}) \geq \epsilon\right) \leq \lim_{n \to \infty}\sum_{b=1}^B \sum_{k=1}^K P\left(D_{KL}(\mathbb{P}, \mathbb{Q}_{\hat{\theta}_b}^{(k)}) \geq \frac{\epsilon}{B \cdot K} \right) = 0{\color{blue},}
\]
by the union bound, which concludes that
\[
D_{KL}(\mathbb{P}, \mathbb{Q}) \xrightarrow{p} 0.
\]
\qed

\subsubsection{Theoretical Discussion on the Convergence Rate}

In this section, we provide additional insights into the convergence rate of the estimation error $	D_{KL}(\mathbb{P}, \mathbb{P}_{\hat{\theta}_{AM}}) - D_{KL}(\mathbb{P}, \mathbb{P}_{\theta^*})$
to zero. Specifically, we explore how fast the population generated by the AM estimated model converges to its optimum. 

Firstly, we define the empirical risk minimizer regarding $\hat{\mathbb{P}}$ as 
\[
 \hat{\theta}^* =  \underset{\theta \in \Theta^{(n)}}{\text{arg min}} E_{(X, Y)\sim \hat{\mathbb{P}}}L(\theta|X, Y).
\]
By writing the estimation error defined in Definition 3 as
% Following \cite{Shen2024}, which is a work utilizing the synthetic data for effectively performing statistical tasks under a different setting, we assume that the imputed population $\mathbb{Q}$ is effective in the sense that
% \begin{equation}\label{sp:eq-convergence-rate-1}
%     D_{KL}(\mathbb{P}, \mathbb{Q}_{\tilde{\theta}^*}) \leq D_{KL}(\mathbb{P}, \mathbb{P}_{\hat{\theta}^*})
% \end{equation}
% where 
% \[
%  \tilde{\theta}^* =  \underset{\theta \in \Theta^{(n)}}{\text{arg min}} E_{(X, Y)\sim \mathbb{Q}}L(\theta|X, Y)
% \]
% \[
%  \hat{\theta}^* =  \underset{\theta \in \Theta^{(n)}}{\text{arg min}} E_{(X, Y)\sim \hat{\mathbb{P}}}L(\theta|X, Y)
% \]
% are the ERM on $\mathbb{Q}$ and $\hat{\mathbb{P}}$, correspondingly. (\ref{sp:eq-convergence-rate-1}) suggests that the model estimation conducted on the imputed population $\mathbb{Q}$ provides closer population to the true population than the model estimation conducted on the empirical samples. By viewing $\mathbb{Q}$ as the synthetic data generated from the empirical population $\hat{\mathbb{P}}$, such gain in effectiveness from the synthetic data has been studied in the literature \citep{Shen2024}.  
\begin{align*}
& \ \ \ \ D_{KL}(\mathbb{P}, \mathbb{P}_{\hat{\theta}_{AM}}) - D_{KL}(\mathbb{P}, \mathbb{P}_{\theta^*})\\
&=E_{(X, Y)\sim \mathbb{P}}L(\hat{\theta}_{AM}|X, Y) - E_{(X, Y)\sim \mathbb{P}}L(\theta^*|X, Y) \\
&= E_{(X, Y)\sim \mathbb{P}}L(\hat{\theta}_{AM}|X, Y) - E_{(X, Y)\sim \mathbb{P}}L(\hat{\theta}^*|X, Y) + E_{(X, Y)\sim \mathbb{P}}L(\hat{\theta}^*|X, Y) - E_{(X, Y)\sim \mathbb{P}}L(\theta^*|X, Y)\\
&= \left(E_{(X, Y)\sim \mathbb{P}}L(\hat{\theta}^*|X, Y) - E_{(X, Y)\sim \mathbb{P}}L(\theta^*|X, Y) \right) - \left( E_{(X, Y)\sim \mathbb{P}}L(\hat{\theta}^*|X, Y) - E_{(X, Y)\sim \mathbb{P}}L(\hat{\theta}_{AM}|X, Y)\right),
\end{align*}
we observe that the first term
\[
E_{(X, Y)\sim \mathbb{P}}L(\hat{\theta}^*|X, Y) - E_{(X, Y)\sim \mathbb{P}}L(\theta^*|X, Y)
\]
represents the effectiveness gap between the empirical risk minimizer and the population risk minimizer. Meanwhile, the second term
\[
E_{(X, Y)\sim \mathbb{P}}L(\hat{\theta}^*|X, Y) - E_{(X, Y)\sim \mathbb{P}}L(\hat{\theta}_{AM}|X, Y)
\]
reflects the gain from the AM estimator compared to the empirical risk minimizer. The convergence rate of the first term has been extensively studied in literature, with a classic lower bound of $\tilde{O}(\sqrt{d/n})$ established under mild assumptions \citep{Liu2018}. The second term regarding the gain of the AM estimator depends on  specific scenarios and the imputation strategy used. A more robust imputation strategy designed to a specific scenario will yield greater gains. This brief discussion sets the stage for conducting more detailed convergence rate analyses in specific scenarios.

\clearpage
\subsection{Imputation with Finite-sample Valid Predictive Coverage} \label{ss:finite-sample-valid-coverage}
\subsubsection{Theoretical Support} \label{ss:finite-sample-valid-coverage-theory}
Here, we study the theoretical properties of the resampling-based imputation scheme in the finite-sample scenario. As proposed in Section~2.3, we utilize an adaptive $m$-out-of-$n$ boostrap resampling scheme. Here, $m$ is selected to ensure uniformly distributed cumulative distribution function (CDF) values of the data, as detailed in (11) and (12). Controlling the effectiveness of the data distribution in this manner can lead to desirable theoretical properties, which is summarized into the following theorem.
% \addtocounter{theorem}
\begin{theorem}[Valid Predictive Coverage] \label{theorem:3}
%Following the definition of Corollary~\ref{corollary:1}, assume $|\mathcal{A}(\theta^*)| < \infty$ for any $\theta^* \in \Theta_*^{(n)}$.
Consider an imputation model with the distribution function $\hat{F}(\cdot|x)$. Suppose that for an unseen data point $(x_*, y_*) \sim \mathbb{P}$, it holds that
\(
\hat{F}(y_* | x_*) \sim \text{Uniform}(0,1).
\)
Then, for any $\alpha \in (0,1)$ and an interval $[a_0,b_0]$ such that
$ \hat{F}(b_0|x_*) - \hat{F}(a_0|x_*) = 1-\alpha$, 
we have
\[
Prob(a_0 \leq y_* \leq b_0) = 1-\alpha.
\]
\end{theorem}
\begin{proof}
    % This follows directly from the condition of the standard uniform distribution.
Without loss of generality, assume $x$ is a random variable with a continuous distribution function $G(\cdot)$. We have
\begin{align*}
    Prob(a_0 \leq y_* \leq b_0) &= \int \int_{a_0}^{b_0}d\hat{F}(y_*|x_*) dG(x_*) \\
    &= (1-\alpha)  \int dG(x_*) \\
    &=  1- \alpha
\end{align*}
given that \(
\hat{F}(y_* | x_*) \sim \text{Uniform}(0,1).
\)
\end{proof}

Thus, imposing (12) ensures a $(1-\alpha)$ coverage rate of the observed data (unseen by the imputation model) using the $(1-\alpha)$ predictive interval of the imputation model for any $\alpha \in (0,1)$. That is,  suppose $(x, y) \sim \mathbb{\hat{P}}$, consider an interval $[a_0,b_0]$ such that
\[ 
\sum_{k=1}^K \left(\hat{F}^{(k)}_{\hat{\theta}_b}(b_0 | x)  - \hat{F}^{(k)}_{\hat{\theta}_b}(a_0 | x)\right) \cdot \mathbbm{1}\left(x \in \{\bx_{b}^{(k)}\}\right) = 1 - \alpha
\]
where $\mathbbm{1}(\cdot)$ denotes the indicator function and $\{\bx_b^{(k)}\}$ denotes the holdout set of $x$ in the $k$-th fold. Then, we have
\[
Prob(a_0 \leq y \leq b_0) = 1-\alpha.
\]
The coverage rate is an unbiased estimate of the true predictive coverage rate due to the use of the holdout data unseen by the imputation model, and the variance decreases with both the observed sample size $n$ and the number of imputation models $B\cdot K$. This property provides a strong theoretical foundation for its similarity to the true but unknown data distribution. 

\subsubsection{Empirical Evidence}

In this section, we present empirical evidence to support the finite-sample predictive coverage theory discussed above in Section~\ref{ss:finite-sample-valid-coverage-theory}. We consider the $n < p$ linear regression scenario from Section~5.2, with different true model sparsity and signal-to-noise ratios (SNR) controlled by $\alpha$ and $\tau$. For the first repetition in each setting, we generate $1,000$ new data points from the underlying true model and assess the coverage rate of the mixture of $B\cdot K$ imputation models.

Algorithm~3 is used to impose (12). We first check the condition in Theorem~\ref{theorem:3} with the $1,000$ new data points. That is, for each new data point $(x_*, y_*) \sim \mathbb{P}$, it is required that
\(
\hat{F}(y_* | x_*) \sim \text{Uniform}(0,1)
\), where $\hat{F}(\cdot|x)$ is the distribution function of the mixture of $B \cdot K$ imputation models. The results are demonstrated in the Q-Q plots in Figure~\ref{fig:finite-sample-valid-coverage}. It can be seen that Algorithm~3 is effective in approximating this target. 

To construct the $(1-\alpha_0)$ predictive interval where $\alpha_0 \in (0,1)$,  for each $x_*$, we sample $N = 10,000$ values of $\hat{y}_*$ from the mixture of $B \cdot K$ imputation models. Then, we take the $\alpha_0/2$-th and $(1-\alpha_0/2)$-th quantiles of the $\hat{y}_*$ values as the lower bound and upper bound of the predictive interval respectively. Table~{\ref{tab:finite-sample-valid-coverage}} displays the coverage rates of the $1,000$ new data points with these constructed $(1- \alpha_0)$ predictive intervals across different values of $\alpha_0$. It can be seen that the coverage is satisfactory across various settings.

    \ifthenelse{1=0}{
    %       \centerline{--- Fig. \ref{fig:procedure} ---}
    }{

    \begin{figure}[!htbp]
    \centering
    \includegraphics[width=14cm]{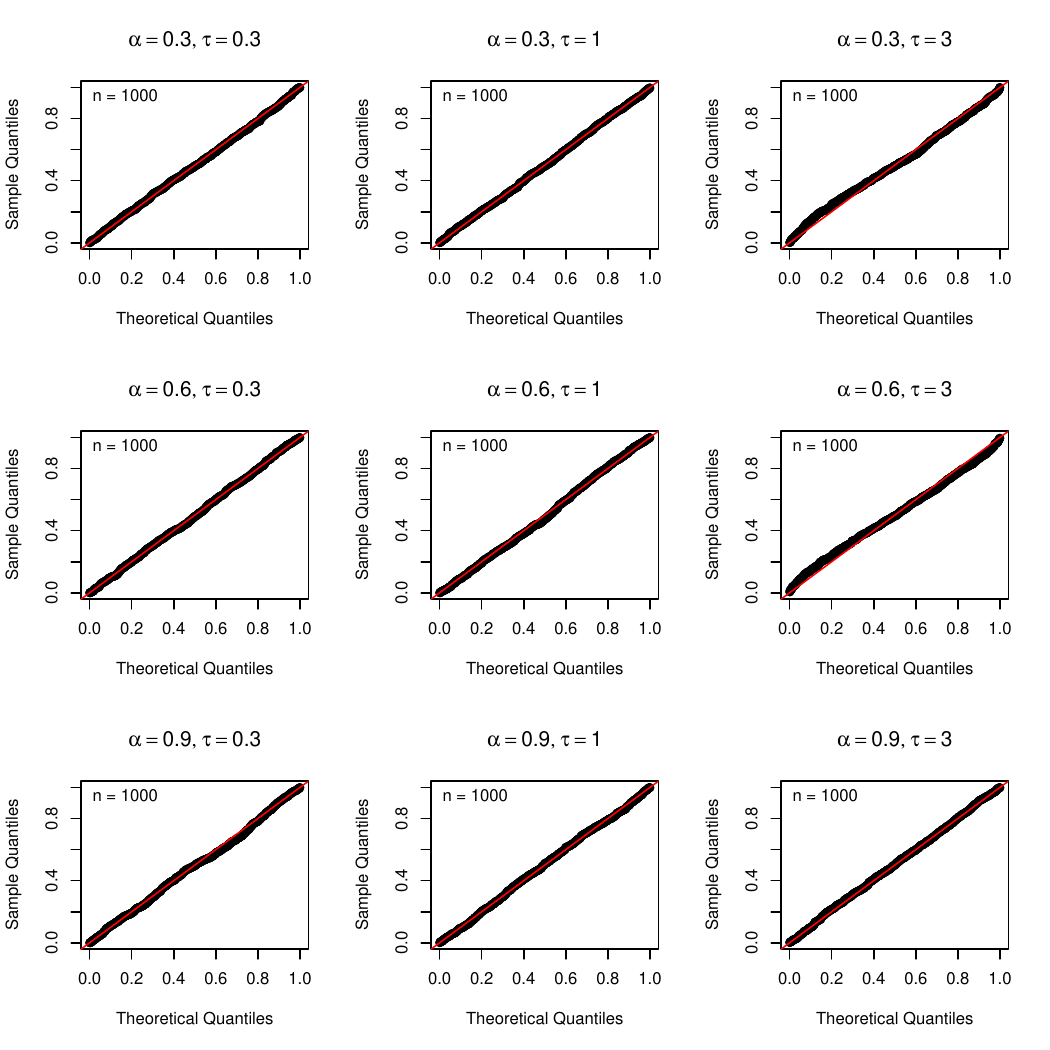}
    \caption{Q-Q plots of the estimated CDF values of the 1,000 new data points with the mixture of $B\cdot K$ imputation models obtained using Algorithm~2, against the standard uniform distribution, in the linear regression example (Section~5.2). Presented here are the results from the first repetition across various settings, using the (unweighted) $L_1$ duality function.}\label{fig:finite-sample-valid-coverage}
    \end{figure}
    }

\begin{table}[!htbp]
\caption{Coverage rate of the 1,000 new data points with the $(1-\alpha_0)$ predictive interval (estimated by Monte-Carlo sampling) of the mixture of $B\cdot K$ imputation models, with different values of $\alpha_0$, in the linear regression example (Section~5.2). Presented here are the results from the first repetition across various settings, using the (unweighted) $L_1$ duality function. \label{tab:finite-sample-valid-coverage}}
\centering
% \tiny
\scriptsize
\begin{tabular}{cccccccccc}
\hline
\textit{Setting}           & {\ul 0.90} & {\ul 0.80} & {\ul 0.70} & {\ul 0.60} & {\ul 0.50} & {\ul 0.40} & {\ul 0.30} & {\ul 0.20} & {\ul 0.10} \\
$\alpha = 0.3, \tau = 0.3$ & 0.909      & 0.816      & 0.722      & 0.621      & 0.518      & 0.423      & 0.313      & 0.219      & 0.106      \\
$\alpha = 0.3, \tau = 1$   & 0.919      & 0.812      & 0.705      & 0.607      & 0.497      & 0.39       & 0.295      & 0.198      & 0.105      \\
$\alpha = 0.3, \tau = 3$   & 0.936      & 0.857      & 0.766      & 0.669      & 0.561      & 0.45       & 0.338      & 0.238      & 0.129      \\
$\alpha = 0.6, \tau = 0.3$ & 0.889      & 0.789      & 0.697      & 0.604      & 0.508      & 0.404      & 0.298      & 0.195      & 0.085      \\
$\alpha = 0.6, \tau = 1$   & 0.883      & 0.786      & 0.681      & 0.587      & 0.485      & 0.379      & 0.285      & 0.181      & 0.084      \\
$\alpha = 0.6, \tau = 3$   & 0.946      & 0.876      & 0.78       & 0.68       & 0.566      & 0.461      & 0.351      & 0.24       & 0.115      \\
$\alpha = 0.9, \tau = 0.3$ & 0.894      & 0.804      & 0.708      & 0.602      & 0.502      & 0.424      & 0.327      & 0.226      & 0.13       \\
$\alpha = 0.9, \tau = 1$   & 0.905      & 0.796      & 0.703      & 0.61       & 0.492      & 0.384      & 0.299      & 0.201      & 0.098      \\
$\alpha = 0.9, \tau = 3$   & 0.901      & 0.791      & 0.707      & 0.618      & 0.51       & 0.409      & 0.301      & 0.208      & 0.097      \\ \hline
\end{tabular}
\end{table}

\clearpage
\subsection{Implementation Details in the Many-Normal-Means Example }\label{appendix:mnm}
\subsubsection{Technical Details}\label{appendix:mnm-technical}
The problem of many normal means in this paper uses the model
\[
	Y_i|\mu_i \sim N(\mu_i, 1), \qquad i = 1,\dots,n,
\]
with
\(
\Prob{\mu_i =\eta_j} =\alpha_j, %\qquad k=1,...,l,
\)
$k=1,...,l,$
%for $i = 1,\dots, n$. Additional constraints are 
where $\eta_1\leq \eta_2 \leq \ldots\leq \eta_l$. The \dual~function takes the form
\[
	\pi(\eta,\lambda)
	=\sum_{j=2}^{l}\lambda_k \left(\eta_j - \eta_{j-1}\right) \qquad(\lambda_j\geq 0 \mbox{ for }j=2,\dots,l).
\]
For convenience, let $\lambda_1 = \lambda_{l+1} = 0$.
The loss function takes the form
\[
L(\alpha, \eta | Y = y) = - \frac{1}{n}\ln \left(\sum_{i=1}^n\sum_{j=1}^l\alpha_j e^{-\frac{(y_i-\eta_j)^2}{2}}\right).
\]
%Denote $\lambda_1 = \lambda_{l+1} = 0$. 

In this section, we elaborate on the derivation of estimation of imputation models. First, we obtain the data represented by $\hat{\mathbb{P}}^{(k)}_b$ and $\tilde{\mathbb{P}}^{(k)}_b$ following the data-splitting and resampling scheme in Algorithm~2 at iteration $(b, k)$, for any arbitrary $b = 1,\dots, B$ and $k = 1,\dots,K.$ For simplicity, we omit the index $(b, -k)$ on each indexed observation $(x, y)$. The samples $\tilde{\mathbb{P}}^{(k)}_b:= \{(\tilde{x}_i, \tilde{y}_i) : i = 1,\dots,m\}$ are treated as the current observations and the samples $\hat{\mathbb{P}}^{(k)}_b:= \{(x_i, y_i) : i = 1,\dots,n_k\}$ are treated as the future observations, where the sample sizes in $\hat{\mathbb{P}}^{(k)}_b$ and $\tilde{\mathbb{P}}^{(k)}_b$ are denoted as $n_k$ and $m$ respectively. Using the matrix notation, the two samples are denoted by $(\bx, \by)$ and $(\tilde{\bx}, \tilde{\by})$, omitting the $(b, -k)$ indexing used in Algorithm~2.

% For ease of explanation, we simplify our notation, using $\hat{\mathbb{P}}$ to denote $\hat{\mathbb{P}}^{(k)}_b$ and $\tilde{\mathbb{P}}$ for $\tilde{\mathbb{P}}^{(k)}_b$, as explained in Algorithm~1, since the derivation applies to any $k = 1,\dots, K$ and $b = 1, \dots, B$.  Here, the sample from $\hat{\mathbb{P}}$ is represented as $Y$ with a sample size of $n$, and the sample from $\tilde{\mathbb{P}}$ is denoted as $\tilde{Y}$, with a sample size of $m$.

The estimation of $(\alpha, \eta, \lambda)$ should satisfy
\begin{align*}
    0=
\frac{ \partial G_{\tilde{\mathbb{P}}^{(k)}_b}(\alpha, \eta,  \lambda)}{\partial \alpha_j} 
%&= \frac{\partial L(\alpha, \eta | \tilde{Y})}{\partial \alpha_i}\\
= \frac{\partial L(\alpha, \eta | \tilde{\by})}{\partial \alpha_j}
 &= -\frac{1}{m} \sum_{i=1}^m \frac{\exp\left(-\frac{1}{2}(\tilde{y}_i - \eta_j)^2\right)}{\sum_{j=1}^l \alpha_j \exp\left(-\frac{1}{2}(\tilde{y}_i - \eta_j)^2\right)}
\end{align*}
and
\begin{align*}
    0=
\frac{ \partial G_{\tilde{\mathbb{P}}^{(k)}_b}(\alpha, \eta,  \lambda)}{\partial \eta_j} 
&= \frac{\partial L(\alpha, \eta | \tilde{\by})}{\partial \eta_j} + \frac{\partial \pi(\eta,\lambda)}{\partial \eta_j}\\
 &= -\frac{1}{m} \sum_{i=1}^m \frac{\alpha_j \exp\left(-\frac{1}{2}(\tilde{y}_i - \eta_j)^2\right) (\tilde{y}_i - \eta_j)}{\sum_{j=1}^l \alpha_j \exp\left(-\frac{1}{2}(\tilde{y}_i - \eta_j)^2\right)} + \left(\lambda_j -\lambda_{j+1}\right)
\end{align*}
for $j = 1,\dots, l$ and
\begin{align*}
   \lambda %&
   =  \min_\lambda \left\|\frac{ \partial V_{\hat{\mathbb{P}}^{(k)}_b,\tilde{\mathbb{P}}^{(k)}_b}(\alpha, \eta, \lambda)}{\partial \eta}
\right\|_2^2 %\\
&=  \min_{\tilde{\lambda}} \left\| d_\eta - \tilde{d}_\eta - M \tilde{\lambda} \right\|_2^2,
\end{align*}
where
\[
d_\eta = \left(\frac{\partial L(\alpha, \eta | \by)}{\partial \eta_1}, \dots, \frac{\partial L(\alpha, \eta | \by)}{\partial \eta_l} \right)^T
\]
\[
\tilde{d}_\eta = \left(\frac{\partial L(\alpha, \eta | \tilde{\by})}{\partial \eta_1}, \dots, \frac{\partial L(\alpha, \eta | \tilde{\by})}{\partial \eta_l} \right)^T
\]
\begin{equation}\label{eq:appendix-mnm-M}
M = \begin{bmatrix}
1 & 0 & 0 & \cdots & 0 & 0 \\
-1 & 1 & 0 & \cdots & 0 & 0 \\
0 & -1 & 1 & \cdots & 0 & 0 \\
\vdots & \vdots & \vdots & \ddots & \vdots & \vdots \\
0 & 0 & 0 & \cdots & 1 & 0 \\
0 & 0 & 0 & \cdots & -1 & 1 \\
0 & 0 & 0 & \cdots & 0 & -1 \\
\end{bmatrix}_{l \times (l-1)}
\end{equation}
\[
\tilde{\lambda} = \left(\lambda_2, \dots, \lambda_l\right).
\]
Following Section~3, the estimation can be done with the projected gradient descent approach \citep{boyd2003subgradient} to deal with the constraint on $\alpha$, $\eta$ and $\lambda$. Specifically, for $\alpha$, we have
\[
\alpha^{(t+1/2)}_j = \alpha^{(t)}_j - \gamma_1 \cdot \left(-\frac{1}{m} \sum_{i=1}^m \frac{\exp\left(-\frac{1}{2}(\tilde{y}_i - \eta_j^{(t)})^2\right)}{\sum_{j=1}^l \alpha_j^{(t)} \exp\left(-\frac{1}{2}(\tilde{y}_i - \eta_j^{(t)})^2\right)} \right)
\]
for $i = 1,\dots,l$, where $\gamma_1$ is the step size, followed with solving
\begin{equation}\label{eq:appendix-mnm-1}
\begin{aligned}
& \underset{\alpha^{(t+1)}}{\text{minimize}}
& & \frac{1}{2} \left\| \alpha^{(t+1)} - \alpha^{(t + \frac{1}{2})} \right\|_2^2 \\
& \text{subject to}
& & 0 \leq \alpha^{(t+1)} \leq 1, \\
&&& \mathbf{1}^T \alpha^{(t+1)} = 1.
\end{aligned}
\end{equation}
Note that (\ref{eq:appendix-mnm-1}) is essentially the problem of projecting a vector onto the standard or probability simplex, where many existing solutions are available and we use the method in \cite{Blondel2014}. For $\eta$, we have
\[
\eta^{(t+1/2)}_j = \eta^{(t)}_j - \gamma_2 \cdot \left(-\frac{1}{m} \sum_{i=1}^m \frac{\alpha_j^{(t)} \exp\left(-\frac{1}{2}(\tilde{y}_i - \eta_j^{(t)})^2\right) (\tilde{y}_i - \eta_j^{(t)})}{\sum_{j=1}^l \alpha_j^{(t)} \exp\left(-\frac{1}{2}(\tilde{y}_i - \eta_j^{(t)})^2\right)} + \left(\lambda_j^{(t)} -\lambda_{j+1}^{(t)}\right) \right)
\]
for $j = 1,\dots,l$, where $\gamma_2$ is the step size, followed with finding
\begin{equation*}\label{eq:appendix-mnm-2}
\begin{aligned}
& \underset{\eta^{(t+1)}}{\text{minimize}}
& & \frac{1}{2} \left\| \eta^{(t+1)} - \eta^{(t + \frac{1}{2})} \right\|_2^2 \\
& \text{subject to}
& & \eta_1^{(t+1)} \leq \eta_2^{(t+1)} \leq \dots \leq \eta_l^{(t+1)},
\end{aligned}
\end{equation*}
or equivalently,
\begin{equation}\label{eq:appendix-mnm-3}
\begin{aligned}
& \underset{\eta^{(t+1)}}{\text{minimize}}
& & \frac{1}{2} \eta^{(t+1)}{}^T\eta^{(t+1)} - \eta^{(t + \frac{1}{2})}{}^T\eta^{(t+1)} \\
& \text{subject to}
& & M^T\eta^{(t+1)} \leq \mathbf{0},
\end{aligned}
\end{equation}
where $M$ is defined in (\ref{eq:appendix-mnm-M}). Note that (\ref{eq:appendix-mnm-3}) is a canonical quadratic programming problem with inequality constraint. For $\lambda$, we have 
\[
\lambda^{(t+1)} = \lambda^{(t)} - \gamma_3 \cdot \left(2 M^TM \lambda^{(t)} - 2 M^T \left(d_{\eta^{(t+1)}} - \tilde{d}_{\eta^{(t+1)}}\right)\right),
\]
where $\gamma_3$ is the step size.

The estimation procedure for an imputation model is considered complete once both $\eta^{(t)}$ and $\alpha^{(t)}$ have converged. The resulting model, characterized by the parameters $(\hat{\eta}^{(k)}_b, \hat{\alpha}^{(k)}_b)$ is then applied on the hold-out set $\by^{(k)}_b$. Depending on the objective, this can be used for the purpose of imputing future observations using Algorithm~2, or for assessing the efficiency of the imputation model to select the resampling parameter with Algorithm~3.

Performing Algorithm~3 for selecting resampling parameter $\tilde{\alpha}$ requires the calculation of CDF values with a single imputation model estimation iteration ($B = 1$). The CDF value can be calculated with
\[
F(y^{(k)}_{bi}) = \sum_{j = 1}^{l} \hat{\alpha}^{(k)}_{bj} \Phi \left(y^{(k)}_{bi} - \hat{\eta}^{(k)}_{bj} \right),
\]
for $i = 1,\dots, n-n_k$, where $b = 1$ and $\Phi(\cdot)$ represents the standard normal CDF. The hold-out set is denoted as $(\bx^{(k)}_b, \by^{(k)}_b) := \{(x^{(k)}_{bi}, y^{(k)}_{bi}):i=1,\dots, n-n_k\}$. After iterating through $k = 1,\dots,K$, we can obtain $n$ such CDF values. This collection of CDF values is then tested against the standard uniform distribution with the KS-test for the selection of the best $\tilde{\alpha}$.

Note that performing Algorithm~3 already yields an imputation model obtained with the selected $\tilde{\alpha}$. Thus, Algorithm~2 only requires $B - 1$ iterations. In Algorithm~2, future observations are created with the mixture of normal distributions %expressed as
\[
y^{(k)}_{*bi} \sim \sum_{j=1}^l \hat{\alpha}^{(k)}_{bj} N(\hat{\eta}^{(k)}_{bj}, 1)
\]
for $i = 1,\dots, n_k$.  By iteratively performing the estimation of the imputation model and the imputation process for $b = 1,\dots, B$ and $k = 1,\dots, K$, we construct the imputation distribution $\mathbb{Q}$, which consists of $n\cdot B$ samples.

The estimation process with Algorithm~1 follows the same implementation of estimation of imputation models as described above, with a straightforward modification: $\hat{\mathbb{P}}^{(k)}_b$ is replaced with $\mathbb{Q}$ and $\tilde{\mathbb{P}}^{(k)}_b$ with $\hat{\mathbb{P}}$, where the latter represents the full observations.

\subsubsection{Implementation Details of AM}\label{appendix:mnm-implementation}
We implemented AM for the many-normal-means example in R. The optimization problem for quadratic programming (\ref{eq:appendix-mnm-2}) is solved with the R package $\texttt{quadprog}$. 
Algorithm~2 was implemented with $B = 5$ and $K = 5$. Algorithm~3 is used to select the resampling parameter $\tilde{\alpha}$, with the candidate set $\{0.5, 0.8, 1.0, 1.2\}$. The optimization-related parameters that apply to all experiments are selected as: $\gamma_1 = 0.05, \ \gamma_2 = 0.5$, and $\gamma_3 = 0.1$.

\subsubsection{Implementation Details of Other Methods}
We implemented the James-Stein estimator in its original form. For $g$-modeling, we use the relevant function in the R package $\texttt{deconvolveR}$ \citep{Narasimhan2020}, with the default settings. The discrete support points were selected to range from the minimum to the maximum observed values of $y$, with intervals of $0.1$ between consecutive points. Additionally, we included the point at $0$ among the support points.

\clearpage
\subsection{Implementation Details in the Linear Regression Example}\label{appendix:linear}
\subsubsection{Technical Details} \label{appendix:linear-technical}
For the observed data $(\bx, \by)$, we assume that $\bx$ is standardized to have mean 0 and variance 1 for each column. The response vector $\by$ is also standardized to have mean 0 and variance 1. The loss function can be written as
\[
L(\beta, \sigma |X = x, Y = y) = \frac{1}{2}\log(\sigma^2) + \frac{1}{2\sigma^2n}\sum_{i=1}^n(y_i-x_i'\beta)^2.
\]
% \[
% L(\beta, \sigma |X, Y) = \frac{n}{2}\log(\sigma^2) + \frac{1}{2\sigma^2}\left(Y - X \beta\right)^T\left(Y - X \beta\right)
% \]
The duality function takes the form 
\(
\pi(\beta, \sigma, \lambda) = \lambda_0 \log\left(\frac{1}{\sigma^2}\right) +  \sum_{i = 1}^p \lambda_i | \beta_i | \quad (\lambda_i \geq 0, i = 0,1,\dots,p)
\)
or
\(
\pi(\beta, \sigma, \lambda) = \lambda_0 \log\left(\frac{1}{\sigma^2}\right) +  \sum_{i = 1}^p \lambda_i \beta_i^2  \quad (\lambda_i \geq 0, i = 0,1,\dots,p).
\)
Note that the estimation of $\beta$ is independent of the estimation of $\sigma$. For this reason, the estimation of $\sigma$ can be done after the estimate of $\beta$ converges to the solution. We denote the reparameterization $\phi = \log\left(\frac{1}{\sigma^2}\right)$.

In this section, we elaborate on the derivation for estimating an imputation model. First, we obtain the data represented by $\hat{\mathbb{P}}^{(k)}_b$ and $\tilde{\mathbb{P}}^{(k)}_b$ following the data-splitting and resampling scheme in Algorithm~2 at iteration $(b, k)$, for any arbitrary $b = 1,\dots, B$ and $k = 1,\dots,K.$ For simplicity, we omit the index $(b, -k)$ on each indexed observation $(x, y)$. The samples $\tilde{\mathbb{P}}^{(k)}_b:= \{(\tilde{x}_i, \tilde{y}_i) : i = 1,\dots,m\}$ are treated as the current observations and the samples $\hat{\mathbb{P}}^{(k)}_b:= \{(x_i, y_i) : i = 1,\dots,n_k\}$ are treated as the future observations, where the sample sizes in $\hat{\mathbb{P}}^{(k)}_b$ and $\tilde{\mathbb{P}}^{(k)}_b$ are denoted as $n_k$ and $m$ respectively. Using the matrix notation, the two samples are denoted by $(\bx, \by)$ and $(\tilde{\bx}, \tilde{\by})$, omitting the $(b, -k)$ indexing used in Algorithm~2.

Denotes the AM estimate of the parameters by $(\hat{\beta}^{(k)}_b, \hat{\sigma}^{(k)}_b, \hat{\lambda}^{(k)}_b)$. Denote the reparameterized estimate of the standard deviation parameter by $\hat{\phi}^{(k)}_b$. When $\beta = \hat{\beta}^{(k)}_b$, $\sigma = \hat{\sigma}^{(k)}_b$,  
and
\begin{align*}
\frac{ \partial G_{\tilde{\mathbb{P}}^{(k)}_b}(\beta, \sigma, \lambda)}{\partial \phi} = 0.
\end{align*}
Thus, we have 
\begin{align*}
   \hat{\lambda}^{(k)}_{b0} &=  \min_{\lambda_0} \left\|\frac{ \partial V_{\hat{\mathbb{P}}^{(k)}_b,\tilde{\mathbb{P}}^{(k)}_b}(\hat{\beta}^{(k)}_b, \hat{\sigma}^{(k)}_b, \lambda)}{\partial \hat{\phi}^{(k)}_b}
\right\|_2^2\\
&=  \min_{\lambda_0} \left\| \frac{\partial L(\hat{\beta}^{(k)}_b, \hat{\sigma}^{(k)}_b | \tilde{\bx}, \tilde{\by})}{\partial \hat{\phi}^{(k)}_b} \right\|_2^2,
\end{align*}
resulting in
\[
\left(\hat{\sigma}^{(k)}_b\right)^2 = \frac{1}{\left(1-2\hat{\lambda}^{(k)}_{b0}\right)m}\sum_{i=1}^m(\tilde{y}_i-\tilde{x}_i'\hat\beta^{(k)}_b)^2
\] 
and
\begin{equation*}
\hat{\lambda}^{(k)}_{b0} =
\left\{
\begin{array}{lcl}
\frac{1}{2}\left(1 - \frac{\sum_{i=1}^m(\tilde{y}_i-\tilde{x}_i'\hat{\beta}^{(k)}_b)^2/m}{\sum_{i=1}^{n_k}(y_i-x_i'\hat{\beta}^{(k)}_b)^2/n_k}\right) && \mbox{if 
%$ \bar{y} \le \bar{\tilde{y}} \le 0$ or $ \bar{y} \ge \bar{\tilde{y}} \ge 0$, {\it i.e.},
$\sum_{i=1}^m(\tilde{y}_i-\tilde{x}_i'\hat{\beta}^{(k)}_b)^2/m \leq \sum_{i=1}^{n_k}(y_i-x_i'\hat{\beta}^{(k)}_b)^2/n_k$}\\
0 && \mbox{if 
%$ \bar{y} \le \bar{\tilde{y}} \le 0$ or $ \bar{y} \ge \bar{\tilde{y}} \ge 0$, {\it i.e.},
$\sum_{i=1}^m(\tilde{y}_i-\tilde{x}_i'\hat{\beta}^{(k)}_b)^2/m > \sum_{i=1}^{n_k}(y_i-x_i'\hat{\beta}^{(k)}_b)^2/n_k$}.\\
\end{array}
\right.
\end{equation*}
%For this reason, 
This further implies that
\begin{equation}\label{eq:appendix-lr-sigma}
    \left(\hat{\sigma}^{(k)}_b\right)^2 = \max\left(\frac{1}{m}\sum_{i=1}^m(\tilde{y}_i-\tilde{x}_i'\hat{\beta}^{(k)}_b)^2, \ \frac{1}{n_k}\sum_{i=1}^{n_k}(y_i-x_i'\hat{\beta}^{(k)}_b)^2\right).
\end{equation}

% Accordingly, the final estimate of $\sigma$ is calculated as the mean squared error (MSE) of the residuals obtained with the estimated $\beta$, based on either the predictions on the current observations $\hat{\mathbb{P}}^{(k)}_b$ or the predictions on the future observations $\tilde{\mathbb{P}}^{(k)}_b$, whichever is greater.

% We update $\theta$ first by minimizing the loss function over $\beta$ and then minimizing over $\phi$, where the solution is independent of $\sigma^2$, with $\beta$ fixed at its (current) minimizer $\hat\beta$. Note that the minimizer $\hat\phi$ is given by
% \[
% \frac{n}{2} + \lambda_0 - \frac{e^{-\phi}}{2}\sum_{i=1}^n(y_i-x_i'\hat\beta)^2 = 0, 
% \]
% that is,
% \[
% \hat{\sigma}^2 = \frac{1}{n+2\lambda_0}\sum_{i=1}^n(y_i-x_i'\hat\beta)^2.
% \] 
% Also note that $\lambda_0$ can be updated along with the update of $\lambda_1, ..., \lambda_p$.

Next, we focus on the estimation of $\beta$ with the modified loss function obtained by including only the terms related to $\beta$. This leads to the canonical mean squared error (MSE) loss function taking the form
\[
L(\beta | X = x, Y = y) = \frac{1}{2n}(y - x\beta)^T(y - x\beta).
\]
% \subsubsection{Weighted Duality Function}
% \subsubsection{Single-parameter Duality Function}
The duality function in its general form can be rewritten as:
\(
\pi(\beta,\lambda) = \lambda\sum_{i=1}^{p}|\beta_i|
\)
and
\(
\pi(\beta,\lambda) = \lambda\sum_{i=1}^{p}\beta_i^2,
\)
where $\lambda$ can be considered as either a vector or a scalar, depending on whether the duality function is weighted. We first consider the $L_1$ duality function. In this case, the estimate of $(\beta, \lambda)$ should satisfy
\begin{align*}
    0=
\frac{ \partial G_{\tilde{\mathbb{P}}^{(k)}_b}(\beta, \lambda)}{\partial \beta}
&=  \frac{\partial L(\beta| \tilde{\bx}, \tilde{\by})}{\partial \beta}
	+ \frac{\partial \pi(\beta, \lambda)}{\partial \beta}\\
 &= -\frac{1}{m}\tilde{\bx}^T(\tilde{\by} - \tilde{\bx}\beta) + \lambda \cdot\text{sign}(\beta) 
\end{align*}
and
\begin{align*}
   \lambda &=  \min_\lambda \left\|\frac{ \partial V_{\hat{\mathbb{P}}^{(k)}_b,\tilde{\mathbb{P}}^{(k)}_b}(\beta, \lambda)}{\partial \beta}
\right\|_2^2\\
&=  \min_\lambda \left\| \left(\frac{1}{n_k} \bx^{T}\bx - \frac{1}{m}\tilde{\bx}^{T}\tilde{\bx}\right)\beta - \left(\frac{1}{n_k}\bx^{T}\by - \frac{1}{m}\tilde{\bx}^{T}\tilde{\by}\right) - \lambda \cdot\text{sign}(\beta) \right\|_2^2
\end{align*}
with the subgradient method \citep{Shor1985}. Thus, the stochastic update is given by
\[
\beta^{(t+1)} = \beta^{(t)} - \gamma_1 \cdot \left(-\frac{1}{m}\tilde{\bx}^T(\tilde{\by} - \tilde{\bx}\beta^{(t)}) + \lambda^{(t)} \cdot\text{sign}(\beta^{(t)})\right)
\]
and
\[
\lambda^{(t+1)} = \left(\lambda^{(t)} - \gamma_2 \cdot \left(2 v_1^Tv_1 \lambda^{(t)} - 2 v_1^Tv_2\right)\right)_+
\]
if $\lambda$ is a scalar and
\[
\lambda^{(t+1)} = \left(\lambda^{(t)} - \gamma_2 \cdot 2 (v_1  - \lambda^{(t)} \odot v_1) \odot (-v_1)\right)_+
\]
otherwise, where
%\begin{align*}
\(v_1 = \text{sign}(\beta^{(t+1)}), \) %\\
\(v_2 = \left(\frac{1}{n_k} \bx^T\bx - \frac{1}{m}\tilde{\bx}^T\tilde{\bx}\right) \beta^{(t+1)} - \left(\frac{1}{n_k}\bx^T\bx - \frac{1}{m}\tilde{\bx}^T\tilde{\by}\right),\)
%\end{align*}
and $\gamma_1$ and $\gamma_2$ are the chosen step sizes. For faster convergence, the update of $\beta$ can be replaced with the update scheme of the proximal gradient method \citep{Boyd2014}, that is,
\[
\beta^{(t+1/2)} = \beta^{(t)} - \gamma_1 \cdot \left(-\frac{1}{m}\tilde{\bx}^T(\tilde{\by} - \tilde{\bx}\beta^{(t)})\right)
\]
\[
\beta^{(t+1)} = \mbox{sign}\left(\beta^{(t+1/2)} \right)
\left(|\beta^{(t+1/2)} | - \gamma_1\lambda^{(t)}\right)_+,
\]
where
\begin{equation*}
\mbox{sign}\left(z\right)\left(|z| - \gamma\right)_+ =
\left\{
\begin{array}{lcl}
z - \gamma && \mbox{if 
%$ \bar{y} \le \bar{\tilde{y}} \le 0$ or $ \bar{y} \ge \bar{\tilde{y}} \ge 0$, {\it i.e.},
$z > 0$ and $\gamma < |z|$}\\
z + \gamma && \mbox{if 
%$ \bar{y} \le \bar{\tilde{y}} \le 0$ or $ \bar{y} \ge \bar{\tilde{y}} \ge 0$, {\it i.e.},
$z < 0$ and $\gamma < |z|$}\\
0 && \mbox{if 
%$ \bar{y} \le \bar{\tilde{y}} \le 0$ or $ \bar{y} \ge \bar{\tilde{y}} \ge 0$, {\it i.e.},
$\gamma \geq |z|$.}\\
\end{array}
\right.
\end{equation*}
For the $L_2$ duality function, we should have
\begin{align*}
    0=
\frac{ \partial G_{\tilde{\mathbb{P}}^{(k)}_b}(\beta, \lambda)}{\partial \beta}
&=  \frac{\partial L(\beta| \tilde{\bx}, \tilde{\by})}{\partial \beta}
	+ \frac{\partial \pi(\beta, \lambda)}{\partial \beta}\\
 &= -\frac{1}{m}\tilde{\bx}^T(\tilde{\by} - \tilde{\bx}\beta) +  2\lambda\beta
\end{align*}
and
\begin{align*}
   \lambda &=  \min_\lambda \left\|\frac{ \partial V_{\hat{\mathbb{P}}^{(k)}_b,\tilde{\mathbb{P}}^{(k)}_b}(\beta, \lambda)}{\partial \beta}
\right\|_2^2\\
&=  \min_\lambda \left\| \left(\frac{1}{n_k} \bx^T\bx - \frac{1}{m}\tilde{\bx}^T\tilde{\bx}\right)\beta - \left(\frac{1}{n_k}\bx^T\by - \frac{1}{m}\tilde{\bx}^T\tilde{\by}\right) - 2\lambda\beta \right\|_2^2.
\end{align*}
Thus, the stochastic update is given by
\[
\beta^{(t+1)} = \beta^{(t)} - \gamma_1 \cdot \left(-\frac{1}{m}\tilde{\bx}^T(\tilde{\by} - \tilde{\bx}\beta^{(t)}) + 2 \lambda^{(t)} \beta^{(t)}\right)
\]
and
\[
\lambda^{(t+1)} = \left(\lambda^{(t)} - \gamma_2 \cdot \left(2 v_1^Tv_1 \lambda^{(t)} - 2 v_1^Tv_2\right)\right)_+
\]
if $\lambda$ is a scalar and
\[
\lambda^{(t+1)} = \left(\lambda^{(t)} - \gamma_2 \cdot 2 (v_1  - \lambda^{(t)} \odot v_1) \odot (-v_1)\right)_+
\]
otherwise, where
%\begin{align*}
\(v_1 = 2\beta^{(t+1)}, \) %\\
\(v_2 = \left(\frac{1}{n_k} \bx^T\bx - \frac{1}{m}\tilde{\bx}^T\tilde{\bx}\right) \beta^{(t+1)} - \left(\frac{1}{n_k}\bx^T\by - \frac{1}{m}\tilde{\bx}^T\tilde{\by}\right),\)
%\end{align*}
and $\gamma_1$ and $\gamma_2$ are chosen step sizes.

To accelerate convergence, the update process for $\lambda$ with both duality functions can be replaced by the ADAM-type update \citep{kingma2014adam} which incorporates momentum. That is, 
\begin{align*}
\mu^{(t+1)} &= \tilde{\beta}_1 \mu^{(t)}+ (1 - \tilde{\beta}_1) \cdot \nabla_\lambda \left\|\frac{ \partial V_{\hat{\mathbb{P}}^{(k)}_b,\tilde{\mathbb{P}}^{(k)}_b}(\beta^{(t+1)}, \lambda^{(t)})}{\partial \beta^{(t+1)}}
\right\|_2^2 \\
\nu^{(t+1)} &= \tilde{\beta}_2 \nu^{(t)} + (1 - \tilde{\beta}_2) \left(\nabla_\lambda \left\|\frac{ \partial V_{\hat{\mathbb{P}}^{(k)}_b,\tilde{\mathbb{P}}^{(k)}_b}(\beta^{(t+1)}, \lambda^{(t)})}{\partial \beta^{(t+1)}}
\right\|_2^2\right)^2 \\
\lambda^{(t+1)} &= \left( \lambda^{(t)} - \gamma_2 \frac{\mu_t^{(t+1)}/(1 + \tilde{\beta}_1)}{\sqrt{\nu_t^{(t+1)} / (1-\tilde{\beta}_2)} + \epsilon} \right)_+,
\end{align*}
where $\mu$ and $\nu$ represent the first and second momentum vectors or scalars initialized as zero, the two scalars $\tilde{\beta}_1$ and $\tilde{\beta}_2$ are momentum parameters with $\tilde{\beta}_1 \in [0,1)$ and $\tilde{\beta}_2 \in [0,1)$, the hyper-parameter $\gamma_2$ represents the learning rate, and $\epsilon$ is a small constant added for numerical stability.

An estimation procedure for an imputation model is considered complete once $\beta^{(t)}$ has converged. The estimate of $\sigma$ can then be obtained by (\ref{eq:appendix-lr-sigma}) with the converged $\beta^{(t)}$ using the argument above in this section. The resulting model, characterized by the parameters $(\hat{\beta}^{(k)}_b, \hat{\sigma}^{(k)}_b)$ is then applied to the hold-out set $(\bx^{(k)}_b, \by^{(k)}_b)$. Depending on the objective, this can be used for the purpose of imputing future observations using Algorithm~2, or for assessing the efficiency of the imputation model to select the resampling parameter with Algorithm~3.

Performing Algorithm~3 for selecting resampling parameter $\tilde{\alpha}$ requires the calculation of CDF values with a single imputation model estimation iteration ($B = 1$). The CDF value can be calculated with
\[
F(y^{(k)}_{bi} | x^{(k)}_{bi}) = \Phi\left(\frac{y^{(k)}_{bi} - x^{(k)}_{bi} \hat{\beta}^{(k)}_b}{\hat{\sigma}^{(k)}_b} \right)
\]
for $i = 1,\dots, n-n_k$, where $b = 1$ and $\Phi(\cdot)$ represents the standard normal CDF. The hold-out set is denoted as $(\bx^{(k)}_b, \by^{(k)}_b) := \{(x^{(k)}_{bi}, y^{(k)}_{bi}):i=1,\dots, n-n_k\}$. After iterating through $k = 1,\dots,K$, we can obtain $n$ such CDF values. This collection of CDF values is then tested against the standard uniform distribution with KS-test for the selection of best $\tilde{\alpha}$.

Note that performing Algorithm~3 already yields an imputation model obtained with the selected $\tilde{\alpha}$. Thus, Algorithm~2 only requires $B - 1$ iterations. In Algorithm~2, future observations are created by the normal distribution
\[
y^{(k)}_{*bi} \sim N \left(x^{(k)}_{bi} \hat{\beta}^{(k)}_b, \left(\hat{\sigma}^{(k)}_b\right)^2\right)
\]
for $i = 1,\dots, n_k$.  By iteratively performing the estimation of the imputation model and the imputation process for $b = 1,\dots, B$ and $k = 1,\dots, K$, we construct the imputation distribution $\mathbb{Q}$, which consists of $n\cdot B$ samples.

The estimation process with Algorithm~1 follows the same implementation of estimation of imputation models as described above, with a straightforward modification:  $\hat{\mathbb{P}}^{(k)}_b$ is replaced with $\mathbb{Q}$ and $\tilde{\mathbb{P}}^{(k)}_b$ with $\hat{\mathbb{P}}$, where the latter represents the full empirical observations.

\subsubsection{Implementation Details of AM} \label{appendix:linear-implementation}
We implemented AM for the linear regression example from scratch in R, with no contributed R packages utilized for the implementation.
Algorithm~2 was implemented with $B = 5$ and $K = 5$. Algorithm~3 was used to select the resampling parameter $\tilde{\alpha}$, with the candidate set $\{0.1, 0.2, 0.3, 0.5\}$. The optimization related parameters that apply to all experiments are listed as follows:
\begin{itemize}
    \item Imputation process: $\gamma_1 = 1e-5, \ \gamma_2 = 1.0, \ \tilde{\beta}_1 = 0.9, \ \tilde{\beta}_2 = 0.999,$ and  $ \ \epsilon = 1e-7$;
    \item Estimation process: $\gamma_1 = 1e-6, \ \gamma_2 = 1.0, \ \tilde{\beta}_1 = 0.9, \ \tilde{\beta}_2 = 0.999,$ and  $\ \epsilon = 1e-7$.
\end{itemize}

\subsubsection{Implementation Details of Other Methods}
We utilized the $\texttt{cv.glmnet}$ and $\texttt{glmnet}$ function from the R package $\texttt{glmnet}$ \citep{Friedman2010} for other methods, applying the default settings. For the elastic net, we chose candidate $\alpha$ values, which balance the $L_1$ and $L_2$ penalties, ranging from $0$ to $1$ in increments of $0.01$. The residuals for the adaptive lasso were obtained with the ridge estimate generated by $\texttt{cv.glmnet}$ and $\texttt{glmnet}$. The method of \cite{Reid2016}, implemented in the R package $\texttt{natural}$ \citep{Yu2019}, is used for estimating the standard deviation.

\clearpage
\subsection{Implementation Details in the Neural Network Example}\label{appendix:nn}
\subsubsection{Technical Details} \label{appendix:nn-technical}
In the context of classification using neural networks, consider the observed data $(x,y)$  and the model parameters $\theta \in \mathbb{R}^p$. A general loss function for this setting is defined as follows:
\[
L(\theta |X = x, Y = y) = \frac{1}{n}\sum_{i=1}^n H\left(y_{i},\hat{y}\left(x_i;\theta\right)\right),
\]
where $y_i$ is the observed response or category and $\hat{y}(x_i;\theta)$ represents the vector of predicted output probabilities for input $x_i$. The function $H(\cdot, \cdot)$ denotes the cross-entropy loss function, commonly used in machine learning.  It takes the same form as the negative log-likelihood function used in (multinomial) logistic regression. For a multi-class classification problem with $C$ classes (for instance, $C = 10$ in the MNIST dataset), the loss function is expressed as:
\[
H\left(y_{i},\hat{y}\left(x_i;\theta\right)\right) = -\sum_{c = 0}^{C-1} y_{ic} \log \hat{y}_{c}\left(x_i;\theta\right),
\]
where $\hat{y}_{c}\left(x_i;\theta\right)$ denotes the estimated probability of observing the class $c$ given the input $x_i$, $y_{ic} = 1$ if $y_i = c$, and $y_{ic} = 0$ otherwise. 

In this section, we elaborate on the derivation for estimating an imputation model. First, we obtain the data represented by $\hat{\mathbb{P}}^{(k)}_b$ and $\tilde{\mathbb{P}}^{(k)}_b$ following the data-splitting and resampling scheme in Algorithm~2 at iteration $(b, k)$, for any arbitrary $b = 1,\dots, B$ and $k = 1,\dots,K.$ For simplicity, we omit the index $(b, -k)$ on each indexed observation $(x, y)$. The samples $\tilde{\mathbb{P}}^{(k)}_b:= \{(\tilde{x}_i, \tilde{y}_i) : i = 1,\dots,m\}$ are treated as the current observations and the samples $\hat{\mathbb{P}}^{(k)}_b:= \{(x_i, y_i) : i = 1,\dots,n_k\}$ are treated as the future observations, where the sample sizes in $\hat{\mathbb{P}}^{(k)}_b$ and $\tilde{\mathbb{P}}^{(k)}_b$ are denoted as $n_k$ and $m$ respectively. Using the matrix notation, the two samples are denoted by $(\bx, \by)$ and $(\tilde{\bx}, \tilde{\by})$, omitting the $(b, -k)$ indexing used in Algorithm~2.

The duality function in its general form can be rewritten as:
\(
\pi(\theta,\lambda) = \lambda\sum_{i=1}^{p}|\theta_i|
\)
and
\(
\pi(\theta,\lambda) = \lambda\sum_{i=1}^{p}\theta_i^2,
\)
where $\lambda$ can be considered as a vector or scalar, depending on whether the duality function is weighted. We first consider the $L_1$ duality function. In this case, the estimation of $(\theta, \lambda)$ should satisfy
\begin{align*}
    0=
\frac{ \partial G_{\tilde{\mathbb{P}}^{(k)}_b}(\theta, \lambda)}{\partial \theta}
%&
= \frac{\partial L(\theta| \tilde{\bx}, \tilde{\by})}{\partial \theta}
	+ \frac{\partial \pi(\theta, \lambda)}{\partial \theta}%\\
 &= \tilde{g}(\theta;\tilde{\bx}, \tilde{\by})+ \lambda \cdot\text{sign}(\theta) 
\end{align*}
and
\begin{align*}
   \lambda %&
   =  \min_\lambda \left\|\frac{ \partial V_{\hat{\mathbb{P}}^{(k)}_b,\tilde{\mathbb{P}}^{(k)}_b}(\theta, \lambda)}{\partial \theta}
\right\|_2^2 %\\
&=  \min_\lambda \Big\| \hat{g}(\theta; \bx, \by) - \tilde{g}(\theta; \tilde{\bx}, \tilde{\by}) - \lambda \cdot\text{sign}(\theta) \Big\|_2^2
\end{align*}
with the subgradient method \citep{Shor1985}, where
\[
\hat{g}(\theta; \bx, \by) = \frac{\partial L(\theta| \bx, \by)}{\partial \theta}
%\]
\qquad\mbox{ and }\qquad
%\[
\tilde{g}(\theta;  \tilde{\bx}, \tilde{\by}) = \frac{\partial L(\theta| \tilde{\bx}, \tilde{\by})}{\partial \theta}
\]
are the gradients of the loss function with respect to $\theta$ on  $\hat{\mathbb{P}}^{(k)}_b$ and $\tilde{\mathbb{P}}^{(k)}_b$, respectively.  These gradients are calculated using the chain rule, known as backpropagation, based on the specific structure of the neural network. Both $\hat{g}(\theta; \bx, \by)$ and $\tilde{g}(\theta;  \tilde{\bx}, \tilde{\by})$ can be computed over a small batch of data using the batch-specific loss function
\[
L_{\text{batch}}^{(b')} (\theta) = \frac{1}{n_{\text{batch}}} \sum_{i=1}^{n_{\text{batch}}} H \left(y_i^{(b')}, \hat{y}\left(x^{(b')}_i;\theta\right)\right),
\]
where $n_{\text{batch}}$ denotes the batch size, typically chosen as $64, 128,$ or $256$. There are $B_{\text{batch}} = \text{floor} (m/n_{\text{batch}})$ batches, and the data in the $b'$-th batch ($b' = 1,\dots, B_{\text{batch}}$) is denoted as $\{(x_i^{(b')}, y_i^{(b')}): i = 1,\dots,n_{\text{batch}}\}$.

The required gradient for the optimizers to optimize $\theta$ and $\lambda$ can be expressed as
\[
\nabla_\theta G_{\tilde{\mathbb{P}}^{(k)}_b}(\theta, \lambda) = \tilde{g}(\theta; \tilde{\bx}, \tilde{\by})+ \lambda \cdot\text{sign}(\theta)
\]
and
\[
\nabla_\lambda \left\|\frac{ \partial V_{\hat{\mathbb{P}}^{(k)}_b,\tilde{\mathbb{P}}^{(k)}_b}(\theta, \lambda)}{\partial \theta}
\right\|_2^2 = -2 \cdot\left(\hat{g}(\theta; \bx, \by) - \tilde{g}(\theta;  \tilde{\bx}, \tilde{\by}) - \lambda \odot \text{sign}(\theta) \right) \cdot \text{sign}(\theta),
\]
when $\lambda$ is weighted, and 
\[
\nabla_\lambda \left\|\frac{ \partial V_{\hat{\mathbb{P}}^{(k)}_b,\tilde{\mathbb{P}}^{(k)}_b}(\theta, \lambda)}{\partial \theta}
\right\|_2^2 = 2\lambda  \cdot \left[\text{sign}(\theta)\right]^T\left[\text{sign}(\theta)\right] - 2\cdot \left[\text{sign}(\theta)\right]^T\left(\hat{g}(\theta; \bx, \by) - \tilde{g}(\theta;  \tilde{\bx}, \tilde{\by}) \right)
\]
otherwise, with the subgradient method \citep{Shor1985}. When the $L_2$ duality function is used, we have
\[
\nabla_\theta G_{\tilde{\mathbb{P}}^{(k)}_b}(\theta, \lambda) = \tilde{g}(\theta;  \tilde{\bx}, \tilde{\by}) + 2\lambda \theta
\]
and
\[
\nabla_\lambda \left\|\frac{ \partial V_{\hat{\mathbb{P}}^{(k)}_b,\tilde{\mathbb{P}}^{(k)}_b}(\theta, \lambda)}{\partial \theta}
\right\|_2^2 = -2 \cdot\left(\hat{g}(\theta; \bx, \by) - \tilde{g}(\theta;  \tilde{\bx}, \tilde{\by}) - 2\lambda \theta \right) \cdot \text{sign}(\theta)
\]
when $\lambda$ is weighted, and 
\[
\nabla_\lambda \left\|\frac{ \partial V_{\hat{\mathbb{P}}^{(k)}_b,\tilde{\mathbb{P}}^{(k)}_b}(\theta, \lambda)}{\partial \theta}
\right\|_2^2 = 2\lambda  \cdot \left(2\theta\right)^T\left(2 \theta\right) - 2\cdot \left(2\theta\right)^T\left(\hat{g}(\theta; \bx, \by) - \tilde{g}(\theta;  \tilde{\bx}, \tilde{\by})\right)
\]
otherwise. 
% The implementation of optimizers will be introduced in Section~\ref{appendix:nn-implementation} in detail.
The parameter $\theta$ is updated using the stochastic gradient descent with momentum (SGDM) optimizer with the updates:
\begin{align*}
v^{(t+1)} &= \rho v^{(t)} + \nabla_\theta G_{\hat{\mathbb P}}(\theta^{(t)} , \lambda^{(t)} ) \\
\theta^{(t+1)} &= \theta^{(t)} - \eta_1 v^{(t+1)},
\end{align*}
where $v$ represents the velocity vector initialized as zero,
$\rho$ is the momentum parameter with $\rho \in [0,1)$, and $\eta_1$ represents the learning rate.

Following the update of $\theta$, $\lambda$ is updated with the ADAM optimizer \citep{kingma2014adam}, with a slightly modified implementation. Specifically,
\begin{align*}
\mu^{(t+1)} &= \beta_1 \mu^{(t)}+ (1 - \beta_1) \cdot \nabla_\lambda \left\|\frac{ \partial V_{\hat{\mathbb{P}}^{(k)}_b,\tilde{\mathbb{P}}^{(k)}_b}(\theta^{(t+1)}, \lambda^{(t)})}{\partial \theta^{(t+1)}}
\right\|_2^2 \\
\nu^{(t+1)} &= \beta_2 \nu^{(t)} + (1 - \beta_2) \left(\nabla_\lambda \left\|\frac{ \partial V_{\hat{\mathbb{P}}^{(k)}_b,\tilde{\mathbb{P}}^{(k)}_b}(\theta^{(t+1)}, \lambda^{(t)})}{\partial \theta^{(t+1)}}
\right\|_2^2\right)^2 \\
\lambda^{(t+1)} &= \left( \lambda^{(t)} - \eta_2 \frac{\mu_t^{(t+1)}/(1 + \beta_1)}{\sqrt{\nu_t^{(t+1)} / (1-\beta_2)} + \epsilon} \right)_+,
\end{align*}
where $\mu$ and $\nu$ represent the first and second momentum vectors initialized as zero, the two scalars $\beta_1$ and $\beta_2$ are the momentum parameters with $\beta_1 \in [0,1)$ and $\beta_2 \in [0,1)$, the hyper-parameter $\eta_2$ represents the learning rate and $\epsilon$ is a small constant added for numerical stability.
Additionally, the learning rates $\eta_1$ and $\eta_2$ are controlled by a scheduler and are updated every epoch with
%\begin{align*}
$\eta_1 = \eta_1 \cdot \gamma$ and %\\
$\eta_2 = \eta_2 \cdot \gamma,$
%\end{align*}
where $\gamma$ is the pre-specified learning rate-decay rate. 

The procedure of estimating one imputation model is considered complete once $\theta^{(t)}$ or the training loss has converged. The resulting model, characterized by the parameter vector $\hat{\theta}^{(k)}_b$, is then applied on the hold-out set $(\bx_b^{(k)}, \by^{(k)}_b)$. Depending on the objective, this can be used for the purpose of imputing future observations using Algorithm~2, or for assessing the efficiency of the imputation model to select the resampling parameter with Algorithm~3.

Performing Algorithm~3 for selecting the resampling parameter $\tilde{\alpha}$ requires obtaining surrogate CDF values. The details are provided in Section~5.3 and Section~\ref{appendix:nn-surrogate}. After iterating through $k = 1,\dots,K$, we obtain $n$ such surrogate CDF values. This collection of surrogate CDF values is then tested against the standard uniform distribution with the KS-test for the selection of the best $\tilde{\alpha}$.

Note that performing Algorithm~3 already yields an imputation model obtained with the selected $\tilde{\alpha}$, thus Algorithm~2 only requires $B - 1$ iterations. When generating imputed future observations, new labels are created by generating
\[
y^{(k)}_{*bi} \sim \text{Multinomial} \left(\hat{y}\left(x^{(k)}_{bi};\hat{\theta}_b^{(k)}\right)\right)
\]
for $i = 1,\dots, n - n_k$. In this expression, $\hat{\theta}_b^{(k)}$ represents the estimated parameters of the imputation model, and $\hat{y}(x^{(k)}_{bi};\hat{\theta}_b^{(k)})$ denotes the estimated probability vector for the 10 digits, given the image $x^{(k)}_{bi}$ as the input. By iteratively conducting the imputation model estimation and the imputation process for $b = 1,\dots, B$ and $k = 1,\dots, K$, we construct the imputation distribution $\mathbb{Q}$. The imputation distribution $\mathbb{Q}$ consists of $60,000 \times B$ images, each with its newly imputed label.

The estimation process with Algorithm~1 follows the same implementation for estimating an imputation model as described above, with a straightforward modification:  $\hat{\mathbb{P}}$ is replaced with $\mathbb{Q}$ and $\tilde{\mathbb{P}}^{(k)}_b$ with $\hat{\mathbb{P}}$, where the latter represents the full empirical observations.

\subsubsection{Implementation Details of AM} \label{appendix:nn-implementation}
% We first introduce the implementation of the optimizers as mentioned in Section~\ref{appendix:nn-technical} for optimizing $\theta$ and $\lambda$.
We implemented AM using the PyTorch framework \citep{Paszke2019}, with CUDA being used for faster training. 
During the data preprocessing phase, the images are normalized to have a mean of 0 and a standard deviation of 1. Beyond this normalization, no additional techniques, including data augmentation, are used. Before training, the neural network model is initialized using the default setting in Pytorch.

The batch training process introduced in Section~\ref{appendix:nn-technical} is implemented in the following way. An epoch is defined as a complete pass over the current observations through the neural network. That is, within each epoch, there are $B_{\text{batch}} = \text{floor} (m/n_{\text{batch}})$ batches for the imputation process and $B_{\text{batch}}= \text{floor} (n/n_{\text{batch}})$ batches for the estimation process. The future observations are similarly shuffled and divided into mini-batches of size $n_{\text{batch}}$, indexed as $1, \dots, B_{\text{batch}}'$. During the $b'$-th iteration, the $b'$-th mini-batch from the current observations is used to calculate the current gradient (referred to as $\tilde{g}(\theta; \tilde{\bx}, \tilde{\by})$ in Section~\ref{appendix:nn-technical}) and the $b$-th mini-batch from the future observations is used for the future gradient (denoted as $\hat{g}(\theta; \bx, \by)$ in Section~\ref{appendix:nn-technical}). If the number of future observations is smaller than that of the current observations, which occurs when using a resampling parameter $\alpha > 1$, the data in the future observations are duplicated before creating mini-batches. For the estimation process, the imputed data from the $B$ imputation iterations are cyclically used as future observations for each epoch. %\CLdel{This ensures that the quantity of current and future observations \CLrep{CAN THIS BE MADE CLEARER?}{matches} in each training epoch.}

For faster convergence, a ``warm-up'' training phase is conducted before the estimation process, utilizing only the observed data $\mathbb{\hat{P}}$. The ``warm-up" training phase is carried out as follows. We implement the AM estimation, as detailed in Section~\ref{appendix:nn-technical}, by utilizing $\hat{\mathbb{P}}$ as both the current and future observations. This phase employs the weighted-$L_1$ duality function. Despite $\hat{\mathbb{P}}$ serving a dual role, the use of the duality function remains meaningful due to the inherent randomness in the mini-batch creation. This warm-up training lasts 50 epochs.  The values of neural network parameters obtained at the end of this phase are then used as the initial values for the estimation process.

The training-related hyper-parameters are specified as follows. The imputation process is performed by applying Algorithm~2 with $B = 5$ and $K = 2$. Algorithm~3 is used to select the resampling parameter $\tilde{\alpha}$, with the candidate set $\{0.5, 0.8, 1.0, 1.2\}$. The optimization related parameters are listed as follows (apply to all experiments):
\begin{itemize}
    \item Imputation process: $\eta_1 = 0.01, \ \rho = 0.95,\ \eta_2 = 0.001, \ \beta_1 = 0.99, \ \beta_2 = 0.9999, \ \epsilon = 1e-8,$ and $ \ \gamma = 0.98$. Training stops after 50 epochs;
    \item ``Warm-up" training: $\eta_1 = 0.01, \ \rho = 0.95,\ \eta_2 = 0.001, \ \beta_1 = 0.99, \ \beta_2 = 0.9999, \ \epsilon = 1e-8,$ and $ \ \gamma = 1$. Training stops after 50 epochs;
    \item Estimation process: $\eta_1 = 0.001, \ \rho = 0.95,\ \eta_2 = 0.0001 \ 
 (\text{weighted duality functions}), \ \eta_2 = 0.00001 \ 
 (\text{unweighted duality functions}), \ \beta_1 = 0.99, \ \beta_2 = 0.9999, \ \epsilon = 1e-8,$ and $ \ \gamma = 0.99$. Training stops after 50 epochs; and
    \item Batch size is chosen to be $n_{\text{batch}} = 64$.
\end{itemize}

\subsubsection{Implementation Details of Other Methods} 
We implemented other methods using the PyTorch framework \citep{Paszke2019}, with CUDA being used for faster training. Before training, the neural network model is initialized using the default setting in Pytorch.

For $L_1$ and $L_2$ regularized training, the model was trained using $\lambda$ values from the set $\{1e-1, 1e-2, 1e-3, 1e-4, 1e-5, 1e-6\}$, with the best results reported. For early-stopping,  we allocated $10,000$ of the training images as a random validation set. In the case of Dropout, it was implemented after the input layer and before each fully connected layer, using dropout ratios selected from $\{0.2,0.5,0.8\}$, reflecting common choices in the literature \citep{Srivastava2014}. All methods employ SGDM for training with a learning rate of $0.01$ and a momentum parameter of $0.95$. This learning rate was chosen from the set $\{0.1,0.01,0.001,0.0001\}$ based on optimal performance.

\subsubsection{Surrogate CDF Values} \label{appendix:nn-surrogate}

When applying Algorithm~3 to choose $m$ for the $m$-out-of-$n$ resampling scheme, a continuous CDF value of the data distribution within the range of $0$ to $1$ is necessary. Here, we adopt and extend the randomization approach proposed by \cite{dunn1996randomized}. For an observed pair of $x$ and $y$, let $p=(p_0, ..., p_9)'$ represent the vector of predictive probabilities for the digits 0 through 9. If $y$ is drawn from the multinomial distribution $\mbox{Multinomial}_{10}(1, p)$, the data generation process can be described as follows
\[
\mbox{Sample } U\sim \mbox{Unif}(0,1) \quad \mbox{and find $y$ such that:}\quad 
\sum_{k=0}^{y-1} p_k \leq U <\sum_{k=0}^y p_k.
\]
where we define $\sum_{k=0}^{-1} p_k =0$. Reversing this data generation process requires finding $U$ given the observed $y$, resulting the interval $[\sum_{k=0}^{y-1} p_k, \sum_{k=0}^y p_k]$. From this interval, We draw a random sample, $U_{x,y}$, uniformly distributed over this interval. This $U_{x,y}$ serves as the surrogate value of the original $U$, allowing for the application of one-sample testing methods that maintain desired frequency properties for goodness-of-fit testing with induced noise; see, {\it i.e.}, \cite{liu2018residuals,cheng2021surrogate,yang2021assessment,yang2024double,gerber2023} and references therein. Consequently, $U_{x,y}$ also functions as a surrogate value of $F(y | x)$, which can be tested against the standard uniform distribution for selecting an appropriate resampling scheme. The properties outlined in Section~\ref{ss:finite-sample-valid-coverage} remain valid with similar arguments, given that
\[
U_{x,y} \sim \mbox{Uniform}(0,1).
\]
Notably, reordering the indices for alternative intervals $\left[\sum_{k=0}^{y-1} p_k, \sum_{k=0}^y p_k\right]$, independent of the observed $y$, can be considered elsewhere for improved efficiency of the underlying goodness-of-fit test.

\subsubsection{Estimation Process of AM} \label{appendix:nn-training-path}
The details of the estimation process and the resultant parameters are visually depicted in Figure~\ref{fig:convergence}. This illustration clearly demonstrates how AM effectively shrinks the parameters toward zero.

\ifthenelse{1=0}{
%       \centerline{--- Fig. \ref{fig:procedure} ---}
}{

\begin{figure}[!htbp]
\centering
\includegraphics[width=14cm]{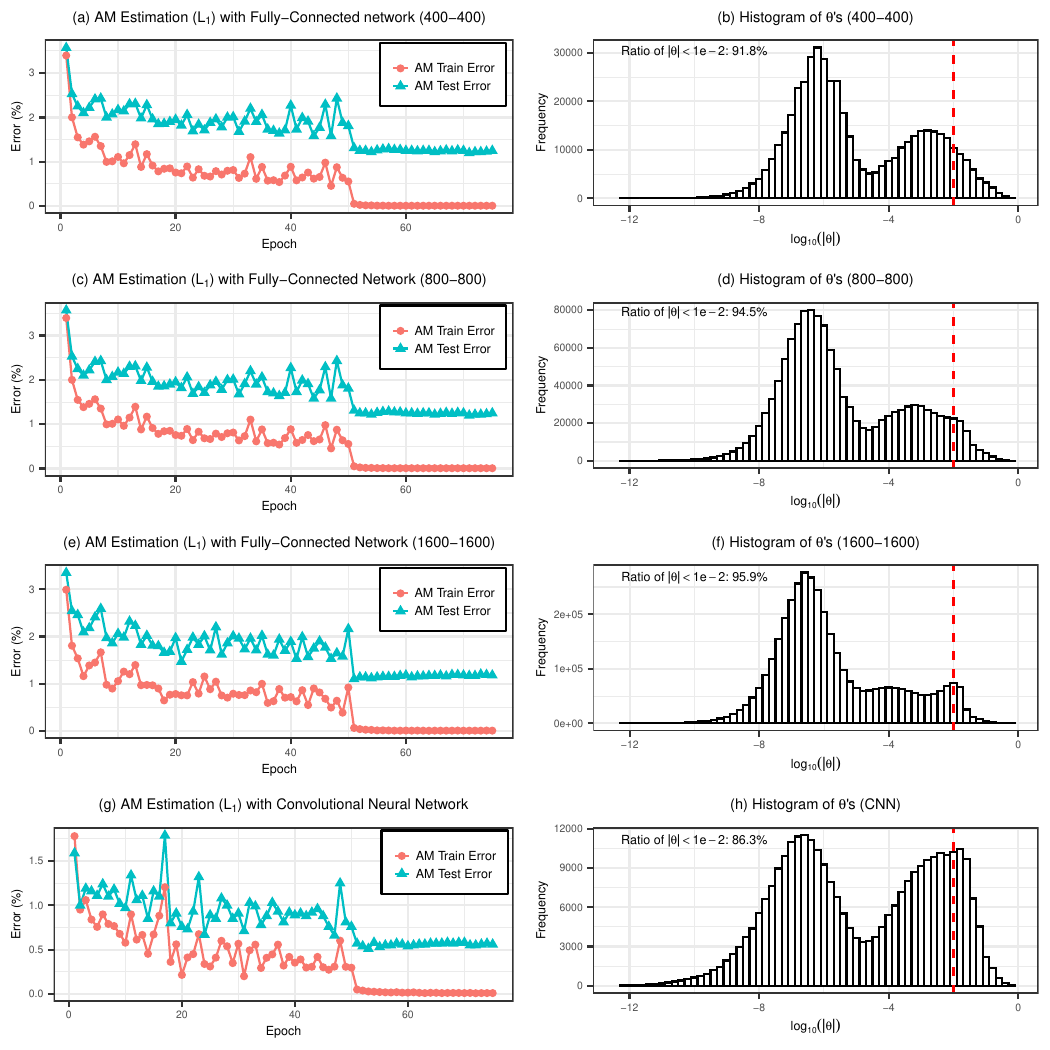}
\caption{The learning curves in the estimation process and the histograms of the estimated parameters using AM (weighted $L_1$ %simplicity-preference 
	\dual~function) with different model structures.}\label{fig:convergence}
\end{figure}
}

\subsubsection{Questionable Labels Found by AM} \label{appendix:nn-questionable-labels}
 An intriguing aspect of the AM estimator is its tendency to achieve nearly, but not quite, 100\% prediction accuracy on training images. By aggregating the training images that were incorrectly predicted across all experiments (as detailed in Table~4, we identified 62 unique images. Closer examination of these images, as showcased in Figure~\ref{fig:questionable-labels-2}, revealed that many seem to possess incorrect labels. Notably, some of the labels we identified as questionable also align with findings from recent research by \cite{Northcutt2021}, which focused on detecting label inaccuracies. 
 % Our method, however, uncovered a greater number of potential errors. 
 Therefore, AM shows promise as a tool for effectively detecting incorrectly labeled data, serving as an ancillary benefit to its primary purpose.
 
 % All 62 training images detected by AM as having potential questionable labels are shown in Figure~\ref{fig:questionable-labels-2}.

    \ifthenelse{1=0}{
    %       \centerline{--- Fig. \ref{fig:procedure} ---}
    }{

    \begin{figure}[!htbp]
    \centering
    \includegraphics[width=14cm]{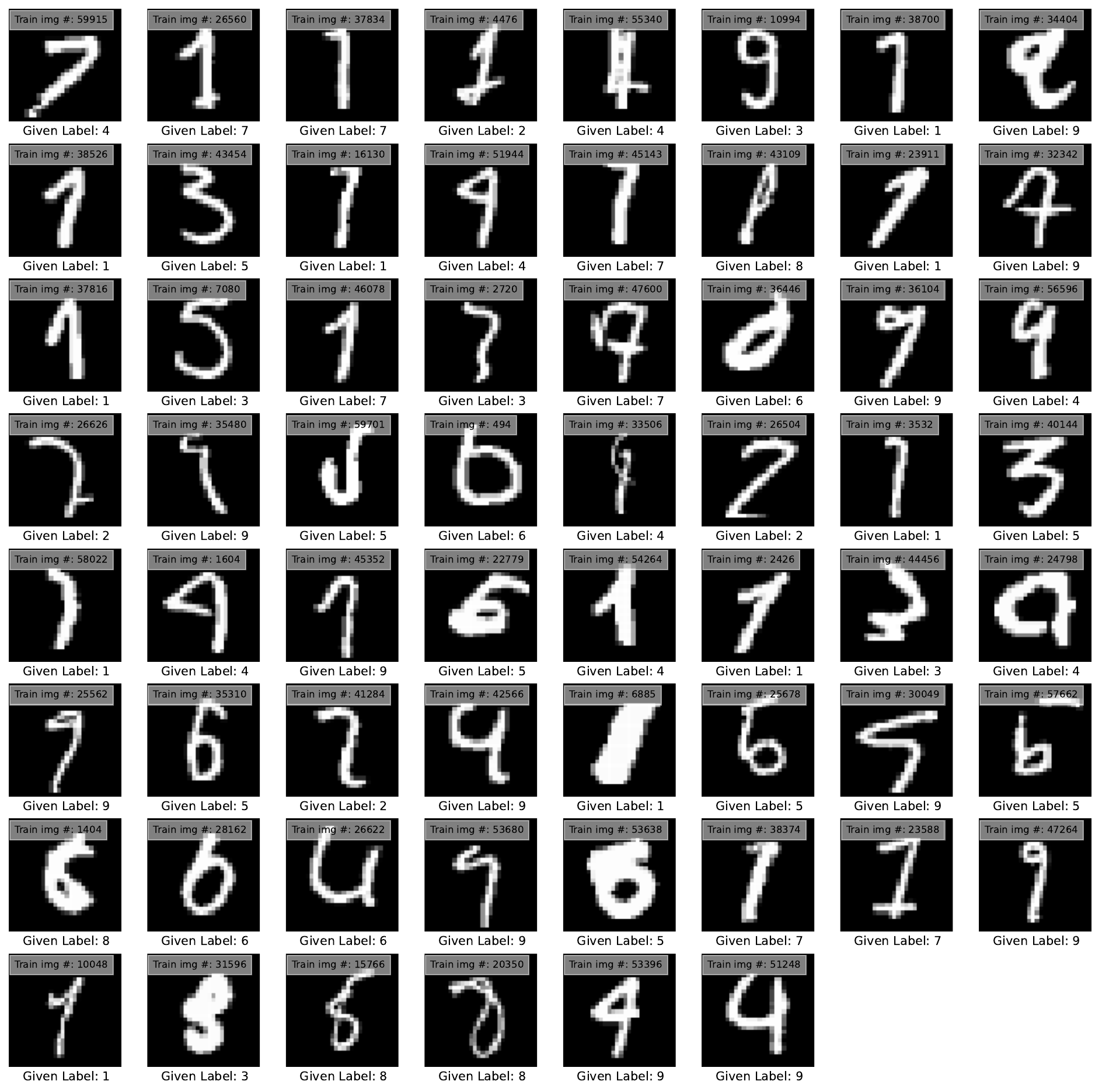}
    \caption{All images detected by AM as having questionable labels, displayed in a left-to-right, top-to-bottom sequence, and sorted by the decreasing frequency of their identification.}\label{fig:questionable-labels-2}
    \end{figure}
    }

\clearpage
\subsection{Imputation Efficiency Evaluation with Q-Q Plots}\label{appendix:qq-plots}
\subsubsection{The Effect of $m$ on Imputation Efficiency} \label{appendix:qq-plots-various-m}
The imputation efficiency is measured by how good each imputation model fits its holdout sample (12). Besides the KS-test approach used in Algorithm~3, such efficiency can be further visually inspected by Q-Q plots, with the estimated CDF values against the standard uniform distribution. Note that the Q-Q plots here can also be viewed as Probability-Probability (P-P) plots. Alternatively, one can also visualize Q-Q plots with normal quantile transformation.

\begin{figure}[!htbp]
    \centering
    \includegraphics[width=10cm]{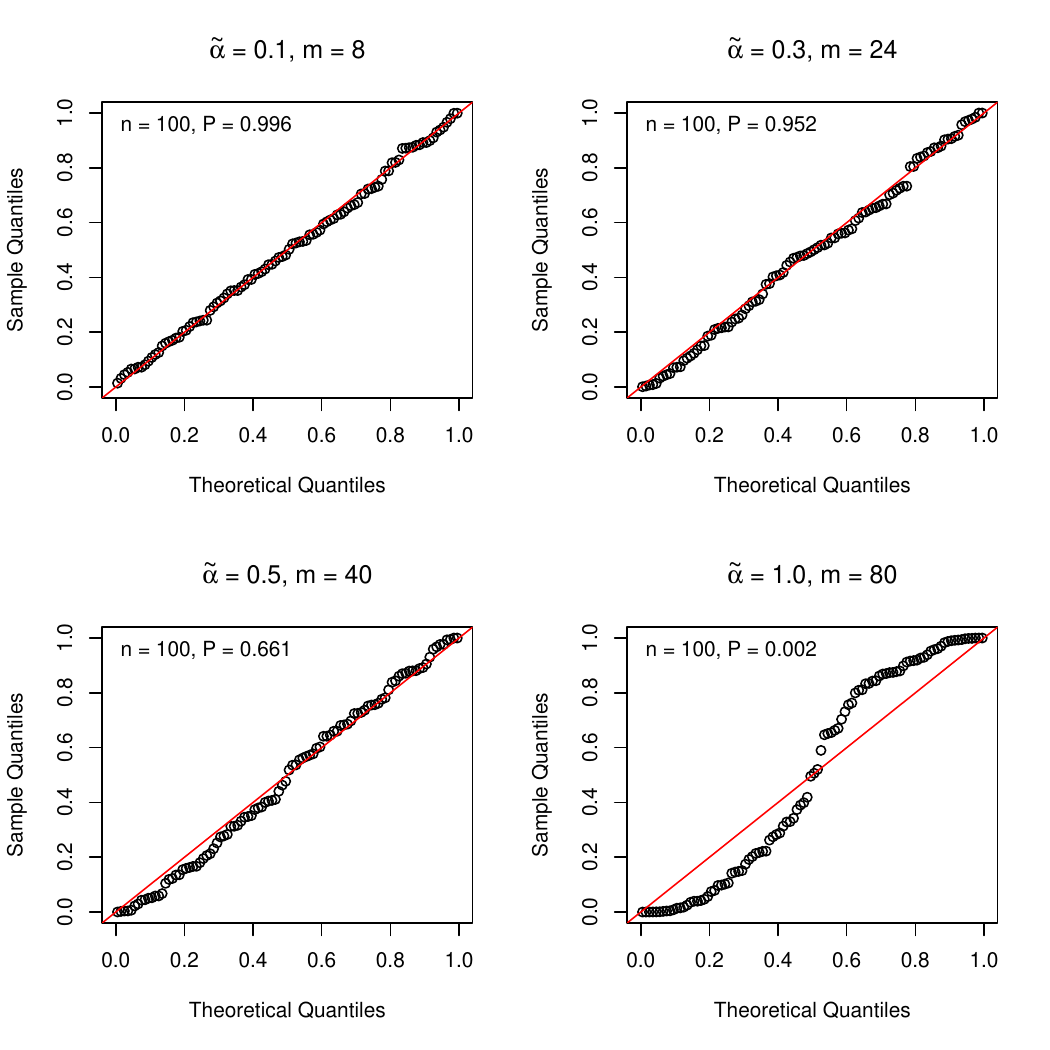}
    \caption{Q-Q plots of the estimated CDF values from the imputation model obtained using the $m$-out-of-$n$ bootstrap with varying $m =\lceil \tilde{\alpha}n_k \rceil$ values, against the standard uniform distribution, in a typical linear regression simulation example in Section~5.2. The CDF values are obtained on the holdout set, following the $K$-fold data-splitting method introduced in Algorithm~3. Here $n_k$ denotes the sample size of the future observations used to fit the $k$-th imputation model, and $n_k = \lceil n(K-1)/K\rceil = 80 \text{ for } k = 1,\dots, K$. In this example we have $K = 5$ and $n_1 =\dots=n_5$, resulting in a uniform $m$ value across $K$ imputation models for each $\tilde{\alpha}$. The observed sample size and the $p$-value of the KS-test are indicated as $n$ and $P$, respectively, in the top-left corner of each plot. Notably, the case with $\tilde{\alpha} = 1$ and $m = 80$ corresponds to the standard bootstrap and shows that the resulting imputation model is severely underdispersed.}\label{fig:qq-various-m} 
    \end{figure}

The Q-Q plots in Figure~\ref{fig:qq-various-m} demonstrate the effect of $m$ on the imputation efficiency in a typical linear regression simulation study in Section~5.2. It can be seen that when the standard bootstrap does not work well in estimating the data distribution, the $m$-out-of-$n$ bootstrap with the optimal $m$ can result in a much better approximation to the true data distribution with regard to (12). Furthermore, $\tilde{\alpha}$ values spanning a wide range, from $0.1$ to $0.5$, consistently yield satisfactory results. This demonstrates the robustness of Algorithm~3 to the grid density of the candidate set for $\tilde{\alpha}$.

\subsubsection{Imputation Efficiency with the Optimal $m$} \label{appendix:qq-plots-optimal-m}
\begin{figure}[!htbp]
    \centering
    \includegraphics[width=14cm]{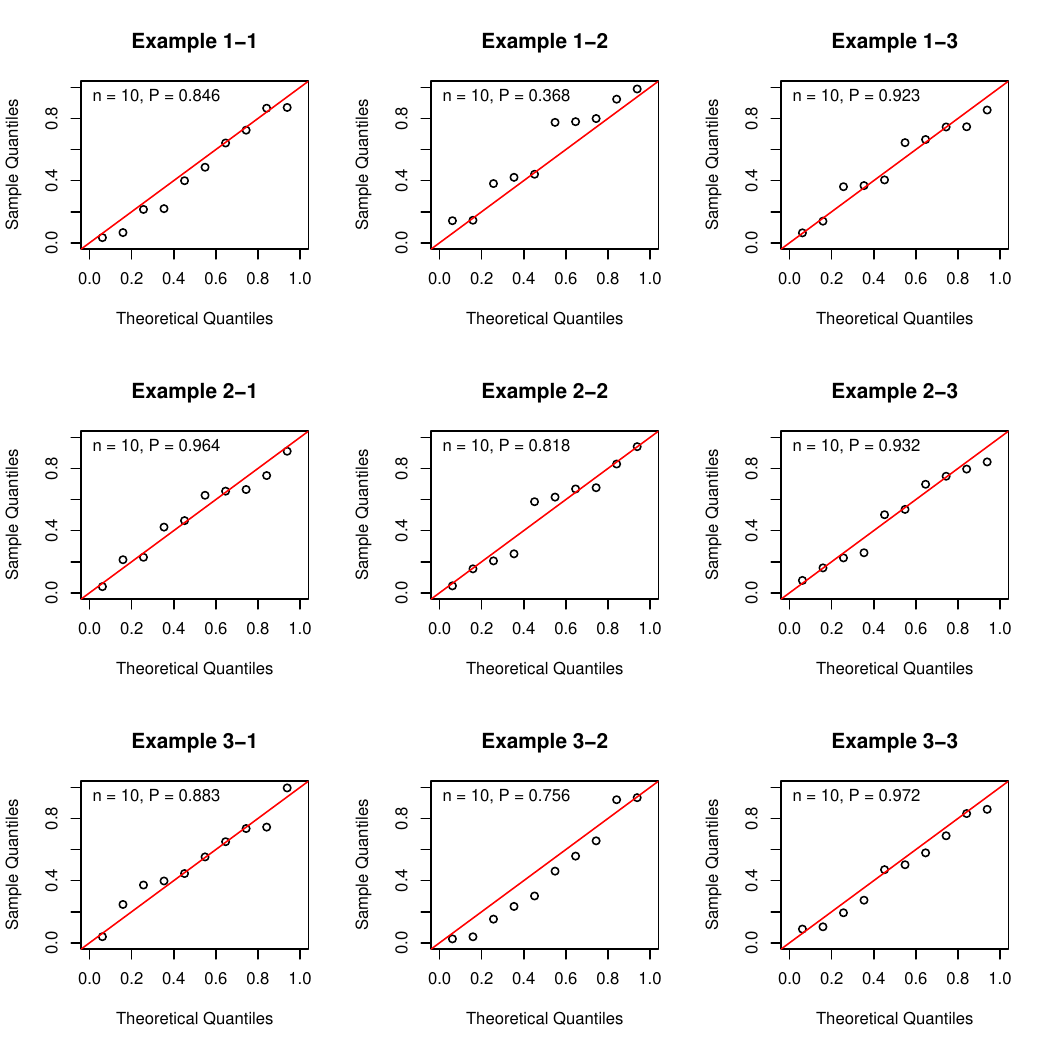}
    \caption{Q-Q plots of the estimated CDF values from the imputation model obtained using Algorithm~3, against the standard uniform distribution, in the many-normal-means simulation study (Section~5.1) with sample size $n = 10$. The $p$-value of the KS-test is indicated as $P$ in the top-left corner of each plot. Presented here are the results from the first three repetitions across the three settings.}\label{fig:qq-1}
    \end{figure}

\begin{figure}[!htbp]
    \centering
    \includegraphics[width=14cm]{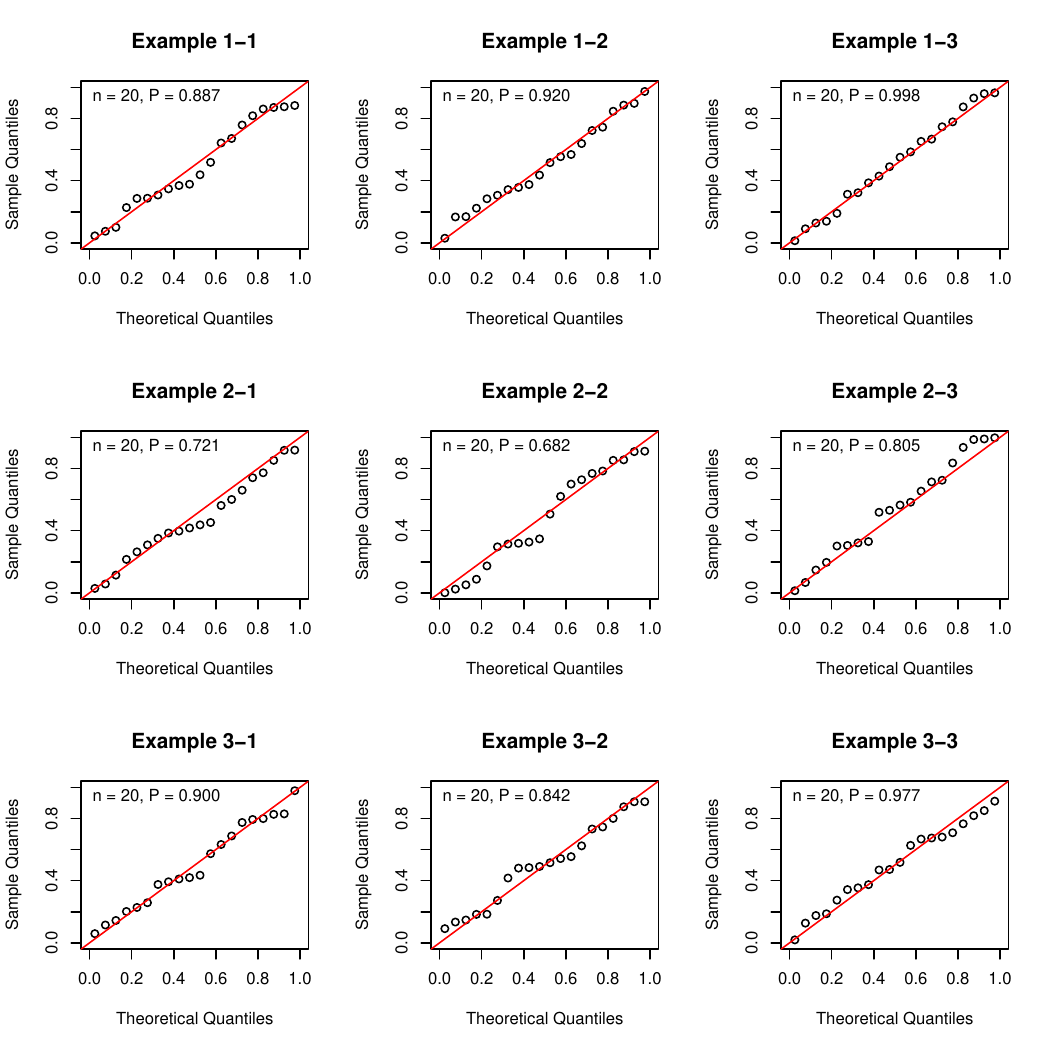}
%    \caption{Q-Q plots of the estimated CDF values from the imputation model obtained using Algorithm~2, against the standard uniform distribution, in the many-normal-means simulation study (Section~\ref{s:mnm}) with sample size $n = 20$. The $p$-value of the KS-test is indicated as $P$ in the top-left corner of each plot. Presented here are the results from the first three repetitions across the three settings.}\label{fig:qq-2}
     \caption{The same legend as Figure \ref{fig:qq-1}, but for $n=20.$}\label{fig:qq-2}
    \end{figure}

\begin{figure}[!htbp]
    \centering
    \includegraphics[width=14cm]{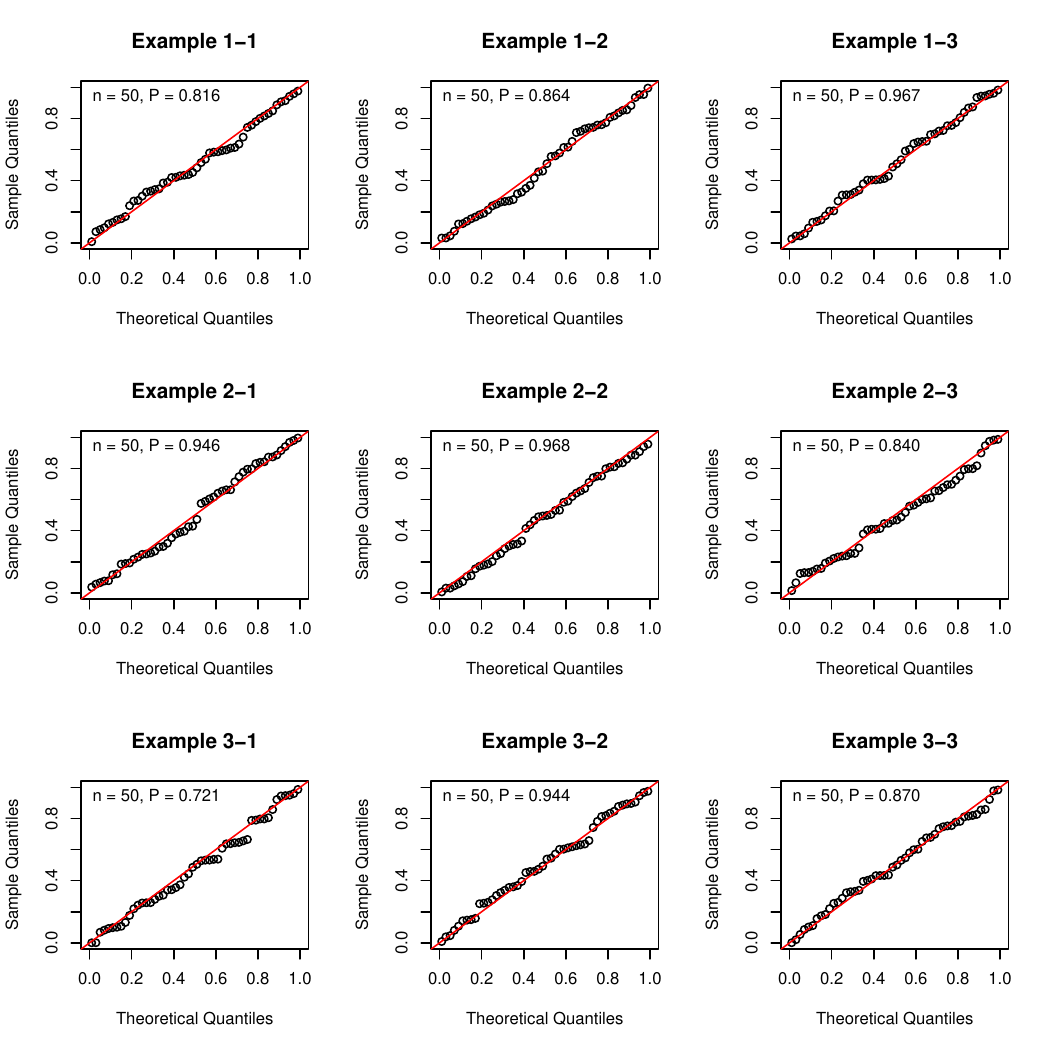}
%    \caption{Q-Q plots of the estimated CDF values from the imputation model obtained using Algorithm~2, against the standard uniform distribution, in the many-normal-means simulation study (Section~\ref{s:mnm}) with sample size $n = 50$. The $p$-value of the KS-test is indicated as $P$ in the top-left corner of each plot. Presented here are the results from the first three repetitions across the three settings.}\label{fig:qq-3}
    \caption{ The same legend as Figure \ref{fig:qq-1}, but for $n=50.$}\label{fig:qq-3}
    \end{figure}

\begin{figure}[!htbp]
    \centering
    \includegraphics[width=14cm]{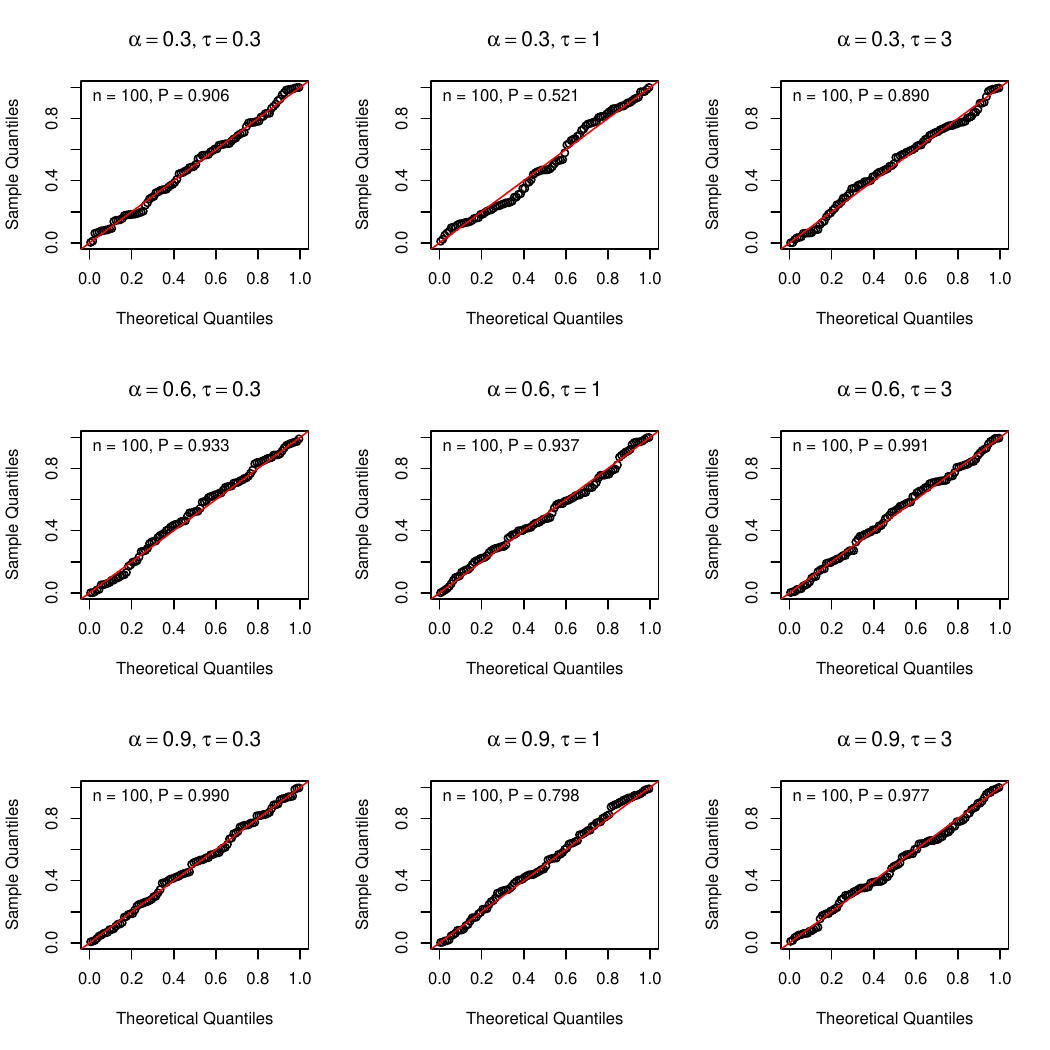}
    \caption{Q-Q plots of the estimated CDF values from the imputation model obtained using Algorithm~3, against the standard uniform distribution, in the linear regression example (Section~5.2). The $p$-value of the KS-test is indicated as $P$ in the top-left corner of each plot. Presented here are the results from the first repetition across various settings, using the (unweighted) $L_1$ duality function. }\label{fig:qq-4}
    \end{figure}

\begin{figure}[!htbp]
    \centering
    \includegraphics[width=14cm]{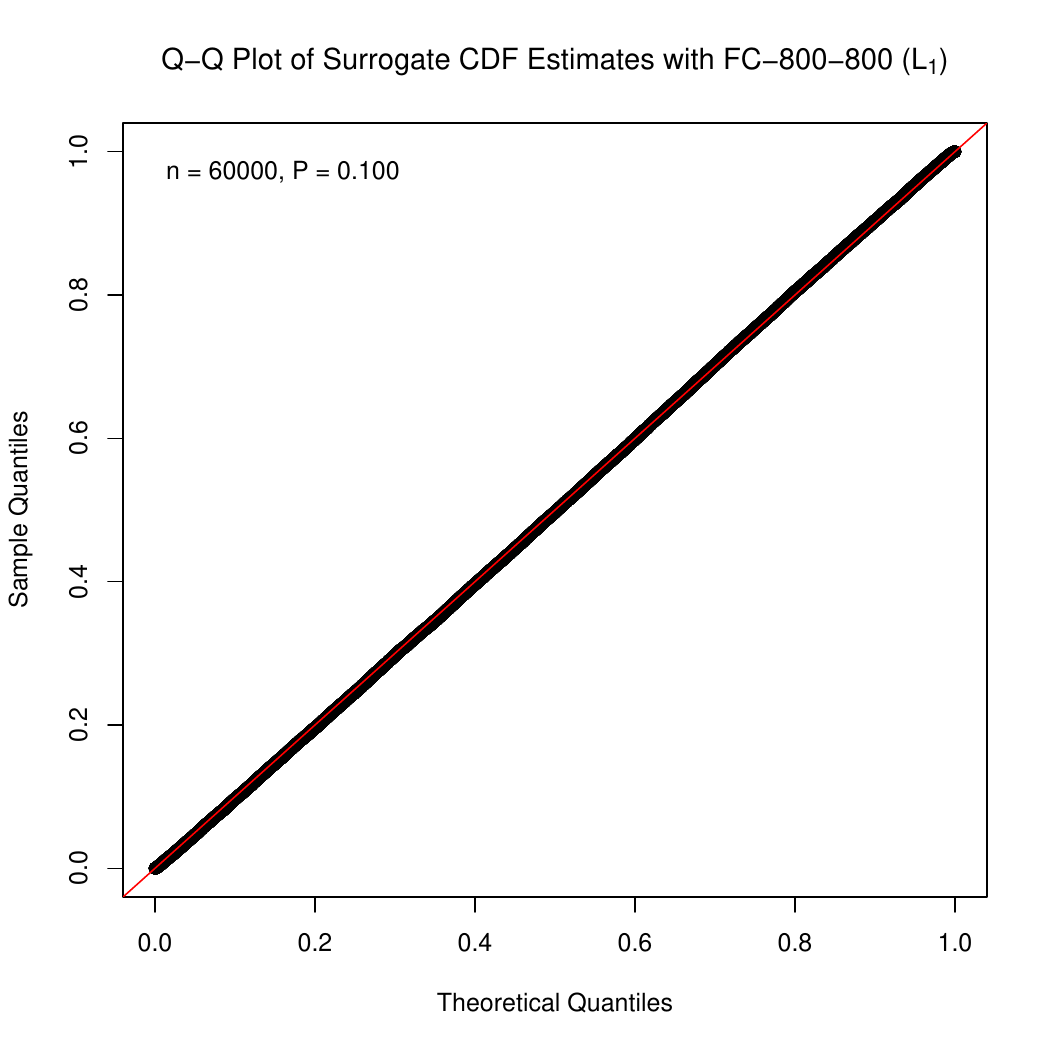}
    \caption{Q-Q plots of the estimated (surrogate) CDF values from the imputation model obtained using Algorithm~3, against the standard uniform distribution, in the neural network example (Section~5.3). The observed sample size and the $p$-value of the KS-test are indicated as $n$ and $P$, respectively, in the top-left corner of each plot. Presented here are the results obtained using the (weighted) $L_1$ duality function in fully connected networks featuring 800 hidden nodes.}\label{fig:qq-5}
    \end{figure}

The Q-Q plots in Figure~\ref{fig:qq-1} to \ref{fig:qq-5} demonstrate the imputation efficiency with the optimal $m$ selected by Algorithm~3. These plots cover all three application examples discussed in Section~5, with observed sample sizes ranging from $n = 10$ to $n = 60,000$. It can be seen that Algorithm~3 effectively approximates the target specified in (12) across all scenarios.

\clearpage

\subsection{A Comparison between AM and the Standard Regularization Approach}
In this section, we aim to outline the advantages of the AM framework compared to standard regularization techniques:
\begin{enumerate}
    \item The AM estimator offers significantly more flexibility in the choice of model forms compared to standard regularization techniques.
    \item The AM objective formulation potentially overcomes the limitations associated with CV error in cross-validation (CV).
    \item The AM objective formulation provides better upper-bound performance than the standard regularization objective, assuming the true population $\mathbb{P}$ is known.
    \item The AM framework is conceptually advanced and compatible with alternative modern synthetic data generation methods. As these techniques continue to evolve, the performance of the AM method is expected to improve correspondingly.
\end{enumerate}

Arguments 2 and 3 are further reinforced by argument 4. Note that for ease of explanation, we assume that the standard regularization techniques are accompanied by CV for the estimation of hyperparameters. However, many of the arguments (1,3, and 4) remain valid irrespective of this assumption. 

To facilitate a comparison with standard regularization techniques, we first formalize the expression of the standard regularization method. This method typically solves for $\theta$:
\begin{equation}\label{eq:3-3-1}
    \hat{\theta}_\lambda =  \mbox{arg} \min_{\theta} \frac{1}{n} \sum_{i = 1}^n L(\theta| x_i,y_i) + P(\theta, \lambda)
\end{equation}
where $P(\theta, \lambda)$ represents the penalty function, and $\lambda$ is the tuning parameter. For comparison, we assume the penalty function $P(\theta,\lambda)$ takes the same form as the duality function $\pi(\theta, \lambda)$ used in AM. The tuning parameters are typically determined through methods such as cross-validation (CV). To select $\lambda$ using the $K-$fold CV, consider the $K$ partitions $\mathcal{T} = \mathcal{T}_1 \cup \mathcal{V}_1 = \dots = \mathcal{T}_K \cup \mathcal{V}_K$, where $\mathcal{T}$ denotes the observed samples or training data and $\mathcal{V}_k$ the validation data for each $k$-th partition $\mathcal{T}_k \cup \mathcal{V}_k$. At the same time, $\mathcal{V}_k$ ($k = 1,\dots,K$) should have approximately equal sample size and $\mathcal{T}_k \cap \mathcal{V}_k = \emptyset$. The $K-$fold CV error or risk is introduced as \citep[{\it c.f.},][]{darren2017}:
\begin{equation}\label{eq:3-3-2}
    \hat{R}_{\mathcal{V}}(\lambda) = \hat{R}_{\mathcal{V}}\left( \hat{\theta}_\lambda^{(\mathcal{V}_1)}, \dots, \hat{\theta}_\lambda^{(\mathcal{V}_k)}\right) := \frac{1}{K} \sum_{v \in \mathcal{V}} \frac{1}{|v|} \sum_{r\in v} L(\hat{\theta}_\lambda^{(v)}| x_r,y_r)
\end{equation}
where $\mathcal{V} = \{ \mathcal{V}_1, \dots, \mathcal{V}_K\}$ is a set of validation sets and $\hat{\theta}_\lambda^{(v)}$ is the estimator in (\ref{eq:3-3-1}) with the observations in the validation set $v \in \mathcal{V}$ removed. The optimal tuning parameter $\lambda$ is identified by minimizing the CV-risk (\ref{eq:3-3-2}),
\begin{equation}\label{eq:3-3-2.3}
    \hat{\lambda} = \mbox{arg} \min_{\lambda \in \Lambda_s} \hat{R}_{\mathcal{V}}(\lambda)
\end{equation}
where $\Lambda_s$ is a predefined set of candidate parameters. We will next discuss the advantages of the AM estimator over standard regularization techniques from several perspectives.

\subsubsection{Modeling Flexibility}
In the standard regularization technique, the choice of candidate hyperparameter set $\Lambda_s$ is crucial, which is typically structured through a grid-search approach when no prior information being available. This approach determines the grid density and range of hyperparameters considered during optimization. However, the computational cost increases exponentially with the increase in the dimension of hyperparameters. Conversely, the AM estimator offers more flexibility. It allows various choices of the duality function with minimal impact on computational demands, enabling more adaptable modeling.

For example, when AM is used, the many-normal-means (MNM) problem can be modeled in a much more flexible over-parameterized model structure, as detailed in Section 5.1. Attempting a similar modeling approach with standard regularization would require the dimension of the hyperparameter $\lambda$ to match the sample size $n$, rendering the computation impractical for even modest sample sizes (e.g., $n = 10$).

Furthermore, the discrete nature of the candidate hyperparameter set in standard regularization methods restricts flexibility. Both the range of the search and the density of the grid are critical factors that need careful selection, significantly influencing the outcomes \citep{darren2017}.

\subsubsection{Potential Limitations of CV Error}
Standard regularization approaches are often validated through the bias-variance trade-off framework. Typically, the mean squared error (MSE) is decomposed into bias and variance components. Incorporating a regularization term generally increases the bias but decreases the variance, ideally leading to a reduction in overall error. The true population error is commonly approximated by the CV error (\ref{eq:3-3-2}), which provides guidance to the choice of hyperparameter $\lambda$. However, this approach presents notable.

Firstly, recent research \cite{Bates2023} indicates that the CV error may not effectively approximate the desired prediction error. Specifically, \cite{Bates2023} proves that in the linear regression model fitted with OLS, CV \textit{estimates the average prediction error of models fit on other unseen training sets drawn from the same population.} Secondly, the standard regularization approach separate the model selection and model estimation process, presuming that the ``best'' $\lambda$ will yield optimal model fitting across different datasets (this is because in $K$-fold cross-validation, the same tuned hyperparameters are evaluated across $K$ models, each fitted on different observations, as well as the final model using the full observation data). However, the optimal hyperparameters are inherently dependent on the specific observations used to fit the model, and may not even lead to a consistent estimation, especially in the over-parameterized models.

Given these challenges, our AM formulation (indexed by (8) in the main manuscript)
\begin{equation} \label{eq:3-4}
\underset{\theta, \lambda}{\text{min}}
\; G_{\hat{\mathbb{P}}}(\theta, \lambda) + V_{\mathbb{P}, \hat{\mathbb{P}}}(\theta, \lambda),  \quad \text{subject to}  \quad
\theta = \underset{\tilde\theta}{\text{arg min}} \, G_{\hat{\mathbb{P}}}(\tilde\theta, \lambda),
\end{equation}
which directly aims at filling the generalization gap, provides a natural alternative to the existing standard regularization approach. Moreover, the imputation method for creating an approximation of $\mathbb{P}$ is theoretically supported in both asymptotic and finite-sample scenarios.

\subsubsection{Better Upper-Bound Performance of AM}
Now, assuming the true population $\mathbb{P}$ is known, we compare the performance of the standard regularization approach and the AM framework. This is essentially the upper-bound performance of both methods in real scenarios. In this ideal scenario, the use of CV will no longer be necessary for the standard regularization, as the prediction error can be directly calculated from the known $\mathbb{P}$. The objective becomes:
\begin{equation}\label{eq:3-5}
\hat{\lambda} = \underset{\lambda \in \Lambda_s}{\text{min}} \ E_{(X,Y) \sim \mathbb{P}} L(\hat{\theta}_{\lambda} | X, Y),
\end{equation}
where $\hat{\theta}_{\lambda}$ is defined in (\ref{eq:3-3-1}). The final estimation of $\theta$ uses $\hat{\theta}_{\hat{\lambda}}$. However, this formulation is generally less effective than the AM formulation in (\ref{eq:3-4}) due to the discreteness of the candidate set $\Lambda_s$ and the potential non-uniqueness of solutions to (\ref{eq:3-3-1}) with given $\lambda$. Conversely, for AM, the formulation in (\ref{eq:3-4}) effectively combines the model selection and model estimation procedure, and an efficient numerical optimization method is developed for (\ref{eq:3-4}) in Section~3 of the main manuscript.

\begin{figure}[!htbp]
\centering
\includegraphics[width=10cm]{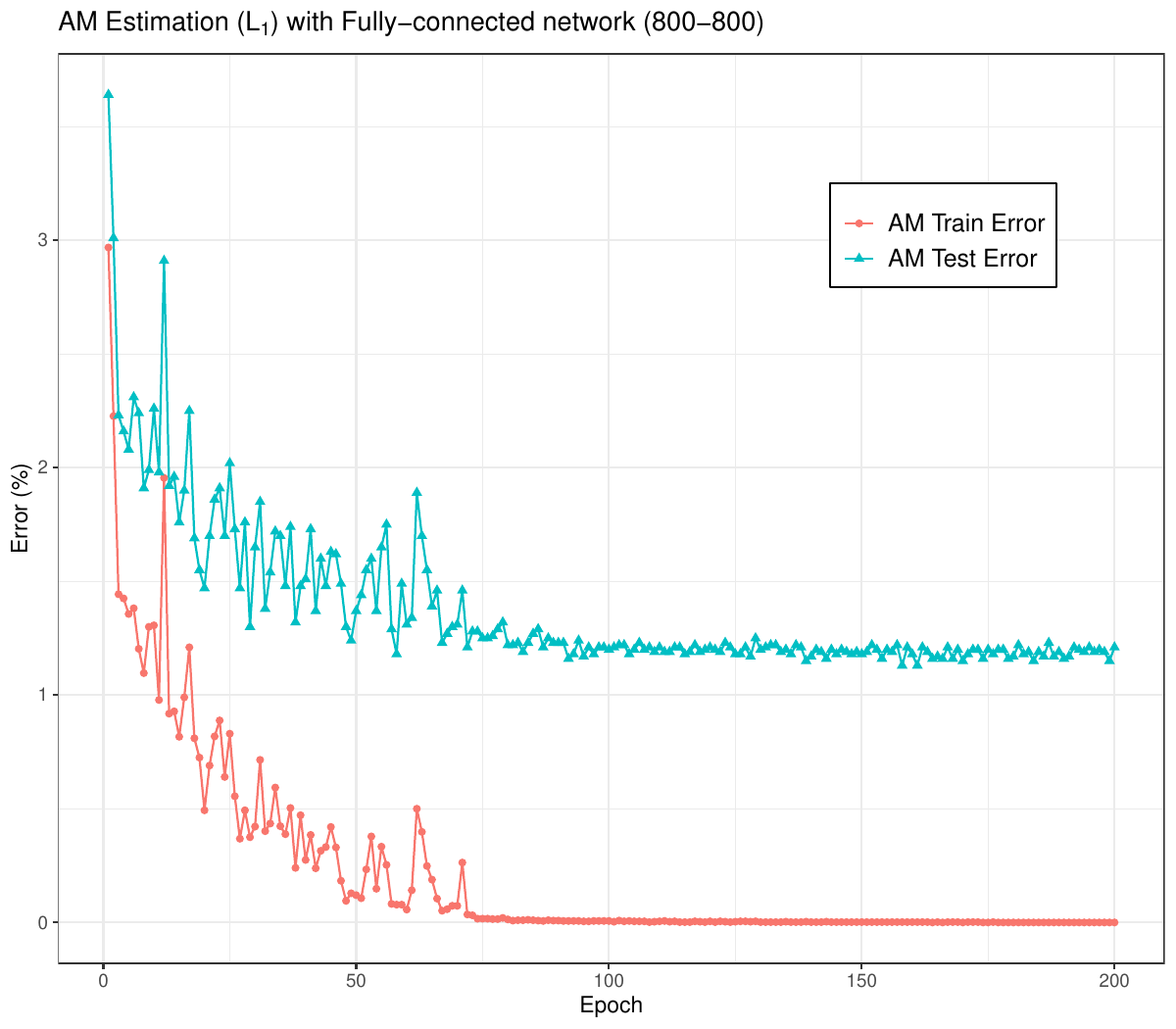}
\caption{The learning curve in the estimation process of AM on the MNIST experiment. A bootstrap sample of 37,950 unique images from the 60,000 training data points serves as $\hat{\mathbb{P}}$, with the entire 60,000 training images considered as $\mathbb{P}$.  Training errors are calculated using the 60,000 training images, and test errors are evaluated using the 10,000 test images. }\label{sp-8:fig-1}
\end{figure}

For illustration purpose, we conducted additional experiments using the neural network model on the MNIST data. Assuming the 60,000 training images and their labels represent
 $\mathbb{P}$, and a bootstrap sample from the training set, containing 37,950 unique images, serves as $\hat{\mathbb{P}}$ in (\ref{eq:3-4}). Surprisingly, yet plausibly given the theoretical underpinnings, by solving the formulated optimization problem, the AM estimator achieves an exact $100\%$ accuracy on the full training data viewed as $\mathbb{P}$ (see Figure~\ref{sp-8:fig-1}). Interestingly, when trained in this manner, the model also provides near state-of-the-art test accuracy, when compared to methods using the same model structure.
 
 In contrast, it is computationally infeasible for the standard regularization approach in (\ref{eq:3-5}) to attain such ``best possible'' results. This difficulty arises particularly because neural network estimation is heavily over-parameterized, making the control of $\hat{\theta}_\lambda$ through a discrete candidate set of $\lambda$ both less efficient and less flexible. 
 % Additionally, the estimation $\hat{\theta}_\lambda$ is not unique for each $\lambda$, increasing such difficulty. 
 Therefore, finding a $\hat{\theta}_\lambda$ that reaches 100\% accuracy on the ``new'' data $\mathbb{P} \setminus \hat{\mathbb{P}}$ (in this case, the remaining 22,050 images) through a candidate set of $\lambda$ values can be extremely challenging.    

\subsubsection{Recent Trends in Synthetic Data Generation}
A key distinction between the standard regularization approach and the proposed AM framework lies in the use of newly generated data. The significance of synthetic data generation methods has grown considerably in recent years. Recent studies \citep{hinton2015distillingknowledgeneuralnetwork, Angelopoulos2023, Zrnic2024} have demonstrated that utilizing model-predicted labels in inference and estimation tasks can substantially enhance performance.  Although it significantly diverges from these existing methods, the imputation framework employed in the proposed AM aligns with these trends of leveraging synthetic data generation, making it a conceptually advanced alternative to traditional methods like standard regularization techniques.

Furthermore, the AM framework is compatible with various synthetic data generation techniques for creating imputation populations. Recent synthetic data generation methods such as normalizing flows \citep{Kingma2018}, GPT \citep{Brown2020}, diffusion models \citep{Ho2020} and GANs \citep{goodfellow2020generative} have shown its promise in various applications. These methods offer viable options for constructing an imputation population that closely approximates the true population $\hat{\mathbb{P}}$. See Supplementary~S.12 for additional discussion. As these synthetic data generation techniques continue to evolve, the performance of the AM method is expected to improve correspondingly. This further underscores the earlier arguments about AM's potential to surpass traditional methods by mitigating the limitations of CV error and achieving upper-bound performance in practical applications. However, since the primary focus of this paper remains on the foundational aspects of modeling, we have opted not to use these more complex models for imputation in this paper, in the hope to avoid a logically circular modeling process or a chicken-and-egg type of situation.

\clearpage
\subsection{Limitations of the Standard Bootstrap Imputation Approach}

In addition to the imputation approach using an adaptive $m$-out-of-$n$ bootstrap accompanied with data-splitting, we might also consider the simple standard bootstrap imputation approach in certain scenarios. The imputation algorithm is detailed below.

\begin{alg}[Standard Bootstrap Imputation]
\rm
With observed data $(\bx, \by)$, the future observations $(\bx, \by_*)$ is created in the following steps:
% One future observation $(x, y)$ is created in the following three steps:
\begin{quote}
\begin{description}
\item[\it Step 1.]
Take a bootstrap resampled data
$\tilde{S} = \{(\tilde{x}_i, \tilde{y}_i): i=1,...,n\}$. Denote by $\tilde{\mathbb{P}}$ the 
corresponding empirical distribution.
\item[\it Step 2.]

Find the bootstrap estimate $\hat{\tilde{\theta}}_{AM}$ by solving the AM objective (8)  with the future population 
                $\mathbb{P} = (\bx, \by)$
                and the empirical distribution 
                $\hat{\mathbb{P}} = \tilde{\mathbb{P}}$.   
		%obtained by simultaneously minimizing $G_{\tilde{\mathbb{P}}}(\theta, \lambda)$ over $\theta$ and $\sum_{i=1}^p\left|\frac{\partial V_{\hat{\mathbb{P}}, \tilde{\mathbb{P}}}(\theta, \lambda)}{\partial \theta_i}\right|$ over $\lambda$ for $i = 1,\dots,p$.
%for all $i=1,...,p$.
\item[\it Step 3.]
Simulate $\by_*$ from the fitted predictive model obtained in Step 2 with $\bx$.
\end{description}
\end{quote}
\end{alg}

Compared to the new imputation algorithm, this simple imputation algorithm is conceptually straightforward, and can be useful in dealing with low-dimensional data. However, the standard bootstrap method used here exhibits less robustness in high-dimensional, finite-sample scenarios, necessitating the use of a more adaptive $m$-out-of-$n$ bootstrap. The use of the $m$-out-of-$n$ bootstrap provides much more flexibility compared with the standard bootstrap as it provide a better control of the imputation population through the selection of $m$. Meanwhile, the use of data splitting allows for checking model fitness and helps with the adaptive selection of resampling size $m$. However, the standard bootstrap does not have such benefits due to its fixed resampling size $m = n$.

\begin{figure}[!htbp]
\centering
\includegraphics[width=10cm]{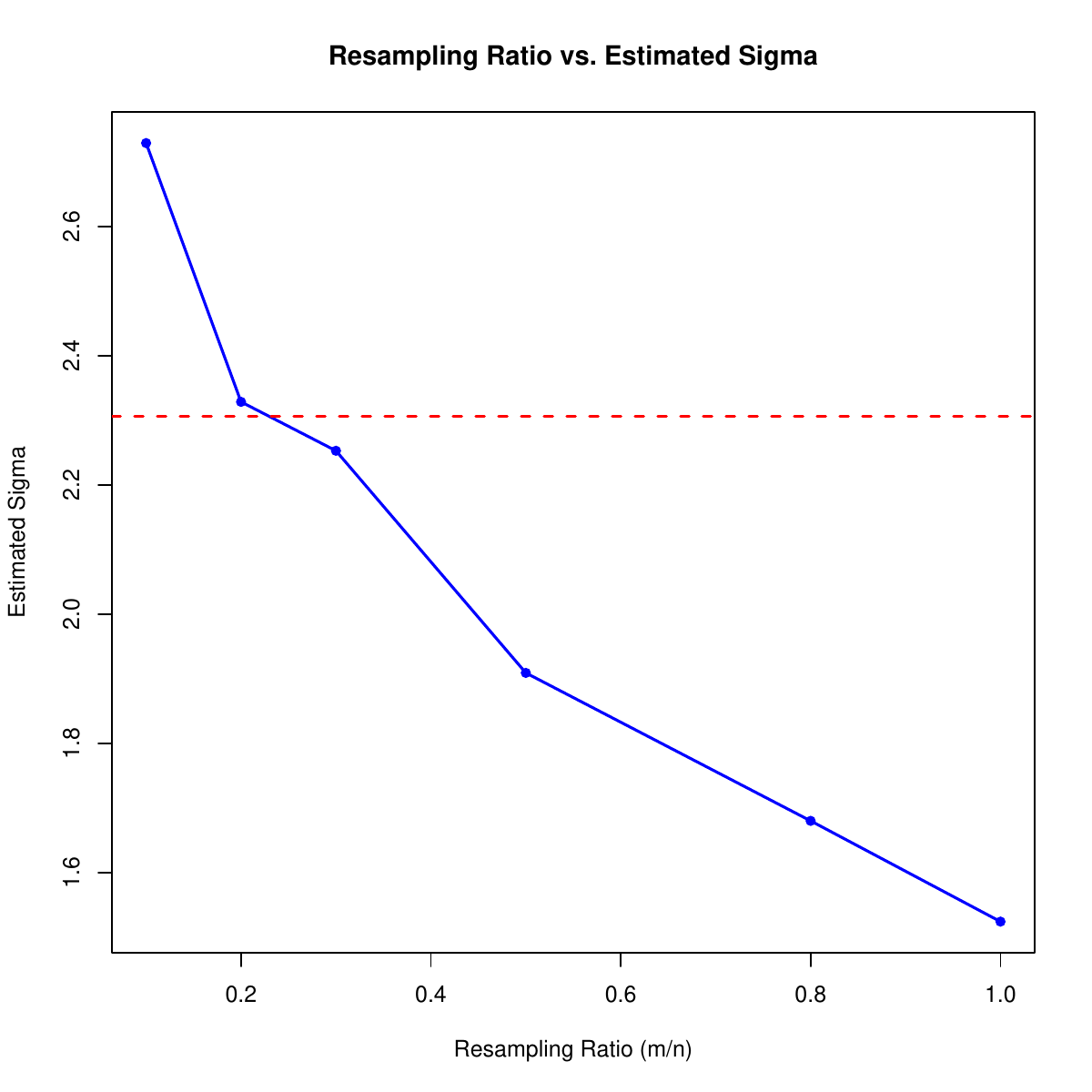}
\caption{Resampling ratio versus the mean value of $\hat{\sigma}$ obtained in the imputation step of the first simulated data set in the linear regression study. The true $\sigma$ is indicated as the red dotted horizontal line. }\label{sp-9:fig-1}
\end{figure}

As an illustrative example, we revisited the $n < p$ linear regression scenario. As is shown in Figure~\ref{sp-9:fig-1}, the value of the estimated variance parameter used for imputation, $\hat{\sigma}$, decreases as the sampling ratio $m/n$ increases. In this specific case, applying the standard bootstrap ($m/n = 1$)  for imputation results in underestimation of the variance parameter, leading to undercoverage in the constructed parametric confidence intervals. However, the proposed imputation framework in Section~2.3 utilizes an adaptive $m$-out-of-$n$ bootstrap, which allows for the selection of an optimal, smaller resampling ratio. With such flexibility, the new imputation method gives an estimation of $\sigma$ that provides a robust and reliable confidence interval at $95\%$ or other levels (see Section 5.2 and Supplementary~S.8.2).

\clearpage
\subsection{Additional Discussion on the Many-Normal-Means Example}

In this section, we first review the generalized version of the James-Stein Estimator that supports multiple shrinkage targets \citep{George1987}. Additionally, we present results from further experiments conducted using this generalized estimator, demonstrating that without meaningful prior information, the accuracy gains may be negligible. We also describe an feasible adaptive implementation of such generalized estimator for obtaining the results of MJS estimator reported in Section 5.1, using prior information obtained from a certain additional modeling procedure. Furthermore, we report the results of $g$-modeling using increased grid density of the knots, showing that increased modeling flexibility does not lead to improved performance for $g$-modeling.  Based on these findings, we then take a retrospective look at the proposed AM estimator in light of these insights. 

\subsubsection{Multiple-Shrinkage James-Stein Estimator}
Recall that the normal means problem is about making inference
on the unknown means $\mu_1, ..., \mu_n$ from the sample $y_1,..., y_n$
with the model
%\[
$	Y_i|\{\mu_1,...,\mu_n\} \sim N(\mu_i, 1)$, $i=1,...,n$,
%\]
where $y_1,..., y_n$ are independent of each other. Denote $\by = (y_1, \dots, y_n)$ as the observed vector and $\bar{y} = \frac{1}{n}\sum_{i = 1}^n y_i$. The classical James-Stein (JS) estimator, as is introduced in \cite{Stein1956,js1961}, aims to shrink the MLEs of the mean parameters toward their mean value, and is expressed as (equivalent to the expression on page 20 of the main manuscript)
\begin{equation}\label{eq:mnm-1}
    \hat{\mu}^{JS} = \by - (n- 3) \frac{\by - \bar{\by}}{\| \by - \bar{\by}\|},
\end{equation}
where $\bar{\by}$ denotes the vector of $n$ $\bar{y}$s.

 The multiple-shrinkage James-Stein (MJS) estimator proposed in \cite{George1987} (see also \cite{George1986a, George1986b}) considers $K$ distinct choices of the target subspace used for shrinkage, denoted by $V_1, \dots, V_K$, which results in $K$ specific JS estimators
 \[
    \hat{\mu}_{k}^{JS} = \by - (q_k- 2) \frac{Q_k\by}{\|Q_k\by\|}, \quad k = 1,\dots,K,
 \]
 where $V_k$ is a subspace of $\mathbb{R}^n$ of dimension $n - q_k$, with $q_k\ge 3$, and $Q_k\by$ denotes the projection of $\by$ onto the orthogonal complement of $V_k$. It can be seen that \eqref{eq:mnm-1} is a special case of this generalization with $K = 1$ and the projection vector $(1, \dots, 1)$. \cite{George1987} suggests that these $K$ targets are based on {\it ``the results of several previous related experiments suggested as reasonable targets''}.  For our study, we consider $V_1, \dots, V_K$ to be $K$ fixed vectors in $\mathbb{R}^n$, that is, $V_k := (v_{k1}, \dots, v_{kn})'$, for $k = 1,\dots,K$. This allows us to express the $K$ estimators more specifically as:%\CLdel{$(n-2)$ or $(n-3)$, because $\mbox{dim}(V_k)=1$? Also, is the numeric correct? We have $Q_kY = Y - (V_k'V_k)^{-1}V_k V_k'Y = Y - \frac{V_k'Y}{V_k'V_k}V_k$. Do you assume $\frac{V_k'Y}{V_k'V_k}=$?} 
%{\color{blue} Thanks for the concern. From George's paper, (S.39) has $V = [1_n]$, a subspace spanned by the $n \times 1$ vector, and thus $\mbox{dim}(V) = 1$. In this case, $V$ is a fixed vector, which gives $\mbox{dim} (V) = 0$. Verified again from the Wikipedia page of James-Stein estimator about $n - 2$. }
 \[
    \hat{\mu}_{ik}^{JS} = y_i - (n- 2) \frac{y_i - v_{ki}}{\sum_{i=1}^n (y_i - v_{ki})^2}, \quad k = 1,\dots,K,
 \]
 for $i = 1,\dots,n$. \cite{George1987} further expresses the combination of these $K$ estimators in the following form as the multiple shrinkage estimator:
 \begin{equation}\label{eq:mnm-2}
 \hat{\mu}_{i *}^{JS} = \sum_{k = 1}^K \rho_k \hat{\mu}_{ik}^{JS}, \quad i = 1, \dots, n,
\end{equation}
 where%\CLdel{Again, please check the difference $y_i - v_{ki}$. Check also the power $\frac{n-2}{2}$. Should it be $\frac{n-3}{2}$ instead?}
  \[
 \rho_k = \frac{w_k \left(\sum_{i=1}^n \left(y_i - v_{ki}\right)^2 \right)^{-\frac{n-2}{2}}}{\sum_{l=1}^K w_l \left(\sum_{i=1}^n \left(y_i - v_{li}\right)^2 \right)^{-\frac{n-2}{2}}}, \quad k = 1,\dots,K,
 \]
 and $w_k$ is some pre-defined weight as the probability for choosing the $k$-th JS estimator and, thereby,
 \[
 w_k > 0 \quad \text{ and } \quad \sum_{k = 1}^K w_k = 1.
 \]
Besides the theoretical investigation, \cite{George1987} also demonstrates the minimaxity of this estimator through a simulation study considering the model $\by \sim N_{10}(\theta, I)$, with shrinkage towards two predefined targets $V_1$ and $V_2$ equally weighted at $w_1 = w_2 = 0.5$. The minimaxity is illustrated by varying $\theta$ as $(1-\lambda) V_1 + \lambda V_2$ with $\lambda = -0.5, 1.5,$ and $ 0.25$.

\subsubsection{Advantages and Limitations of the Generalized Estimator}

The MJS estimators address a key limitation of the original JS estimator, which only allows for shrinkage towards a single target. This adaptability is especially beneficial in scenarios where multiple plausible target regions in the parameter space have been identified for prior experiments. However, a significant limitation emerges when prior information about these targets is limited or absent. Under such conditions, the advantages of multiple shrinkage over the traditional JS estimator may become marginal. For instance, in the simulation study described in Section 5.1, there was no prior information about the true parameters, ensuring an arguably fair comparison among all methods.

To deepen our understanding and for further illustration purpose, we conducted additional experiments using the MJS estimators in the simulation study. We consider a ``flat prior'', selecting a grid of equally spaced target points within the range of observed $\by$. The results, presented in Table~\ref{tab:1}, indicate that the performance of the MJS is very similar to the original JS estimator in \eqref{eq:mnm-1}.

\begin{table}[!htbp]
\caption{%Summarized 
MPE results under three simulation settings with the James-Stein estimator and its multiple shrinkage generalization. The standard deviation of each value (estimated with bootstrap) is given in parentheses. \label{tab:1}}
\scriptsize
\centering
\begin{tabular}{cccccccccccc}
\hline
\multirow{2}{*}{\textit{Method}}                                 & \multicolumn{3}{c}{$\mu \sim N(0,0.01)$}                                                                                                                                    &                      & \multicolumn{3}{c}{\begin{tabular}[c]{@{}c@{}}$\mu^1 \sim N(-2,0.01)$\\ $\mu^2 \sim N(2,0.01)$\end{tabular}}                                                                &                      & \multicolumn{3}{c}{\begin{tabular}[c]{@{}c@{}}$\mu^1 = 0$\\ $\mu^2 \sim N(-3,1)$\end{tabular}}                                                                              \\ \cline{2-4} \cline{6-8} \cline{10-12} 
                                                                 & {\ul $n = 10$}                                          & {\ul $n = 20$}                                          & {\ul $n = 50$}                                          &                      & {\ul $n = 10$}                                          & {\ul $n = 20$}                                          & {\ul $n = 50$}                                          &                      & {\ul $n = 10$}                                          & {\ul $n = 20$}                                          & {\ul $n = 50$}                                          \\
James-Stein                                                      & \begin{tabular}[c]{@{}c@{}}0.293\\ (0.009)\end{tabular} & \begin{tabular}[c]{@{}c@{}}0.164\\ (0.005)\end{tabular} & \begin{tabular}[c]{@{}c@{}}0.068\\ (0.002)\end{tabular} &                      & \begin{tabular}[c]{@{}c@{}}0.881\\ (0.012)\end{tabular} & \begin{tabular}[c]{@{}c@{}}0.865\\ (0.008)\end{tabular} & \begin{tabular}[c]{@{}c@{}}0.830\\ (0.005)\end{tabular} &                      & \begin{tabular}[c]{@{}c@{}}0.554\\ (0.012)\end{tabular} & \begin{tabular}[c]{@{}c@{}}0.500\\ (0.008)\end{tabular} & \begin{tabular}[c]{@{}c@{}}0.485\\ (0.005)\end{tabular} \\
\multicolumn{1}{l}{}                                             & \multicolumn{1}{l}{}                                    & \multicolumn{1}{l}{}                                    & \multicolumn{1}{l}{}                                    & \multicolumn{1}{l}{} & \multicolumn{1}{l}{}                                    & \multicolumn{1}{l}{}                                    & \multicolumn{1}{l}{}                                    & \multicolumn{1}{l}{} & \multicolumn{1}{l}{}                                    & \multicolumn{1}{l}{}                                    & \multicolumn{1}{l}{}                                    \\
\begin{tabular}[c]{@{}c@{}}James-Stein\\ (Multiple)\end{tabular} & \begin{tabular}[c]{@{}c@{}}0.285\\ (0.009)\end{tabular} & \begin{tabular}[c]{@{}c@{}}0.159\\ (0.005)\end{tabular} & \begin{tabular}[c]{@{}c@{}}0.065\\ (0.002)\end{tabular} &                      & \begin{tabular}[c]{@{}c@{}}0.880\\ (0.012)\end{tabular} & \begin{tabular}[c]{@{}c@{}}0.864\\ (0.008)\end{tabular} & \begin{tabular}[c]{@{}c@{}}0.829\\ (0.005)\end{tabular} &                      & \begin{tabular}[c]{@{}c@{}}0.549\\ (0.012)\end{tabular} & \begin{tabular}[c]{@{}c@{}}0.498\\ (0.008)\end{tabular} & \begin{tabular}[c]{@{}c@{}}0.485\\ (0.005)\end{tabular} \\ \hline
\end{tabular}
\end{table}

One might wonder if the superior performance of the AM method compared to the JS method is due to its distinctive model structure. However, the discussion above clarifies that even though the generalized JS method allows for flexible modeling, effective estimation still requires meaningful prior information, which is often absent in practical scenarios.

\subsubsection{An Adaptive Implementation of the Multiple-Shrinkage James-Stein Estimator}

To address the lack of meaningful prior information for the MJS estimator, we employ a data-driven approach, in favor of MJS, to determine the necessary shrinkage target information, specifically utilizing Dirichlet process mixture models (DPMM). After running DPMM, we obtain a $C \times 1$ shrinkage target vector, $(\eta_1, \dots \eta_C)' \in \mathbb{R}^C$ and an $n \times C$ membership matrix \( \mathbf{P} \in \mathbb{R}^{n \times C} \), where \( P_{ij} \ (i = 1,\dots, n; j = 1, \dots, C)\) denotes the probability of assigning the \( i \)-th data point to component with center \( \eta_j \) for $j=1,..., C$ and, thereby, each row of \( \mathbf{P} \) sums up to 1.

In calculating the MJS estimator, the membership information from DPMM (or similar methods) can be considered as adaptive prior information for probabilistic shrinkage of each observation to multiple targets. Intuitively, each component of $\by$ can be shrunk to any of the $C$ centers. Thus, there are $C^n$ such fixed target vectors, each of which takes the form of $(\eta_{(1)},\dots, \eta_{(n)})'$, where $\eta_{(k)}$ ($k = 1,\dots, n$) can be any of the $C$ centers $\eta_1, \dots, \eta_C$. Ideally, we would construct these $C^n$ subspaces $V_1, \dots, V_{C^n}$ for MJS, where each subspace corresponds to a unique permutation of membership assignments for the $n$ observations, with associated weights $w_i \ (i = 1, \dots, C^n)$ calculated as products of the corresponding probabilities from $\mathbf{P}$. 

However, the above approach is computationally infeasible for large $n$ and $C$, so we use an approximation through Monte-Carlo sampling \citep{robert2004monte}.
In our experiments, we perform $M = 1,000$ Monte-Carlo iterations. For each iteration $m = 1, \dots, M$, we sample a shrinkage target vector $V^{(m)} = (v^{(m)}_1, \dots, v^{(m)}_n)'$ from the membership matrix $\mathbf{P}$ and calculate the weight $w^{(m)}$ by taking the product of the corresponding probabilities. To prevent numerical floating error issues, we use log-probabilities here. Then, we normalize $w^{(1)}, \dots, w^{(M)}$ so they sum to $1$. Finally, we compute the MJS estimators using $V^{(1)}, \dots V^{(M)}$ and the normalized weights $w^{(1)}, \dots, w^{(M)}$ as defined in (\ref{eq:mnm-2}).

\subsubsection{$g$-modeling with Increased Grid Density of Knots}

For $g$-modeling, when no prior information is available, \cite{efron2016empirical} and \cite{Narasimhan2020} recommend selecting equally spaced discrete support points or knots, $\eta_1, \dots, \eta_l$, within a reasonable range. In our results for $g$-modeling presented in Section~5.1, we chose knots ranging from the minimum to the maximum observed values of $y$, with a spacing of $d = 0.1$ between consecutive points, and included the point at 0 as an additional knot. A natural question arises about the adequacy of this knot spacing with an interval length of $d = 0.1$, as AM does not rely on fixed knots, and its advantage over $g$-modeling could stem from modeling flexibility rather than estimation efficiency. To address this, we conducted additional experiments for $g$-modeling across the three examples, varying the interval length at $d \in \{0.1, 0.05, 0.01\}$.

\begin{table}[!htbp]
\caption{%Summarized 
MPE results under three simulation settings of the $g$-modeling with varying interval length $d$. The standard deviation of each value (estimated with bootstrap) is given in parentheses. \label{tab:2}}
\scriptsize
\centering
\begin{tabular}{cccccccccccc}
\hline
\multirow{2}{*}{\textit{Method}}                                    & \multicolumn{3}{c}{$\mu \sim N(0,0.01)$}                                                                                                                                    &  & \multicolumn{3}{c}{\begin{tabular}[c]{@{}c@{}}$\mu^1 \sim N(-2,0.01)$\\ $\mu^2 \sim N(2,0.01)$\end{tabular}}                                                                &  & \multicolumn{3}{c}{\begin{tabular}[c]{@{}c@{}}$\mu^1 = 0$\\ $\mu^2 \sim N(-3,1)$\end{tabular}}                                                                              \\ \cline{2-4} \cline{6-8} \cline{10-12} 
                                                                    & {\ul $n = 10$}                                          & {\ul $n = 20$}                                          & {\ul $n = 50$}                                          &  & {\ul $n = 10$}                                          & {\ul $n = 20$}                                          & {\ul $n = 50$}                                          &  & {\ul $n = 10$}                                          & {\ul $n = 20$}                                          & {\ul $n = 50$}                                          \\
\begin{tabular}[c]{@{}c@{}}$g$-modeling\\ ($d = 0.1$)\end{tabular}  & \begin{tabular}[c]{@{}c@{}}0.394\\ (0.014)\end{tabular} & \begin{tabular}[c]{@{}c@{}}0.390\\ (0.010)\end{tabular}  & \begin{tabular}[c]{@{}c@{}}0.170\\ (0.004)\end{tabular} &  & \begin{tabular}[c]{@{}c@{}}0.771\\ (0.018)\end{tabular} & \begin{tabular}[c]{@{}c@{}}0.731\\ (0.012)\end{tabular} & \begin{tabular}[c]{@{}c@{}}0.729\\ (0.008)\end{tabular} &  & \begin{tabular}[c]{@{}c@{}}0.548\\ (0.015)\end{tabular} & \begin{tabular}[c]{@{}c@{}}0.538\\ (0.013)\end{tabular} & \begin{tabular}[c]{@{}c@{}}0.378\\ (0.007)\end{tabular} \\
\begin{tabular}[c]{@{}c@{}}$g$-modeling\\ ($d = 0.05$)\end{tabular} & \begin{tabular}[c]{@{}c@{}}0.397\\ (0.014)\end{tabular} & \begin{tabular}[c]{@{}c@{}}0.430\\ (0.011)\end{tabular} & \begin{tabular}[c]{@{}c@{}}0.235\\ (0.005)\end{tabular} &  & \begin{tabular}[c]{@{}c@{}}0.765\\ (0.018)\end{tabular} & \begin{tabular}[c]{@{}c@{}}0.724\\ (0.012)\end{tabular} & \begin{tabular}[c]{@{}c@{}}0.734\\ (0.008)\end{tabular} &  & \begin{tabular}[c]{@{}c@{}}0.559\\ (0.015)\end{tabular} & \begin{tabular}[c]{@{}c@{}}0.603\\ (0.013)\end{tabular} & \begin{tabular}[c]{@{}c@{}}0.417\\ (0.007)\end{tabular} \\
\begin{tabular}[c]{@{}c@{}}$g$-modeling\\ ($d = 0.01$)\end{tabular} & \begin{tabular}[c]{@{}c@{}}0.400\\ (0.014)\end{tabular} & \begin{tabular}[c]{@{}c@{}}0.440\\ (0.011)\end{tabular} & \begin{tabular}[c]{@{}c@{}}0.505\\ (0.007)\end{tabular} &  & \begin{tabular}[c]{@{}c@{}}0.760\\ (0.018)\end{tabular} & \begin{tabular}[c]{@{}c@{}}0.718\\ (0.012)\end{tabular} & \begin{tabular}[c]{@{}c@{}}0.729\\ (0.008)\end{tabular} &  & \begin{tabular}[c]{@{}c@{}}0.565\\ (0.015)\end{tabular} & \begin{tabular}[c]{@{}c@{}}0.658\\ (0.013)\end{tabular} & \begin{tabular}[c]{@{}c@{}}0.598\\ (0.008)\end{tabular} \\ \hline
\end{tabular}
\end{table}

The results summarized in Table~\ref{tab:2} clearly indicate that increasing the density of the grid leads to a performance decline in many examples. Consequently, we report the results for $g$-modeling using an interval length of $d = 0.1$ in Section~5.1.

\subsubsection{A Retrospective Look on AM}

The numerical results for the adaptive implementation of the MJS estimator, together with other comparable methods reported in Section~5.1,  validates the AM estimator's effectiveness in model estimation in the normal means problem. In other words, the robustness and flexibility of the AM approach in estimating normal means stem not from any inherent structure in the means but from the innovative framework itself. Specifically, AM uses its prediction-based imputation-estimation framework for generating observations from the imputation population and use these observations for estimating flexible models. This framework can effectively handle practical scenarios lacking prior information.

%\CLdel{Furthermore, as discussed in Section~5.1, $g$-modeling, which appears in the comparison in Section~5.1, uses a similar empirical Bayes model structure approach as AM, aimed at recovering the prior distribution $g(\theta)$. However, unlike AM ,which impute future observations, $g$-modeling is constrained to penalized likelihood maximization using only empirical samples, thus offering less flexible choices of means $\mu_1, \dots, \mu_n$. The promising results of AM in Section~5.1 further demonstrates the superiority of the AM's key idea to impute and fit models to future observations.}

\subsubsection{Additional Experiments: $\mu$ Following Uniform Distribution}

Here, we conduct additional experiments by letting $\mu$ follow a non-Gaussian distribution. In the first simulation study, we assume $\mu_i$ ($i = 1, \dots, n$) follows a uniform distribution in $(-3.1, -2.9)$. The second simulation study considers the multi-modal case: half of the $n$ unknown means are drawn from a uniform distribution in $(-2.1, -1.9)$ and the other half from a uniform distribution in $(1.9, 2.1)$. These two settings serve as counterparts to the unimodal and bimodal cases used in Section~5.1, but with $\mu$ following a non-Gaussian distribution.

The results are summarized in Table~\ref{tab:3}. They show that, in both examples, all methods except for DPMM and AM perform similarly to their performance in Section~5.1. However, under the unimodal setting, AM demonstrates an improvement compared to Section~5.1, while DPMM exhibits a decline in performance. These numerical results indicate that in this setting, DPMM does not outperform AM, even in the simplest case, which contrasts with the observations in Section~5.1.

\begin{table}[!htbp]
\caption{%Summarized 
MPE results under two simulation settings with different methods. Each entry is taken as the average value obtained with $500$ repetitions and the standard deviation of each value (estimated with bootstrap) is given in parentheses. The best result in each setting is highlighted in boldface.\label{tab:3}}
\scriptsize
\centering
\begin{tabular}{cccccccc}
\hline
\multirow{2}{*}{\textit{Method}} & \multicolumn{3}{c}{$\mu \sim \text{Unif} \ (-3.1, -2.9)$}                                                                                                                                                       &  & \multicolumn{3}{c}{\begin{tabular}[c]{@{}c@{}}$\mu^1 \sim \text{Unif} \ (-2.1, -1.9)$\\ $\mu^2 \sim \text{Unif} \ (1.9, 2.1)$\end{tabular}}                                                                              \\ \cline{2-4} \cline{6-8} 
                                 & {\ul $n = 10$}                                                   & {\ul $n = 20$}                                                   & {\ul $n = 50$}                                                   &  & {\ul $n = 10$}                                                   & {\ul $n = 20$}                                                   & {\ul $n = 50$}                                                   \\
MLE                              & \begin{tabular}[c]{@{}c@{}}0.992\\ (0.021)\end{tabular}          & \begin{tabular}[c]{@{}c@{}}0.995\\ (0.015)\end{tabular}          & \begin{tabular}[c]{@{}c@{}}0.990\\ (0.009)\end{tabular}          &  & \begin{tabular}[c]{@{}c@{}}1.007\\ (0.020)\end{tabular}          & \begin{tabular}[c]{@{}c@{}}1.014\\ (0.013)\end{tabular}          & \begin{tabular}[c]{@{}c@{}}1.002\\ (0.009)\end{tabular}          \\
JS                               & \begin{tabular}[c]{@{}c@{}}0.323\\ (0.025)\end{tabular}          & \begin{tabular}[c]{@{}c@{}}0.150\\ (0.007)\end{tabular}          & \begin{tabular}[c]{@{}c@{}}0.062\\ (0.003)\end{tabular}          &  & \begin{tabular}[c]{@{}c@{}}0.894\\ (0.017)\end{tabular}          & \begin{tabular}[c]{@{}c@{}}0.858\\ (0.011)\end{tabular}          & \begin{tabular}[c]{@{}c@{}}0.813\\ (0.007)\end{tabular}          \\
MJS                              & \begin{tabular}[c]{@{}c@{}}0.384\\ (0.019)\end{tabular}          & \begin{tabular}[c]{@{}c@{}}0.211\\ (0.010)\end{tabular}           & \begin{tabular}[c]{@{}c@{}}0.106\\ (0.005)\end{tabular}          &  & \begin{tabular}[c]{@{}c@{}}0.744\\ (0.025)\end{tabular}          & \begin{tabular}[c]{@{}c@{}}0.629\\ (0.02)\end{tabular}           & \begin{tabular}[c]{@{}c@{}}0.492\\ (0.015)\end{tabular}          \\
DPMM                             & \begin{tabular}[c]{@{}c@{}}0.213\\ (0.012)\end{tabular}          & \textbf{\begin{tabular}[c]{@{}c@{}}0.107\\ (0.006)\end{tabular}} & \textbf{\begin{tabular}[c]{@{}c@{}}0.047\\ (0.003)\end{tabular}} &  & \begin{tabular}[c]{@{}c@{}}0.771\\ (0.027)\end{tabular}          & \begin{tabular}[c]{@{}c@{}}0.546\\ (0.018)\end{tabular}          & \begin{tabular}[c]{@{}c@{}}0.375\\ (0.010)\end{tabular}          \\
$g$-modeling                     & \begin{tabular}[c]{@{}c@{}}0.414\\ (0.014)\end{tabular}          & \begin{tabular}[c]{@{}c@{}}0.416\\ (0.010)\end{tabular}          & \begin{tabular}[c]{@{}c@{}}0.176\\ (0.004)\end{tabular}          &  & \begin{tabular}[c]{@{}c@{}}0.772\\ (0.017)\end{tabular}          & \begin{tabular}[c]{@{}c@{}}0.740\\ (0.012)\end{tabular}          & \begin{tabular}[c]{@{}c@{}}0.736\\ (0.009)\end{tabular}          \\
Auto-modeling                    & \textbf{\begin{tabular}[c]{@{}c@{}}0.187\\ (0.013)\end{tabular}} & \begin{tabular}[c]{@{}c@{}}0.109\\ (0.007)\end{tabular}          & \textbf{\begin{tabular}[c]{@{}c@{}}0.047\\ (0.003)\end{tabular}} &  & \textbf{\begin{tabular}[c]{@{}c@{}}0.643\\ (0.028)\end{tabular}} & \textbf{\begin{tabular}[c]{@{}c@{}}0.498\\ (0.018)\end{tabular}} & \textbf{\begin{tabular}[c]{@{}c@{}}0.360\\ (0.011)\end{tabular}} \\ \hline
\end{tabular}
\end{table}

\clearpage
\subsection{Computational Time of the Application Examples}

We present the computational times of AM recorded for the application examples discussed in the paper. These times are based on the first repetition of each application example. The evaluations were performed on a personal computer equipped with an Intel Core i7-12700KF CPU and an NVIDIA GeForce RTX 3070 Ti GPU. We conducted the computations without employing any parallel processing techniques. The detailed computational results are summarized in Table~\ref{tab:computation}.
Notably, the application of AM in both the many-normal-means and linear regression scenarios was implemented with R. It is important to note that a further acceleration in performance is achievable through implementation in C. The result presented in Table~\ref{tab:computation} effectively demonstrates the scalability of AM when applied to large-scale problems.

Additionally, the computation comparisons for the three examples are presented in Tables~\ref{tab:computation-1} to \ref{tab:computation-3}. The relatively higher computation times observed in Tables~\ref{tab:computation-1} and \ref{tab:computation-2} are primarily due to AM being implemented from scratch in R for clarity, whereas most other methods are optimized for efficiency using C. The results in Table~\ref{tab:computation-3} are consistent with our computation cost analysis in Remark 4, as both AM and the comparable methods are implemented within the PyTorch framework.

\begin{table}[!htbp]
\caption{Computation times required for AM to estimate a single imputation model and complete the entire imputation-estimation scheme. The results are obtained with the first repetition under the first setting across various application examples, and are measured in seconds. It is observed that both weighted and unweighted duality functions exhibit similar computation times. Furthermore, the execution times for neural networks, employing various duality functions and structures as described in the paper, are approximately equivalent, due to the use of CUDA.\label{tab:computation} }

\centering
% \tiny
\scriptsize
\begin{tabular}{ccc}
\hline
                                    & Time Per Imputation Model (s) & Total Time (s) \\
Many-Normal-Means ($n = 10$)        & 0.02                          & 0.87           \\
Many-Normal-Means ($n = 20$)        & 0.06                          & 2.63           \\
Many-Normal-Means ($n = 50$)        & 0.29                          & 15.71          \\
Linear Regression-$L_1$ ($n = 100$) & 2.25                          & 118.74         \\
Linear Regression-$L_2$ ($n = 100$) & 0.23                          & 34.88          \\
Neural Network ($n = 60,000$)       & 53.78                         & 1127.56        \\ \hline
\end{tabular}
\end{table}

\begin{table}[!htbp]
\caption{Computation time for AM compared with other methods in the many-normal-means problem, as presented in Section~5.1. Note that cross-validation is excluded due to prohibitively high computation costs in this example. \label{tab:computation-1} }
\centering
% \tiny
\scriptsize
\begin{tabular}{ccccccc}
\hline
                             & AM    & DPMM & $g$-modeling & MJS  & JS   & MSE  \\
Computation Time ($n = 10$) & 0.87  & 0.46 & $<0.1$         & 0.50 & $<0.1$ & $<0.1$ \\
Computation Time ($n = 20$) & 2.63  & 0.79 & $<0.1$         & 0.87 & $<0.1$ & $<0.1$ \\
Computation Time ($n = 50$) & 15.71 & 1.83 & $<0.1$         & 2.09 & $<0.1$ & $<0.1$ \\ \hline
\end{tabular}
\end{table}

\begin{table}[!htbp]
\caption{Computation time of AM compared to other methods in the linear regression example presented in Section~5.2.\label{tab:computation-2} }
\centering
% \tiny
\scriptsize
\begin{tabular}{ccccccc}
\hline
                             & AM-$L_1$ & AM-$L_2$ & Ridge & Adaptive Lasso & Lasso & Elastic Net \\
Computation Time ($n = 100$) & 118.74   & 34.88    & 0.28  & 0.65           & 0.09  & 8.15        \\ \hline
\end{tabular}
\end{table}

\begin{table}[!htbp]
\caption{Computation time of AM compared to other methods in the neural network example presented in Section~5.3.\label{tab:computation-3} }
\centering
% \tiny
\scriptsize
\begin{tabular}{cccc}
\hline
                                & AM      & Cross-validation with Penalty & Early-Stopping \\
Computation Time ($n = 60,000$) & 1127.56 & 622.65                        & 23.21          \\ \hline
\end{tabular}
\end{table}

\clearpage
\subsection{Discussion on Modern Synthetic Data Generation Methods}

\subsubsection{The Iterative Nature of Modeling}

According to \cite{box1980sampling}, scientific learning and statistical modeling are ongoing processes; no statistical model can be definitively assumed adequate. In general scientific learning, the dual processes of induction and deduction continually contribute to the evolution of new knowledge. Specifically, in statistical modeling and inference, the processes of \textit{criticism} (model checking) and \textit{estimation} (model fitting) occur iteratively, allowing models to evolve and adapt over time.

In the context of AM, the model checking and estimation are combined and conducted by using the \dual~function. Given the focus of this paper on the foundational aspects of modeling, we consider AM as a unified process starting from a base model. As a result, we consider the base scenario that the imputation model used is the same as the base model and conducted a resampling-based imputation method developed for this fundamental process. 

Due to the iterative nature of modeling, it is plausible, and indeed expected, that models will evolve. Therefore, the initial imputation model might be replaced by more sophisticated models that could incorporate knowledge transferred from other studies. This potential for model evolution and improvement has inspired us to review modern synthetic data generation methods, exploring alternative approaches for imputation that may offer enhanced capabilities.

\subsubsection{Review of Modern Synthetic Data Generation Methods}
In this section, we review two main categories of advancements: first, the methods for generating synthetic data, and second, the applications of synthetic data to enhance performance across various statistical tasks. We then explore the connections between these application methods and our proposed AM method, aiming to provide insights into how these approaches can be integrated to improve the efficacy of statistical analyses.

Synthetic data generation techniques, such as our proposed imputation method discussed in Section~2.3, strive to produce a synthetic data distribution $\tilde{\mathbb{P}}$ that replicates the true data distribution $\mathbb{P}$. The related methods developed in recent years include normalizing flows \citep{Kingma2018}, GPT \citep{Brown2020}, diffusion models \citep{Ho2020} and GANs \citep{goodfellow2020generative}, which have shown promise in a variety of applications. Theoretical insights about the efficiency, measured by the total variance measure $TV(\tilde{\mathbb{P}}, \mathbb{P})$, is also provided, for example, by \cite{oko2023} for the diffusion model. However, the overall theoretical understanding of these methods generally remains limited.

The traditional resampling-based approaches, such as the standard bootstrap \citep{efron1979}, face challenges in handling high-dimensional data, prompting the development of an adaptive $m$-out-of-$n$ bootstrap with data-splitting as described in Section~2.3. One of the key advantages of modern synthetic data generation methods over traditional resampling-based approaches, such as bootstrap, is their capability to handle high-dimensional data \citep{Liu2024}. This attribute is particularly valuable for addressing challenges like the estimation of over-parameterized models, as discussed in this paper, presenting them as promising alternatives to our proposed approach. However, the method introduced in Section 2.3 stands out for its simplicity and strong foundational potential, providing distinct advantages over more complex synthetic data generation techniques.

 Several recent methods have adapted modern synthetic data generation techniques to enhance statistical task performance. For instance, \cite{Shen2024} introduced the Synthetic Data Generation for Analytics (Syn) framework, which uses transfer learning to increase the precision of statistical methods applied to high-fidelity synthetic data. Similarly, \cite{Liu2024} developed the Perturbation-Assisted Inference (PAI) framework, which employs synthetic data generated by the Perturbation-Assisted Sample Synthesis (PASS) method for uncertainty quantification in complex models, offering statistical guarantees and justification.
 
Despite the differences from these existing methods, the imputation-estimation framework of our proposed AM method aligns with these trends in leveraging synthetic data generation to enhance statistical task performance. It offers a conceptually advanced alternative to traditional methods like standard regularization for estimating over-parameterized models. Additionally, the simplicity and foundational potential of our proposed imputation method in Section 2.3, together with AM's compatibility with various generative models for dealing with more complex problems, make it a flexible and promising approach for further exploration.

\clearpage
\subsection{
Beyond the Conventional Penalization: Fitting Tree Models from the AM Perspective
}

\subsubsection{Introduction}

The three %\CLdel{application examples}
applications presented in Section~5,
the many-normal-means, %\CLdel{ problem,} 
$n < p$ linear regression, and neural network 
models, utilize the duality function in the form of canonical penalty functions. 
The superior performance of AM can be primarily attributed to three factors: valid data-driven imputation, the ability to enable highly flexible regularization using high-dimensional hyperparameters, and efficient hyperparameter estimation.
To further demonstrate the innovative idea of AM and its broad applicability across different modeling scenarios, here we introduce an example of using AM to effectively fit tree models. In this case, the duality function, which does not have an explicit form, is used to control the model structure.

Tree models, such as classification and regression trees (CART) and random forests (RF), have consistently played an essential role in modern machine learning and statistics \citep{hastie2009book}, even as more sophisticated models such as neural networks have emerged and developed. A key reason for their enduring popularity is the interpretability they provide compared to more complex models. For instance, tree models have proven useful in identifying important genetic variants in genome-wide association studies (GWAS) within biostatistics \citep{Song2014, Hu2020}.

Similar to the models used in the three examples in Section~5, a primary challenges in fitting tree models is tuning hyperparameters, such as the maximum depth of any node of the final tree for CART algorithm, to avoid over-fitting the training data. However, to our knowledge, the most wide-acceptable and effective method for hyperparameter tuning in tree models remains the CV approach \citep{Gomes2024, Loh2014}. In this section, we consider a simple and effective AM alternative for hyperparameter tuning in building CART models and present a numerical example to illustrate its advantages over the CV approach. While further exploration of AM’s application in other tree models would be valuable, we will limit our discussion here due to the scope of this supplement section.

Recall that when the duality function $\pi(\theta, \lambda)$ takes the form of the canonical penalty function, we have $G_{\hat{\mathbb{P}}}(\theta, \lambda)
= E_{(X, Y)\sim\hat{\mathbb{P}}} L(\theta|X, Y)
  + \pi(\theta, \lambda)$, and $\theta = \underset{\tilde\theta}{\text{arg min}} \, G_{\hat{\mathbb{P}}}(\tilde\theta, \lambda)$ in the objective function (8) is equivalent to optimizing penalized ERM. Here, we let $\lambda$ be the maximum depth of any node of the final tree for the CART algorithm. In this way, the estimation process for finding $\theta = \underset{\tilde\theta}{\text{arg min}} \, G_{\hat{\mathbb{P}}}(\tilde\theta, \lambda)$ is equivalent to fitting the tree with maximum depth set to $\lambda$. Consequently, AM can be implemented in two steps: 1. Impute future observations $\mathbb{Q}$ using the observed data $\hat{\mathbb{P}}$; and 2. Select the optimal maximum depth hyperparameter $\lambda^*$ such that, after fitting CART on $\hat{\mathbb{P}}$ with $\lambda^*$, the loss calculated on the imputed population $\mathbb{Q}$ is minimized.

We implement the bootstrap imputation approach for AM. Specifically, we grow $M$ trees (without limiting the maximum depth), each fitted on a bootstrap sample of $\hat{\mathbb{P}}$, denoted by $\tilde{\mathbb{P}}^{(m)} \ (m = 1,\dots,M)$. With each tree model, we impute new responses $\by_*^{(m)}$ using the covariate $X$ from the out-of-bag samples (observations not included in $\tilde{\mathbb{P}}^{(m)}$). We then add these covariates and their imputed responses to the imputed population $\mathbb{Q}$ as future observations.

Notably, growing $M$ trees in this way closely resembles the process of growing RF. This suggests that when dealing with high-dimensional input, we could, at each candidate split in the tree-fitting process, use a random subset of features, as in RF. This approach could improve imputation effectiveness for high-dimensional data, given the well-documented advantage of RF in handling high-dimensional data. For simplicity, however, this process is not used in our simulation study.

\subsubsection{Simulation Study}

To demonstrate the effectiveness of the AM framework for growing CART, we consider a simulation setup that is similar to that used in \cite{Zhang2009}. Specifically, each dataset contains 100 observations, with one response variable and 50 predictors generated from a Bernoulli distribution with a success probability of 0.5. We randomly select $\nu$ of the 50 predictors, labeled as $X_1, \ldots, X_\nu$, to determine the response variable. Without loss of generality, we define the response variable as:
\[
y = 
\begin{cases} 
1, & \text{if } \frac{1}{\nu} \sum_{i=1}^{\nu} X_i + \epsilon > 0.5; \\
0, & \text{otherwise},
\end{cases}
\]
where $\epsilon$ is a random variable following a normal distribution with mean zero and variance $\sigma^2$. We vary $\nu$ over $\{3,5,10,20\}$ and $\sigma$ over $\{0.1,0.3\}$. To approximate the true population, a sufficiently large test set is generated independently. This procedure is repeated to produce $M = 500$ simulated datasets.

To compare AM with the CV method, we implement CV as $K$-fold cross-validation. Since different values of $K$ yield very similar results, we report the findings using $K = 5$ in this section.
To assess the effectiveness of hyperparameter tuning methods, we define a dataset-specific oracle depth, which is the maximum depth value that yields the highest test accuracy when fitting a CART model to the training data. The corresponding test accuracy, referred to as the oracle test accuracy, represents the best possible test accuracy a method can achieve through hyperparameter tuning.

As both AM and CV involve tuning the maximum depth hyperparameter, we evaluate the methods' performance using two metrics: \textit{regret} and \textit{depth consistency rate}. Regret is defined as the difference between the oracle test accuracy and the test accuracy obtained by the method's model, averaged across all 500 datasets. This metric reflects how closely each method’s performance approaches the oracle test accuracy. Depth consistency rate measures the likelihood that a method selects the oracle depth. Specifically, it is the estimated probability that the chosen maximum depth hyperparameter matches the oracle depth, calculated across the $M = 500$ datasets.

The results are summarized in Table~\ref{tab:rf-1} and Table~\ref{tab:rf-2}. Across all simulation settings, which vary in signal sparsity (controlled by $\nu$) and signal-to-noise ratio (controlled by $\sigma$), AM consistently outperforms CV. Specifically, AM achieves test accuracy that is significantly closer to the oracle test accuracy and exhibits a notably higher probability of selecting the optimal hyperparameter.

\begin{table}[!htbp]
\caption{Regret results under various simulation settings for AM and CV methods, calculated based on 500 data set. The standard deviation of each value (estimated with bootstrap) is given in parentheses. The best result in each setting is highlighted in boldface. \label{tab:rf-1}}
\centering
% \tiny
\scriptsize
\begin{tabular}{cccccccccccc}
\hline
                & \multicolumn{2}{c}{$\nu = 3$}                                                                                                         &           & \multicolumn{2}{c}{$\nu = 5$}                                                                                                         &           & \multicolumn{2}{c}{$\nu = 10$}                                                                                                        &           & \multicolumn{2}{c}{$\nu = 20$}                                                                                                        \\ \cline{2-3} \cline{5-6} \cline{8-9} \cline{11-12} 
\textit{Method} & {\ul $\sigma = 0.1$}                                              & {\ul $\sigma = 0.3$}                                              &           & {\ul $\sigma = 0.1$}                                              & {\ul $\sigma = 0.3$}                                              &           & {\ul $\sigma = 0.1$}                                              & {\ul $\sigma = 0.3$}                                              &           & {\ul $\sigma = 0.1$}                                              & {\ul $\sigma = 0.3$}                                              \\
AM              & \textbf{\begin{tabular}[c]{@{}c@{}}0.004 \\ (0.001)\end{tabular}} & \textbf{\begin{tabular}[c]{@{}c@{}}0.019 \\ (0.002)\end{tabular}} & \textbf{} & \textbf{\begin{tabular}[c]{@{}c@{}}0.026 \\ (0.002)\end{tabular}} & \textbf{\begin{tabular}[c]{@{}c@{}}0.022 \\ (0.002)\end{tabular}} & \textbf{} & \textbf{\begin{tabular}[c]{@{}c@{}}0.019 \\ (0.001)\end{tabular}} & \textbf{\begin{tabular}[c]{@{}c@{}}0.019 \\ (0.001)\end{tabular}} & \textbf{} & \textbf{\begin{tabular}[c]{@{}c@{}}0.020 \\ (0.001)\end{tabular}} & \textbf{\begin{tabular}[c]{@{}c@{}}0.017 \\ (0.001)\end{tabular}} \\
CV              & \begin{tabular}[c]{@{}c@{}}0.007 \\ (0.002)\end{tabular}          & \begin{tabular}[c]{@{}c@{}}0.026 \\ (0.002)\end{tabular}          &           & \begin{tabular}[c]{@{}c@{}}0.033 \\ (0.002)\end{tabular}          & \begin{tabular}[c]{@{}c@{}}0.030 \\ (0.002)\end{tabular}          &           & \begin{tabular}[c]{@{}c@{}}0.026 \\ (0.001)\end{tabular}          & \begin{tabular}[c]{@{}c@{}}0.023 \\ (0.001)\end{tabular}          &           & \begin{tabular}[c]{@{}c@{}}0.022 \\ (0.001)\end{tabular}          & \begin{tabular}[c]{@{}c@{}}0.019 \\ (0.001)\end{tabular}          \\ \hline
\end{tabular}
\end{table}

\begin{table}[!htbp]
\caption{Depth consistency rate results under various simulation settings for AM and CV methods, calculated based on 500 data set. The standard deviation of each value (estimated with bootstrap) is given in parentheses. The best result in each setting is highlighted in boldface. \label{tab:rf-2} }
\centering
% \tiny
\scriptsize
\begin{tabular}{cccccccccccc}
\hline
                & \multicolumn{2}{c}{$\nu = 3$}                                                                                                         &           & \multicolumn{2}{c}{$\nu = 5$}                                                                                                         &  & \multicolumn{2}{c}{$\nu = 10$}                                                                                                        &  & \multicolumn{2}{c}{$\nu = 20$}                                                                                                        \\ \cline{2-3} \cline{5-6} \cline{8-9} \cline{11-12} 
\textit{Method} & {\ul $\sigma = 0.1$}                                              & {\ul $\sigma = 0.3$}                                              &           & {\ul $\sigma = 0.1$}                                              & {\ul $\sigma = 0.3$}                                              &  & {\ul $\sigma = 0.1$}                                              & {\ul $\sigma = 0.3$}                                              &  & {\ul $\sigma = 0.1$}                                              & {\ul $\sigma = 0.3$}                                              \\
AM              & \textbf{\begin{tabular}[c]{@{}c@{}}0.938 \\ (0.011)\end{tabular}} & \textbf{\begin{tabular}[c]{@{}c@{}}0.486 \\ (0.022)\end{tabular}} & \textbf{} & \textbf{\begin{tabular}[c]{@{}c@{}}0.512 \\ (0.022)\end{tabular}} & \textbf{\begin{tabular}[c]{@{}c@{}}0.374 \\ (0.021)\end{tabular}} &  & \textbf{\begin{tabular}[c]{@{}c@{}}0.318 \\ (0.021)\end{tabular}} & \textbf{\begin{tabular}[c]{@{}c@{}}0.306 \\ (0.021)\end{tabular}} &  & \textbf{\begin{tabular}[c]{@{}c@{}}0.268 \\ (0.020)\end{tabular}} & \textbf{\begin{tabular}[c]{@{}c@{}}0.236 \\ (0.019)\end{tabular}} \\
CV              & \begin{tabular}[c]{@{}c@{}}0.602\\ (0.022)\end{tabular}           & \begin{tabular}[c]{@{}c@{}}0.278 \\ (0.020)\end{tabular}          & \textbf{} & \begin{tabular}[c]{@{}c@{}}0.412 \\ (0.022)\end{tabular}          & \begin{tabular}[c]{@{}c@{}}0.270 \\ (0.020)\end{tabular}          &  & \begin{tabular}[c]{@{}c@{}}0.200 \\ (0.018)\end{tabular}          & \begin{tabular}[c]{@{}c@{}}0.228 \\ (0.019)\end{tabular}          &  & \begin{tabular}[c]{@{}c@{}}0.180 \\ (0.017)\end{tabular}          & \begin{tabular}[c]{@{}c@{}}0.172 \\ (0.017)\end{tabular}          \\ \hline
\end{tabular}
\end{table}

\begin{figure}[!htbp]
\centering
\includegraphics[width=14cm]{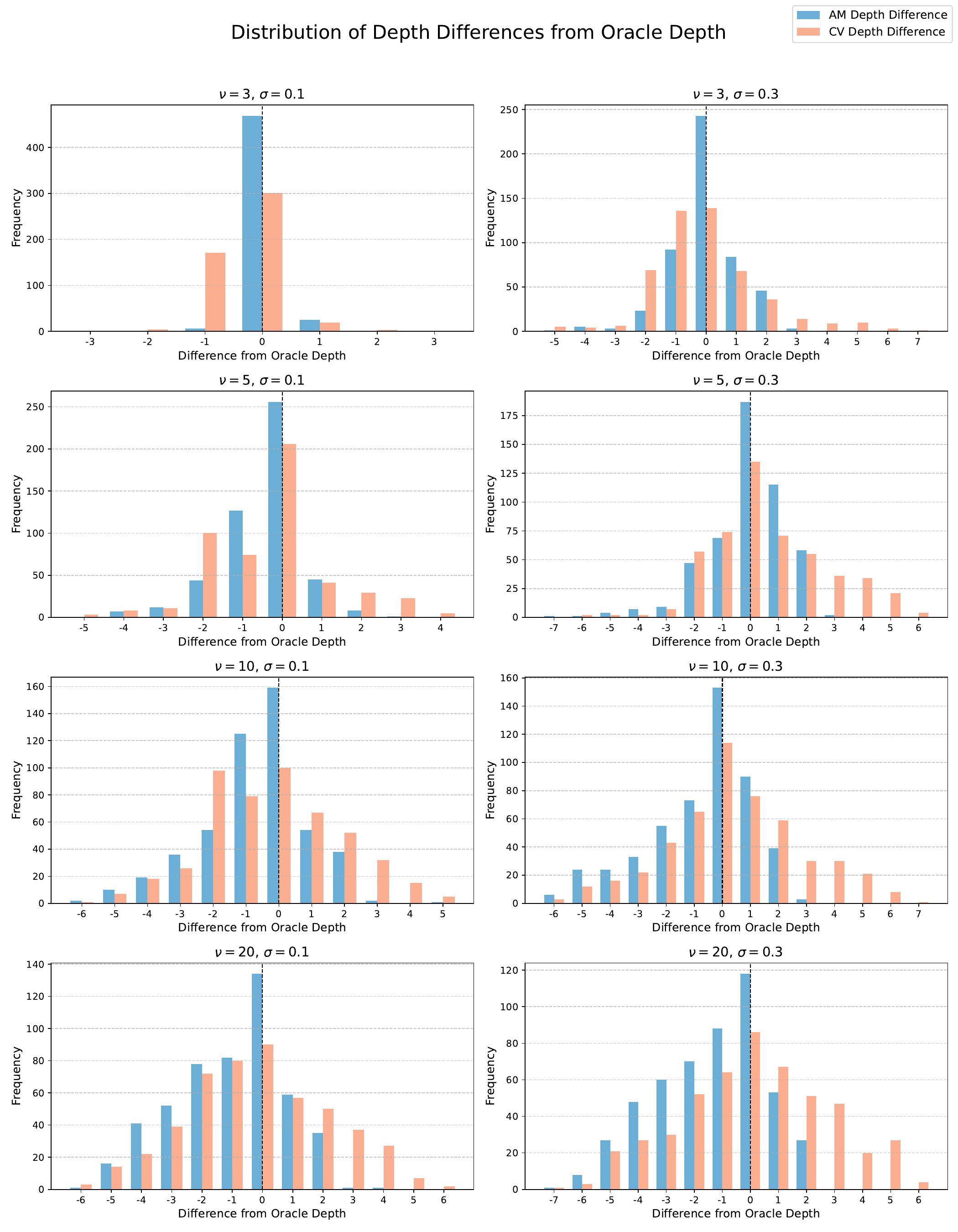}
\caption{Distribution of Depth Differences from Oracle Depth for AM and CV across $M = 500$ datasets under various simulation settings. The vertical dashed line at zero represents the oracle depth. Values closer to zero indicate better alignment with the oracle, with negative values showing a preference for simpler models and positive values indicating a preference for more complex models.\label{fig:rf-1}}
\end{figure}

Additionally, we examine the distribution of the difference between the selected depth and the oracle depth for both methods. Since the maximum depth hyperparameter controls model complexity, this analysis provides insights into the relative complexity of the models selected by each method compared to the optimal model complexity. The results in Figure~\ref{fig:rf-1} clearly show that, compared to CV, AM tends to select simpler models. This can be a significant advantage of AM, as tree models are often chosen for their interpretability, and simpler models generally enhance interpretability.

Overall, these findings demonstrate the effectiveness of AM over the widely-used CV in fitting CART, highlighting its potential for various applications. %\CLadd{AM's framework could also be adapted to fit more advanced machine learning models that, unlike tree models, may have lower interpretability but higher predictive power. Beyond the CNN model discussed in Section~5.3, it would be valuable to explore the application of AM in recent architectures, such as transformers \citep{Vaswani2017Attention}. However, a deeper exploration of these models is beyond the scope of this work.}

\subsubsection{Discussion}
This section provides an additional example, substantially distinct from the three in Section~5, demonstrating the potential of AM in fitting tree models and illustrating its effectiveness across a different modeling scenario.

There are several promising directions for future research. In this example, we focus on the maximum depth hyperparameter; however, another relevant hyperparameter is the maximum number of leaves, which also controls model complexity. Additionally, for illustration purposes, we apply AM to fit a single CART model, but similar concepts can be extended to fit a forest of trees. Given AM's promising results with a single CART, we anticipate similar effectiveness in a forest setting, where automated efficient estimation of hyperparameters provided by AM will be especially desirable.
%\CLdel{, where the use of hyper-parameters of high-dimensions would unleash the power of random forest}\CLadd{\color{red}Question: Is this true? Re: I suggest remove the claim, it is not clear for now.}.

AM's framework could also be adapted to fit more advanced machine learning models that, unlike tree models, may have lower interpretability but higher predictive power. Beyond the CNN model discussed in Section~5.3, it would be valuable to explore the application of AM in recent architectures. These include transformers \citep{Vaswani2017Attention}, with the powerful self-attention and cross-attention mechanisms to capture long-range dependencies that have been proven to be challenging in scientific modeling, and foundation models \citep[see, {\it e.g.},][and references therein]{garza2023timegpt} where high-dimensional hyperparameter-based regularization and efficient estimation of its haperparameters are expected to useful. 
However, a deeper exploration of these models is beyond the scope of this work.

% As a remark, the original simulation setting in \cite{Zhang2009} uses 500 observations and 30 predictors. However, both AM and CV reaches near-oracle in all settings, so we decide to adjust to more challenging scenarios.

%\CLadd{\subsubsection{The Attention Mechanism and Transformer Models}\label{ss:transformer}}

\clearpage

%\subsection{\CLdel{Implementation of AM  without Penalty Terms}}
\subsection{Alternative Implementation of AM: An Example Beyond Conventional Penalty}

Although the use of the duality function in AM, which introduces penalty terms, is chosen for our exposition of the idea of `fitting models to future observations' and is often effective in model estimation, it is not inherently the core idea of AM. In other words, the idea of AM can be implemented by going beyond the use of ERM with conventionally used penalty terms. To make the current discussion simple, consider, for example, the method of final estimation, where an imputed $\mathbb{Q}$ is available in one way or another; see the discussion and references in Section~6 of the main paper.  An alternative implementation of AM without utilizing a penalty function can be achieved by creating two statistically exchangeable imputation populations based on $\mathbb{Q}$, denoted by $\mathbb{Q}_1$ and $\mathbb{Q}_2$. In practice, $\mathbb{Q}_1$ and $\mathbb{Q}_2$ can also be different imputations from different sources; see more discussion in our numerical example below where we make $\mathbb{Q}_1$ slightly more over-dispersed than $\mathbb{Q}_2$. The first population, $\mathbb{Q}_1$, plays a role analogous to the duality function and is used for model estimation, whereas the second population, $\mathbb{Q}_2$, plays the role of finding the optimal value of $w$ with respect to the population $\mathbb{Q}_2$.
% ``model selection''
% \footnote{This statement of saying model selection can be questionable. How about : \color{red} whereas the second population, $\mathbb{Q}_2$, plays the role of finding the optimal value of $w$ with respect to the population $\mathbb{Q}_2$.}.

 Specifically, we now define the $G$ function as
\begin{equation}\label{eq:KD}
G_{\hat{\mathbb{P}}}(\theta, w)
= w \cdot E_{(X, Y)\sim\hat{\mathbb{P}}} L(\theta|X, Y)
  + (1-w) \cdot E_{(X, Y)\sim\mathbb{Q}_1} L(\theta|X, Y),
\end{equation}
where the introduced mixture rate hyperparameter $w$ is assumed to be in $[0,1)$ for technical simplicity and
is used to mix the empirical observations and the imputed future observations to obtain an estimator different from ERM. Thus, $w$ serves a similar role to the duality function described in Section~3. Notably, when $w=0$, we estimate the model entirely based on imputed data. This is in fact a simple way of combining information from the imputations, which originally motivated such a technical development and led to the consideration of the definition of the $G$ function in \eqref{eq:KD} as a weighted sum of the empirical risk and imputed-population risk with the weight $w\in [0,1)$.

Now, the AM objective function is then expressed as:
\begin{equation}\label{eq:am-objective-s}
\begin{aligned}
& \underset{\theta, w}{\text{minimize}}
& & E_{(X, Y)\sim \mathbb{Q}_2} L(\theta|X, Y) \\
& \text{subject to}  
& & \theta = \underset{\tilde\theta}{\text{arg min}} \, G_{\hat{\mathbb{P}}}(\tilde\theta, w).
\end{aligned}
\end{equation}
We implemented this alternative AM estimator with objective (\ref{eq:am-objective-s}) for the normal means problem described in Section~5.1. For simplicity, the imputed samples, obtained through the process described in Section~2.3 using the duality function, were split equally at random into two parts serving as $\mathbb{Q}_1$ and $\mathbb{Q}_2$, respectively. A small random noise, $\epsilon \sim N(0,0.01)$, was added to the samples in $\mathbb{Q}_1$. For this simple experiment, the grid search method was employed to search for $w$ from the candidate set $\{0, 0.2, 0.4, 0.6, 0.8, 1.0\}$.

The numerical results are summarized in Table~\ref{tab:4}. It can be observed that this implementation of AM, without using penalty terms, achieves performance close to that of the standard AM reported in Section~5.1. While it does not match the results of the standard AM, it still outperforms other methods in many cases.
% considering split the imputed samples forming the population $\mathbb{Q}$ into two parts: $\mathbb{Q}_1$ and $\mathbb{Q}_2$. 

 \begin{table}[!htbp]
\caption{MPE results under three simulation settings with different methods. Each entry is taken as the average value obtained with $500$ repetitions and the standard deviation of each value (estimated with bootstrap) is given in parentheses. The results, except for the new AM method, are reproduced from Table 1 in the main manuscript for ease of comparison. The best result in each setting is highlighted in  boldface.  \label{tab:4}}
\centering
\scriptsize
\begin{tabular}{cccccccccccc}
\hline
\multirow{2}{*}{\textit{Method}}                                     & \multicolumn{3}{c}{$\mu \sim N(0,0.01)$}                                                                                                                                                               &                      & \multicolumn{3}{c}{\begin{tabular}[c]{@{}c@{}}$\mu^1 \sim N(-2,0.01)$\\ $\mu^2 \sim N(2,0.01)$\end{tabular}}                                                                                           &                      & \multicolumn{3}{c}{\begin{tabular}[c]{@{}c@{}}$\mu^1 = 0$\\ $\mu^2 \sim N(-3,1)$\end{tabular}}                                                                                                         \\ \cline{2-4} \cline{6-8} \cline{10-12} 
                                                                     & {\ul $n = 10$}                                                   & {\ul $n = 20$}                                                   & {\ul $n = 50$}                                                   &                      & {\ul $n = 10$}                                                   & {\ul $n = 20$}                                                   & {\ul $n = 50$}                                                   &                      & {\ul $n = 10$}                                                   & {\ul $n = 20$}                                                   & {\ul $n = 50$}                                                   \\
JS                                                                   & \begin{tabular}[c]{@{}c@{}}0.313\\ (0.017)\end{tabular}          & \begin{tabular}[c]{@{}c@{}}0.155\\ (0.007)\end{tabular}          & \begin{tabular}[c]{@{}c@{}}0.069\\ (0.003)\end{tabular}          &                      & \begin{tabular}[c]{@{}c@{}}0.899\\ (0.018)\end{tabular}          & \begin{tabular}[c]{@{}c@{}}0.85\\ (0.012)\end{tabular}           & \begin{tabular}[c]{@{}c@{}}0.812\\ (0.007)\end{tabular}          &                      & \begin{tabular}[c]{@{}c@{}}0.535\\ (0.016)\end{tabular}          & \begin{tabular}[c]{@{}c@{}}0.514\\ (0.012)\end{tabular}          & \begin{tabular}[c]{@{}c@{}}0.496\\ (0.007)\end{tabular}          \\
MJS                                                                  & \begin{tabular}[c]{@{}c@{}}0.226\\ (0.019)\end{tabular}          & \begin{tabular}[c]{@{}c@{}}0.110\\ (0.007)\end{tabular}          & \begin{tabular}[c]{@{}c@{}}0.049\\ (0.002)\end{tabular}          & \multicolumn{1}{l}{} & \begin{tabular}[c]{@{}c@{}}0.873\\ (0.018)\end{tabular}          & \begin{tabular}[c]{@{}c@{}}0.836\\ (0.012)\end{tabular}          & \begin{tabular}[c]{@{}c@{}}0.806\\ (0.007)\end{tabular}          & \multicolumn{1}{l}{} & \begin{tabular}[c]{@{}c@{}}0.449\\ (0.015)\end{tabular}          & \begin{tabular}[c]{@{}c@{}}0.463\\ (0.013)\end{tabular}          & \begin{tabular}[c]{@{}c@{}}0.464\\ (0.008)\end{tabular}          \\
DPMM                                                                 & \textbf{\begin{tabular}[c]{@{}c@{}}0.125\\ (0.006)\end{tabular}} & \textbf{\begin{tabular}[c]{@{}c@{}}0.081\\ (0.004)\end{tabular}} & \textbf{\begin{tabular}[c]{@{}c@{}}0.043\\ (0.002)\end{tabular}} & \multicolumn{1}{l}{} & \begin{tabular}[c]{@{}c@{}}0.772\\ (0.029)\end{tabular}          & \begin{tabular}[c]{@{}c@{}}0.533\\ (0.018)\end{tabular}          & \begin{tabular}[c]{@{}c@{}}0.392\\ (0.010)\end{tabular}          & \multicolumn{1}{l}{} & \begin{tabular}[c]{@{}c@{}}0.522\\ (0.020)\end{tabular}          & \begin{tabular}[c]{@{}c@{}}0.469\\ (0.014)\end{tabular}          & \begin{tabular}[c]{@{}c@{}}0.385\\ (0.009)\end{tabular}          \\
$g$-modeling                                                         & \begin{tabular}[c]{@{}c@{}}0.394\\ (0.014)\end{tabular}          & \begin{tabular}[c]{@{}c@{}}0.390\\ (0.010)\end{tabular}          & \begin{tabular}[c]{@{}c@{}}0.170\\ (0.004)\end{tabular}          &                      & \begin{tabular}[c]{@{}c@{}}0.771\\ (0.018)\end{tabular}          & \begin{tabular}[c]{@{}c@{}}0.731\\ (0.012)\end{tabular}          & \begin{tabular}[c]{@{}c@{}}0.729\\ (0.008)\end{tabular}          &                      & \begin{tabular}[c]{@{}c@{}}0.548\\ (0.015)\end{tabular}          & \begin{tabular}[c]{@{}c@{}}0.538\\ (0.013)\end{tabular}          & \begin{tabular}[c]{@{}c@{}}0.378\\ (0.007)\end{tabular}          \\
\begin{tabular}[c]{@{}c@{}}Auto-modeling\\ (No Penalty)\end{tabular} & \begin{tabular}[c]{@{}c@{}}0.257\\ (0.014)\end{tabular}          & \begin{tabular}[c]{@{}c@{}}0.141\\ (0.007)\end{tabular}          & \begin{tabular}[c]{@{}c@{}}0.070\\ (0.003)\end{tabular}          &                      & \begin{tabular}[c]{@{}c@{}}0.727\\ (0.029)\end{tabular}          & \begin{tabular}[c]{@{}c@{}}0.540\\ (0.018)\end{tabular}          & \begin{tabular}[c]{@{}c@{}}0.391\\ (0.010)\end{tabular}          & \textbf{}            & \begin{tabular}[c]{@{}c@{}}0.500\\ (0.015)\end{tabular}          & \begin{tabular}[c]{@{}c@{}}0.410\\ (0.013)\end{tabular}          & \begin{tabular}[c]{@{}c@{}}0.329\\ (0.009)\end{tabular}          \\
\begin{tabular}[c]{@{}c@{}}Auto-modeling\\ (Standard)\end{tabular}   & \begin{tabular}[c]{@{}c@{}}0.196\\ (0.013)\end{tabular}          & \begin{tabular}[c]{@{}c@{}}0.116\\ (0.007)\end{tabular}          & \begin{tabular}[c]{@{}c@{}}0.059\\ (0.003)\end{tabular}          &                      & \textbf{\begin{tabular}[c]{@{}c@{}}0.646\\ (0.029)\end{tabular}} & \textbf{\begin{tabular}[c]{@{}c@{}}0.497\\ (0.019)\end{tabular}} & \textbf{\begin{tabular}[c]{@{}c@{}}0.365\\ (0.010)\end{tabular}} &                      & \textbf{\begin{tabular}[c]{@{}c@{}}0.437\\ (0.020)\end{tabular}} & \textbf{\begin{tabular}[c]{@{}c@{}}0.400\\ (0.014)\end{tabular}} & \textbf{\begin{tabular}[c]{@{}c@{}}0.324\\ (0.008)\end{tabular}} \\ \hline
\end{tabular}
\end{table}

It is worth mentioning that this implementation is technically related to the concept of knowledge distillation \citep{hinton2015} in the machine learning literature, where complex models serve as ``teacher'' ($\mathbb{Q}_1$ in our notation) for simpler ``student'' ($\hat{\mathbb{P}}$ in our notation) models. More recent discussions on the understanding and applications of knowledge distillation are provided by \cite{Zhang2020} and \cite{Moslemi2024}, to name a few. For example, \cite{Zhang2020} wrote:
% \footnote{Can we find a better quote along this line. The present one is fine.}
\begin{quote}\it
It has been recently demonstrated that multi-generational self-distillation can improve generalization. Despite this intriguing observation, reasons for the enhancement remain poorly understood.
\end{quote}

\noindent The AM perspective has the potential to provide statistical insights into such methodologies, paving the way for further exploration in future work.

\clearpage